\documentclass[11pt]{article}
\usepackage{latexsym} % to have the ``\Box'' macro,
\newcommand{\nn}{\nonumber}
\def\lagr{{\cal L}}

\def\gstring{g_{\rm s}}

\def\refeq#1{(\ref{#1})}
\newcommand{\be}{\begin{equation}}
\newcommand{\ee}{\end{equation}}
\newcommand{\bea}{\begin{eqnarray}}
\newcommand{\eea}{\end{eqnarray}}
\newcommand{\vv}{{\cal V}}
\renewcommand{\a}{\alpha}
\renewcommand{\b}{\beta}

\renewcommand{\d}{\delta}
\newcommand{\pa}{\partial}
\newcommand{\g}{\gamma}
\newcommand{\G}{\Gamma}
\newcommand{\e}{\epsilon}

\renewcommand{\L}{\Lambda}
\newcommand{\m}{\mu}
\newcommand{\n}{\nu}

\newcommand{\s}{\sigma}

\renewcommand{\o}{\omega}

\newcommand{\ft}[2]{{\textstyle\frac{#1}{#2}}}
\newcommand{\so}[1]{{\rm SO}{$(#1)$}}  
\newcommand{\eqn}[1]{(\ref{#1})}
\def\slash{\llap /}
\begin{document}
%%%%%%%%%%%%%%%%%%%%%%%%%%%%%%%%%%%%%%%%%%%%%%%%%%%
%\thispagestyle{empty}
%%%%%%%%%%%%%%%
\begin{flushright}
ITP-UU-02/56\\
SPIN-02/35\\[1mm]
September 2002
\end{flushright}
%%%%%%%%%%%%%%%
\vspace{8mm}
\begin{center}
{\huge{\bf SUPERGRAVITY}}\\[8mm]
{Bernard de Wit}\\[4mm]
{Institute for Theoretical Physics \& Spinoza Institute, \\
Utrecht University, Utrecht, The Netherlands}\\[8mm]
{\bf Abstract}\\
%\end{center}
%\begin{quotation}
{These notes are based on lectures presented at the 2001 Les
Houches Summerschool ``Unity from Duality: Gravity, Gauge Theory and
Strings''. }
%\end{quotation}
\end{center}
\vspace{15mm}
%%%%%%%%%%%%%%%%%%%%%%%%%%%%%%%%%%%%%%%%%%%%%%%%%%%%%%%%%%%%%%%%%%
%%%%%%%%%%%%%%%%%%%%%%%%%%%%%%%%%%%%%%%%%%%%%%%%%%%%%%%%%%%%%%%%%%
% \newpage
\tableofcontents
%%%%%%%%%%%%%%%%%%%%%%%%%%%%%%%%%%%%%%%%%%%%%%%%%%%%%%%%%%%%%%%%%%
\newpage  %%%%%%%%%%%%%%%%%%%%%%%%%%%%%%%%%%%%%%%%%%%%%%%%%%%%%%%%
%%%%%%%%%%%%%%%%%%%%%%%%%%%%%%%%%%%%%%%%%%%%%%%%%%%%%%%%%%%%%%%%%%
%%%%%%%%%%%%%%%%%%%%%%%%%%%%%%%%%%%%%%%%%%%%%%%%%%%%%%%%%%%%%%%%%%
%%%%%%%%%%%%%%%%%%%%%%%%%%%%%%%%%%%%%%%%%%%%%%%%%%%%%%%%%%%%%%%%%%
%%%%%%%%%%%%%%%%%%%%%%%%%%%%%%%%%%%%%%%%%%%%%%%%%%%%%%%%%%%%%%%%%%
\section{Introduction}
Supergravity plays a prominent role in our ideas about the
unification of fundamental forces beyond the standard model, in our
understanding of many central features of superstring theory, and in
recent developments of the conceptual basis of quantum field theory and
quantum gravity. The advances made have found their place in many 
reviews and textbooks (see, {\it e.g.}, \cite{general-sg}), but the
subject has grown so much and has so many different facets that no
comprehensive treatment is available as of today. Also in these
lectures, which will cover a number of basic aspects of supergravity,
many topics will be left untouched. 

During its historical development the perspective of supergravity has
changed. Originally it was envisaged as an elementary field theory
which should be free of ultraviolet divergencies and thus bring about
the long awaited unification of gravity with the other fundamental
forces in nature. But nowadays supergravity is primarily viewed as an
effective 
field theory describing the low-mass degrees of freedom of a more
fundamental underlying theory. The only candidate for such a theory is
superstring theory (for some reviews and textbooks, see, {\it e.g.},
\cite{superstring-general}), or rather, yet another, somewhat hypothetical,  
theory, called M-theory. Although we know a lot about M-theory, its
underlying principles have only partly been established. String theory
and supergravity in their modern incarnations now represent some of the
many faces of M-theory. String theory is no longer a theory exclusively of
strings but includes other extended objects that emerge in the
supergravity context as solitonic objects. Looking backwards it
becomes clear that there are many reasons why neither superstrings nor
supergravity could account for all the relevant degrees of freedom
and we have learned to appreciate that  M-theory has many different
realizations. 

Because supersymmetry is such a powerful symmetry it plays a central
role in 
almost all these developments. It controls the dynamics and, because
of nonrenormalization theorems, precise predictions can be made in many
instances, often relating strong- to weak-coupling regimes. To
appreciate the implications of supersymmetry, chapter~2 starts with a
detailed discussion of supersymmetry and its
representations. Subsequently supergravity theories are introduced in 
chapter~3, mostly concentrating on the maximally supersymmetric
cases. In chapter~4 gauged nonlinear sigma models with
homogeneous target spaces are introduced, paving the way for the
construction of gauged supergravity. This construction is explained in
chapter~5, where 
the emphasis is on gauged supergravity with 32 supercharges in 4 and 5
spacetime  dimensions. These theories can describe anti-de Sitter
ground states which are fully supersymmetric. This is one of the
motivations for considering anti-de Sitter supersymmetry and the
representations of the anti-de Sitter group in  
chapter~6. Chapter~7 contains a short introduction to
superconformal transformations and superconformally invariant
theories. This chapter is self-contained, but it is of course related to 
the discussion on anti-de Sitter representations in chapter~6, as well
as to the adS/CFT correspondence. 

This school offers a large number of lectures dealing with gravity,
gauge theories and string theory from various perspectives. We
intend to stay within the supergravity perspective and to try and
indicate what the possible implications of supersymmetry and
supergravity are for these subjects. Our hope is that the material
presented below will offer a helpful introduction to and will blend
in naturally with the material presented in other lectures. 
%%%%%%%%%%%%%%%%%%%%%%%%%%%%%%%%%%%%%%%%%%%%%%%%%%%%%%%%%%%%%%%%
%%%%%%%%%%%%%%%%%%%%%%%%%%%%%%%%%%%%%%%%%%%%%%%%%%%%%%%%%%%%%%%%
%%%%%%%%%%%%%%%%%%%%%%%%%%%%%%%%%%%%%%%%%%%%%%%%%%%%%%%%%%%%%%%%%%
%%%%%%%%%%%% CHAPTER II %%%%%%%%%%%%%%%%%%%%%%%%%%%%%%%%%%%%%%%%%%
%%%%%%%%%%%%%%%%%%%%%%%%%%%%%%%%%%%%%%%%%%%%%%%%%%%%%%%%%%%%%%%%%%
\section{Supersymmetry in various dimensions}
An enormous amount of information about supersymmetric theories is 
contained in the structure of the underlying representations of 
the supersymmetry algebra (for some references, see
\cite{general-sg,nahm,multiplets,Strathdee,dewitlouis}. Here we should
make a distinction between a supermultiplet of fields which
transform irreducibly 
under the supersymmetry transformations, and a supermultiplet of states
described by a supersymmetric theory. In this chapter\footnote{%%%%%
%%%%%%%%%%%%%%%%%%%%%%%%%%%%%%%%%%%%%%%%%%%%%%%%%%%%%%%%%%%%%%%%%%%%
   The material presented in this and the following chapter is an
   extension of the second chapter of \cite{dewitlouis}. } %%%%%%%%%
%%%%%%%%%%%%%%%%%%%%%%%%%%%%%%%%%%%%%%%%%%%%%%%%%%%%%%%%%%%%%%%%%%%%
we concentrate on supermultiplets of states, primarily restricting
ourselves to flat Minkowski spacetimes of  
dimension $D$. The relevant symmetries in this 
case form an extension of the Poincar\'e transformations, which consist
of translations and Lorentz transformations. However, many of the
concepts that we introduce will also play a role in the discussion of
other superalgebras, such as the anti-de Sitter (or conformal)
superalgebras. For a recent practical introduction to superalgebras,
see \cite{VanProe}. 
%%%%%%%%%%%%%%%%%%%%%%%%%%%%%%%%%%%%%%%%%%%%%%%%%%%%%%%%%%%%%%%%%
 \subsection{The Poincar\'e supersymmetry algebra}
The generators of the 
super-Poincar\'e algebra comprise the supercharges, transforming as 
spinors under the Lorentz group, the energy and momentum 
operators, the generators of the Lorentz group, and possibly 
additional generators that commute with the supercharges. For the 
moment we ignore these additional charges, often called {\it 
central} charges.\footnote{%%%%%%%%%%%%%%%%%%%%%%%%%%%%%%%%%%%%%%%
   The terminology adopted in the literature is  
   not always very precise. Usually, all charges that commute 
   with the supercharges, but not necessarily with all the 
   generators of the Poincar\'e algebra, are called `central 
   charges'. We adhere to this nomenclature. Observe that the 
   issue of central charges is different when not in flat space, as
   can be seen, for example, in the context of the anti-de Sitter
   superalgebra (discussed in chapter~\ref{susy-ads}).} %%%%%%%%%%%%
There are other relevant superalgebras, such as the 
supersymmetric extensions of the anti-de Sitter (or the conformal) 
algebras. These will be encountered in due course. 

The most important anti-commutation relation of the super-Poincar\'e
algebra is the one of two supercharges,
\be\label{susy-algebra}
\{Q_\a , \bar Q_\b \} = -2 i P_\m\, (\G^\m)_{\a\b}\,,
\ee
where we suppressed the central charges.
Here $\G^\m$ are the gamma matrices that generate the Clifford 
algebra ${\cal C}(D-1,1)$ with Minkowskian metric 
$\eta_{\m\n}= {\rm diag} (-,+,\cdots,+)$.  

The size of a supermultiplet depends exponentially on the number of 
independent supercharge components $Q$. The first step is therefore to 
determine $Q$ for any given number of spacetime 
dimensions $D$. The result is summarized in 
table~\ref{susy-charges}. As shown, there exist five  
different sequences of spinors, corresponding to spacetimes of 
particular dimensions. When this dimension is odd, it is 
possible in certain cases to have Majorana spinors. These cases 
constitute the first sequence.  
The second one corresponds to those odd dimensions where Majorana 
spinors do not exist. The spinors are then Dirac spinors. In even 
dimension one may distinguish three sequences. In the first one, 
where the number of dimensions is a multiple of 4, 
charge conjugation relates positive- with negative-chirality 
spinors. All spinors in this sequence can be restricted to 
Majorana spinors. For the remaining two  
sequences, charge conjugation preserves the chirality of the 
spinor. Now there are again two possibilities, depending on whether 
Majorana spinors can exist or not. The cases where we cannot have 
Majorana spinors, corresponding to $D=6$~mod~8, comprise the fourth
sequence.  For the last sequence with $D=2$~mod~8, Majorana   
spinors exist and the charges can be restricted to so-called
Majorana-Weyl spinors.  

%%%%%%%%%%%%%%%%%%%%%%%%%%%%%%%%%%%%%%%%%%
\begin{table}
\begin{center}
\begin{tabular}{l l l l}\hline
$D$ & $Q_{\rm irr}$ &  ${\rm H}_{\rm R}$   & type \\\hline
~&~&~&~ \\[-3.5mm]
3, 9, 11, mod 8   & $2^{(D-1)/2}$ & ${\rm SO}(N)$ & M\\
5, 7, mod 8          & $2^{(D+1)/2}$ & ${\rm USp}(2N)$ & D\\
4, 8, mod 8          & $2^{D/2}$     & ${\rm U}(N)$ & M \\
6, mod 8             & $2^{D/2}$    & ${\rm USp}(2N_+)\times 
                                       {\rm USp}(2N_-)$ & W\\
2, 10, mod 8         & $2^{D/2-1}$   & ${ \rm SO}(N_+)\times {\rm SO}(N_-)$ 
                                              & MW\\\hline
\end{tabular}
\end{center}
\caption{\small 
The supercharges in flat Minkowski spacetimes of dimension
$D$. In the second column, $Q_{\rm irr}$ specifies the real dimension of 
an irreducible spinor in a $D$-dimensional 
Minkowski spacetime. The third column specifies the group ${\rm H}_{\rm 
R}$ for $N$-extended supersymmetry, defined in the text, acting 
on $N$-fold reducible spinor  
charges. The fourth column denotes the type of spinors: Majorana 
(M), Dirac (D), Weyl (W) and Majorana-Weyl (MW).}\label{susy-charges}
\end{table}
%%%%%%%%%%%%%%%%%%%%%%%%%%%%%%%%%%%%%%%%%%%

One can consider {\it extended supersymmetry}, where the spinor 
charges transform reducibly under the Lorentz group and comprise 
$N$ irreducible spinors. For Weyl charges, one can  
consider combinations of $N_+$ positive- and $N_-$ 
negative-chirality spinors. In all these cases there exists 
a group ${\rm H}_{\rm R}$ of rotations of the spinors which commute
with the Lorentz group  and leave the supersymmetry algebra
invariant. This group, often referred
to as the `R-symmetry' group, is thus defined as the largest subgroup of
the automorphism group of the supersymmetry algebra that  
commutes with the Lorentz group. It is often realized as 
a manifest invariance group of a supersymmetric field theory, but this
is by no means necessary. There are other versions of the R-symmetry
group ${\rm H}_{\rm R}$ which play a role, for instance, in the
context of the 
Euclidean rest-frame superalgebra for massive representations or for
the anti-de Sitter superalgebra. Those will be discussed later in this
chapter. In table~\ref{susy-charges} we have listed the 
corresponding ${\rm H}_{\rm R}$ groups for $N$ irreducible spinor
charges. Here we have assumed that ${\rm H}_{\rm R}$ is compact so
that it preserves a positive-definite metric. In the latter two
sequences of spinor charges shown in table~\ref{susy-charges}, we
allow $N_\pm$ charges of opposite chirality, so that ${\rm H}_{\rm R}$
decomposes into the product of two such groups, one for each chiral
sector. 

Another way to present some of the results above, is shown in  
table~\ref{simple-susy}. Here we list the real dimension of an 
irreducible spinor charge and the corresponding spacetime 
dimension. In addition we include the number of states of the 
shortest\footnote{%
   By the {\it shortest} multiplet, we mean the multiplet with 
   the helicities of the states as low as possible. This is 
   usually (one of) the smallest possible supermultiplet(s). } %
supermultiplet of massless states, written as a sum of 
bosonic and fermionic states. We return to a more general discussion
of the R-symmetry groups and their consequences in
section~\ref{R-symmetry}. 

%%%%%%%%%%%%%%%%%%%%%%%%%%%%%%%%%%%%%%%%%%%%%
\begin{table}
\begin{center}
\begin{tabular}{l l l}\hline
$Q_{\rm irr}$ & $D$  & shortest supermultiplet \\\hline
32   & $D=11$        & $128+128$\\
16   & $D= 10,9,8,7$ & $8+8$\\
8    & $D= 6,5$      & $4+4$\\
4    & $D= 4$        & $2+2$\\
2    & $D=3$         & $1+1$ \\\hline
\end{tabular}
\end{center}
\caption{\small Simple supersymmetry in various dimensions. We present 
the dimension of the irreducible spinor charge with $2\leq Q_{\rm 
irr}\leq 32$ and the corresponding spacetime dimensions $D$. The 
third column represents the number of bosonic + fermionic 
{\it massless} states for the shortest supermultiplet.  }\label{simple-susy}
\end{table}
%%%%%%%%%%%%%%%%%%%%%%%%%%%%%%%%%%%%%%%%%%%%%%
%%%%%%%%%%%%%%%%%%%%%%%%%%%%%%%%%%%%%%%%%%%%%%
\subsection{Massless supermultiplets}
%%%%%%%%%%%%%%%%%%%%%%%%%%%%%%%%%%%%%%%%%%%%%%
Because the momentum operators $P_\m$ commute with the 
supercharges, we may  
consider the states at arbitrary but fixed momentum $P_\m$, which, 
for massless representations, satisfies $P^2 =0$. The matrix 
$P_\m\G^\m$ on the right-hand side of \eqn{susy-algebra} has 
therefore zero eigenvalues. In a positive-definite Hilbert space  
some (linear combinations) of the supercharges must therefore 
vanish. To exhibit this more explicitly, let us rewrite 
\eqn{susy-algebra} as (using $\bar Q= i Q^\dagger\Gamma^0$),
\be
\{Q_\a , Q^\dagger_\b \} = 2 \,(P\!\slash\, \G^0)_{\a\b}\,.
\ee
For light-like $P^\m= (P^0,\vec P)$ the right-hand side is 
proportional to a  
projection operator $({\bf 1} + \G_\parallel \G^0)/2$. Here 
$\G_\parallel$ is the gamma matrix along the spatial momentum 
$\vec P$ of the states.  The supersymmetry anti-commutator can 
then be written as
\be\label{susy-algebra2}
\{Q_\a , Q^\dagger_\b \} = 2 \, P^0\Big({\bf 1} + \tilde \G_D 
\tilde \G_\perp \Big)_{\a\b}  \,.
\ee
Here $\tilde \G_D$ consists of the product of all $D$ independent 
gamma matrices, and $\tilde\G _\perp$ of the product of all $D-2$ 
gamma matrices in the transverse directions (i.e., perpendicular 
to $\vec P$), with phase factors such that
\be
(\tilde \G_D)^2 = ( \tilde \G_\perp)^2 = {\bf 1}\,,\qquad [ 
\tilde \G_D , \tilde \G_\perp] = 0\,.
\ee
This shows that the right-hand side of \eqn{susy-algebra2} is 
proportional to a projection operator, which projects out half of 
the spinor space. Consequently, half the spinors must vanish on 
physical states, whereas the other ones generate a Clifford algebra. 

Denoting the real dimension of the supercharges by $Q$, the 
representation space of the charges decomposes into the two chiral 
spinor representations of SO$(Q/2)$. When confronting these results 
with the last column in table~\ref{simple-susy}, it turns out 
that the dimension of the shortest supermultiplet is not just 
equal to $2^{Q_{\rm irr}/4}$, as one might naively expect. For 
$D=6$, this is so because the representation is complex. For $D=3,
4$ the representation is twice  
as big because it must also accommodate fermion number (or, 
alternatively, because it must be CPT self-conjugate). The derivation
for $D=4$ is presented in many  places (see, for instance,
\cite{general-sg,multiplets,Strathdee}). For $D=3$ we refer to
\cite{DWTN}. 

The two chiral spinor subspaces correspond to the bosonic and 
fermionic states, respectively. For the massless 
multiplets, the dimensions are shown in table~\ref{simple-susy}. 
Bigger supermultiplets can be obtained by combining irreducible
multiplets by requiring them to transform nontrivially 
under the Lorentz group. We shall demonstrate 
this below in three relevant cases, corresponding to 
$D=11, 10$ and 6 spacetime dimensions. 
Depending on the number of spacetime dimensions, many supergravity theories
exist. Pure supergravity theories with spacetime dimension $4\leq D
\leq 11$ can exist with  $Q=32,24,20,16,12,8$ and 4
supersymmetries.\footnote{%%%%%%%%%%%%%%%%%%%%%%%%%%%%%%%%%%%%% 
  In $D=4$  there exist theories with $Q=12,20$ and 24;  
  in $D=5$  there exists a theory with $Q=24$ \cite{cremmer2}. In
  $D=6$ there are three theories with $Q=32$ and one with $Q=24$. So
  far these    supergravities have played no role in string
  theory. For a more   recent discussion, see \cite{Hull2}.} %%%%
Some of these theories will be discussed later in more detail (in  
particular supergravity in $D=11$ and 10 spacetime dimensions). 

%%%%%%%%%%%%%%%%%%%%%%%%%%%%%%%%%%%%%%%%%%%%%%%%%%%%%%%%%%%%%
%%%%%%%%%%%%%%%%%%%%%%%%%%%%%%%%%%%%%%%%%%%%%%%%%%%%%%%%%%%%% 
 \subsubsection{$D=11$ Supermultipets}
In 11 dimensions we are dealing with 32 independent real 
supercharges.  
In odd-dimensional spacetimes irreducible spinors are subject to 
the eigenvalue condition $\tilde\G_D=\pm {\bf 1}$. Therefore 
\eqn{susy-algebra2} simplifies and shows that the 16 
nonvanishing spinor charges transform according to a single  
spinor representation of the helicity group SO(9). 

On the other hand, when regarding the 16 spinor charges as gamma 
matrices, it follows that the representation space constitutes 
the spinor representation of SO(16), which 
decomposes into two chiral subspaces, one corresponding to the 
bosons and the other one to the fermions. To determine the 
helicity content of the bosonic and fermionic states,  
one considers the embedding of the SO(9) spinor representation in 
the SO(16) vector representation. It then turns out that one of 
the ${\bf 128}$ representations branches into helicity representations 
according to  
${\bf 128}\to {\bf 44} + {\bf 84}$, while the second one 
transforms irreducibly according to the ${\bf 128}$  
representation of the helicity group.

The above states comprise precisely the massless states 
corresponding to $D=11$ supergravity \cite{cjs}. The graviton states 
transform in the ${\bf 44}$, the antisymmetric tensor states in the 
${\bf 84}$ and the gravitini states in the ${\bf 128}$ representation 
of SO(9). 
Bigger supermultiplets consist of multiples of 256 states. For 
instance, without central charges, the smallest massive 
supermultiplet comprises $32768+32768$ states. These multiplets 
will not be considered here.

%%%%%%%%%%%%%%%%%%%%%%%%%%%%%%%%%%%%%%%%%%%%%%%%%%%%%%%%%%%%%%%
 \subsubsection{$D=10$ Supermultiplets}
In 10 dimensions the supercharges are both Majorana and Weyl 
spinors. 
The latter means that they are eigenspinors of $\tilde \G_D$. 
According to \eqn{susy-algebra2}, when we have simple (i.e.,  
nonextended) supersymmetry with 16 charges, the nonvanishing charges 
transform in a chiral spinor representation of the SO(8) helicity 
group. With 8 nonvanishing supercharges we are dealing with an 
8-dimensional Clifford  
algebra, whose irreducible representation space corresponds to the 
bosonic and fermionic states, each transforming according to a 
chiral spinor representation. Hence we are dealing with three 
8-dimensional representations of SO(8), which are inequivalent. One is the    
representation to which we assign the supercharges, which we will 
denote by ${\bf 8}_s$; to the other two, denoted as the ${\bf 
8}_v$ and ${\bf 8}_c$ representations, we  
assign the bosonic and fermionic states, respectively. 
The fact that SO(8) representations  
appear in a three-fold variety is known as {\it triality}, which 
is a characteristic property of the group SO(8). With the 
exception of certain  
representations, such as the adjoint and the 
singlet representation, the three types of representation are 
inequivalent. They are traditionally distinguished by labels $s$,
$v$ and $c$ (see, for instance, \cite{slansky}).\footnote{%%%%%%%
%%%%%%%%%%%%%%%%%%%%%%%%%%%%%%%%%%%%%%%%%%%%%%%%%%%%%%%%%%%%%%%%%
  The representations can be characterized according to the four
  different   conjugacy classes of the ${\rm SO}(8)$ weight vectors,
  denoted by   $0$, $v$, $s$ and $c$. In this context one uses the
  notation ${\bf 1}_0$, ${\bf 28}_0$, and ${\bf 35}_0$, ${\bf
  35}^\prime_0$,  ${\bf 35}^{\prime\prime}_0$  for ${\bf 35}_v$,
  ${\bf 35}_s$, ${\bf 35}_c$, respectively. } %%%%%%%%%%%%%%%%%%%
%%%%%%%%%%%%%%%%%%%%%%%%%%%%%%%%%%%%%%%%%%%%%%%%%%%%%%%%%%%%%%%%%

The smallest massless supermultiplet has now been constructed 
with 8 bosonic and 8 fermionic states and corresponds to the 
vector multiplet of supersymmetric Yang-Mills theory in 10 
dimensions \cite{10D-YM}.
Before constructing the supermultiplets that are relevant for 
$D=10$ supergravity, let us first discuss some other properties of 
SO(8) representations. One way to distinguish the inequivalent 
representations, is to investigate how they decompose into 
representations of an SO(7) subgroup. Each of the 8-dimensional 
representations leaves a different SO(7) subgroup
of SO(8) invariant. Therefore there is an 
SO(7) subgroup under which the ${\bf 8}_v$ representation 
branches into 
$$
{\bf 8}_v \longrightarrow {\bf 7} + {\bf 1} . 
$$
Under this SO(7) the other two 8-dimensional representations branch 
into
$$
{\bf 8}_s \longrightarrow {\bf 8} \, , \qquad {\bf 8}_c 
\longrightarrow {\bf 8} \,,
$$
where ${\bf 8}$ is the spinor representation of SO(7). Corresponding 
branching rules for the 28-, 35- and 56-dimensional representations are
\be
\begin{array}{rcl}
{\bf 28} & \longrightarrow  & {\bf 7}   + {\bf 21} \,, \\
{\bf 35}_v & \longrightarrow  & {\bf 1}  +  {\bf 7}  +  {\bf 27} 
\,, \\
{\bf 35}_{c,s} & \longrightarrow  & {\bf 35} \,,
\end{array}\qquad
\begin{array}{rcl}
~          &  ~               &   ~  \\
{\bf 56}_v & \longrightarrow  & {\bf 21} + {\bf 35}  \,,\\
{\bf 56}_{c,s} & \longrightarrow  & {\bf 8}   +  {\bf 48} \,.
\end{array} 
\ee

%%%%%%%%%%%%%%%%%%%%%%%%%%%%%%%%%%%%%%%%%%%%%
\begin{table}
\begin{center}
\begin{tabular}{l l l}\hline
supermultiplet & bosons & fermions \\\hline
vector multiplet    & ${\bf 8}_v$            & ${\bf 8}_c$\\
graviton multiplet      & ${\bf 1} +{\bf  28} + {\bf 35}_v$  & 
${\bf 8}_s +{\bf 56}_s$ \\
gravitino multiplet     &  ${\bf 1} +{\bf  28} + {\bf 35}_c$ & 
${\bf 8}_s +{\bf 56}_s$ \\
gravitino multiplet     &  ${\bf 8}_v +{\bf 56}_v$ &  ${\bf 8}_c 
+{\bf 56}_c$ \\ \hline
\end{tabular}
\end{center}
\caption{\small Massless $N=1$ supermultiplets in $D=10$ spacetime 
dimensions containing $8+8$ or $64 + 64$ bosonic and fermionic 
degrees of freedom.  }\label{D10N1-susy-mult}
\end{table}
%%%%%%%%%%%%%%%%%%%%%%%%%%%%%%%%%%%%%%%%%%%%%%

In order to obtain the supersymmetry representations relevant for 
supergravity we consider tensor products of the smallest 
supermultiplet consisting of ${\bf 8}_v + {\bf 8}_c$, with one 
of the 8-dimensional
representations. There are thus three different possibilities, each
leading to a 128-dimensional supermultiplet. Using the multiplication 
rules for SO(8) representations,
\be\label{so8-multiplication}
\begin{array}{rcl}
{\bf 8}_v  \times  {\bf 8}_v & = & {\bf 1}  + {\bf 28} + {\bf 
35}_v\,, \\
{\bf 8}_s \times  {\bf 8}_s & = & {\bf 1} +  {\bf 28}   + 
{\bf 35}_s \,, \\
{\bf 8}_c  \times  {\bf 8}_c & = & {\bf 1}  +  {\bf 28}  + 
{\bf 35}_c \,,
\end{array}
\qquad
\begin{array}{rcl}
{\bf 8}_v  \times {\bf 8}_s & = & {\bf 8}_c +  {\bf 56}_c \,, \\
{\bf 8}_s  \times  {\bf 8}_c & = & {\bf 8}_v  +  {\bf 56}_v \,, \\
{\bf 8}_c  \times  {\bf 8}_v & = & {\bf 8}_s  +  {\bf 56}_s \,,
\end{array}
\ee
it is straightforward to obtain these new multiplets. Multiplying 
${\bf 8}_v$ with ${\bf 8}_v + {\bf 8}_c$ yields ${\bf 8}_v\times 
{\bf 8}_v$ bosonic and ${\bf 8}_v\times{\bf 8}_c$ fermionic 
states, and leads to the second supermultiplet shown in 
table~\ref{D10N1-susy-mult}. This supermultiplet contains the
representation ${\bf 35}_v$, which can be associated with the 
states of the graviton in $D=10$ dimensions (the field-theoretic 
identification of the various states has  been clarified in many
places; see {\it e.g.} the appendix in \cite{dewitlouis}). Therefore
this supermultiplet will be called the  
{\em graviton multiplet.} Multiplication with ${\bf 8}_c$ or 
${\bf 8}_s$ goes in the same fashion, except that we will
associate the {\bf 8}$_c$ and {\bf 8}$_s$ representations with 
fermionic quantities (note that these are the representations to which
the fermion states of the Yang-Mills multiplet and the supersymmetry
charges are assigned).
Consequently, we interchange the boson and fermion assignments in 
these 
products. Multiplication with ${\bf 8}_c$ then leads to ${\bf 
8}_c\times {\bf 8}_c$ bosonic and ${\bf 8}_c\times{\bf 8}_v$ 
fermionic states, whereas multiplication with ${\bf 8}_s$ gives 
${\bf 8}_s\times{\bf 8}_c$
bosonic and ${\bf 8}_s\times{\bf 8}_v$ fermionic states. These
supermultiplets contain fermions transforming according to the ${\bf 
56}_s$ and ${\bf 56}_c$ representations, respectively, which can 
be associated  with gravitino states, but no graviton states as 
those transform in the  ${\bf 35}_v$ representation. Therefore 
these two supermultiplets are called {\em  gravitino
multiplets}. We have thus established the existence of two 
inequivalent gravitino multiplets. The explicit SO(8) 
decompositions of the vector,
graviton and gravitino supermultiplets are shown in 
table~\ref{D10N1-susy-mult}.
         
By combining a graviton and a gravitino multiplet it is possible to 
construct an $N=2$ supermultiplet of 128 + 128 bosonic and fermionic states. 
However, since there are two inequivalent gravitino multiplets, 
there will also be two inequivalent $N=2$ supermultiplets 
containing the states corresponding to a
graviton and two gravitini. According to the construction presented 
above, one $N=2$  
supermultiplet may be be viewed as the tensor product of two identical
supermultiplets (namely ${\bf 8}_v + {\bf 8}_c$). Such a multiplet 
follows if one starts from a supersymmetry algebra based on {\em 
two} Majorana-Weyl spinor charges $Q$ with the {\em same} 
chirality. The states of this multiplet decompose as follows:
%%%%%%
\begin{eqnarray} \label{eq:IIB}
\mbox{\it Chiral $N=2$ supermultiplet {\rm (IIB)}}\!\!\!&& 
\nonumber\\[-.55cm]
({\bf 8}_v+ {\bf 8}_c)\times({\bf 8}_v+{\bf 8}_c)&\Longrightarrow&  
\left\{\begin{array}{l}
\mbox{\bf bosons}: \\
{\bf 1}+ {\bf 1} +  {\bf 28} +  {\bf 28}+ {\bf 35}_v + {\bf 35}_c \\[.5cm] 
\mbox{\bf fermions}: \\
{\bf 8}_s+ {\bf 8}_s +  {\bf 56}_s+ {\bf 56}_s 
\end{array}
\right. 
\end{eqnarray}
%%%%%%%%
This is the multiplet corresponding to IIB supergravity 
\cite{SH}.
Because the supercharges have the same chirality, one can perform rotations
between these spinor charges which leave the supersymmetry algebra 
unaffected. Hence the automorphism group $H_{\rm R}$ is equal to 
SO(2). This feature reflects itself in the multiplet 
decomposition, where the $\bf 1$, ${\bf 8}_s$, $\bf 28$ and ${\bf 
56}_s$ representations are degenerate and constitute doublets 
under this SO(2) group. 

A second supermultiplet may be viewed as the tensor  
product of a (${\bf 8}_v + {\bf 8}_s$) supermultiplet with a 
second supermultiplet
(${\bf 8}_v + {\bf 8}_c$). In this case the supercharges 
constitute two Majorana-Weyl spinors of opposite chirality. Now the 
supermultiplet decomposes as follows: 
\pagebreak
%%%%%%
\begin{eqnarray} \label{eq:IIA}
\mbox{\it Nonchiral $N=2$ supermultiplet {\rm (IIA)}}\!\!\!&& 
\nonumber\\[-.55cm]
({\bf 8}_v+ {\bf 8}_s)\times({\bf 8}_v+{\bf 8}_c)&\Longrightarrow&  
\left\{\begin{array}{l}
\mbox{\bf bosons}: \\
{\bf 1} +  {\bf 8}_v+ {\bf 28} +{\bf 35}_v + {\bf 56}_v \\[.5cm] 
\mbox{\bf fermions}: \\
{\bf 8}_s+ {\bf 8}_c +  {\bf 56}_s+ {\bf 56}_c 
\end{array}
\right. 
\end{eqnarray}
%%%%%%%%
This is the multiplet corresponding to IIA supergravity \cite{IIASG}.
It can be obtained by a straightforward reduction of $D=11$ 
supergravity. The 
latter follows from the fact that two $D=10$ Majorana-Weyl 
spinors with opposite 
chirality can be combined into a single $D=11$ Majorana spinor.
The formula below summarizes the massless states 
of IIA supergravity from an 11-dimensional perspective. The 
massless states of 11-dimensional supergravity transform according
to the ${\bf 44}$, ${\bf 84}$ and ${\bf 128}$ representation of the 
helicity  group SO(9).  They correspond to the degrees of freedom 
described by the metric, a 3-rank antisymmetric gauge field and 
the gravitino field, respectively. We also show how the 
10-dimensional states can subsequently be branched into 
9-dimensional states, characterized in terms of representations of the
helicity group ${\rm SO}(7)$:  
%%%%%%
\begin{eqnarray} \label{eq:11-IIA-9}
{\bf 44} &\Longrightarrow&  
\left\{\begin{array}{lcl}
{\bf 1}   & \longrightarrow &  {\bf 1} \\
{\bf 8}_v & \longrightarrow & {\bf 1} + {\bf 7}\\
{\bf 35}_v& \longrightarrow & {\bf 1} + {\bf 7} + {\bf 27}
\end{array}
\right. 
\nonumber\\
{\bf 84} &\Longrightarrow&  
\left\{\begin{array}{lcl}
{\bf 28}   & \longrightarrow &   {\bf 7}+{\bf 21}  \\
{\bf 56}_v & \longrightarrow & {\bf 21} + {\bf 35}
\end{array}
\right. 
\\
{\bf 128} &\Longrightarrow&  
\left\{\begin{array}{lcl}
{\bf 8}_s   & \longrightarrow &   {\bf 8}  \\
{\bf 8}_c   & \longrightarrow &  {\bf 8}   \\
{\bf 56}_s  & \longrightarrow &  {\bf 8} +{\bf 48} \\
{\bf 56}_c  & \longrightarrow &  {\bf 8}+{\bf 48}  \\
\end{array}
\right. 
\nonumber
\end{eqnarray}
%%%%%%%%
Clearly, in $D=9$ we have a degeneracy of states, associated with the
group ${\rm H}_{\rm R} = {\rm SO}(2)$. We note the presence of graviton and 
gravitino states, transforming in the ${\bf 27}$ and ${\bf 48}$ 
representations of the ${\rm SO}(7)$ helicity group. 

One could also take the states of the IIB supergravity and 
decompose them into $D=9$ massless states. This leads to 
precisely the same supermultiplet as the reduction of the states 
of IIA supergravity. Indeed, the reductions of IIA and IIB 
supergravity to 9 dimensions, yield the same theory 
\cite{DHS,d-brane1,berg}. However, the massive states are
still characterized in terms of the group SO(8), which in $D=9$
dimensions comprises the rest-frame rotations. Therefore 
the Kaluza-Klein states that one obtains when compactifying the
ten-dimensional theory on a circle remain {\it inequivalent} for the IIA and
IIB theories (see \cite{ADLN} for a discussion of this phenomenon
and its consequences). It turns out that the $Q=32$ supergravity
multiplets are unique in all spacetime  dimensions $D>2$, except for 
$D=10$. Maximal supergravity will be introduced in chapter~3. 
The field content of the maximal $Q=32$ 
supergravity theories for dimensions $3\leq D\leq 11$ will be 
presented in  two tables ({\it c.f.} table~\ref{maximal-sg-bosons} and 
\ref{maximal-sg-fermions}). 

%%%%%%%%%%%%%%%%%%%%%%%%%%%%%%%%%%%%%%%%%%%%%
\begin{table}
\begin{center}
\begin{tabular}{l l l l l}\hline
SU$_+$(2)&$N_+=1$&$N_+=2$& $N_+=3$ & $N_+=4$ \\\hline
{\bf 5} & ~      &    ~  &  ~      &   1    \\
{\bf 4} & ~      &    ~  &  1      &   8    \\
{\bf 3} & ~      &    1  &  6      &   27   \\
{\bf 2} & 1      &    4  &  14     &   48   \\
{\bf 1} & 2      &    5  &  14     &   42   \\ \hline
~      & $(2+2)_{\bf C}$ & $(8+8)_{\bf R}$ & $(32+32)_{\bf C}$ & 
$(128+128)_{\bf R}$ \\ \hline
\end{tabular}
\end{center}
\caption{\small 
Shortest massless supermultiplets of $D=6$ $N_+$-extended 
chiral supersymmetry. The states transform both in the SU$_+$(2) 
helicity group and under a USp(2$N_+$) group. For odd values of 
$N_+$ the representations are complex, for even $N_+$ they can be chosen 
real. Of course, an identical table can be given for 
negative-chirality spinors. }\label{D6-chiral-susy-mult} 
\end{table}
%%%%%%%%%%%%%%%%%%%%%%%%%%%%%%%%%%%%%%%%%%%%%%

%%%%%%%%%%%%%%%%%%%%%%%%%%%%%%%%%%%%%%%%%%%%%%%%%%%%%%%%%%%%%%%% 
%
 \subsubsection{$D=6$ Supermultiplets}
In 6 dimensions we have chiral spinors, which are not Majorana. 
Because the charge conjugated spinor has the same chirality, the  
chiral rotations of the spinors can be extended to the group
USp$(2N_+)$, for $N_+$ chiral spinors. Likewise $N_-$ 
negative-chirality spinors transform under USp$(2N_-)$. This feature
is already incorporated in  
table~\ref{susy-charges}. In principle we have $N_+$ positive- 
and $N_-$ negative-chirality charges, but almost all 
information follows from first considering the purely chiral 
case. In table~\ref{D6-chiral-susy-mult} we present the  
decomposition of the various helicity representations of 
the smallest supermultiplets based on $N_+=1,2,3$ or 4 
supercharges. In $D=6$ dimensions the helicity group SO(4) 
decomposes into the product of two SU(2) groups: SO(4)$\cong ({\rm 
SU}_+(2) \times {\rm SU}_-(2))/{{\bf Z}_2}$. When we have 
supercharges of only one chirality, the smallest supermultiplet will 
only transform under one SU(2) factor of the helicity group, as 
is shown in table~\ref{D6-chiral-susy-mult}.\footnote{%%%%%%
  The content of this table also specifies the shortest 
  {\em massive} supermultiplets in four dimensions as well as with the  
  shortest {\it massless} multiplets in five dimensions. The SU(2) 
  group is then associated with spin or with
  helicity, respectively. 
}%%%%%%%%%%%%%%%%%%%%%%%%%%%%%%%%%%%%%%%%%%%%%%%%%%%%%%%%%%%% 

Let us now turn to specific supermultiplets. Let us recall that the  
helicity assignments of the states describing gravitons, 
gravitini, vector and (anti)selfdual tensor gauge fields, and spinor
fields are $(3,3)$, $(2,3)$ or $(3,2)$, $(2,2)$, $(3,1)$ or $(1,3)$,
and $(2,1)$ or $(1,2)$. Here $(m,n)$ denotes that the dimensionality
of the reducible representations of the two SU(2) factors of the
helicity group are of dimension $m$ and $n$. For the derivation of these
assignments, see for instance one of the appendices in \cite{dewitlouis}.

In the following we will first restrict ourselves to helicities that
correspond to at most the three-dimensional representation of either
one of the SU(2) factors. Hence we have only $(3,3)$, $(3,2)$,
$(2,3)$, $(3,1)$ or $(1,3)$ representations, as well as the
lower-dimensional ones. When a supermultiplet contains $(3,2)$ or
$(2,3)$ representations, we insist that it will also contain a single
$(3,3)$ representation, because gravitini without a graviton are not
expected to give rise to a consistent interacting field theory. The
multiplets of this type are shown in table~\ref{D6-susy-mult}. There
are no such multiplets for more than $Q=32$ supercharges. 

There are supermultiplets with higher SU(2)
helicity representations, which contain neither gravitons nor
gravitini. Some of these multiplets are shown in
table~\ref{D6-chiral} and we will discuss them in due
course. 

We now elucidate the construction of the supermultiplets listed in
table~\ref{D6-susy-mult}. The simplest case is
$(N_+, N_-)=(1,0)$, where the smallest supermultiplet is the (1,0) {\it 
hypermultiplet}, consisting of a complex doublet of spinless 
states and a chiral 
spinor. Taking the tensor product of the 
smallest supermultiplet with the $(2,1)$ helicity representation 
gives the (1,0) {\it tensor multiplet}, with a selfdual tensor, a 
spinless state and a doublet of chiral spinors. The tensor product with 
the $(1,2)$ helicity representation yields the (1,0) {\it vector 
multiplet}, with a vector state, a doublet of chiral spinors and a scalar. 
Multiplying the hypermultiplet with the (2,3) helicity representation,
one obtains  the states of (1,0) {\it supergravity}. Observe that the
selfdual tensor fields in the tensor and supergravity supermultiplet
are of opposite selfduality phase. 

%%%%%%%%%%%%%%%%%%%%%%%%%%%%%%%%%%%%%%%%%%%%%
\begin{table}
\begin{center}
\begin{tabular}{l l l l}\hline
multiplet        &\#& bosons        & fermions  \\ \hline
(1,0) hyper      &$4+4$  & $(1,1;2,1) + {\rm h.c.} $ & $(2,1;1,1)$ \\[1mm]
(1,0) tensor     &$4+4$  & $(3,1;1,1) + (1,1;1,1)$   & $(2,1;2,1)$ \\[1mm]
(1,0) vector     &$4+4$  & $(2,2;1,1)$               & $(1,2;2,1)$ \\[1mm]
(1,0) graviton   &$12+12$& $(3,3;1,1) + (1,3;1,1)$   & $(2,3;2,1)$ \\[1mm]
(2,0) tensor     &$8+8$  & $(3,1;1,1) + (1,1;5,1)$   & $(2,1;4,1)$ \\[1mm]
(2,0) graviton   &$24+24$& $(3,3;1,1) + (1,3;5,1)$   & $(2,3;4,1)$ \\[1mm]
(1,1) vector     &$8+8$  & $(2,2;1,1) + (1,1;2,2)$   & $(2,1;1,2)$ \\ 
~                &~      &~                          & $+(1,2;2,1)$ \\[1mm]
(1,1) graviton   &$32+32$& $(3,3;1,1)$               & $(3,2;1,2)$ \\
~                &~ & $+(1,3;1,1) + (3,1; 1,1)$    &$+(2,3;2,1)$ \\
~                &~ & $+(1,1;1,1) + (2,2;2,2)$     &$+(1,2;1,2)$ \\ 
~                &~      &~                          & $+ (2,1;2,1)$ \\[1mm]
(2,1) graviton   &$64+64$& $(3,3;1,1)$               & $(3,2;1,2)$\\ 
~                  &~ & $+(1,3;5,1) + (3,1; 1,1)$ & $+ (2,3;4,1)$\\

~                  &~ & $+(2,2;4,2)+(1,1;5,1)$  & $+(1,2;5,2)$\\
~                  &~ &~                        &$+(2,1;4,1)$ \\[1mm]
(2,2) graviton   &$128+128$& $(3,3;1,1)$              & $(3,2;4,1)$\\ 
~                  &~ & $+(3,1;1,5)+(1,3;5,1)$    & $+ (2,3;1,4)$ \\
~                  &~ & $+(2,2;4,4) + (1,1;5,5)$  & $+(2,1;4,5)$\\
~                  &~ &~                          & $+ (1,2;5,4)$ \\ 
\hline
\end{tabular}
\end{center}
\caption{\small Some relevant $D=6$ supermultiplets with $(N_+,N_-)$ 
supersymmetry. The states $(m,n;\tilde m,\tilde n)$ are assigned 
to $(m,n)$ representations of  
the helicity group SU$_+(2)\times {\rm SU}_-(2)$ and $(\tilde m,
\tilde n)$ representations of  
USp$(2N_+)\times {\rm USp}(2N_-)$. The second column lists 
the number of bosonic + fermionic states for each 
multiplet. }\label{D6-susy-mult}
\end{table}
%%%%%%%%%%%%%%%%%%%%%%%%%%%%%%%%%%%%%%%%%%%%%%

Next consider $(N_+,N_-)=(2,0)$ supersymmetry. The smallest 
multiplet, shown in table~\ref{D6-chiral-susy-mult}, then 
corresponds to the (2,0) {\it tensor multiplet}, with the bosonic states 
decomposing into a selfdual tensor and a five-plet of spinless 
states, and a four-plet of chiral fermions. Multiplication with 
the (1,3) helicity representation yields the (2,0) supergravity 
multiplet, consisting of the graviton, four chiral gravitini and 
five selfdual tensors  \cite{townsend1}. Again, the selfdual 
tensors of the tensor   
and of the supergravity supermultiplet are of opposite selfduality  
phase.

Of course, there exists also a nonchiral version with 16 supercharges, 
namely the one corresponding to $(N_+,N_-)=(1,1)$. The smallest 
multiplet is now given by the tensor product of the 
supermultiplets with (1,0) and (0,1) supersymmetry. This yields 
the vector multiplet, with the vector state and four scalars, 
the latter transforming with respect to the (2,2) representation of 
USp$(2) \times {\rm USp}(2)$. There are two doublets of chiral 
fermions with opposite chirality, each transforming as a doublet 
under the corresponding USp(2) group.
Taking the tensor product of the vector multiplet with the (2,2) 
representation of the helicity group yields the states of the (1,1)
{\it supergravity} multiplet. It consists of 32 bosonic states,  
corresponding to a graviton, a tensor, a 
scalar and four vector states, where the latter transform under the 
(2,2) representation of USp$(2) \times {\rm USp}(2)$. The 32 
fermionic states comprise two doublets of chiral gravitini and two 
chiral spinor doublets, transforming as doublets under the 
appropriate USp(2) group.

%%%%%%%%%%%%%%%%%%%%%%%%%%%%%%%%%%%%%%%%%%%%%%%%%%%%%%%%%%%%%%%%%
\begin{table}
\begin{center}
\begin{tabular}{l l l l}\hline
supersymmetry          &\#& bosons        & fermions  \\ \hline
(1,0) &$8+8$  & $(5;1) + (3;1)$   & $(4;2)$           \\[1mm]
(2,0) &$24+24$& $(5;1)+(1,1)$               & $(4;4)+(2;4)$     \\
~                &~ & $+(3;1) + (3;5)$ & ~   \\[1mm]
(3,0) &$64+64$& $(5;1)+(1;14)$               & $(4;6)$    \\
~                  &~ & $+(3;1) + (3;14)$ & $+(2;6) + (2;14)$ \\[1mm]
(4,0)  &$128+128$& $(5;1)+ (3;27)$              & $(4;8) + (2;48)$ \\
~                  &~ & $+(1;42)$    & ~      \\ \hline
\end{tabular}
\end{center}
\caption{\small 
$D=6$ supermultiplets without gravitons and gravitini with
$(N,0)$  supersymmetry, a single $(5;1)$ highest-helicity state and at
most 32 supercharges. The theories based on these multiplets 
have only rigid supersymmetry. The multiplets are identical to those
that underly the five-dimensional 
$N$-extended supergravities. They are all chiral, so that the
helicity group in six dimensions is restricted to SU$(2)\times {\bf
1}$ and the states are characterized as representations of USp$(2N)$.  
The states $(n;\tilde n)$ are assigned to the $n$-dimensional
representation of  SU(2) and the $\tilde n$-dimensional representation
of USp$(2N)$. The second column lists 
the number of bosonic + fermionic states for each 
multiplet. }\label{D6-chiral}
\end{table}
%%%%%%%%%%%%%%%%%%%%%%%%%%%%%%%%%%%%%%%%%%%%%%%%%%%%%%%%%%%%%%

Subsequently we discuss the case $(N_+,N_-)=(2,1)$. Here a
supergravity multiplet exists \cite{DauriaKF} and can be obtained from
the product of the 
states of the $(2,0)$ tensor multiplet with the $(0,1)$ tensor
multiplet. There is in fact a smaller supermultiplet, which we
discard because it contains gravitini but no graviton states. 

Finally, we turn to the case of $(N_+,N_-)=(2,2)$. The smallest 
supermultiplet is given by the tensor product of the smallest 
(2,0) and (0,2) supermultiplets. This yields the 128 + 128 states 
of the (2,2) {\it supergravity} multiplet. These states transform 
according to representations of USp$(4) \times {\rm USp}(4)$. 

In principle, one can continue and classify representations for 
other values of $(N_+,N_-)$. As is obvious from the construction 
that we have presented, this will inevitably lead to states 
transforming in higher-helicity representations. Some of these
multiplets will suffer from the fact that they have more than one
graviton state, so that we expect them to be inconsistent at the
nonlinear level. However, there are the chiral theories which contain
neither graviton nor gravitino states. Restricting ourselves to 32
supercharges and requiring the highest helicity to be a
five-dimensional representation of one of the SU(2) factors, there are
just four theories, summarized in table~\ref{D6-chiral}. For
a recent discussion of one of these theories, see \cite{Hull2}. 
%
%%%%%%%%%%%%%%%%%%%%%%%%%%%%%%%%%%%%%%%%%%%%%%%%%%%%%%%%%%%%%%
 \subsection{Massive supermultiplets}
 \label{massive-supermultiplets}
%%%%%%%%%%%%%%%%%%%%%%%%%%%%%%%%%%%%%%%%%%%%%%%%%%%%%%%%%%%%%%
Generically massive supermultiplets are bigger than massless ones
because the number of supercharges that generate the multiplet is not
reduced, unlike for massless supermultiplets where one-half of the
supercharges vanishes. However, in the presence of mass parameters
the superalgebra may also contain central charges, which could give
rise to a shortening of the representation in a way similar to what
happens for the massless supermultiplets. This only happens for
special values of these charges. The shortened supermultiplets are
known as BPS multiplets. Central charges and multiplet shortening are
discussed in subsection~\ref{central-charges-shortening}. In this
section we assume that the central charges are absent. 

The analysis of massive supermultiplets takes place in the
restframe. The states then organize themselves into representations
of the rest-frame rotation group, ${\rm SO}(D-1)$, associated with
spin. The supercharges transform as spinors under this group, so that
one obtains a Euclidean supersymmetry algebra, 
\begin{equation}
  \label{eq:nr-susy}
 \{Q_\a, Q^\dagger_\b\} = 2 M \,\d_{\a\b}\,. 
\end{equation}
Just as before, the spinor charges transform under the automorphism
group of the supersymmetry algebra that commutes with the spin
rotation group. This group will also be denoted by ${\rm H}_{\rm R}$;
it is the nonrelativistic variant of the R-symmetry group that was
introduced previously. Obviously the nonrelativistic group can be
bigger than its relativistic counterpart, as it is required to commute
with a smaller group. For instance, in $D=4$ spacetime dimensions, the
relativistic R-symmetry group is equal to ${\rm U}(N)$,
while the nonrelativistic one is the group ${\rm USp}(2N)$, which 
contains ${\rm U}(N)$ as a subgroup according to ${\bf 2 N} = {\bf N}
+ \overline {\bf N}$. Table~\ref{massive-4d} shows the smallest massive 
representations for $N\leq 4$ in $D=4$ dimensions as an
illustration. Clearly the states of given spin can be assigned to 
representations of the nonrelativistic group ${\rm H}_{\rm R}= {\rm
USp}(2N)$ and decomposed in terms of irreducible representations of the
relativistic R-symmetry group ${\rm U}(N)$. More explicit
derivations can be found in \cite{multiplets}. 

Knowledge of the relevant groups ${\rm H}_{\rm R}$ is 
important and convenient in writing down the supermultiplets. It can
also reveal certain relations between supermultiplets, even between
supermultiplets living in spacetimes of different dimension. Obviously,
supermultiplets living in higher dimensions can always be decomposed
into supermultiplets living in lower dimensions, and massive
supermultiplets can be decomposed in terms of massless ones, but
sometimes there exists a relationship that is less trivial. For
instance, the $D=4$ {\it massive} multiplets shown in
table~\ref{massive-4d} coincide with the {\it massless}
supermultiplets of chirally extended supersymmetry in $D=6$ dimensions
shown in table~\ref{D6-chiral-susy-mult}. In particular the $N=4$
supermultiplet of table~\ref{massive-4d} appears in many places and
coincides with the massless $N=8$ supermultiplet in $D=5$
dimensions, which is shown in tables~\ref{maximal-sg-bosons} and
\ref{maximal-sg-fermions}. The reasons for this are clear. The $D=5$
and the chiral $D=6$ massless supermultiplets are subject to the same
helicity group ${\rm SU}(2)$,  which in turn coincides with the spin
rotation group for $D=4$. Not surprisingly, also the relevant
automorphism groups ${\rm H}_{\rm R}$ coincide, as the reader can
easily verify. Since the number of {\it effective} supercharges is
equal in these cases and given by $Q_{\rm eff}= 16$ (remember that
only half of the charges play a role in building up massless
supermultiplets), the multiplets must indeed be identical.

%%%%%%%%%%%%%%%%%%%%%%%%%%%%%%%%%%%%%%%%%%%%%%%%%%%%%%%%%%%%%%%%%%%%%%
\begin{table}
\begin{center}
\begin{tabular}{c c c }\hline
spin& $N=1$      &  $N=2$   \\ \hline
1   & ~          & $1=1$     \\
1/2 &$1=1$       &$4=2+\bar 2$ \\
0   &$2=1+\bar 1$&$5=3+1+\bar1$ \\ \hline
 {~}&  $N=3$           & $N=4$ \\ \hline
2   &   ~~             & $1=1$  \\
3/2 & $1=1$            & $8=4+ \bar 4$ \\
1   & $6=3+\bar 3$     & $27=15+6+\bar6$\\
1/2 & $14=8+3+\bar3$&$48=20+\overline{20}+4+\bar4$ \\
0   & $14=6+\bar 6+ 1+\bar1$ & $42=20^\prime
+ 10+\overline{10} + 1 +\bar 1$   \\ \hline

\end{tabular}
\end{center}
\caption{\small 
  Minimal $D=4$ massive supermultiplets without central charges for
  $N\leq4$. The states are listed as ${\rm USp}(2N)$ representations
  which are subsequently decomposed into representations of ${\rm
  U}(N)$.}\label{massive-4d} 
\end{table}
%%%%%%%%%%%%%%%%%%%%%%%%%%%%%%%%%%%%%%%%%%%%%%%%%%%%%%%%%%%%%%%%%%

Here we also want to briefly draw the attention to the relation between
off-shell multiplets and massive representations. So far we discussed
supermultiplets consisting of states on which the supercharges
act. These states can be described by a field theory in which the
supercharges generate corresponding supersymmetry variations on the
fields. Very often the transformations on the fields do {\it not} close
according to the supersymmetry algebra unless one imposes the
equations of motion for the fields. Such representations are called
{\it on-shell} representations. The lack of closure has many consequences,
for instance, when determining quantum corrections. In certain cases
one can improve the situation by introducing extra fields which do not
directly correspond to physical fields. These fields are known as 
{\it auxiliary fields}. By employing such fields one may be able to
define an {\it off-shell} representation, where the transformations close
upon (anti)commutation without the need for 
imposing field equations. Unfortunately, many theories do not possess
(finite-dimensional) off-shell representations. Notorious examples are
gauge theories and 
supergravity theories with 16 or more supercharges. This fact makes is
much more difficult to construct an extended variety of actions for
these theories, because the transformation rules are implicitly
dependent on the action. There is an 
off-shell counting argument. according to which the field degrees of
freedom should comprise a {\it massive} supermultiplet (while the states
that are described could be massless). For instance, the off-shell
description of the $N=2$ vector multiplet in $D=4$ dimensions can be
formulated in terms of a gauge field (with three degrees of freedom), a
fermion doublet (with eight degrees of freedom) and a triplet of auxiliary
scalar fields (with three degrees of freedom), precisely in accord with
the $N=2$ entry in  table~\ref{massive-4d}. In fact, this multiplet
coincides with the multiplet of the currents that couple to an $N=2$
supersymmetric gauge theory.

The $N=4$ multiplet in table~\ref{massive-4d} corresponds to the
gravitational supermultiplet of currents \cite{BdRdW}. These are the
currents that 
couple to the fields of $N=4$ conformal supergravity. Extending the
number of supercharges beyond 16 will increase the minimal spin of a
massive multiplet beyond spin-2. Since higher-spin fields can usually
not be coupled, one may conclude that conformal supergravity does not
exist for more than 16 charges. For that reason there can be no
off-shell formulations for supergravity with more than 16
charges. Conformal supergravity will be discussed in chapter~7. 

In section~2.5 we will present a table listing the
various groups ${\rm H}_{\rm R}$ for spinors associated with certain 
Clifford algebras ${\cal C}(p,q)$ with corresponding rotation groups
${\rm SO}(p,q)$. Subsequently we then discuss some further
implications of these results.
%%%%%%%%%%%%%%%%%%%%%%%%%%%%%%%%%%%%%%%%%%%%%%%%%%%%%%%%%%%%%%%%%%
%%%%%%%%%%%%%%%%%%%%%%%%%%%%%%%%%%%%%%%%%%%%%%%%%%%%%%%%%%%%%%%%%%
%%%%%%%%%%%%%%%%%%%%%%%%%%%%%%%%%%%%%%%%%%%%%
\begin{table}
\begin{center}
\begin{tabular}{l l l l l l l l}\hline
$D$ &${\rm H}_{\rm R}$& $p=0$ & $p=1$ & $p=2$ & $p=3$ & $p=4$ & $p=5$
\\ \hline
11  & 1      & ~  & ~           & 1   & ~  & ~  & 1 \\
~  & ~      & ~   & ~           & $[55]$  & ~  & ~  & 
$[462]$ \\[1mm]
10A$\!\!\!$ & 1      & 1  & $1$       & 1   & ~  & 1    & $1+1$ \\
~   & ~      & $[1]$  & $[10]$    &$[45]$   & ~  & $[210]$  &  
$[126]$ \\[1mm]
10B$\!\!\!$ & SO(2)  & ~  & $2$& ~   & 1   & ~  & $1+2$ \\
~   & ~      & ~  & $[10]$          & ~   & $[120]$ & ~  & 
$[126]$          \\[1mm]
9   & SO(2)  & $1+2$ & $2$ & 1 & 1  & $1+2$ & ~ \\
~   &        & $[1]$           & $[9]$          &$[36]$ &$[84]$  
& $[126]$         & ~ \\[1mm]
8   & U(2)   & $3+\bar 3$  & $3$      & $1+\bar 1$& $1+3$ 
&$3+\bar 3$ & ~ \\
~   &        & $[1]$           & $[8]$          & $[28]$        & 
$[56]$    &$[35]$ & ~ \\[1mm]
7   & USp(4) & 10  & $5$ & $1+5$      & 10  & ~  & ~ \\
~   &        & $[1]$   & $[7]$     & $[21]$         & $[35]$  & ~ 
 & ~ \\[1mm]
6   & ${\rm USp}(4)$   & $(4,4)$ & $(1,1)$ & (4,4) & $(10,1)$ & ~ & ~ \\ 
~   & $\times{\rm USp}(4)$  &   ~   & $+(5,1)$ & ~  & $+(1,10)$
&~&~\\
~   & ~          &  ~       & $+(1,5)$ & ~   & ~     & ~ & ~  \\
~   & ~  & $[1]$ & $[6]$ & $[15]$ & $[10]$ & ~ & ~ \\[1mm] 
5   & USp(8) & $1+27$   & $27$ & 36   & ~  & ~  & ~ \\
~   & ~      & $[1]$        & $[5]$      & $[10]$   & ~  & ~  & ~ 
\\[1mm]
4   & U(8) & $28+\overline{28}$ & $63$& $36+\overline{36}$ & ~ & ~ & ~ \\  
~   & ~    & $[1]$                 & $[4]$    & $[3]$      & ~ & 
~ & ~ \\[1mm] 
3   & SO(16) & 120   & $135$ & ~   & ~  & ~  & ~ \\
~   & ~      & $[1]$     & $[3]$   & ~   & ~  & ~  & ~ \\[1mm]
 \hline
\end{tabular}
\end{center}
\caption{\small Decomposition of the central extension in the 
supersymmetry algebra with $Q=32$ supercharge components in terms 
of $p$-rank Lorentz tensors. The second row specifies the number of 
independent components for each $p$-rank tensor charge. The total 
number of central charges is equal to $528-D$, because we have 
not listed the $D$ independent momentum operators}
\label{maximal-central-extension} 
\end{table}
%%%%%%%%%%%%%%%%%%%%%%%%%%%%%%%%%%%%%%%%%%%%%%%%%%%%%%%%%%%%%%%
%%%%%%%%%%%%%%%%%%%%%%%%%%%%%%%%%%%%%%%%%%%%%%%%%%%%%%%%%%%%%%%
\subsection{Central charges and multiplet shortening}
\label{central-charges-shortening}
%%%%%%%%%%%%%%%%%%%%%%%%%%%%%%%%%%%%%%%%%%%%%%%%%%%%%%%%%%%%%%%
The supersymmetry algebra of the maximal supergravities comprises 
general coordinate transformations, local supersymmetry 
transformations and the gauge transformations associated with the 
antisymmetric gauge fields.\footnote{%%%%%%%%%%%%%%%%%%%%%%%%%%
   There may be 
   additional gauge transformations that are of interest to us. 
   As we discuss in due course, it is possible to have 
   (part of) the automorphism group ${\rm H}_{\rm R}$ realized as a 
   local invariance. However, the corresponding gauge fields are 
   composite and do not give rise to physical states (at least, 
   not in perturbation theory).} %%%%%%%%%%%%%%%%%%%%%%%%%%%% 
These gauge transformations usually appear in the anticommutator of
two supercharges, and may be regarded as central charges. In
perturbation theory, the theory does not contain charged fields, so
these central charges simply vanish  
on physical states. However, at the nonperturbative level, there 
may be solitonic or other states that carry charges. An example 
are magnetic monopoles, dyons, or black holes. At the
M-theory level, these charges are associated with certain brane
configurations. On 
such states, some of the central charges may take finite values. 
Without further knowledge about the kind of states  
that may emerge at the nonperturbative level, we can  
generally classify the possible central charges, by considering a
decomposition of the anticommutator.  
This anticommutator carries at least two spinor indices and two 
indices associated with the group ${\rm H}_{\rm R}$. Hence we may
write  
\be
\{Q_\a, Q_\b\} \propto \sum_p \;(\G^{\m_1\cdots \m_p} C)_{\a\b} \, 
Z_{\m_1\cdots \m_p}\,,
\ee
where $\G^{\m_1\cdots \m_p}$ is the antisymmetrized product of 
$p$ gamma matrices, $C$ is the charge-conjugation matrix and 
$Z_{\m_1\cdots \m_p}$ is the central charge, which transforms as 
an antisymmetric $p$-rank Lorentz tensor and depends on 
possible additional ${\rm H}_{\rm R}$ indices attached to the 
supercharges. The central charge must be symmetric or antisymmetric in
these indices, depending on whether the $(\Gamma^{\m_1\cdots\m_p}
C)_{\a\b}$ is antisymmetric or symmetric in $\a,\b$, so that the
product with $Z_{\m_1\cdots\m_p}$ is always symmetric in the combined
indices of the supercharges. For given spacetime dimension all 
possible central charges can be classified.\footnote{%%%%%%%%%%%
   For related discussions see, for example,
   \cite{Mtheory,sergio,ObersPioline} and 
   references therein.} %%%%%%%%%%%%%%%%%%%%%%%%%%%%%%%%%%%%%%%%
For the maximal supergravities in spacetime dimensions $3\leq D\leq11$
this  classification is given in
table~\ref{maximal-central-extension}, where we list all possible
charges and their ${\rm H}_{\rm R}$ representation assignments.
Because we have 32 supercharge components, the sum of the 
independent momentum operators and the central charges must be 
equal to $(32\times 33)/2 = 528$. The results of the table are in
direct correspondence with the eleven-dimensional superalgebra with
the most general central charges, 
\begin{equation}
  \label{eq:11D-algebra}
  \{Q_\a,\bar Q_\b\} = -2i P_M \G^M_{\a\b} + Z_{MN}\G^{MN}_{\a\b} +
  Z_{MNPQR} \G^{MNPQR}_{\a\b}\,. 
\end{equation}
The two central charges, $Z_{MN}$ and $Z_{MNPQR}$ can be associated
with the winding numbers of two- and five-branes. 

In order to realize the supersymmetry algebra in a positive-definite
Hilbert space, the right-hand side of the anticommutator is subject to
a positivity condition, which generically implies that the mass
of the multiplet is larger than or equal to the central
charges. Especially in higher dimensions, the bound may can take a
complicated form. This positivity bound is known as the Bogomol'nyi
bound. When the bound is saturated one speaks of BPS states. For BPS
multiplets some of the supercharges must vanish on the states, in the
same way as half of the charges vanish for the massless
supermultiplets. This vanishing of some of the supercharges leads to a
shortening of the multiplet. Qualitatively, this phenomenon of multiplet
shortening is the same as for massless supermultiplets, but 
here the fraction of the charges that vanishes is not necessarily
equal to 1/2. Hence one speaks of 1/2-BPS, 1/4-BPS supermultiplets,
etcetera, to indicate which fraction of the supercharges vanishes on
the states. The fact that the BPS supermultiplets have a completely
different field content than the generic massive supermultiplets makes
that they exhibit a remarkable stability under `adiabatic' deformations.
This means that perturbative results based on BPS supermultiplets can
often be extrapolated to a nonperturbative regime.

For higher extended supersymmetry the difference in size of BPS
supermultiplets and massive supermultiplets can be enormous in view
of the fact that the number of states depend exponentionally on the
number of nonvanishing central charges. For lower supersymmetry the
multiplets can be comparable in size, but nevertheless they are quite
different. For instance, consider $N=2$ {\it massive} vector
supermultiplets in four spacetime dimensions. Without central charges,
such a multiplet comprises $8+8$ states, corresponding to the three
states of spin-1, the eight states of four irreducible spin-$\ft12$ 
representations, and five states with spin 0. On the other hand there
is another massive vector supermultiplet, which is 
BPS and comprises the three states of spin-1, the four states of two
spin-$\ft12$ representations and two states of spin-0. These
states are subject to a 
nonvanishing central charge which requires that the states are all
doubly degenerate, so that we have again $8+8$ states, but with a
completely different spin content. When
decomposing these multiplets into massless $N=2$ supermultiplets, the
first multiplet decomposes into a massless vector multiplet and a
hypermultiplet. Hence this is the multiplet one has in the Higgs
phase, where the hypermultiplet provides the scalar degree of freedom
that allows the conversion of the massless to massive spin-1
states. This multiplet carries no central charge. The second
supermultiplet, which is BPS, appears as a massive charged vector
multiplet when breaking a nonabelian supersymmetric gauge theory to an
abelian subgroup. This realization is known as the Coulomb phase. 

In view of the very large variety of BPS supermultiplets, we do not
continue this general discussion of supermultiplets with central
charges. In later chapters we will discuss specific BPS
supermultiplets as well as other mechanisms of multiplet shortening in
anti-de Sitter space.  

%%%%%%%%%%%%%%%%%%%%%%%%%%%%%%%%%%%%%%%%%%%%%%%%%%%%%%%%%%%%%%%%%%
%%%%%%%%%%%%%%%%%%%%%%%%%%%%%%%%%%%%%%%%%%%%%%%%%%%%%%%%%%%%%%%%%%
\subsection{On spinors and the R-symmetry group ${\rm H}_{\rm R}$}
\label{R-symmetry} %%%%%%%%%%%%%%%%%%%%%%%%%%%%%%%%%%%%%%%%%%%%%%%
%%%%%%%%%%%%%%%%%%%%%%%%%%%%%%%%%%%%%%%%%%%%%%%%%%%%%%%%%%%%%%%%%%
In this section we return once more to the spinor
representations and the corresponding automorphism group 
${\rm H}_{\rm R}$, also known as the R-symmetry
group. Table~\ref{pq-spinors} summarizes information for 
spinors up to (real) dimension 32 associated with the groups 
${\rm SO}(p,q)$, where we restrict $q\leq2$. From this table we can
gain certain insights into the properties of spinors living in
Euclidean, Minkowski and (anti-)de Sitter spaces as well as the
supersymmetry algebras based on these spinors. Let us first elucidate
the information presented in the table. Subsequently we shall discuss
some correspondences between the various spinors in different
dimensions.  

%%%%%%%%%%%%%%%%%%%%%%%%%%%%%%%%%%%%%%%%%%%%%%%%%%%%%%%%%%%%%%%% 
%%%%%%%%%%%%%%%%%%%%%%%%%%%%%%%%%%%%%%%%%%%%%%%%%%%%%%%%%%%%%%%%
\begin{table}
\begin{center}
\begin{tabular}{l l c c c l}\hline
$d_{\cal C}$ & ${\cal C}(p,q)$ & $r$  & centralizer & 
   $d_{{\rm SO}(p,q)}$ & ${\rm H}_{\rm R}$ \\ \hline
1 & ${\cal C}(1,0)$ &1& $\bf R$ & $1$   & ${\rm SO}(N)$   \\
2 & ${\cal C}(0,1)$ &3& $\bf C$ & $1+1$ & ${\rm SO}(N)$    \\
2 & ${\cal C}(1,1)$ &0& $\bf R$ & $1+1$ & ${\rm SO}(N)$   \\
2 & ${\cal C}(2,0)$ &2& $\bf R$ & $2$   & ${\rm U}(N)$    \\
2 & ${\cal C}(2,1)$ &1& $\bf R$ & $2$   & ${\rm SO}(N)$   \\
4 & ${\cal C}(0,2)$ &2& $\bf H$ & $2+2$ & ${\rm U}(N)$    \\
4 & ${\cal C}(1,2)$ &3& $\bf C$ & $2+2$ & ${\rm SO}(N)$   \\
4 & ${\cal C}(2,2)$ &0& $\bf R$ & $2+2$ & ${\rm SO}(N)$   \\
4 & ${\cal C}(3,0)$ &3& $\bf H$ & $4$   & ${\rm USp}(2N)$ \\
4 & ${\cal C}(3,1)$ &2& $\bf C$ & $4$   & ${\rm U}(N)$    \\
4 & ${\cal C}(3,2)$ &1& $\bf R$ & $4$   & ${\rm SO}(N)$   \\
8 & ${\cal C}(4,0)$ &0& $\bf H$ & $4+4$ & ${\rm USp}(2N)$ \\  
8 & ${\cal C}(4,1)$ &3& $\bf H$ & $8$   & ${\rm USp}(2N)$ \\
8 & ${\cal C}(4,2)$ &2& $\bf C$ & $8$   & ${\rm U}(N)$    \\
8 & ${\cal C}(5,0)$ &1& $\bf H$ & $8$   & ${\rm USp}(2N)$ \\
16& ${\cal C}(5,1)$ &0& $\bf H$ & $8+8$ & ${\rm USp}(2N)$ \\
16& ${\cal C}(5,2)$ &3& $\bf C$ & $16$  & ${\rm USp}(2N)$ \\
16& ${\cal C}(6,0)$ &2& $\bf H$ & $8+8$ & ${\rm U}(N)$    \\
16& ${\cal C}(6,1)$ &1& $\bf H$ & $16$  & ${\rm USp}(2N)$ \\
16& ${\cal C}(7,0)$ &3& $\bf C$ & $8+8$ & ${\rm SO}(N)$   \\
16& ${\cal C}(8,0)$ &0& $\bf R$ & $8+8$ & ${\rm SO}(N)$   \\ 
16& ${\cal C}(9,0)$ &1& $\bf R$ & $16$  & ${\rm SO}(N)$   \\
32& ${\cal C}(6,2)$ &0& $\bf H$ & $16+16$ & ${\rm USp}(2N)$\\
32& ${\cal C}(7,1)$ &2& $\bf C$ & $16+16$ & ${\rm U}(N)$   \\
32& ${\cal C}(7,2)$ &3& $\bf H$ & $32$    & ${\rm USp}(2N)$  \\
32& ${\cal C}(8,1)$ &3& $\bf C$ & $16+16$ & ${\rm SO}(N)$  \\
32& ${\cal C}(9,1)$ &0& $\bf R$ & $16+16$ & ${\rm SO}(N)$  \\
32& ${\cal C}(10,0)$&2& $\bf R$ & $32$  & ${\rm U}(N)$     \\
32& ${\cal C}(10,1)$&1& $\bf R$ & $32$    & ${\rm SO}(N)$  \\ 
64& ${\cal C}(8,2)$ &2& $\bf H$ & $32+32$ & ${\rm U}(N)$   \\
64& ${\cal C}(9,2)$ &3& $\bf C$ & $32+32$ & ${\rm SO}(N)$  \\
64& ${\cal C}(10,2)$&0& $\bf R$ & $32+32$ & ${\rm SO}(N)$  \\
\hline
\end{tabular}
\end{center}
\caption{\small Representations of the Clifford algebras 
    ${\cal C}(p,q)$ with $q\leq2$ and their centralizers, and the
    ${\rm SO}(p,q)$ spinors of maximal real dimension 32 and their
    R-symmetry group. We also list the dimensions of the Clifford
    algebra and spinor representation, as well as $r=p-q \;\rm{mod}\; 
    4$. }   
\label{pq-spinors}
\end{table}
%%%%%%%%%%%%%%%%%%%%%%%%%%%%%%%%%%%%%%%%%%%%%%%%%%%%%%%%%%%%%%%%%%%
%%%%%%%%%%%%%%%%%%%%%%%%%%%%%%%%%%%%%%%%%%%%%%%%%%%%%%%%%%%%%%%%%%

We consider the Clifford algebras ${\cal C}(p,q)$ based on  $p+q$
generators, denoted by $e_1,e_2,\ldots, e_{p+q}$, with a nondegenerate
metric of signature $(p,q)$. This means that $p$ 
generators square to the identity and $q$ to minus the identity. We
list the real dimension of the irreducible Clifford algebra
representation, denoted by $d_{\cal C}$, and the values $r$ (equal to
$0,\ldots, 3$), where $r$ is defined by $r\equiv p-q$~mod~4. The value
for $r$ determines the square of the matrix  
built from forming the product $e_1\cdot e_2\cdots e_{p+q}$ of all the
Clifford algebra generators. For 
$r=0,1$ this square equals the identity, while for $r=2,3$ the square
equals minus the identity. Therefore, for $r=0$, the subalgebra 
${\cal C}_+(p,q)$ generated by products of {\it even}
numbers of generators is not simple and breaks into two simple ideals,
while for $r=1$, the {\it full} Clifford algebra ${\cal C}(p,q)$
decomposes into two simple ideals. 

We also present the centralizer of the irreducible representations
of the Clifford algebra, which, according to Schur's lemma, must form
a division algebra and is thus isomorphic to the real numbers 
(${\bf R}$), the complex numbers (${\bf C}$), or the quaternions 
(${\bf H}$). This means that the irreducible representation commutes 
with the identity and none, one or three complex structures, 
respectively, which generate the corresponding division
algebra. Table~\ref{pq-spinors} reflects also the 
so-called periodicity theorem \cite{periodicity}, according to which
there exists an isomorphism between the Clifford algebras 
${\cal C}(p+8,q)$ (or ${\cal C}(p,q+8)$) and ${\cal C}(p,q)$ times the
$16\times16$ real matrices. Therefore, the dimension of the
representations of ${\cal C}(p+8,q)$ (or ${\cal C}(p,q+8)$) and 
${\cal C}(p,q)$ differs by a factor 16.   

Finally the table lists the branching of the Clifford algebra
representation into ${\rm SO}(p,q)$ spinor representations. When
$r=0$ the Clifford algebra representation decomposes into two
chiral spinors. Observe that for $r=2$ we can also have chiral
spinors, but they are complex so that their real dimension remains
unaltered. For $r=2,3$ there are no chiral spinors, but nevertheless
in certain cases the Clifford algebra representation can still
decompose into two  irreducible 
spinor  representations. The last column gives the compact group ${\rm
H}_{\rm R}$, consisting of the linear transformations that commute with the
group ${\rm SO}(p,q)$ and act on $N$ irreducible spinors, leaving a
positive-definite metric invariant. For $r=0$, a group 
${\rm H}_{\rm R}$ should be assigned to each of the chiral sectors
separately. Again, 
according to Schur's lemma the centralizer of ${\rm SO}(p,q)$ must
form a division algebra for irreducible spinor
representations. Correspondingly, the group ${\rm SO}(p,q)$ commutes
with the identity and none, one or three 
complex structures, which leads to ${\rm H}_{\rm R}= {\rm SO}(N)$,
${\rm U}(N)$, or ${\rm USp}(2N)$, respectively. We note that the
results of table~\ref{pq-spinors} are in accord with the results
presented earlier in tables~\ref{susy-charges} and \ref{simple-susy}.

We now  discuss and clarify a number of correspondences between
spinors living in different dimensions. The
first correspondence is between spinors of ${\rm SO}(p,1)$ and  
${\rm SO}(p-1,0)$. According to the table, for any $p>1$, the
dimensions of the corresponding spinors differ by a factor two, while
their respective groups  
${\rm H}_{\rm R}$ always coincide. From a physical perspective, this
correspondence can be 
understood from the fact that ${\rm  SO}(p-1,0)$ is the {\it
helicity} group of massless spinor states in flat Minkowski space of
dimension $D=p+1$. In a field-theoretic context the reduction of the
spinor degrees of freedom is effected by the massless Dirac equation
and the automorphism groups ${\rm H}_{\rm R}$ that commute with the
Lorentz transformations and the transverse helicity rotations,
respectively, simply coincide. The two algebras \eqn{susy-algebra} and 
\eqn{susy-algebra2} thus share the same automorphism group. From a
mathematical viewpoint, this correspondence is related to the
isomorphism 
\be
\label{iso-pq}
{\cal C}(p,q) \cong {\cal C}(p-1,q-1)\otimes {\cal C}(1,1)\,,
\ee
where we note that ${\cal C}(1,1)$ is isomorphic with the real
$2\times2$ matrices. 

Inspired by the first correspondence one may investigate a second one
between spinors of ${\rm SO}(p,1)$ and   
${\rm SO}(p,0)$ with $p>1$. Physically this correspondence is relevant
when considering relativistic massive spinors in flat Minkowski
spacetime of dimension $D=p+1$, 
which transform in the restframe under $p$-dimensional spin
rotations. As the table shows, this correspondence is less systematic
and, indeed, an  underlying isomorphism for the corresponding Clifford
algebras is lacking. The results of the table should therefore be
applied with care. In a number of cases the relativistic spinor
transforms irreducibly under the nonrelativistic rotation group. In
that case the dimension of the 
automorphism group ${\rm H}_{\rm R}$ can increase, as it does for
$p=3$ and 10, but not for $p=5$ and 9. For $p=3$ (or, equivalently,
$D=4$) the implications of the fact that the nonrelativistic
automorphism group ${\rm USp}(2N)$ is bigger than the relativistic
one, have already been discussed in section~\ref{massive-supermultiplets}.  
In the remaining cases, $p=4,6,7,8$ (always
modulo 8), the relativistic spinor decomposes into two nonrelativistic
spinors. Because the number of irreducible spinors is then doubled, the
nonrelativistic automorphism group has a tendency to increase, but one
should consult the table for specific cases.

The third correspondence relates spinors of ${\rm SO}(p,2)$ and  
${\rm SO}(p,1)$ with $p>1$. Again the situation depends
sensitively on the value for $p$. In a number of cases ({\it i.e.}, 
$p=2,3,4,6$) the spinor dimension is the same for both groups. This
can be understood from the fact that the Clifford algebra
representations are irreducible with respect to ${\rm SO}(p,1)$, so
that one can always extend the generators of ${\rm SO}(p,1)$, which
are  proportional to $\G^{[a}\G^{b]}$, to those of ${\rm
SO}(p,2)$ by including the gamma matrices $\G^a$. However, the
R-symmetry group is not necessarily the same. For $p=4$ the
${\rm SO}(4,2)$ spinors allow the R-symmetry group ${\rm U}(N)$, while
for ${\rm SO}(4,1)$ the R-symmetry group is larger and equal to ${\rm
USp}(2N)$. Therefore theories formulated in flat Minkowski
spacetime of dimension $D=3,4,5,6$ can in principle be elevated to
anti-de Sitter space. For $D=5$ the R-symmetry reduces to ${\rm
U}(N)$, while for $D=3,4,6$ the R-symmetry remains the same. In the
remaining dimensions, $D=2,7,8,9,10$, a single Minkowski spinor can not
be elevated to anti-de Sitter space, and one 
must at least start from an even number of flat Minkowski spinors (so that
$N$ is even). For these cases, it is hard to make general statements
about the fate of the R-symmetry when moving to anti-de Sitter space
and one has to consult table~\ref{pq-spinors}.

The fourth correspondence is again more systematic, as it is based on
the isomorphism \eqn{iso-pq}. The correspondence relates spinors of
${\rm SO}(p,2)$ and of ${\rm SO}(p-1,1)$. For all $p>1$ the spinor
dimension differs by one-half while the R-symmetry group remains the
same. Observe that ${\rm SO}(p,2)$ can be regarded as the group of
conformal symmetries in a Minkowski space of $p$ dimensions, or as
the isometry group of an anti-de Sitter space of dimension
$p+1$. This correspondence extends this statement to the level of
spinors. It implies that the extension of the Poincar\'e
superalgebra in $D=p$ spacetime dimensions 
to a superconformal algebra requires a doubling of the number of 
supercharges. This feature is well known \cite{FKTvN} and the 
two supersymmetries are called $Q$- and $S$-supersymmetry. The
anticommutator of two $S$-supersymmetry charges yields the conformal
boosts. Both set of charges transform under the R-symmetry group of
the Poincar\'e algebra, which plays a more basic role in the
superconformal algebra as its generators appear in the anticommutator
of a $Q$-supersymmetry and an $S$-supersymmetry charge. In the anti-de
Sitter context, the spinor charge is irreducible but has simply twice
as many components. We return to the superconformal invariance
and related aspects in chapter~7.

It is illuminating to exploit some of the previous correspondences and
the relations between various supersymmetry representations in the context
of the so-called adS/CFT correspondence \cite{Mald}. We close this
chapter by exhibiting a chain of relationships between various
supermultiplets. We start with an $N=4$ supersymmetric gauge theory in
$D=4$ spacetime dimensions, whose massless states are characterized as 
representations of the ${\rm SO}(2)$ helicity group and the
R-symmetry group ${\rm SU}(4)$.\footnote{%%%%%%%%%%%%%%%%%%%%%%%%%%%
%%%%%%%%%%%%%%%%%%%%%%%%%%%%%%%%%%%%%%%%%%%%%%%%%%%%%%%%%%%%%%%%%%%%
  Because this multiplet is CPT self-conjugate, the ${\rm U}(1)$
  subgroup of ${\rm U}(4)$ coincides with the helicity group and plays
  no independent role here.} %%%%%%%%%%%%%%%%%%%%%%%%%%%%%%%%%%%%%%%
%%%%%%%%%%%%%%%%%%%%%%%%%%%%%%%%%%%%%%%%%%%%%%%%%%%%%%%%%%%%%%%%%%%%
Hence we have a field theory with $Q=16$ supersymmetries, of which only
8 are realized on the massless supermultiplet. This supermultiplet
decomposes as follows, 
\be
\label{gauge-mult}
(\pm1,{\bf 1})_{\bf p}+ (0,{\bf 6})_{\bf p}+ (\ft 12,{\bf 4})_{\bf p}
+ (-\ft12, \overline {\bf 4})_{\bf p} \,,
\ee 
where ${\bf p}$ indicates the three-momentum, $\vert{\bf p}\vert$ the
energy, and the entries in the parentheses denote the helicity and the
${\rm SU}(4)$ representation of the states. 

Multiplying this multiplet with a similar one, but now with opposite
three-momentum $-{\bf p}$, yields a multiplet with zero
momentum and with mass $M= 2\vert{\bf p}\vert$. As it turns out the
helicity states can now be assembled into states that transform under
the 3-dimensional rotation group, so that they can be characterized by 
their spin. The resulting supermultiplet consists of $128 + 128$
degrees of freedom. While the states of the original multiplet
\eqn{gauge-mult} were only subject to the helicity group and 8
supersymmetries, the composite multiplet is now a full supermultiplet
subject to 16 supersymmetries and the rotation (rather than the
helicity) group. Indeed, inspection shows that this composite
multiplet is precisely the $N=4$ massive multiplet shown in
table~\ref{massive-4d}. In this form the relevant R-symmetry group is
extended to ${\rm USp}(8)$. 

It is possible to cast the above product of states into a product of
fields of the 4-dimensional gauge theory. One then finds that the
spin-2 operators correspond to the energy-momentum tensor, which is
conserved ({\it i.e.} divergence-free) and traceless, so that it has
precisely the 5 independent components appropriate for spin-2. The
spin-1 operators decompose into 15 conserved vectors, associated with the
currents of ${\rm SU}(4)$, and 6 selfdual antisymmetric
tensors. The spin-0 operators are scalar composite
operators. Furthermore there are 4 chiral and 4 antichiral
vector-spinor operators, which are conserved and traceless (with
respect to a contraction with gamma matrices) such that 
each of them correspond precisely to the 4 components appropriate for
spin-$\ft32$. These are the supersymmetry currents. Finally there are
20 and 4 chiral 
and antichiral spin-$\ft12$ operators. This is precisely the
supermultiplet of currents \cite{BdRdW}, which couples to the fields
of conformal supergravity. Because neither the currents nor the
conformal supergravity fields are subject to any field equations
(unlike the supersymmetric gauge multiplet from 
which we started, which constitutes only an on-shell supermultiplet),
it forms the basis for a proper off-shell theory of $N=4$
conformal supergravity \cite{BdRdW}. The presence of the traceless and
conserved energy-momentum tensor and  supersymmetry currents, and of
the ${\rm SU}(4)$ conserved currents, is a consequence of the 
superconformal invariance of the underlying 4-dimensional gauge
theory. The $N=4$ 
conformal supergravity theory couples consistently to the $N=4$
supersymmetric gauge theory. Chapter~7 will further explain the
general setting of superconformal theories that is relevant in this
context. 

The off-shell $N=4$ conformal supergravity multiplet in 4 dimensions
can also be 
interpreted as an on-shell {\it massless} supermultiplet in 5 dimensions
with 32 supersymmetries. Because of the masslessness, the states are
annihilated by half the supercharges and are still classified
according to ${\rm SO}(3)$,  which now acts as the helicity group; the
R-symmetry group coincides with 
the ${\rm USp}(8)$ R-symmetry of the relativistic 5-dimensional
supersymmetry algebra. Hence this is the same multiplet that describes
$D=5$ maximal supergravity. This theory has a nonlinearly
realized ${\rm E}_{6(6)}$ invariance whose linearly realized subgroup
(which is relevant for the spectrum) equals ${\rm USp}(8)$. 

The latter theory can be gauged (we refer to chapter~5 for a
discussion of this) in which case it can possess an anti-de Sitter ground
state. According to table~\ref{pq-spinors}, a fully supersymmetric
ground state leads to a ${\rm U}(4)$ R-symmetry group. As we
will discuss in chapter~6, anti-de Sitter space leads to 'remarkable
representations'. These are the singletons, which do not have a smooth
Poincar\'e limit because they are associated with possible degrees of
freedom living on a 4-dimensional boundary. Because the 5-dimensional
anti-de Sitter superalgebra coincides with the 4-dimensional
superconformal algebra, the 4-dimensional boundary theory must be
consistent with  superconformal invariance. Hence it does not come as
a surprise that these singleton representations coincide with the
supermultiplet of 4-dimensional $N=4$ gauge theory. This set-up requires
the gauge group of  5-dimensional supergravity to be chosen such as to
preserve the relevant automorphism group. Therefore the gauge group
must be equal to ${\rm SO}(6)\cong {\rm SU}(4)/{\bf Z}_2$. Indeed,
this gauging allows for an anti-de Sitter maximally supersymmetric
ground state \cite{GunaRomansWarner}. Thus, the circle closes.

We stress that the above excursion, linking the various 
supermultiplets in different dimensions by a series of arguments, is 
purely based on symmetries. It does not capture the dynamical aspects
of the adS/CFT correspondence and has no bearing on the nature of the
gauge group in 4 dimensions. At this stage we thus  
have to content ourselves with the existence of this remarkable chain 
of correspondences. Many aspects of these correspondences will
reappear in later chapters. 
%%%%%%%%%%%%%%%%%%%%%%%%%%%%%%%%%%%%%%%%%%%%%%%%%%%%%%%%%%%%%%%%
%%%%%%%%%%%%%%%%%%%%%%%%%%%%%%%%%%%%%%%%%%%%%%%%%%%%%%%%%%%%%%%%

%%%%%%%%%%%%%%%%%%%%%%%%%%%%%%%%%%%%%%%%%%%%%%%%%%%%%%%%%%%%%%%%
%%%%%%%%%%%%%%%%%%%%%%%%%%%%%%%%%%%%%%%%%%%%%%%%%%%%%%%%%%%%%%%%
%\newpage %%%%%%%%%%%%%%%%%%%%%%%%%%%%%%%%%%%%%%%%%%%%%%%%%%%%%%%
%%%%%%%%%%%%%%%%%%%%%%%%%%%%%%%%%%%%%%%%%%%%%%%%%%%%%%%%%%%%%%%%
%%%%%%%%%%%%%%%%%%%%%%%%%%%%%%%%%%%%%%%%%%%%%%%%%%%%%%%%%%%%%%%%
%%%%%%%%%%%%%%%%%%%%%%%%%%%%%%%%%%%%%%%%%%%%%%%%%%%%%%%%%%%%%%%%
%%%%%%%%%%%%%%%%%%%%% CHAPTER III %%%%%%%%%%%%%%%%%%%%%%%%%%%%%%
%%%%%%%%%%%%%%%%%%%%%%%%%%%%%%%%%%%%%%%%%%%%%%%%%%%%%%%%%%%%%%%%
\section{Supergravity}
\label{supergravity}
\setcounter{equation}{0}
%%%%%%%%%%%%%%%%%%%%%%%%%%%%%%%%%%%%%%%%%%%%%%%%%%%%%%%%%%%%%%%%
In this chapter we discuss field theories that are invariant under
local supersymmetry. Because of the underlying supersymmetry algebra,
the invariance under local supersymmetry implies the invariance under
spacetime diffeomorphisms. Therefore these theories are necessarily
theories of gravity. We exhibit the 
initial steps in the construction of a supergravity theory, with and
without a cosmological term. Then we concentrate on  
maximal supergravity theories in various dimensions, their symmetries,
and dimensional compactifications on tori. At the end we briefly
discuss some of the nonmaximal theories 
%%%%%%%%%%%%%%%%%%%%%%%%%%%%%%%%%%%%%%%%%%%%%%%%%%%%%%%%%%%%%%%
\subsection{Simple supergravity}
\label{simple-supergravity}
The first steps in the construction of any supergravity theory are
usually based on the observation that local supersymmetry
implies the invariance under 
general coordinate transformation. Therefore one must introduce the
fields needed to describe general relativity, namely a vielbein field
$e_\m^{\,a}$ and a spin-connection field $\omega_\m^{\,ab}$. The
vielbein field is nonsingular and its inverse is denoted by
$e_a^{\,\m}$. The vielbein defines a local set of tangent frames
of the spacetime manifold, while the 
spin-connection field is associated with (local) Lorentz transformations 
of these frames. The world indices, $\m, \n,\ldots$, and the
tangent space indices, $a,b,\ldots$, both run from 0 to $D-1$. For an 
introduction to the vielbein formalism we refer to
\cite{Trieste-dW}. Furthermore one needs one or several gravitino 
fields, which carry both a world index and a
spinor index and which act as the gauge fields associated with local
supersymmetry. For simplicity we only consider a single Majorana gravitino
field, denoted by $\psi_\m$, but this restriction is not 
essential.\footnote{%%%%%%%%%%%%%%%%%%%%%%%%%%%%%%%%%%%%%%%%%%
   For definiteness we consider a generic supergravity theory with one
   Majorana 
   gravitino with an antisymmetric charge-conjugation matrix $C$ and 
   gamma matrices $\Gamma_a$ satisfying $C\Gamma^{~}_a C^{-1} = -
   \Gamma_a^{\rm T}$. Furthermore $\G^{~}_\m =
   e_\m^{\;a}\Gamma_a$. This is the case for $D=3,4,10,11$~mod~8. For
   $D=8,9$~mod~8, the charge-conjugation matrix is symmetric and
   $C\Gamma^{~}_a C^{-1} = \Gamma_a^{\rm T}$. For
   $D=5,6,7$~mod~8, Majorana spinors do not exist. } %%%
%%%%%%%%%%%%%%%%%%%%%%%%%%%%%%%%%%%%%%%%%%%%%%%%%%%%%%%%%%%%% 
Hence any supergravity Lagrangian is expected to contain the
Ein\-stein-Hilbert Lagrangian of general relativity 
and the Rarita-Schwinger Lagrangian for the gravitino field, 
\be
\kappa^2\,\lagr= -\ft12e \, R(\o) - \ft12e
\bar\psi_\m\,\G^{\m\n\rho}D_\n(\o) \psi_\rho + \cdots\,,
\ee 
where the covariant derivative on a spinor $\psi$ reads 
\be
D_\m(\o)\psi  = \left( \pa_\m - \ft14 \o_\m{}^{\!\!ab}\, \G_{ab}\right)\psi \,, 
\ee
and $\o_\m{}^{ab}$ is the spin-connection field whose definition will
be discussed in a sequel. The matrices $\ft12 \G_{ab} =
\ft14[\G_a,\G_b]$ are the generators of the Lorentz transformations in
spinor space, $\kappa^2$ is related to Newton's constant
and $e=\det(e_\m^{\;a})$. 
Observe that the spinor covariant derivative on $\psi_\m$ contains no
affine connection, as it should not \cite{Trieste-dW}.

We note the existence of two covariant tensors, namely the curvature
associated with the spin connection $R_{\m\n}^{\,ab}(\o)$ and the torsion
tensor $R_{\m\n}^{\,a}(P)$, which carries this name because it is
proportional to the antisymmetric part of the affine connection,
$\Gamma_{[\m\n]}^{\,\rho}$, upon using the vielbein postulate, 
\bea
R_{\m\n}^{\,ab}(\o)   &=& \pa_\m\omega^{\,ab}_\n- \pa_\n\omega^{\,ab}_\m +
\omega^{\,ac}_\m \,\omega_\n^{\,b}{}^{~}_c - \omega^{\,ac}_\n
\,\omega_\m^{\,b}{}^{~}_c \,,    \nonumber\\ 
R_{\m\n}^{\,a}(P) &=& D^{~}_\m(\omega) e_\n^{\,a} - D^{~}_\n(\omega)
e_\m^{\,a}  \,. 
\eea
We note that these tensors satisfy the Bianchi identities,
\be
D^{~}_{[\m} (\o) R_{\n\rho]}^{ab}(\o)= 0\,,  \qquad 
D^{~}_{[\m} (\o) R_{\n\rho]}^a(P) +
R_{[\m\n}^{ab}(\o)\,e^{~}_{\rho]\,b} = 0\,.  
\ee
It is suggestive to regard $e_\m^{\;a}$ and $\o_\m^{\;ab}$ as the
gauge fields of the Poincar\'e group. In that context $R(\o)$ is
written as $R(M)$, so that $P$ and $M$ denote the translation and the
Lorentz generators of the Poincar\'e algebra. We will use this
notation in later chapters when discussing the anti-de Sitter and the
conformal algebras. Here we will just use the notation $R(\o)$ and
define its contractions with the inverse vielbeine (related to the
Ricci tensor and Ricci scalar) by 
\be
R_\m^{\,a}(e,\o) = e_b^{\;\n}\, R_{\m\n}^{\,ab}(\omega) \,,\qquad 
R(e,\o) = e_a^{\;\m}e_b^{\;\n}\, R_{\m\n}^{\,ab}(\omega)\,.
\ee

The spin connection can be treated as an independent field
(first-order formalism),
which is then solved in terms of its field equations, or it can be
fixed from the beginning (second-order formalism), for instance, by
imposing the constraint, 
\be
R_{\m\n}^{\,a}(P) = 0\,. \label{eq:torsion-constraint}
\ee
Such a constraint is called `conventional' because it expresses one
field in terms of other fields in an algebraic fashion. For pure
gravity the first- and the second-order formalism lead to the
same result. The constraint \eqn{eq:torsion-constraint} can be solved
algebraically and leads to,   
\be
\o_{\m}^{\,ab}(e)= \ft12 e_\m^{\;c} (\Omega^{ab}{}_c -\Omega^b{}_c{}^a
 -\Omega_c{}^{ab})\,, \label{eq:spin-connection}
\ee
where the $\Omega_{ab}{}^c$ are the {\it objects of anholonomity},
\be
\Omega_{ab}{}^c= e_a^{\,\m}e_b^{\,\n}\,(\pa^{~}_\m e_\n^{\,c} -\pa^{~}_\m
e_\n^{\,c}) \,.
\ee
From the spin connection one defines the affine connection by
$\Gamma_{\m\n}{}^{\!\rho}= e_a^{\;\rho}\,D_\m(\o) 
e_\n^{\;a}$, which ensures the validity of the vielbein postulate. 
With the zero-torsion value \eqn{eq:spin-connection} for the spin
connection, the affine connection becomes equal to the Christoffel
symbols and $R_{\m\n\rho}{}^\s = 
R_{\m\n}^{ab}(\o)\, e^{~}_{\rho\,a}\,e_b^{\;\s}$ coincides with the
standard Riemann tensor. 

The action corresponding to the above Lagrangian is locally
supersymmetric up to terms cubic in the gravitino field. The
supersymmetry transformations contain the terms, 
\be 
\d e_\m{}^{a} = \ft12 \bar\e \,\G^a\psi_\m\,,\qquad \d\psi_\m =
D_\m(\o) \e\,, \label{flat-susy-grav} 
\ee
where the gravitino variation is the extension  to curved spacetime of
the spinor gauge invariance of a Rarita-Schwinger field. 
Extending this Lagrangian to a fully supersymmetric one is not always
possible. Usually it requires additional fields of lower spin, whose
existence can be inferred from the knowledge of the possible underlying 
(massless) supermultiplets of states. When 
the spacetime dimension exceeds eleven, conventional supergravity no
longer exists, as we shall discuss in the next section. 

Let us now include a cosmological term into the above Lagrangian as
well as a suitably chosen masslike term for the gravitino field, 
\bea
\lagr&=& -\ft12e\, R(e,\o) - \ft12 e \bar\psi_\m\,\G^{\m\n\rho}D_\n(\o)
\psi_\rho   \nonumber\\
&&+ \ft14 g(D-2) e \,\bar\psi_\m\G^{\m\n}\psi_\n + \ft12 g^2 (D-1)(D-2)
\,e + \cdots\,. \label{cosm-term-lagr}
\eea
As it turns out the corresponding action is still locally
supersymmetric, up to terms 
that are cubic in the gravitino field, provided that we introduce an
extra term to the transformation rules,  
\be 
\d e_\m{}^{a} = \ft12 \bar\e \,\G^a\psi_\m\,,\qquad \d\psi_\m =
\left(D_\m(\o) + \ft12 g \G_\m\right) \e\,. \label{ads-susy-grav} 
\ee
The Lagrangian \eqn{cosm-term-lagr} was first written down in
\cite{townsend} in four space-time dimensions and the correct
interpretation of the masslike term was given in \cite{deser-zumino}. 
Observe that the variation for $\psi_\m$ may be regarded as a
generalized covariant derivative, where $\o_\m^{ab}$ and $e_\m^{\;a}$
act as gauge fields,\footnote{%%%%%%%%%%%%%%%%%%%%%%%%%%%%%%%%%%%%%
  The masslike term in \eqn{cosm-term-lagr} is consistent with that
  interpretation as it can be generated from the Rarita-Schwinger
  Lagrangian by the same change of the covariant derivative, {\it i.e.}, 
  $$
  -\ft12 e \bar \psi_\m \G^{\m\n\rho} (D_\n(\o) + \ft12 g \G_\n)
  \psi_\rho \,.
  $$  } %%%%%%%%%%%%%%%%%%%%%%%%%%%%%%%%%%%%%%%%%%%%%%%%%%%%%%%%% 

Consistency requires that $g\G_\m \e$ satisfies the same Majorana
constraint as $\psi_\m$ and $\e$. With the conventions that we have
adopted this implies that $g$ is real. The reality of $g$ has
important consequences, as it implies that the cosmological term is of
definite sign. Hence supersymmetry does not a priori forbid a
cosmological term, but it must be  of definite sign (at least, if the
ground state is to preserve supersymmetry). This example does not
cover all cases, as one does not always have a single Majorana spinor
with the specified charge conjugation properties. Nevertheless the
conclusion that the cosmological term must have this particular sign
remains, unless one accepts `ghosts': fields whose kinetic terms are
of the wrong sign. For an early discussion, see
\cite{Ferrara,deWitZwartk} and references therein.  We should point
out that there are
situations where a cosmological term is not consistent with
supersymmetry. Assuming that the theory has an anti-de Sitter or de 
Sitter ground state, one may verify whether the
Minkowski spinors have the right dimension to enable them to live in
these spaces. For instance, a Majorana-Weyl spinor in $D=10$
spacetime dimensions has only half the number of components as
a spinor in (anti-)de Sitter space of the same
dimension. Therefore, 
simple supergravity in $D=10$ dimensions cannot possibly have 
(anti-)de Sitter ground states. Such a counting argument does not
exclude anti-de Sitter ground states in $D=11$ spacetime
dimensions, because $D=11$ Lorentz spinors can exist in anti-de
Sitter space. Here the argument may be invoked that no relevant
supersymmetric extension of the anti-de Sitter algebra exists beyond
$D=7$ dimensions \cite{nahm}, but there are also explicit studies
ruling out supersymmetric cosmological terms in 11 dimensions
\cite{deser}.  

The Einstein equation corresponding to \eqn{cosm-term-lagr} reads
(suppressing the gravitino field),
\be 
R_{\m\n} -\ft12 g_{\m\n}\,R + \ft12 g^2(D-1)(D-2)\, g_{\m\n} =0\,,
\ee
which implies, 
\be 
\label{Einstein-ads}
R_{\m\n} = g^2 (D-1)\,g_{\m\n} \,,\qquad R= g^2 D(D-1)\,. 
\ee
Hence we are dealing with a $D$-dimensional  Einstein
space. The maximally symmetric 
solution of this equation is an anti-de Sitter 
space, whose Riemann curvature equals 
\be
R_{\m\n}{}^{\!ab} = 2 g^2 \, e_\m{}^{\![a}\,e_\n{}^{b]}\,.
\label{ads-curv}
\ee
This solution leaves all supersymmetries intact just as flat
Minkowski space does. One can verify this directly by considering the
supersymmetry variation of the gravitino field and by requiring that
it vanishes in the bosonic background. This happens for spinors
$\e(x)$ satisfying  
\be
\left(D_\m(\o) + \ft12 g \G_\m\right) \e=0\,. \label{killing-spinor}
\ee
Spinors satisfying this equation are called Killing spinors. Since
\eqn{killing-spinor} is a first-order differential equation, one
expects that it can be solved provided some integrability condition is
satisfied. To see 
this one notes that also $(D_\m(\o) + \ft12 g \G_\m)(D_\n(\o) + \ft12 g
\G_\n)\e$ must vanish. Antisymmetrizing this expression in $\m$ and
$\n$ then yields the (algebraic) integrability condition, 
\be 
\Big (-\ft14 R_{\m\n}{}^{\!ab} \,\G_{ab} + \ft12 g^2 \,\G_{\m\n} \Big)
\e = 0\,.
\ee
Multiplication with $\G^\n$ yields,
\be
\Big(R_{\m\n} -g^2(D-1)\,g_{\m\n}\Big) \G^\n\e =0\,,
\ee
from which one derives that the Riemann tensor satisfies
\eqn{Einstein-ads}. Therefore supersymmetry requires an Einstein
space. Requiring full supersymmetry, so that \eqn{killing-spinor} holds
for any spinor $\e$, implies \eqn{ads-curv} so that the spinor $\e$
must live in anti-de Sitter space. 

Hence we have seen that supersymmetry can be realized in anti-de
Sitter space. We will return to this issue later in
chapter~\ref{susy-ads}, where we discuss the (super)multiplet
structure in anti-de Sitter space. We stress once more that, in this
section, we have restricted ourselves to the graviton-gravitino
sector. To construct the full theory usually requires more fields and 
important restrictions arise on the dimensionality of spacetime. For
instance, while minimal supergravity in $D=4$ dimensions does not require
additional fields, in $D=11$ dimensions an additional
antisymmetric gauge field is necessary. The need for certain extra
fields can be readily deduced from the underlying massless
supermultiplets, which were extensively discussed in the previous 
chapter. 
%%%%%%%%%%%%%%%%%%%%%%%%%%%%%%%%%%%%%%%%%%%%%%%%%%%%%%%%%%%%%%%%%
 \subsection{Maximal supersymmetry and supergravity} % : $Q\leq32$}
%%%%%%%%%%%%%%%%%%%%%%%%%%%%%%%%%%%%%%%%%%%%%%%%%%%%%%%%%%%%%%%%%
In chapter~2 we restricted ourselves to supermultiplets 
based on $Q\leq32$ supercharge components. {From} the general analysis  
it is clear that increasing the number of supercharges leads to 
higher and higher helicity representations. For instance, the maximal
helicity, $|\lambda_{\rm max}|$, of a massless supermultiplet in $D=4$
spacetime dimensions is larger than or equal to $\ft
1{16}Q$. Therefore, when $Q>8$ we have 
$|\lambda_{\rm max}|\geq1$, so that theories for these multiplets must
include vector gauge fields. When $Q>16$ we have 
$|\lambda_{\rm max}|\geq \ft32$, so that the theory should contain
Rarita-Schwinger fields. In view of the supersymmetry algebra an
interacting supersymmetric theory of this type should contain gravity,
so that in this case we must include $\lambda=2$ states for the
graviton. Beyond $Q=32$ one is dealing with states of helicity
$\lambda>2$. Those are described by gauge fields that are {\it
symmetric} Lorentz tensors.  Symmetric tensor gauge fields for 
arbitrary helicity states can be constructed (in $D=4$ dimensions, 
see, for instance,  \cite{fronsdal}). However, it turns out that 
symmetric gauge fields cannot consistently couple, neither to
themselves nor to other fields. An exception is the graviton field,
which can interact with itself as well as with low-spin matter, but
not with other fields of the same spin \cite{DesAr}. By consistent,
we mean that the respective gauge invariances of the higher-spin
fields (or appropriate deformations thereof) cannot be  
preserved at the interacting level. 
Most of the search for interacting higher-spin fields was 
performed in 4 spacetime dimensions \cite{higher-spin}, but in higher
dimensional spacetimes one expects to arrive at the same 
conclusions, because otherwise, upon dimensional reduction, 
these theories would give rise to theories that are consistent 
in $D=4$. There is also direct  
evidence in $D=3$, where graviton and gravitini fields do not 
describe dynamic degrees of freedom. Hence, one can write down 
supergravity theories based on a graviton field and an arbitrary 
number of gravitino fields, which are topological. However, when 
coupling matter to this theory in the form of scalars and spinors, 
the theory does not support more than 32 supercharges. 
Beyond $Q=16$ there are four unique theories with $Q=18,20, 
24$~and~32 \cite{DWTN}. 

Hence the conclusion is that there is a restriction on the
number $Q$ of independent supersymmetries, as for $Q>32$ no
interacting field theories seem to exist. 
There have been many efforts to circumvent this bound of $Q=32$
supersymmetries. It seems clear that one needs a combination of the 
following ingredients in order to do this (for a review, 
see {\it e.g.} \cite{vasiliev}): (i) an  
infinite tower of higher-spin gauge fields; (ii) interactions 
that are inversely proportional to the cosmological constant; 
(iii) extensions of the super-Poincar\'e or the super-de~Sitter 
algebra with additional fermionic and bosonic charges. Indeed explicit
theories have been constructed which demonstrate this. 
However, conventional supergravity theories are {\it not} of 
this kind. This is the reason why we avoided ({\it i.e.} in 
table~\ref{D6-susy-mult}) to  
list supermultiplets with states transforming in higher-helicity 
representations. The fact that an infinite number of fields can cure
certain inconsistencies is by itself not new. While a massive spin-2
field cannot be coupled to gravity, the coupling of an infinite number
of them can be consistent, as can be seen in Kaluza-Klein theory. 

 %%%%%%%%%%%%%%%%%%%%%%%%%%%%%%%%%%%%%%%%%%%%%
\begin{table}
\begin{center}
\begin{tabular}{l l l l l l l l }\hline
$D$ &${\rm H}_{\rm R}$& graviton & $p=-1$ & $p=0$ & $p=1$ & $p=2$ & 
$p=3\!$  \\   \hline
11  & 1      & 1  & 0   & 0   & 0  & 1  & 0 \\[.5mm]
10A$\!\!\!$ & 1      & 1  & 1   & 1   & 1  & 1  & 0 \\[.5mm]
10B$\!\!\!$ & SO(2)  & 1  & 2   & 0   & 2  & 0  & $1^\ast$ \\[.5mm]
9   & SO(2)  & 1  &$2+1$&$2+1$& 2  & 1  & ~ \\[.5mm]
8   & U(2)   & 1  & $5+1+\bar 1\!$ & $3+\bar 3$ &3 &$[1]$&~\\[.5mm]
7   & USp(4) & 1  & 14  & 10  & 5  & ~  & ~ \\[.5mm]
6   &USp$(4)$& 1  &(5,5)&(4,4)&$(5,1)$ & ~ & ~\\
~   & $\times{\rm USp}(4)$&~&~&~&$+(1,5)$& ~ & ~ \\[.5mm]
5   & USp(8) & 1   & 42 & 27   & ~  & ~  & ~ \\[.5mm]
4   & U(8)   & 1   & $35+\overline{35}$ & $[28]$ & ~ & ~ & ~ \\[.5mm]
3   & SO(16) & 1   & 128 & ~   & ~  & ~  & ~ \\ \hline
\end{tabular}
\end{center}
\caption{\small
Bosonic field content for maximal supergravities. The 
$p=3$ gauge field in $D=10$B has a self-dual field strength. The 
representations [1] and [28] (in $D=8,4$, respectively) are 
extended to U(1) and SU(8) representations through duality transformations 
on the field strengths. These transformations can not be 
represented on the vector potentials. In $D=3$ dimensions, the 
graviton does not describe propagating degrees of freedom. For $p>0$
the fields can be assigned to representations of a bigger group than
${\rm H}_{\rm R}$. This will be discussed in due course.
}\label{maximal-sg-bosons} 
\end{table}
%%%%%%%%%%%%%%%%%%%%%%%%%%%%%%%%%%%%%%%%%%%%%%%%%%%%%%%%%%%%%%
%
%%%%%%%%%%%%%%%%%%%%%%%%%%%%%%%%%%%%%%%%%%%%%%%%%%%%%%%%%%%%%%

In this chapter we review the  maximal supergravities in various 
dimensions. These theories have $Q=32$ supersymmetries and we restrict
our discussion to $3\leq D\leq 11$.  
The highest dimension $D=11$ is motivated by the fact that spinors
have more that 32 components in flat Minkowski space for spacetime
dimensions $D>11$. Observe, however, that this argument assumes 
$D$-dimensional Lorentz invariance. As was stressed in
\cite{strings00,istanbul}, there are scenarios based on
spacetime dimensions higher than $D=11$, where the extra dimensions
can not uniformly decompactify so that the no-go theorem is
avoided. The fact that no uniform decompactification is possible
is closely related to the T-duality between winding
and momentum states that one knows from string theory.

The bosonic fields always comprise the metric tensor for the 
graviton and a number of $(p+1)$-rank antisymmetric gauge fields. For 
the antisymmetric gauge fields, it is a priori unclear whether to 
choose a $(p+1)$-rank gauge field or its dual $(D-3-p)$-rank  
partner, but it turns out that the interactions often prefer  
the rank of the gauge field to be as small as possible. 
Therefore, in table~\ref{maximal-sg-bosons}, 
we restrict ourselves to $p\leq 3$, as in $D=11$ dimensions, 
$p=3$ and $p=4$ are each other's dual conjugates. This table presents 
all the field configurations for maximal supergravity in various 
dimensions. Obviously, the problematic higher-spin fields are 
avoided, because the only symmetric gauge field is the one 
describing  the graviton. In table~\ref{maximal-sg-fermions} we also 
present the fermionic fields, always consisting of gravitini and 
simple spinors. All these fields are classified as 
representations of the R-symmetry group ${\rm H}_{\rm R}$. Note that
the simplest versions of supergravity (which depend on no other
coupling constant than Newton's constant) are  
manifestly invariant under ${\rm H}_{\rm R}$. Actually, as we will 
explain in a sequel, the maximal supergravity theories have 
symmetry groups that are much larger than ${\rm H}_{\rm R}$.

%%%%%%%%%%%%%%%%%%%%%%%%%%%%%%%%%%%%%%%%%%%%%%%%%%%%%%%%%
%%%%%%%%%%%%%%%%%%%%%%%%%%%%%%%%%%%%%%%%%%%%%%%%%%%%%%%%%
\begin{table}
\begin{center}
\begin{tabular}{l l l l}\hline
$D$ &${\rm H}_{\rm R}$& gravitini & spinors   \\ \hline
11  & 1      &  1  & 0    \\[.5mm]
10A & 1      & 1+1 & 1+1    \\[.5mm]
10B & SO(2)  &  2  & 2    \\[.5mm]
9   & SO(2)  &  2  &$2+2$ \\[.5mm]
8   & U(2)   &  $2+\bar 2$  & $2+\bar 2+ 4 +\bar 4$  \\[.5mm]
7   & USp(4) &  4  & 16  \\[.5mm]
6   & USp$(4)\!\times\! {\rm USp}(4)$  & $(4,1)+(1,4)$ 
   & $(4,5) + (5,4)$  \\[.5mm]
5   & USp(8) & 8   & 48  \\[.5mm]
4   & U(8)   & $8+\bar 8$   & $56 +\overline{56}$ \\[.5mm]  
3   & SO(16) & 16  & 128 \\ \hline
\end{tabular}
\end{center}
\caption{\small 
Fermionic field content for maximal supergravities. For 
$D=5,6,7$ the fermion fields are counted as symplectic Majorana 
spinors. For $D=4,8$ we include both chiral and antichiral spinor 
components, which transform in conjugate representations of 
$H_{\rm R}$. In $D=3$ dimensions the gravitino does not describe
propagating degrees of freedom. }\label{maximal-sg-fermions} 
\end{table}
%%%%%%%%%%%%%%%%%%%%%%%%%%%%%%%%%%%%%%%%%%%%%%%%%%%%%%%%%%
%
 \subsection{$D=11$ Supergravity}
Supergravity in 11 spacetime dimensions is based on an ``elfbein" 
field $E_M^{\;A}$, a Majorana gravitino field $\Psi_M$  
and a 3-rank antisymmetric gauge field $C_{MNP}$. With chiral 
(2,0) supergravity in 6 dimensions, it is the 
only $Q\geq16$ supergravity theory without a scalar field. Its 
Lagrangian can be written as follows \cite{cjs},  
\bea\label{D11-lagrangian}
{\cal L}_{11}\!\!\!\!&=&\!\!\!\! {1\over \kappa_{11}^2} \bigg[ 
-\ft12 E\, R(E,\Omega)  
-\ft12 E\bar\Psi_M\G^{MNP}D_N(\Omega)\Psi_P -\ft1{48}E 
(F_{MNPQ})^2 \nonumber\\   
&& \hspace{6mm}- \ft1{3456} \sqrt{2} \, \varepsilon^{MNPQRSTUVWX} 
\,F_{MNPQ} \,F_{RSTU} \,C_{VWX}  \\
&&\hspace{6mm} - \ft1{192}\sqrt{2} E \Big(\bar\Psi_R \G^{MNPQRS} 
\Psi_S + 12 \,\bar\Psi^M  \G^{NP} \Psi^Q\Big) F_{MNPQ} + 
\cdots\bigg] \,, \nonumber
\eea
where the ellipses denote terms of order $\Psi^4$, $E= \det 
E_M^{\;A}$ and $\Omega_M{}^{\!AB}$ denotes the spin connection. 
The supersymmetry transformations are  
\bea
\d E^{\;A}_M   &\!=\!& \ft12 \,\bar \e\,\G^A\Psi_M\,,\nonumber \\
\d C_{MNP} &\!=\!& -\ft18\sqrt{2} \,\bar \e\,\G_{[MN}\Psi_{P]} \,,
 \\
\d\Psi_M   &\!=\!&  D_M(\hat\Omega)\,\e + \ft1{288} \sqrt{2} \,
\Big(\G_M{}^{\!NPQR}- 8\, \d_M^{\,N}\, \G^{PQR} \Big)\,\e\, \hat 
F_{NPQR}\,.  \nonumber 
\eea
Here the derivative $D_M$ is covariant with respect to local 
Lorentz transformations, 
\be
D_M(\Omega)\,\e = \Big(\pa_M -\ft14\Omega_M{}^{AB} \G_{AB}\Big) \e\, ,
\ee
and $\hat F_{MNPQ}$ is the supercovariant field strength
\be
\hat F_{MNPQ} = 24\, \pa_{[M}C_{NPQ]} + \ft32 \sqrt{2} \,\bar 
\Psi_{[M}\G_{NP}\Psi_{Q]} \,.
\ee
The supercovariant spin connection is the 
solution of the following equation,
\be
D_{[M} (\hat\Omega) \,E_{N]}^{\;A} - \ft14 \bar \Psi_M \G^A\Psi_N 
=0\,.
\ee
The left-hand side is the supercovariant torsion tensor.  

Note the presence of a Chern-Simons-like term  
$F\wedge F \wedge C$ in the Lagrangian, so that the action is only
invariant under tensor gauge transformations up to 
surface terms. We also wish to point out that the 
quartic-$\Psi$ terms can be included into the Lagrangian 
\eqn{D11-lagrangian} by replacing the spin-connection field 
$\Omega$ by $(\Omega+\hat \Omega)/2$ in the 
covariant derivative of the gravitino kinetic term and by 
replacing $F_{MNPQ}$ in the last line by $(\hat 
F_{MNPQ}+F_{MNPQ})/2$. These substitutions ensure that the field 
equations corresponding to \eqn{D11-lagrangian} are 
supercovariant.  The Lagrangian is derived in the context of the 
so-called ``1.5-order'' formalism, in which the spin connection is 
defined as a dependent field determined by its (algebraic) 
equation of motion, whereas its supersymmetry variation in the 
action is treated as if it were an independent field 
\cite{1.5-order}. 

We have the following bosonic field equations and Bianchi identities, 
\bea
\label{D11-field-eq}
R_{MN} &=& \ft1{72}g_{MN} \, F_{PQRS}F^{PQRS} -\ft16 
F_{MPQR}\,F_N{}^{\!PQR}\,, \nonumber\\ 
\pa_{M}\Big(E\,  F^{MNPQ}\Big) &=& \ft1{1152} \sqrt{2}\, 
\varepsilon^{NPQRSTUVWXY} F_{RSTU}\,F_{VWXY}\,, \nonumber \\
\pa_{[M}F_{NPQR]}&=&0\,,
\eea
which no longer depend explicitly on the antisymmetric gauge field. 
An alternative form of the second equation is \cite{page}
\be\label{D11-field-eq2}
\pa_{[M}H_{NPQRSTU]} =0\,,
\ee
where $H_{MNPQRST}$ is the dual field strength,
\be
H_{MNPQRST} = {1\over 7!}E\, \varepsilon_{MNPQRSTUVWX} F^{UVWX} 
-\ft12 \sqrt{2}  \, F_{[MNPQ}\,C_{RST]}\,.
\ee
One could imagine that the third equation of \eqn{D11-field-eq} and  
\eqn{D11-field-eq2} receive contributions from charges that would 
give rise to source terms on the right-hand side of the equations. 
These charges are associated with the `flux'-integral of $H_{MNPQRST}$ 
and $F_{MNPQ}$ over the boundary of an 8- and a 5-dimensional 
spatial volume, respectively. In analogy with the Maxwell theory, the
integral $\oint H$ may be associated with electric flux and the
integral $\oint F$ with magnetic flux. The spatial 
volumes are orthogonal to a $p=2$ and a $p=5$ brane configuration,
respectively, and the corresponding charges are 
2- and 5-rank Lorentz tensors. These are just the charges that can
appear as  central charges in the supersymmetry algebra
\eqn{eq:11D-algebra}. Solutions of 
11-dimensional supergravity that contribute to these charges were 
considered in \cite{duffstelle,gueven,solirev}. 

Finally, the constant $1/\kappa_{11}^2$ in front of the Lagrangian 
\eqn{D11-lagrangian}, which carries dimension $[{\rm 
length}]^{-9}\sim [{\rm mass}]^9$, is 
undetermined and depends on fixing some length scale.
To see this consider a continuous  rescaling of the fields, 
\be  
E_M^{\;A}\to {\rm e}^{-\a} E_M^{\;A}\,, \qquad \Psi_M\to {\rm 
e}^{-\a/2}\Psi_M\,, \qquad C_{MNP} \to {\rm e}^{-3\a}C_{MNP} \,.  
\label{scale-11}
\ee
Under this rescaling the Lagrangian changes 
according to
\be
{\cal L}_{11} \to {\rm e}^{-9\a}{\cal L}_{11}\,.
\ee
This change can then be absorbed into a redefinition of 
$\kappa_{11}$,\footnote{%
  Note that the rescalings also leave the supersymmetry 
  transformation rules unchanged, provided the supersymmetry 
  parameter $\e$ is changed accordingly.} 
\be \label{scale2-11}
\kappa_{11}^2\to {\rm e}^{-9\a} \kappa_{11}^2\,.
\ee
This simply means that the Lagrangian depends on only one dimensional
coupling constant, namely $\kappa_{11}$. The same situation is present
in many other supergravity theories. Concentrating on the
Einstein-Hilbert action in $D$ spacetime dimensions, the corresponding  
scaling property is 
\be\label{scaleD}
g_{\m\n}^D \to {\rm e}^{-2\a}g_{\m\n}^D\, ,\qquad
{\cal L}_{D} \to {\rm e}^{(2-D)\a}{\cal L}_{D}\,,\qquad
\kappa_{D}^2\to {\rm e}^{(2-D)\a} \kappa_{D}^2\,.
\ee
Of course, this implies that the physical value of Newton's constant, 
does not necessarily coincide with the parameter 
$\kappa_{D}^2$ in the Lagrangian but it also depends on the precise
value  adopted for the (flat) metric in the ground state of the theory. 

%%%%%%%%%%%%%%%%%%%%%%%%%%%%%%%%%%%%%%%%%%%%%%%%%%%%%%%%%%%%%%%%
\subsection{Dimensional reduction and hidden symmetries}
%%%%%%%%%%%%%%%%%%%%%%%%%%%%%%%%%%%%%%%%%%%%%%%%%%%%%%%%%%%%%%%%
The maximal supergravities in various dimensions are related by 
dimensional reduction. In this reduction some of the spatial
dimensions are  compactified on a hypertorus and one retains only the
fields that do not depend on the torus coordinates. This corresponds
to the theory one obtains when the size of the torus is shrunk to
zero. A subset of the gauge symmetries associated with the
compactified dimensions survive as internal symmetries. The aim
of the present discussion here is to elucidate a number of features
related to these symmetries, mainly in the context of the reduction
of $D=11$ supergravity to $D=10$  dimensions. 

We denote the compactified coordinate by $x^{10}$ which now 
parameterizes a circle of length $L$.\footnote{%
   Throughout these lectures we enumerate spacetime coordinates 
   by $0,1,\ldots, D-1$. } %  
The fields are thus decomposed in a Fourier series as periodic
functions in $x^{10}$ on the interval $0\leq  x^{10}\leq L$. This
results in a spectrum of massless modes and an infinite tower of
massive modes with masses inversely proportional to the circle length
$L$. The massless modes form the basis of the 
lower-dimensional supergravity theory. Because a toroidal background
does not break supersymmetry, the resulting supergravity has  
the same number of supersymmetries as the original one. For 
compactifications on less trivial spaces than the hypertorus, 
this is usually not the case and the number of independent
supersymmetries will be reduced. Fully supersymmetric
compactifications are rare. For instance,  11-dimensional
supergravity can be compactified to a 4-dimensional maximally
symmetric spacetime in only two ways such that all
supersymmetries remain unaffected \cite{biran}. One is the
compactification on a torus $T^7$, the other one the compactification
on a sphere $S^7$. In the latter case the resulting 4-dimensional 
supergravity theory acquires a cosmological term. 

In the formulation of the compactified theory, it is important to 
decompose the higher-dimensional fields in such a way that they 
transform covariantly under the lower-dimensional gauge 
symmetries and under diffeomorphisms of the 
lower-dimensional spacetime. This ensures that various  
complicated mixtures of massless modes with the tower of massive 
modes will be avoided. It is a key element in ensuring that 
solutions of the lower-dimensional theory remain solutions of the 
original higher-dimensional one, which is an obvious requirement for
having consistent truncations to the massless states. Another point of
interest concerns the  
nature of the massive supermultiplets. Because these originate 
from supermultiplets that are massless in higher dimensions, 
they are 1/2-BPS multiplets which are  shortened by
the presence of central charges corresponding to the momenta
in the compactified dimension. Implications of these BPS
supermultiplets will be discussed in more detail in
section~\ref{BPSextended}. 

The emergence of new internal symmetries in theories that 
originate from a higher-dimensional setting, is a standard feature of 
Kaluza-Klein theories \cite{KK}. Following the discussion in \cite{DWVVP} 
we distinguish between symmetries that have a direct explanation 
in terms of the higher-dimensional  symmetries, and symmetries 
whose origin is obscure from a higher-dimensional viewpoint. 
Let us start with the symmetries associated with the metric  
tensor. The 11-dimensional metric can be decomposed according 
to 
\be \label{KKmetric}
{\rm d}s^2 = g_{\m\n} \,{\rm d}x^\m{\rm d}x^\n  + {\rm 
e}^{4\phi/3} 
({\rm d}x^{10}+ V_\m{\rm d}x^\m)({\rm d}x^{10}+ V_\n{\rm 
d}x^\n)\,,
\ee
where the indices $\mu,\nu$ label the 10-dimensional coordinates 
and the factor multiplying $\phi$ is for convenience later on. 
The massless modes correspond to the $x^{10}$-independent parts of the 
10-dimensional metric $g_{\m\n}$, the vector field $V_\m$ and 
the scalar $\phi$. 
Here the $x^{10}$-independent component of $V_\m$ acts as a gauge 
field associated with reparametrizations of the circle coordinate 
$x^{10}$ with an arbitrary function $\xi(x)$ of the 10 remaining 
spacetime coordinates $x^\mu$. Specifically, we have $x^{10} \to 
x^{10} - \xi(x)$ and $x^\m\to x^\m$, leading to 
\be 
V_\m(x)\to V_\m(x)+ \pa_\m \xi(x)\,.
\ee
The massive modes, which correspond to the nontrivial Fourier modes in
$x^{10}$, couple to this gauge field with a charge that is a  
multiple of 
\be 
e_{\scriptscriptstyle\rm KK} = {2\pi\over L}\,. \label{KK-charge}
\ee

Another symmetry of the lower-dimensional theory is more subtle 
to identify.\footnote{%
  There are various discussions of this symmetry in the 
  literature. Its existence in 10-dimensional supergravity was 
  noted long ago (see, e.g. \cite{cremmer2,cremmer3}) and an extensive 
  discussion can be found in \cite{berg}. Our derivation here was 
  alluded to in \cite{DWVVP}, which deals with isometries in 
  $N=2$ supersymmetric Maxwell-Einstein theories in $D=5,4$ 
  and 3 dimensions.} %
In the previous subsection we noted the existence of certain 
scale transformations of the $D=11$ fields, which did not leave 
the theory invariant but could be used to adjust the coupling 
constant $\kappa_{10}$. In the compactified situation we can also 
involve the compactification length into the dimensional 
scaling. The integration over $x^{11}$ introduces an overall 
factor $L$ in the action (we do not incorporate any $L$-dependent 
normalizations in the Fourier sums, so that the 10-dimensional 
and the 11-dimensional fields are directly proportional). Therefore, 
the coupling constant that emerges in the 10-dimensional theory 
equals
\be\label{def-L}
{1\over \kappa_{10}^2}= {L\over \kappa_{11}^2}\,,
\ee
and is of dimension $[{\rm mass}]^8$. 
However, because of the invariance under diffeomorphisms, $L$ 
itself has no intrinsic meaning. It simply expresses the length of the 
$x^{10}$-periodicity interval, which depends on the
coordinatization. Stated differently, we can  
reparameterize $x^{10}$ by some diffeomorphism, as long as we 
change $L$ accordingly. In particular, we may rescale $L$ according to
\be
L \to {\rm e}^{-9\a} L \,,
\ee
corresponding to a reparametrization of the 11-th coordinate,
\be
x^{10}\to {\rm e}^{-9\a} x^{10} \,,\label{diff-11}
\ee 
so that $\kappa_{10}$ remains invariant. Consequently we are then
dealing with a {\it symmetry} of the Lagrangian.  

In the effective 10-dimensional theory, the scale transformations 
\eqn{scale-11} are thus suitably combined with the diffeomorphism 
\eqn{diff-11} to yield an invariance of the Lagrangian. For the 
fields corresponding to the 11-dimensional metric, these 
combined transformations are given by\footnote{%%%%%%%%%%%%%%%%%%%%%%
   Note that these transformations  apply uniformly to all Fourier
   modes, as those depend on  $x^{10}/L$ which is insensitive to the scale
   transformation. This does not imply that the Lagrangian remains 
   invariant when retaining the higher Fourier modes, because the
   Kaluza-Klein charges \eqn{KK-charge} depend explicitly on $L$. This
   issue will be relevant in section~\ref{BPSextended}.} %%%%%%%%%%%%%
%%%%%%%%%%%%%%%%%%%%%%%%%%%%%%%%%%%%%%%%%%%%%%%%%%%%%%%%%%%%%%
\be \label{scale-10x}
e_\m^a \to  {\rm e}^{-\a}e_\m^a \,,\qquad \phi \to  \phi + 12\a 
\,,\qquad V_\m\to {\rm e}^{-9\a} V_\m \,.
\ee
The tensor gauge field $C_{MNP}$ decomposes into a 3- and a 
2-rank tensor in 10 dimensions, which transform according to
\be \label{scale-10y}
C_{\m\n\rho}\to {\rm e}^{-3\a} C_{\m\n\rho}\,, \qquad 
C_{11\m\n} \to  {\rm e}^{6\a}  C_{11\m\n}\,.
\ee

The presence of the above scale symmetry is confirmed by the 
resulting 10-dimensional Lagrangian for the massless  
({\it i.e.}, $x^{10}$-independent) modes. Its purely bosonic terms
read 
\bea \label{D10-lagrangian}
{\cal L}_{10}&\!\!=\!\!& {1\over \kappa_{10}^2} \bigg[ -\ft12 e\,{\rm 
e}^{2\phi/3} R(e,\omega)  -\ft18 e\,{\rm e}^{2\phi}(\pa_\m 
V_\n-\pa_\n V_\m)^2 \\ 
&&\hspace{9.5mm} -\ft1{48}e\,{\rm e}^{2\phi/3}(F_{\m\n\rho\s})^2  
 -\ft34 e\, {\rm e}^{- 2\phi/3}(H_{\m\n\rho})^2 \nonumber  \\
&& \hspace{9.5mm}+ \ft1{1152} \sqrt{2}\,
\varepsilon^{\m_1
%\m_2\m_3\m_4\m_5\m_6\m_7\m_8\m_9
- \m_{10}} 
\,C_{11\m_1\m_2} \,F_{\m_3\m_4\m_5\m_6} \,
F_{\m_7\m_8\m_9\m_{10}}\bigg] \,, \nonumber 
\eea
where $H_{\m\n\rho}= 6\,\pa_{[\m}C_{\n\rho]11}$ is the field 
strength tensor belonging to the 2-rank tensor gauge field. 

The above example exhibits many of the characteristic 
features of dimensional reduction and of the symmetries that emerge as
a result. When reducing to lower dimension one can follow the same
procedure a number of times, consecutively reducing the dimension by
unit steps, or one can reduce at once to lower dimensions.
Before continuing our general discussion, let us briefly discuss an
example of the latter based on 
gravity coupled to an antisymmetric tensor gauge field in $D+n$
spacetime dimensions,  
\be 
\label{gravity-tensor}
\lagr \propto -\ft12  E\, R - \ft94 E\, (\pa_{[M} B_{NP]})^2\,.
\ee
After compactification on a torus $T^n$, the fields that are
independent of the torus coordinates remain massless 
fields in $D$ dimensions: the
graviton, one tensor gauge field, $2n$ abelian vector gauge fields,
and $n^2$ scalar fields. The scalar fields originate from the metric
and the antisymmetric gauge field with both indices taking values in
$T^n$, so that they are parametrized by a symmetric tensor
$g_{ij}$ and an antisymmetric tensor $B_{ij}$. The diffeomorphisms
acting on the torus coordinates $x^i$ which are linear in $x^i$, {\it
i.e.}, $x^i\to {\sf O}^i{}_j\,x^j$, act on $g_{ij}$ and $B_{ij}$
according to $g\to {\sf O}^{\rm T} g \,{\sf O}$ and $B\to {\sf O}^{\rm
T} B\, {\sf O}$.  The matrices $\sf O$ generate the group ${\rm
GL}(n)$, which can be regarded as a
generalization of the scale transformations \eqn{scale-10x} and
\eqn{scale-10y}. The group ${\rm GL}(n)$ contains the rotation group
${\rm SO}(n)$; its remaining part depends on
$\ft12n(n+1)$ parameters, exactly equal to the number of independent
fields $g_{ij}$. Special tensor gauge
transformations with parameters proportional to $\Lambda_{ij}\,x^j$
induce a shift of the massless scalars $B_{ij}$ proportional to the
constants $\Lambda_{[ij]}$. There are thus $\ft12n(n-1)$ independent shift
transformations, so that, in total,  we have now identified $\ft12n(3n-1)$
isometries, which act transitively on the manifold ({\it i.e.} they
leave no point on the manifold invariant). Therefore the manifold is 
homogeneous (for a discussion of such manifolds, see 
section~\ref{sec:nonl-real-symm}). However, it turns out that there
exist $\ft12n(n-1)$ additional isometries, whose origin is {\it not}
directly related to the higher-dimensional context, and which
combine with the 
previous ones to generate the group ${\rm SO}(n,n)$. The homogeneous
space can then be identified as the coset space 
${\rm SO}(n,n)/({\rm SO}(n)\times{\rm SO}(n))$. 

According to the above, 11-dimensional supergravity reduced on a
hypertorus thus leads to a Lagrangian for the massless sector in
lower dimensions (the massive sector is discussed in
section~\ref{BPSextended}), which exhibits a number of invariances 
that find their origin in the diffeomorphisms and gauge
transformations related to the torus coordinates. As already
explained, one must properly account for the periodicity intervals of
the torus coordinates $x^i$, but the action for the massless fields
remains invariant under continuous ${\rm GL}(n)$ transformations. 
Furthermore, all the scalars that emerge from dimensional
reduction of gauge fields are subject to constant shift
transformations. These scalars and the scalars originating 
from the metric transform transitively under the isometry group.
Since 11-dimensional supergravity has itself no scalar fields, the
rank of the resulting symmetry group in lower dimensions is equal to
the rank of ${\rm GL}(n)$, and thus to the number of compactified
dimensions, {\it i.e.}, ${\bf r}= 11-D$, where $D$ is the spacetime
dimension to which 
we reduce. In general these extra symmetries are not necessarily
symmetries of the full action. In even dimensions, the symmetries may
not leave the Lagrangian, but only the field  equations,
invariant. The reason for this is that the isometries may act by means 
of duality transformations on field 
strengths associated with antisymmetric tensor gauge fields of
rank $\ft12 D-1$ which cannot be implemented on the gauge fields  
themselves. In 4 dimensions this phenomenon is known as electric-magnetic
duality (for a recent review, see \cite{susy30}); for $D=6$ we
refer to \cite{Tanii}. For supergravity, it
is easy to see that the scalar manifold (as well as the rest of the 
theory) possesses additional symmetries beyond the ones
that follow from higher dimensions, because the latter do not yet 
incorporate the full R-symmetry group of the underlying  
supermultiplet. We expect  
that ${\rm H}_{\rm R}$ is also realized as a symmetry, because the maximal 
supergravity theory that one obtains from compactification on a
hypertorus has no additional coupling constants (beyond Newton's
constant) which could induce R-symmetry breaking. Therefore we
expect that the target space for the scalar fields is an homogeneous
space, with an isometry group whose generators 
belong to a solvable subalgebra associated with the shift transformations,
to the subalgebra of ${\rm GL}(11-D)$ scale  
transformations and/or to the subalgebra associated with 
${\rm H}_{\rm R}$. Of course, these subalgebras will partly
overlap. Usually a counting argument (of the type first used in 
\cite{cremmer}) then readily indicates what the structure is of the
corresponding homogeneous space that is parametrized by the scalar
fields. In  
table~\ref{maximal-cosets} we list the isometry group G and the
isotropy group ${\rm H}_{\rm R}$ of  
these scalar manifolds for maximal supergravity in dimensions $3\leq 
D\leq 11$. Earlier versions of such tables can, for instance,  be 
found in \cite{cremmer2,cremmer3}. 

A more recent discussion of 
these isometry groups from the perspectives of string theory and
M-theory can be found in, for example,
\cite{hull,ObersPioline}. Here we merely stress a number of
characteristic features of the group 
${\rm G}$. One of them is that ${\rm H}_{\rm R}$ is always the maximal
compact subgroup of $\rm G$. As we mentioned already, another
(noncompact) subgroup is the group 
${\rm GL}{(11-D)}$,  associated with the 
reduction on an $(11-D)$-dimensional torus. Yet another subgroup is
${\rm SL}(2)\times {\rm SO}(n,n)$, where $n= 10-D$. This group, which
emerges for $D<10$  can be understood within the string perspective;
${\rm SL}(2)$ is the S-duality group and ${\rm SO}(n,n)$ is the
T-duality group. It also follows from the toroidal compactifications
of IIB supergravity, which has a manifest ${\rm SL}(2)$ in $D=10$
dimensions. The group ${\rm SO}(n,n)$ is associated with the
invariance of toroidal compactifications that involve the metric and an
antisymmetric tensor field ({\it c.f.} \eqn{gravity-tensor}). 

%%%%%%%%%%%%%%%%%%%%%%%%%%%%%%%%%%%%%%%%%%%%%%%%%%%%%%%%%%%%%%%%%%
%%%%%%%%%%%%%%%%%%%%%%%%%%%%%%%%%%%%%%%%%%%%%%%%%%%%%%%%%%%%%%%%%%%
\begin{table}
\begin{center}
\begin{tabular}{l l l l}\hline
$D$ &G        & H   & ${\rm dim}\,[{\rm G}]-{\rm dim}\,[{\rm H}]$ \\ 
\hline
11  & 1       & 1   & $0-0=0$      \\
10A & SO$(1,1)/{\bf Z}_2$   & 1   & $1-0=1$  \\
10B & SL(2)    & SO(2) &  $3-1=2$  \\
9   & GL(2)    & SO(2) &  $4-1=3$ \\
8   & E$_{3(+3)}\sim {\rm SL}(3)\! \times \!{\rm SL}(2)$    &  
U(2) & $11- 4=7$  \\
7   & E$_{4(+4)}\sim {\rm SL}(5)$  & USp(4) &$24-10=14$   \\
6   & E$_{5(+5)}\sim {\rm SO}(5,5)$ & 
    USp$(4)\!\times\! {\rm USp}(4)$ &$45-20=25$   \\ 
5   & E$_{6(+6)}$  & USp(8) & $78-36= 42$    \\
4   & E$_{7(+7)}$  &  SU(8) & $133-63= 70$    \\  
3   & E$_{8(+8)}$  & SO(16) & $248 - 120= 128$   \\ \hline
\end{tabular}
\end{center}
\caption{\small 
Homogeneous scalar manifolds G/H for maximal 
supergravities in various dimensions. The type-IIB theory 
cannot be obtained from reduction of 11-dimensional supergravity
and is included for completeness. The difference of the 
dimensions of G and H equals the number of scalar fields, 
listed in table~10.%\ref{maximal-sg-bosons}. 
}\label{maximal-cosets} 
\end{table}
%%%%%%%%%%%%%%%%%%%%%%%%%%%%%%%%%%%%%%%%%%%%%%%%%%%%%%%%%%%%%
%%%%%%%%%%%%%%%%%%%%%%%%%%%%%%%%%%%%%%%%%%%%%%%%%%%%%%%%%%%%%

Here we should add that it is generally possible to realize the group 
${\rm H}_{\rm R}$ as a {\it local} symmetry of the Lagrangian. The 
corresponding connections are then composite connections, 
governed by the Cartan-Maurer equations. In such a formulation 
most fields (in particular, the fermions) do not transform under 
the group G, but only under the local ${\rm H}_{\rm R}$ group. 
The scalars transform linearly under both the rigid duality group as well 
as under the local ${\rm H}_{\rm R}$ group; the gauge fields cannot
transform under the local group ${\rm H}_{\rm R}$, as this would be in
conflict with their own gauge invariance. After fixing a gauge, the 
G-transformations become realized nonlinearly (we discuss such
nonlinear realizations in detail in chapters~4 and 5). The fields which 
initially transform only under the local ${\rm H}_{\rm R}$ group, will 
now transform under the duality group G through field-dependent 
${\rm H}_{\rm R}$ transformations. This phenomenon is also realized for 
the central charges, which transform under the group ${\rm H}_{\rm R}$   
as we have shown in table~\ref{maximal-central-extension}. We discuss
some of the consequences for the central charges and the BPS states in
section~\ref{BPSextended}. 

%%%%%%%%%%%%%%%%%%%%%%%%%%%%%%%%%%%%%%%%%%%%%%%%%%%%%%%%%%%%%%%
 \subsection{Frames and field redefinitions}
%%%%%%%%%%%%%%%%%%%%%%%%%%%%%%%%%%%%%%%%%%%%%%%%%%%%%%%%%%%%%%%
The Lagrangian \eqn{D10-lagrangian} does not contain the 
standard Einstein-Hilbert term for gravity, while a standard 
kinetic term for the scalar field $\phi$ is lacking. This does 
not pose a serious problem. In this form the 
gravitational field and the scalar field are entangled and one 
has to deal with the scalar-graviton system as a whole. To 
separate the scalar and gravitational 
degrees of freedom, one applies a so-called Weyl rescaling of 
the metric $g_{\m\n}$ by an appropriate function of $\phi$. In 
the case that we include the massive modes, this rescaling may 
depend on the extra coordinate $x^{10}$. In the context of 
Kaluza-Klein theory this factor is known as the `warp 
factor'. For these lectures two different Weyl rescalings are 
particularly relevant, which lead  
to the so-called Einstein and to the string frame, respectively. 
They are defined by
\be \label{weyl-rescaling}
e_\m^a= {\rm e}^{-\phi/12}\, [e_\m^a]^{\scriptscriptstyle\rm 
Einstein}\,,\qquad e_\m^a=  {\rm e}^{-\phi/3}\,
[e_\m^a]^{\scriptscriptstyle\rm string}\,.
\ee
We already stressed that the the compactification length $L$ is just a
parameter length with no intrinsic meaning as a result of the fact
that one can always apply general
coordinate transformations which involve $x^{10}$. Of course,
one may also consider the {\it geodesic length}, which in the metric
specified by \eqn{KKmetric} is equal to
$L\,\exp[{2\langle\phi\rangle}/3]$. In the Einstein frame, the
geodesic length of the 11-th dimension is invariant under the 
${\rm SO}(1,1)$ transformations. 

After applying the first rescaling \eqn{weyl-rescaling} to the 
Lagrangian  
\eqn{D10-lagrangian} one obtains the Lagrangian in the Einstein 
frame. This frame is characterized by a standard Einstein-Hilbert 
term and by a graviton field that is invariant under the scale 
transformations (\ref{scale-10x},~\ref{scale-10y}). The corresponding 
Lagrangian reads\footnote{%%%%%%%%%%%%%%%%%%%%%%%%%%%%%%%%%%%%%%%%
  Note that under a local scale transformation  
  $e_\m^a\to {\rm  e}^\L e_\m^a$, the Ricci scalar in $D$ 
  dimensions changes according to  
$$
R\to {\rm e}^{-2\L} \Big[ R + 2 (D-1) D^\m\partial_\m \L + 
(D-1)(D-2) g^{\m\n} \,\partial_\m\L\, \partial_\n\L\Big]\,.
$$
  Observe that gauge fields cannot be redefined by these local scale
  transformations because this would interfere with their own 
  gauge invariance. }  %%%%%%%%%%%%%%%%%%%%%%%%%%%%%%%%%%%%%%%%%%%%
%%%%%%%%%%%%%%%%%%%%%%%%%%%%%%%%%%%%%%%%%%%%%%%%%%%%%%%%%%%%%%%%%%%
\bea
\label{D10-lagrangian-E}
{\cal L}_{10}^{\scriptscriptstyle\rm Einstein}&\!\!=\!\!&
 {1\over \kappa_{10}^2} \bigg[ e\,  
\Big[-\ft12 R(e,\omega)  -\ft14 (\pa_\m\phi)^2 \Big]  -\ft18 e\,{\rm 
e}^{3\phi/2}(\pa_\m  V_\n-\pa_\n V_\m)^2 \nonumber\\ 
&&\hspace{8.5mm}   
 -\ft34 e\, {\rm e}^{- \phi}(H_{\m\n\rho})^2 -\ft1{48}e\,{\rm 
e}^{\phi/2}(F_{\m\n\rho\s})^2 \nonumber \\ 
&& \hspace{8.5mm}+ \ft1{1152} \sqrt{2}\,
\varepsilon^{\m_1%
%\m_2\m_3\m_4\m_5\m_6\m_7\m_8\m_9%
-\m_{10}} 
\,C_{11\m_1\m_2} \,F_{\m_3\m_4\m_5\m_6} \,
F_{\m_7\m_8\m_9\m_{10}} \bigg] \,.  
\eea
Supergravity theories are usually formulated in this frame, 
where the isometries of the scalar fields do not act on the 
graviton.  

The second rescaling \eqn{weyl-rescaling} leads to the Lagrangian 
in the string frame,
\bea\label{D10-lagrangian-S}
{\cal L}_{10}^{\scriptscriptstyle\rm string} &\!\!=\!\!& 
{1\over\kappa_{10}^2} \bigg[ e\,{\rm  
e}^{-2\phi} \Big[-\ft12 R(e,\omega)  +2(\pa_\m  \phi)^2 -\ft34 
(H_{\m\n\rho})^2\Big]\nonumber  \\ 
&&\hspace{7.5mm} -\ft18 e\,(\pa_\m 
V_\n-\pa_\n V_\m)^2 -\ft1{48}e\,(F_{\m\n\rho\s})^2   \nonumber \\
&& \hspace{7.5mm}+ \ft1{1152} \sqrt{2}\, 
\varepsilon^{\m_1%
%\m_2\m_3\m_4\m_5\m_6\m_7\m_8\m_9%
-\m_{10}} 
\,C_{11\m_1\m_2} \,F_{\m_3\m_4\m_5\m_6} \,
F_{\m_7\m_8\m_9\m_{10}}\bigg] \,. \;{~~}
\eea
This frame is characterized by the fact that 
$R$ and $(H_{\m\n\rho})^2$ have the same coupling 
to the scalar $\phi$, or, equivalently, that 
$g_{\m\n}$ and $C_{11\m\n}$ transform with equal weights under the
scale transformations (\ref{scale-10x},~\ref{scale-10y}). 
In string theory $\phi$ coincides with the dilaton field that couples  
to the topology of the worldsheet and whose vacuum-expectation value
defines the string coupling constant according to
$\gstring=\exp(\langle\phi\rangle)$. The significance of the dilaton
factors in the Lagrangian above is well known. 
The metric $g_{\mu\nu}$, the antisymmetric tensor
$C_{\mu\nu 11}$ and the dilaton $\phi$ 
always arise in the Neveu-Schwarz sector and  
couple universally to  ${\rm e}^{-2\phi}$.
On the other hand the vector $V_\mu$ and the 3-form 
$C_{\mu\nu\rho}$ describe 
Ramond-Ramond (R-R) states and the specific
form of their vertex operators forbids
any tree-level coupling to the dilaton \cite{various,berg}.
In particular the Kaluza-Klein gauge field 
$V_\m$ corresponds in the string context to the R-R  
gauge field of type-II string theory. The infinite tower of 
massive Kaluza-Klein states carry a  
charge quantized in units of $e_{\scriptscriptstyle\rm KK}$, 
defined in \eqn{KK-charge}. In the context of 10-dimensional 
supergravity,  states with R-R charge are solitonic. In string theory,
R-R charges are carried by the D-brane states. 
 
For later purposes let us note that the above discussion can be
generalized to arbitrary spacetime dimensions.
The Einstein frame in any dimension is defined by a 
gravitational action that is just the 
Einstein-Hilbert action, whereas in the string frame 
the Ricci 
scalar is multiplied by a dilaton term $\exp (-2\phi_D)$, as in 
\eqn{D10-lagrangian-E} and \eqn{D10-lagrangian-S}, respectively.
The Weyl rescaling which connects the two frames 
is given by,
\be
[e_\m^a]^{\scriptscriptstyle\rm string}= {\rm e}^{2\phi_D/(D-2)} \,
[e_\m^a]^{\scriptscriptstyle\rm Einstein}\,.
\ee

Let us now return to 11-dimensional supergravity with the 11-th 
coordinate compactified to a circle so that $0\leq x^{10}\leq L$. 
As we stressed  
already, $L$ itself has no intrinsic meaning and it is better to 
consider the geodesic radius of the 11-th dimension, which reads
\be\label{Rphi}
R_{10}= {L\over 2\pi} \, {\rm e}^{2 \langle\phi\rangle/3}\, .
\ee
This result applies to the frame specified by the 11-dimensional 
theory\footnote{This is the frame specified by the 
   metric given in \refeq{KKmetric}, which leads to the 
   Lagrangian \refeq{D10-lagrangian}.}. % 
In the string frame, the above result reads
\be
(R_{10})^{\scriptscriptstyle\rm string}= {L\over 2\pi} \, {\rm 
e}^{\langle\phi\rangle} \, . 
\ee
It shows that a small 11-th dimension corresponds to small 
values of $\exp\langle \phi\rangle$ which in turn 
corresponds to a weakly coupled string theory. Observe that $L$ is fixed 
in terms of $\kappa_{10}$ and $\kappa_{11}$ 
({\it c.f.}  (\ref{def-L})).
 
{}From the 11-dimensional expressions, 
\be
E_a^{\,M}\pa_M = e_a^{\,\m} (\pa_\m - V_\m \,\pa_{10})\,, \qquad 
E_{10}^{\,M}\pa_M = {\rm e}^{-2\phi/3}\, \pa_{10}\,, 
\ee
where $a$ and $\m$ refer to the 10-dimensional Lorentz and world 
indices, we infer that, in the frame specified by the 11-dimensional 
theory, the Kaluza-Klein masses are multiples of 
\be
M^{\scriptscriptstyle\rm KK} = {1\over R_{10}}\,.
\ee
Hence Kaluza-Klein states have a mass and a Kaluza-Klein charge  
(cf. \eqn{KK-charge}) related by   
\be 
M^{\scriptscriptstyle\rm KK}= 
\vert e_{\scriptscriptstyle\rm KK} \vert\, {\rm e}^{-2 \langle
\phi\rangle/3} \,.  
\ee
In the string frame, this result becomes
\be \label{Mphi}
(M^{\scriptscriptstyle\rm KK})^{\scriptscriptstyle\rm string}= 
\vert e_{\scriptscriptstyle\rm KK} \vert\, {\rm e}^{-
\langle \phi\rangle} \,. 
\ee
Massive Kaluza-Klein states are always BPS states, meaning that 
they are contained in supermultiplets that are `shorter' than the 
generic massive supermultiplets because of nontrivial central 
charges. The central charge here is just the 10-th component of 
the momentum, which is proportional to the Kaluza-Klein charge. 

The surprising insight that emerged, is that the Kaluza-Klein features
of 11-dimensional supergravity have a precise counterpart in string
theory \cite{hull,Townsend,various}. There one has nonperturbative (in
the string coupling constant) states which carry R-R charges. On the
supergravity side these states often appear as solitons. 
%%%%%%%%%%%%%%%%%%%%%%%%%%%%%%%%%%%%%%%%%%%%%%%%%%%%%%%%%%%%%%%%%
%%%%%%%%%%%%%%%%%%%%%%%%%%%%%%%%%%%%%%%%%%%%%%%%%%%%%%%%%%%%%%%%%
\subsection{Kaluza-Klein states and BPS-extended supergravity}
\label{BPSextended}
%%%%%%%%%%%%%%%%%%%%%%%%%%%%%%%%%%%%%%%%%%%%%%%%%%%%%%%%%%%%%%%%%
In most of this chapter we restrict ourselves to pure
supergravity. However, when compactifying dimensions one also
encounters massive Kaluza-Klein states, which couple to the
supergravity theory as massive matter supermultiplets. The presence of
these BPS states introduces a number of qualitative changes to the
theory which we discuss in this section. The most conspicuous change
is that the continuous nonlinearly realized symmetry group $\rm G$ is 
broken to an arithmetic subgroup, known as the
U-duality group. This U-duality group has been conjectured 
to be the exact symmetry group of (toroidally compactified) M-theory
\cite{hull}. The BPS states (which are contained in M-theory) should
therefore be assigned to 
representations of the U-duality group. Here one naively assumes that the
U-duality group acts on the central charges of the BPS states and it
is simply defined as the arithmetic subgroup of $\rm G$ that leaves the
central-charge lattice invariant. However, central charges are in
principle assigned to representations of the group ${\rm H}_{\rm R}$
and not of the group $\rm G$ (although the central charges will
eventually, upon gauge fixing, 
transform nonlinearly under G, via field-dependent ${\rm H}_{\rm R}$
transformations). In most cases, these ${\rm H}_{\rm R}$ 
representations can be elevated to representations of $\rm G$, by 
multiplying with the representatives of the 
coset space ${\rm G}/{\rm H}_{\rm R}$ (representatives of coset spaces
will be discussed in chapters~4 and 5). In this way, the pointlike
(field-dependent) central charges can be assigned to representations
of $\rm G$ for spacetime dimensions $D\geq4$. Similar observations
exist for stringlike and membranelike central charges except that
in these cases the dimension must be restricted even further
\cite{istanbul}.   

Another aspect of the coupling of the BPS states to supergravity is
that the central charges should be related to {\it local} symmetries,
in view of the fact that they appear in the anticommutator of two
supercharges and supersymmetry is realized locally. Therefore nonzero
central charges must couple to appropriate gauge fields in the 
supergravity theory. These gauge fields transform (with minor
exceptions) linearly with constant matrices under the group $\rm
G$. Inspection of the 
tables that we have presented earlier, shows that the gauge fields
usually appear in the $\rm G$-representation required for
gauging the corresponding central charge. Provided that the central
charges can be assigned to the appropriate representations of the
U-duality group and that the appropriate gauge fields are available,
one may thus envisage a (possibly local field) theory of BPS states
coupled to supergravity that is U-duality invariant. This theory would
exhibit many of the features of M-theory and describe many of the
relevant degrees of freedom.

The Kaluza-Klein states that we encounter in
toroidal compactifications of supergravity are a subset of the 1/2-BPS 
states in M-theory. They carry pointlike central charges and they
couple to the Kaluza-Klein photon fields, {\it i.e.}, the vector gauge
fields that emerge from the 
higher-dimensional metric upon the toroidal compactification. However,
they do {\it not} constitute representations of the U-duality group,
because the central charges that they carry are too restricted. This
is the reason why retaining the Kaluza-Klein states in the dimensional
compactification will lead to a breaking of the U-duality group. 
From the eleven-dimensional perspective it is easy to see why the
central charges associated with the Kaluza-Klein states are too
restricted, because under U-duality the central charges
related to the momentum operator in the compactified dimensions
combine with the two- and five-brane charges
({\it c.f.} \eqn{eq:11D-algebra}) in order to define  
representations of the U-duality group. However, conventional dimensional
compactification does not involve any brane charges. 
Nevertheless, in certain cases one may still be able to extend the
Kaluza-Klein states with other BPS states, so that a U-duality
invariant theory is obtained. Such extended theories are called
BPS-extended supergravity \cite{strings00,istanbul}.   

The fact that some of the central charges are
associated with extra spacetime dimensions ({\it i.e.} the charges
carried by the Kaluza-Klein states) implies that the newly introduced
states (associated with wrapped branes) may also have an
interpretation in terms of extra dimensions. In this way, the number
of spacetime dimensions could exceed eleven, although the theory would 
presumably not be able to decompactify uniformly to a flat spacetime of
more than eleven dimensions. The aim of this section is to 
elucidate some of these ideas in the relatively simple context of
$N=2$ supergravity in $D=9$ spacetime dimensions. 

We start by considering the BPS multiplets that are relevant in
9 spacetime dimensions from the perspective of supergravity, string
theory and (super)membranes. In 9 dimensions the R-symmetry group
and the duality group are equal to ${\rm H}_{\rm R}= {\rm SO}(2)$ and  
${\rm G}={\rm SO}(1,1)\times {\rm SL}(2;{\bf R})$, respectively. It is
well known that the 
massive supermultiplets of IIA and IIB string theory  
coincide, whereas the massless states comprise inequivalent
supermultiplets for the simple reason that they transform according
to different representations of the SO(8) helicity group.
When compactifying the theory on a circle, IIA and IIB states that are
massless in 9 spacetime dimensions, transform according to
identical representations of the ${\rm SO}(7)$ helicity group and
constitute equivalent supermultiplets. The corresponding interacting
field theory  
is the unique $N=2$ supergravity theory in 9 spacetime dimensions. 
However, the BPS supermultiplets which carry momentum
along the circle, remain inequivalent as they remain assigned to
the inequivalent representations of the group SO(8) which is now
associated with the restframe (spin) rotations of the massive states. 
Henceforth the momentum states of the IIA and the IIB theories will be
denoted as KKA and KKB states, respectively. The fact that they 
constitute  inequivalent supermultiplets, has implications for the
winding states in order that T-duality remains valid \cite{ADLN}.

In 9 spacetime dimensions with $N=2$ supersymmetry the
Lorentz-invariant central charges are encoded in a two-by-two
real symmetric  matrix $Z^{ij}$, which can be decomposed as 
\be
Z^{ij} = b\,\d^{ij} + a \,(\cos\theta \,\sigma_3 + \sin \theta\,
\sigma_1)^{ij} \,.
\ee
Here $\sigma_{1,3}$ are the real symmetric Pauli matrices. 
We note that the central charge associated with the parameter $a$
transforms as a doublet under the SO(2) R-symmetry group that rotates
the two  supercharge spinors, while the central charge proportional to 
the parameter $b$ is SO(2)
invariant. Subsequently one shows that BPS states that carry these 
charges must satisfy the mass formula,
\be 
M_{\rm BPS} = \vert a \vert + \vert b\vert \,.
\ee
Here one can distinguish three types of BPS supermultiplets. One type 
has central charges $b=0$ and $a\not=0$. These are 
$1/2$-BPS muliplets, because they are annihilated by half of the
supercharges. The KKA supermultiplets that comprise Kaluza-Klein states of IIA
supergravity are of this type.  Another type of $1/2$-BPS mulitplets
has central charges $a=0$ and $b\not =0$. The KKB supermultiplets
that comprise the Kaluza-Klein states of IIB supergravity are of this
type. Finally there are $1/4$-BPS multiplets 
(annihilated by one fourth of the supercharges) 
characterized by the fact that neither $a$ nor $b$ vanishes. 

For type-II string theory one obtains these central charges in
terms of the left- and right-moving momenta, $p_{\rm L}$, $p_{\rm R}$,
that characterize winding and momentum along $S^1$. However, the
result takes a different form for the IIA and the IIB theory as the
following formula shows,
\be
Z^{ij} = \left\{\begin{array}{lr}
\ft12(p_{\rm L}+p_{\rm R}) \d^{ij} +  \ft12(p_{\rm L}-p_{\rm R})
\sigma_3^{ij} \,,\quad & {\rm (for \;IIB)}\\[3mm]
 \ft12(p_{\rm L}-p_{\rm R}) \d^{ij} +  \ft12(p_{\rm L}+p_{\rm R})
\sigma_3^{ij} \,.\quad & {\rm (for \;IIA)}
\end{array} \right.
\ee
The corresponding BPS mass formula is thus equal to
\be
M_{\rm BPS} = \ft12\vert p_{\rm L}+p_{\rm R}\vert  +  \ft12\vert
p_{\rm L}-p_{\rm R}\vert \,.
\ee
For $p_{\rm L}=p_{\rm R}$ we confirm the original identification of
the momentum states, namely that IIA momentum states constitute KKA
supermultiplets, while IIB momentum states constitute KKB
supermultiplets. For the winding states, where $p_{\rm L}=-p_{\rm R}$,
one obtains the opposite result: IIA winding states constitute KKB
supermultiplets, while IIB winding states constitute KKA 
supermultiplets. The $1/4$-BPS multiplets arise for string states that have
either right- or left-moving oscillator states, so that either $M_{\rm
BPS} = \vert p_{\rm L}\vert$ or $\vert p_{\rm R}\vert$ with $p_{\rm L}^{\,2}
\not= p_{\rm R}^{\,2}$. All of this is entirely consistent with 
T-duality\cite{DHS,d-brane1},
according to which there exists a IIA and a IIB perspective, with 
decompactification radii are that inversely proportional and with an
interchange of winding and momentum states. Observe that the
$1/4$-BPS states will never become massless, so that they don't play a
role in what follows. 

It is also possible to view the central charges from the perspective of the
11-dimensional (super)membrane \cite{BST}. Assuming that the
two-brane charge takes values in the compact coordinates labelled 
by 9 and 10, which can be generated by wrapping the membrane
over the corresponding $T^2$, one readily finds the expression,
\be 
Z^{ij} = Z_{9\,10} \,\d^{ij} - (P_9 \,\sigma_3^{ij} - P_{10} \,
\sigma_1^{ij})\,.
\ee
When compactifying on a torus with modular parameter
$\tau$ and area $A$, the BPS mass formula takes the form  
\bea
M_{\rm BPS} &=& \sqrt{P_9^{\,2} + P_{10}^{\,2} } + \vert
Z_{9\,10}\vert \nonumber\\
&=& {1\over \sqrt{A\,\tau_2}} \vert q_1 + \tau \,q_2\vert + T_{\rm m}
A \,\vert p\vert \,. \label{BPS-membrane}
\eea
Here $q_{1,2}$ denote the momentum numbers on the torus and $p$ is the
number of times the membrane is wrapped over the torus; $T_{\rm m}$
denotes the supermembrane tension. 
Clearly the KKA states correspond to the momentum modes on $T^2$ while
the KKB states are associated with the wrapped membranes on the
torus. Therefore there is a rather natural way to describe the IIA and IIB
momentum and winding states starting from a (super)membrane in eleven
spacetime dimensions. This point was first emphasized in \cite{JHS}. 
%%%%%%%%%%%%%%%%%%%%%%%%%%%%%%%%%%%%%%%%%%%%%%%%%%%%%%%%%%%%%

This suggests to consider $N=2$ supergravity in 9 spacetime
dimensions and couple it to the simplest BPS supermultiplets
corresponding to KKA and KKB states. As shown in
tables~\ref{maximal-central-extension} and \ref{maximal-sg-bosons} 
there are three central charges and 9-dimensional 
supergravity possesses precisely three gauge fields that couple to these
charges. From the perspective of 11-dimensional supergravity
compactified on $T^2$, the  Kaluza-Klein states transform as KKA
multiplets. Their charges transform obviously with respect to an SO(2)
associated with 
rotations of the coordinates labelled by 9 and 10. Hence we have a
``double'' tower of these charges with corresponding 
KKA supermultiplets. On the other hand, from the perspective of IIB 
compactified on $S^1$, the Kaluza-Klein states constitute KKB multiplets
and their charge is SO(2) invariant. Here we have a ``single'' tower of KKB
supermultiplets. However, from the perspective of 9-dimensional
supergravity one is led to couple both towers of KKA and KKB
supermultiplets simultaneously. In that case one obtains some
dichotomic theory\cite{ADLN}, which we refer to as BPS-extended 
supergravity. In the
case at hand this new theory describes the 
ten-dimensional IIA and IIB theories in certain decompactification
limits, as well as eleven-dimensional supergravity. But the theory is
in some sense truly 12-dimensional with three compact coordinates,
although there is no 12-dimensional Lorentz invariance, not even
in a uniform decompactification limit, as the fields never depend on
all the 12 coordinates! Whether this kind of BPS-extended
supergravity offers a viable scheme in a more general context than the
one we discuss here, is not known. In the case at 
hand we know a lot about these couplings from our knowledge of
the $T^2$ compactification of $D=11$ supergravity and the
$S^1$ compactification of IIB supergravity. 

The fields of 9-dimensional $N=2$ supergravity are listed in
table~\ref{dichotomic}, where we also indicate their relation with the
fields of 11-dimensional and 10-dimensional IIA/B supergravity upon
dimensional reduction. It is not necessary to work out all the 
nonlinear field redefinitions here, as the corresponding fields can be
uniquely identified by their scaling weights under SO(1,1), a symmetry
of the massless theory that emerges upon dimensional reduction and  is 
associated with scalings of the internal vielbeine. The scalar field
$\sigma$ is related to $G_{99}$, the IIB metric component in the
compactified dimension, by $G_{99}=\exp(\sigma)$; likewise it
is related to the determinant of the 11-dimensional metric in the
compactified dimensions, which is equal to $\exp(-\ft43\sigma)$. The
precise relationship follows from  comparing the SO(1,1) weights
through the dimensional reduction of IIB and eleven-dimensional
supergravity. In 9 dimensions supergravity has two more scalars, 
which are described by a nonlinear sigma model based on 
${\rm SL}(2,{\bf R})/{\rm SO}(2)$. The coset is described by the
complex doublet of fields $\phi^\a$, which satisfy a constraint
$\phi^\a\phi_\a =1$ and are subject to a local SO(2) invariance, so
that they describe precisely two scalar degrees of freedom
($\a=1,2$). We expect that the local SO(2) invariance 
can be incorporated in the full BPS-extended supergravity theory and
can be exploited in the construction of the couplings of the various
BPS supermultiplets to supergravity. 

We already mentioned  the three abelian vector gauge
fields which couple to the central charges. There are two vector
fields $A_\m^\a$, which are the Kaluza-Klein 
photons from the $T^2$ reduction of eleven-dimensional supergravity
and which couple therefore to the KKA states. From the IIA perspective
these correspond to the Kaluza-Klein states on $S^1$ and the D0
states. From the IIB side they 
originate from the tensor fields, which confirms that they couple to
the IIB (elementary and D1) winding states. These two fields transform 
under SL(2), which can be understood {from} the perspective of the
modular transformation on $T^2$ as well as from the S-duality
transformations that rotate the elementary strings with the D1 strings.
The third gauge field, denoted by $B_\m$, is a singlet under SL(2)
and is the Kaluza-Klein photon on the IIB side, so that it couples
to the KKB states. On the IIA side it originates from the IIA tensor
field, which is consistent with the fact that the IIA winding states
constitute KKB supermultiplets. 

{From} the perspective of the supermembrane, the KKA states are the
momentum states on $T^2$, while the KKB states correspond to the
membranes wrapped around the torus. While it is gratifying to see how
all these correspondences work out, we stress that, from the
perspective of 9-dimensional $N=2$ supergravity, the results follow
entirely from supersymmetry.  

%%%%%%%%%%%%%%%%%%%%%%%%%%%%%%%%%%%%%%%%%%%%%%%%%%%%%%%%%%%%%%%%%%
\begin{table}
\begin{center}
\begin{tabular}{ccccr}
\hline $D=11$ & IIA & $D=9$ & {~}\quad IIB \quad{~} & SO(1,1) \\ 
\hline  \\[-3mm]%%%%%%%%%%%%%%%%%%%%%%%%%%%%%%%%%%%%%%%%%%%%%%%%%%%%%
$\hat G_{\mu \nu}$ & $G_{\m \n}$  & $g_{\m \n}$  & $G_{\m \n}$ & $0\;\,$ \\[1mm]
%\hline \\[-3mm]%%%%%%%%%%%%%%%%%%%%%%%%%%%%%%%%%%%%%%%%%%%%%%%%%%%%%%%%%%%%
$\hat{A}_{\mu {\scriptscriptstyle \,9\,10}}$  & $C_{\mu
{\scriptscriptstyle\,9}}$ &  $B_{\m}$ & $G_{\m {\scriptscriptstyle\,9}}$
& $-4\;\,$\\[1mm] 
%\hline \\[-3mm]%%%%%%%%%%%%%%%%%%%%%%%%%%%%%%%%%%%%%%%%%%%%%%%%%%%%%%%%%%%%%
$\hat G_{\mu {\scriptscriptstyle\,9}}$, $\hat G_{\mu
{\scriptscriptstyle\,10}}$  & $G_{\mu {\scriptscriptstyle\,9}}$ ,
$C_{\mu}$  & $A_{\mu}^{\, \alpha}$ & $A_{\m
{\scriptscriptstyle\,9}}^{\, \alpha}$ & $3\;\,$  \\[1mm]  
%\hline \\[-3mm]%%%%%%%%%%%%%%%%%%%%%%%%%%%%%%%%%%%%%%%%%%%%%%%%%%%%%%%%%%%%
$\hat{A}_{\m \n {\scriptscriptstyle\,9}}$, $\hat{A}_{\m \n
{\scriptscriptstyle\,10}}$ & $C_{\m \n {\scriptscriptstyle\,9}}, C_{\m
\n}$ &  $A_{\m\n}^{\,\a}$ &  $A_{\m\n}^{\,\a}$  & $-1\;\,$  \\ [1mm]
%\hline \\[-3mm]%%%%%%%%%%%%%%%%%%%%%%%%%%%%%%%%%%%%%%%%%%%%%%%%%%%%%%%%%%%%
$\hat A_{\m \n\rho}$  & $C_{\m \n \rho}$  & $A_{\m\n\rho}$  &
$A_{\m\n\rho\sigma}$ & $2\;\,$  \\[1mm] 
%\hline \\[-3mm]%%%%%%%%%%%%%%%%%%%%%%%%%%%%%%%%%%%%%%%%%%%%%%%%%%%%%%%%%%%%%
$\hat G_{{\scriptscriptstyle 9 \,10}}$, $\hat{G}_{{\scriptscriptstyle9\,
9}}$, $\hat G_{{\scriptscriptstyle 10\,  10}}$ & $\phi$,  
$G_{{\scriptscriptstyle 9\,9}}$, $C_{\scriptscriptstyle  9}$ &
$\left\{ \begin{array}{l} \phi^\a  \\[1mm] \exp (\sigma ) \end{array}
\right. $ & $\begin{array}{l} \phi^\a \\[1mm] G_{{\scriptscriptstyle
9\,9}} 
\end{array}$ & $\begin{array}{r} $0$ \\[1mm]  $7$ \end{array}$
\\[1mm]\\[-3mm]  
\hline %%%%%%%%%%%%%%%%%%%%%%%%%%%%%%%%%%%%%%%%%%%%%%%%%%%%%%%%%%%%
\end{tabular}
\end{center}
\caption{\small 
The bosonic fields of the eleven dimensional, type-IIA,  
nine-dimensional $N=2$ and type-IIB  supergravity theories. 
The 11-dimensional and 10-dimensional indices, respectively, are 
split as $\hat{M} =(\mu, 9,10)$ and $M=(\mu,9)$, where $\mu = 0,1,\ldots 8$. 
The last column lists the SO(1,1) scaling weights of the fields.}
\label{dichotomic}
\end{table}
%%%%%%%%%%%%%%%%%%%%%%%%%%%%%%%%%%%%%%%%%%%%%%%%%%%%%%%%%%%%%%%%%%%

The resulting BPS-extended theory incorporates 11-dimensional
supergravity and the two type-II supergravities in special
decompactification limits. But, as we stressed above, we are dealing
with a 12-dimensional theory here, although no field can depend
nontrivially on all of these  
coordinates. The theory has obviously two mass scales associated with
the KKA and KKB states. We return to them in a moment. Both S- and
T-duality are manifest, although the latter has become trivial as
the theory is not based on a specific IIA or IIB perspective. One
simply has the freedom to view the theory from a IIA or a IIB
perspective and interpret it accordingly. 

We should discuss the fate of the group ${\rm G}= {\rm SO}(1,1) \times
{\rm SL}(2,{\bf R})$ of pure supergravity after coupling the theory to
the BPS multiplets. The central charges of the BPS states form a
discrete lattice, which is affected by this group. Hence, after
coupling to the BPS states, we only have a discrete subgroup that
leaves the charge lattice invariant. This is the group ${\rm
SL}(2,{\bf Z})$.  

The KKA and KKB states and their interactions with the massless theory
can be understood from the perspective of compactified  11-dimensional
and IIB supergravity. In this way we are able to deduce the following
BPS mass formula, 
\be
M_{\rm BPS}(q_1,q_2,p) = m_{\scriptscriptstyle\rm KKA} \,{\rm
e}^{3\sigma/7} \,\vert q_\a \phi^\a \vert +  m_{\scriptscriptstyle\rm
KKB} \,{\rm e}^{-4\sigma/7} \,\vert p\vert\,, \label{massformula}
\ee
where $q_\a$ and $p$ refer to the integer-valued KKA and KKB charges,
respectively, and $m_{\scriptscriptstyle\rm KKA}$ and
$m_{\scriptscriptstyle\rm KKB}$ are two independent  mass scales. 
This formula can be compared to the membrane BPS formula
\eqn{BPS-membrane} in the 11-dimensional frame. One then finds
that 
\be
m^2_{\scriptscriptstyle \rm KKA}\, m^{~}_{\scriptscriptstyle\rm KKB}
\propto  T_{\rm m}\,,
\ee
with a numerical proportionality constant. However, the most important
conclusion to draw from \eqn{massformula} is that there is no limit in
which the masses of both KKA and KKB states will tend to zero. In
other words, there is no uniform decompactification limit. Therefore,
in spite of the fact that we have more than 11 dimensions,  
there exists no theory with $Q=32$ supercharges in flat Minkowski
spacetime of dimensions $D>11$. 

%%%%%%%%%%%%%%%%%%%%%%%%%%%%%%%%%%%%%%%%%%%%%%%%%%%%%%%%%%%%%%%%%
 \subsection{Nonmaximal supersymmetry: $Q=16$}
%%%%%%%%%%%%%%%%%%%%%%%%%%%%%%%%%%%%%%%%%%%%%%%%%%%%%%%%%%%%%%%%%
%%%%%%%%%%%%%%%%%%%%%%%%%%%%%%%%%%%%%%%%%%%%%%%%%%%%%%
\begin{table}
\begin{center}
\begin{tabular}{l l l l l l l l}\hline
$D$ &${\rm H}_{\rm R}$& $A_\m$ & $\phi$ & $\chi$   \\ \hline
10 & 1       & 1   & 0     &  1    \\
9  & 1       & 1   & 1     &  1   \\
8  & U(1)    & 1   & $1+\bar1$ &  $1+\bar 1$    \\
7   & USp(2)  & 1  & $3$   &   2  \\
6  & USp$(2)\!\times\! {\rm USp}(2)$  & 1 & $(2,2)$ & $(2,1)+(1,2)$\\ 
5   & USp(4)& 1   & 5     &  4    \\
4   & U(4)  & 1   & $6^\ast$ &  $4+\bar 4$    \\  
3   & SO(8) & ~   &  8   &   8     \\ \hline
\end{tabular}
\end{center}
\caption{\small 
Field content for maximal super-Maxwell theories in 
various dimensions. All supermultiplets contain a gauge field 
$A_\m$, scalars $\phi$ and spinors $\chi$ and comprises $8+8$ degrees
of freedom. In $D=3$ dimensions the vector field is dual to a  
scalar. The $6^\ast$ representation of SU(4) is a selfdual 
rank-2 tensor. }
\label{maximal-Maxwell} 
\end{table}
%%%%%%%%%%%%%%%%%%%%%%%%%%%%%%%%%%%%%%%%%%%%%%%%%%%%%%%%%%%%%%%%%%
For completeness we also summarize a number of  results on nonmaximal
supersymmetric theories with $Q=16$  supercharges, which are now
restricted to dimensions $D\leq 10$.   
Table~\ref{maximal-Maxwell} shows the field representations for the vector  
multiplet in dimension $3\leq D\leq 10$. This multiplet comprises 
$8+8$ physical degrees of freedom. 
We also consider the $Q=16$ supergravity theories. The Lagrangian 
can be obtained by truncation of \eqn{D10-lagrangian}. However, 
unlike in the case of maximal supergravity, we now have the 
option of introducing additional matter fields. For $Q=16$ the 
matter will be in the form of vector supermultiplets, possibly 
associated with some nonabelian gauge group. 
Table~\ref{non-maximal-supergravity} summarizes $Q=16$ supergravity 
for dimensions $3\leq D\leq 10$. In $D=10$ dimensions the bosonic 
terms of the supergravity Lagrangian take the form \cite{BDDV},
\bea 
\label{D10-Q16-lagrangian}
{\cal L}_{10}\!&=\!& {1\over \kappa_{10}^2} \bigg[ -\ft12 e\,{\rm 
e}^{2\phi/3} R(e,\omega)  \nonumber \\
&&\hspace{9mm} -\ft34 e\, {\rm e}^{- 
2\phi/3}(H_{\m\n\rho})^2  - \ft14 e (\pa_\m A_\n- \pa_\n A_\m)^2 
\bigg] \,,  
\eea
where, for convenience, we have included a single vector gauge 
field $A_\m$, belonging to an abelian vector supermultiplet. A 
feature that deserves to be mentioned, is that the field strength 
$H_{\m\n\rho}$ 
associated with the 2-rank gauge field contains a Chern-Simons 
term $A_{[\m} \pa_\n A_{\rho]}$. Chern-Simons terms play an important 
role in the anomaly cancellations of this theory. Note also that 
the kinetic term for the Kaluza-Klein vector field in 
\eqn{D10-lagrangian}, depends on  
$\phi$, unlike the kinetic term for the matter vector field in 
the Lagrangian above.  This reflects itself in the extension of 
the symmetry transformations noted in
(\ref{scale-10x},~\ref{scale-10y}), 
\bea 
\label{scale-10z}
&&e_\m^a \to  {\rm e}^{-\a}e_\m^a \,,\qquad \phi \to  \phi + 12\a 
\,,\nonumber \\
&&  C_{11\m\n} \to  {\rm e}^{6\a}  C_{11\m\n}\,, \qquad A_{\m} 
\to  {\rm e}^{3\a}  A_{\m}\,. 
\eea
where $A_\m$ transforms differently from the Kaluza-Klein vector field 
$V_\m$. 

%%%%%%%%%%%%%%%%%%%%%%%%%%%%%%%%%%%%%%%%%%%%%%%%%%%%%%%%%%%%
%%%%%%%%%%%%%%%%%%%%%%%%%%%%%%%%%%%%%%%%%%%%%%%%%%%%%%%%%%%%
\begin{table}
\begin{center}
\begin{tabular}{l l l l l l l l }\hline
$D$ &${\rm H}_{\rm R}$& \#& graviton & $p=-1$ & $p=0$ & $p=1$  \\ \hline
10  & 1          &64& 1 & $1$ & ~    & 1     \\
9   & 1          &56& 1 & $1$ & 1    & 1     \\
8   & U(1)       &48& 1 & $1$ & $1+\bar1$& $1$ \\
7  & USp(2)      &40& 1 & $1$ &$3$   & 1    \\
6A$\!\!$ & USp$(2)\!\times\! {\rm USp}(2)$ &32& 1 &1&(2,2)&(1,1)\\ 
6B$\!\!$ & USp$(4)$ &24& 1 & ~&~ & $5^\ast$ \\ 
5   & USp(4)        &24& 1 & $1$ & 5+1    & ~    \\
4   & U(4)          &16& 1 & $1+\bar 1$ & $[6]$& ~ \\
3   & SO(8)         &$8k$& 1 & $8k$       & ~    & ~ \\  \hline
\end{tabular}
\end{center}
\caption{\small 
Bosonic fields of nonmaximal supergravity with $Q=16$. 
In 6 dimensions type-A and type-B 
correspond to (1,1) and (2,0) supergravity. In the third column, \#
denotes the number of bosonic degrees of freedom. Note that, with the 
exception of the 6B and the 4-dimensional theory, all these 
theories contain precisely one scalar field. The tensor field in the
6B theory is selfdual. In $D=4$ dimensions, the 
SU(4) transformations cannot be implemented on the vector 
potentials, but act on the (abelian) field strengths by duality 
transformations. In $D=3$ dimensions 
supergravity is a topological theory and can be coupled to scalars and 
spinors. The scalars parametrize the coset space ${\rm SO}(8,
k)/({\rm SO}(8)\times {\rm SO}(k))$, where $k$ is an arbitrary 
integer. }
\label{non-maximal-supergravity} 
\end{table}
%%%%%%%%%%%%%%%%%%%%%%%%%%%%%%%%%%%%%%%%%%%%%%

In this case there are three different Weyl rescalings that are 
relevant, namely 
\bea \label{weyl-rescaling2}
e_\m^a&=& {\rm e}^{-\phi/12}\, [e_\m^a]^{\scriptscriptstyle\rm 
Einstein}\,,\nonumber\\
 e_\m^a&=&  {\rm e}^{-\phi/3}\,
[e_\m^a]^{\scriptscriptstyle\rm string}\,,\nonumber \\
 e_\m^a&=&  {\rm e}^{\phi/6}\,
[e_\m^a]^{\scriptscriptstyle\rm string'}\,.
\eea
It is straightforward to obtain the corresponding Lagrangians. In 
the Einstein frame, the graviton is again invariant under the 
isometries of the scalar field. The bosonic terms read
\bea \label{D10-Q16-lagrangian1}
{\cal L}_{10}^{\scriptscriptstyle\rm Einstein}\!\!\!&=\!\!\!& {1\over 
\kappa_{10}^2} \bigg[ -\ft12 e\,  
R(e,\omega)  -\ft14 e (\pa_\m\phi)^2 \nonumber\\
&& \hspace{8mm} -\ft34 e\, {\rm e}^{- 
\phi}(H_{\m\n\rho})^2  - \ft14 e \,{\rm e}^{-\phi/2} (\pa_\m A_\n- 
\pa_\n A_\m)^2   
\bigg] \,. \quad{~} 
\eea

The second Weyl rescaling leads to the following Lagrangian, 
\bea \label{D10-Q16-lagrangian2}
{\cal L}_{10}^{\scriptscriptstyle\rm string}\!&=\!& {1\over 
\kappa_{10}^2} {\rm e}^{-2\phi}\bigg[ -\ft12 e\, R(e,\omega) + 
2e(\pa_\m\phi)^2 \nonumber  \\
&& \hspace{15mm}  -\ft34 e\, 
(H_{\m\n\rho})^2   
- \ft14 e (\pa_\m A_\n- \pa_\n A_\m)^2  \bigg] \,,  
\eea
which shows a uniform coupling to the dilaton. This is the 
low-energy effective Lagrangian relevant for the heterotic string. 
Eventually the matter gauge field  
has to be part of an nonabelian gauge theory based on the groups
${\rm SO}(32)$ or ${\rm E}_8\times{\rm E}_8$ in order to be
anomaly-free.  

Finally, the third Weyl rescaling yields
\bea \label{D10-Q16-lagrangian3}
{\cal L}_{10}^{\scriptscriptstyle\rm string'}\!&=\!& {1\over 
\kappa_{10}^2} \bigg[e\,{\rm e}^{2\phi}  \Big[-\ft12 R(e,\omega) 
+2(\pa_\m\phi)^2 \Big] \nonumber \\ 
&& \hspace{9mm}  -\ft34 e\, 
(H_{\m\n\rho})^2  - \ft14 e\,{\rm  
e}^{\phi}(\pa_\m A_\n- \pa_\n A_\m)^2  \bigg] \,.
\eea
Now the dilaton seems to appear with the wrong sign. As it turns 
out, this is the low-energy effective action of the type-I 
string, where the type-I dilaton must be associated with $-\phi$. 
This is related to the fact that the ${\rm SO}(32)$ heterotic string
theory is S-dual to type-I string theory \cite{polch-witt}. 

%%%%%%%%%%%%%%%%%%%%%%%%%%%%%%%%%%%%%%%%%%%%%%%%%%%%%%%%%%%%%%%%%%%
%%%%%%%%%%%%%%%%%%%%%%%%%%%%%%%%%%%%%%%%%%%%%%%%%%%%%%%%%%%%%%%%%%%

%%%%%%%%%%%%%%%%%%%%%%%%%%%%%%%%%%%%%%%%%%%%%%%%%%%%%%%%%%%%%%%%
%%%%%%%%%%%%%%%%%%%%%%%%%%%%%%%%%%%%%%%%%%%%%%%%%%%%%%%%%%%%%%%%
%\newpage %%%%%%%%%%%%%%%%%%%%%%%%%%%%%%%%%%%%%%%%%%%%%%%%%%%%%%%
%%%%%%%%%%%%%%%%%%%%%%%%%%%%%%%%%%%%%%%%%%%%%%%%%%%%%%%%%%%%%%%%
%%%%%%%%%%%%%%%%%%%%%%%%%%%%%%%%%%%%%%%%%%%%%%%%%%%%%%%%%%%%%%%%
%%%%%%%%%%%%%%%%%%%%%%%%%%%%%%%%%%%%%%%%%%%%%%%%%%%%%%%%%%%%%%%%
%%%%%%%%%%%%%%%%%%%%%%%%%%%%%%%%%%%%%%%%%%%%%%%%%%%%%%%%%%%%%%%%%
%%%%%%%%%%%% CHAPTER IV %%%%%%%%%%%%%%%%%%%%%%%%%%%%%%%%%%%%%%%%%
%%%%%%%%%%%%%%%%%%%%%%%%%%%%%%%%%%%%%%%%%%%%%%%%%%%%%%%%%%%%%%%%%
\section{Homogeneous spaces and nonlinear sigma models}
\label{homogeneous}
\setcounter{equation}{0}
%%%%%%%%%%%%%%%%%%%%%%%%%%%%%%%%%%%%%%%%%%%%%%%%%%%%%%%%%%%%%%%%%
This chapter offers an introduction to coset spaces and nonlinear
sigma models based on such target spaces with their possible
gaugings. The aim of this introduction is to facilitate the discussion
in the next chapter, where we explain the gauging of maximal
supergravity, concentrating  on the maximal supergravities in $D=4,5$
spacetime dimensions. As discussed earlier, these theories have a
nonlinearly realized symmetry group, equal to ${\rm E}_{7(7)}$ and
${\rm E}_{6(6)}$, respectively. It is possible to elevate the abelian
gauge group associated with the vector gauge fields to a nonabelian 
group, which is a subgroup of these exceptional groups. The
construction of these gaugings makes an essential use of 
the concepts and techniques discussed here. 

We start by introducing the concept of a coset
space G/H, where H is a subgroup of a group G. Most of this material
is standard and can be found in textbooks, such as 
\cite{Gilmore,frebook}. Then we 
discuss the corresponding nonlinear sigma models, based on homogeneous
target spaces, and present their description in a form that
emphasizes a local gauge invariance associated with the group H. Finally
we introduce the so-called {\it gauging} of this class of  nonlinear
sigma models. In this introduction we try to be as general as possible
but in the examples we restrict ourselves to
pseudo-orthogonal groups: ${\rm SO}(n)$ or the noncompact versions
${\rm SO}(p,q)$. The latter enable us to include some material on de
Sitter and anti-de Sitter spacetimes, which we make use of in later
chapters. 
%%%%%%%%%%%%%%%%%%%%%%%%%%%%%%%%%%%%%%%%%%%%%%%%%%%%%%%%%%%%%%%%%%%
\subsection{Nonlinearly realized symmetries}
\label{sec:nonl-real-symm}
As an example consider the $n$-dimensional sphere $S^n$ of unit radius
which we may embed in an $(n+1)$-dimensional real vector space ${\bf
R}^{n+1}$. The sphere is obviously invariant under SO$(n+1)$, the
group of $(n+1)$-dimensional rotations. Such invariances are called
isometries and ${\rm G}= {\rm SO}(n+1)$ is therefore known as the {\it
isometry group}. A {\it homogeneous} space is a space where every two
points can be connected by an isometry transformation. Clearly the
sphere is such a homogeneous space as every two points on $S^n$ can be
related by an ${\rm SO}(n+1)$ isometry. However, the rotation
connecting these two points is not unique as every point on $S^n$ 
is invariant under an ${\rm SO}(n)$ subgroup. This group is called the {\it
isotropy group} (or stability subgroup), denoted by H. Obviously, for a
homogeneous manifold, the isotropy groups for two arbitrary points are 
isomorphic (but not identical as one has to rotate between these
points). It is convenient to choose a certain point on the
sphere (let us call it the north pole) with coordinates in ${\bf R}^{n+1}$
given by 
$(0,\ldots,0,1)$. From the north pole, we can reach each point by a
suitable rotation. However, the north pole itself is invariant under
the \so{n} isotropy group, consisting of the following orthogonal 
matrices embedded into \so{n+1}, 
\be
\label{H-decomposition}
h =   \pmatrix{
\ast&\cdots &\ast & 0 \cr
\vdots&\ast &\vdots & \vdots \cr
\ast&\cdots &\ast & 0 \cr 
0   &\cdots   &0 & 1\cr}\,,\qquad h\in {\rm H}\,.
\ee
Therefore, if a rotation $g_1\in {\rm G}$, with ${\rm G}={\rm
SO}(n+1)$ maps the north pole onto a certain point on the sphere, then
the transformation $g_2= g_1\cdot h$ will do the same. Therefore points on
the sphere can be associated with the class of group elements $g\in G$
that are equivalent up to multiplication by elements $h\in {\rm H}$
from the {\it right}. Such equivalence classes are called cosets, and
therefore the space $S^n$ is a coset space G/H with ${\rm G}={\rm
SO}(n+1)$ and ${\rm H}={\rm SO}(n)$. The sphere is obviously just one
particular example of a homogeneous space. Every such space can be
described in terms of appropriate ${\rm G}/{\rm H}$ cosets  based on
an isometry group G and a isotropy subgroup H.

A parametrization of the cosets of ${\rm SO}(n+1)/{\rm SO}(n)$, which
assigns a single $SO(n+1)$ element to every coset, is thus equivalent
to giving  a parametrization of the sphere. It is not difficult to
find such a parametrization. One first observes that every element of
${\rm SO}(n+1)$ can be decomposed as the product of 
\be
g(\alpha) = \exp 
\pmatrix{ 
    0    &  \alpha^i \cr
\noalign{\vskip2mm} 
-\alpha_j& 0 \cr} \,, \label{coset-representative-alpha}
\ee 
with some element of ${\rm H}$. Here $i,j=1,\ldots,n$. In view of
various applications we extend our notation to noncompact 
versions of the orthogonal group, so that we can also deal with
noncompact spaces. The noncompact groups leave an indefinite metric
invariant. For ${\rm SO}(p,q)$ we have a diagonal metric with $p$
eigenvalues $+1$  and $q$ eigenvalues $-1$ (or vice versa as the overall
sign is not relevant). Using the same decomposition as in
\eqn{H-decomposition}, we choose this metric of the form 
${\rm diag}\,(\eta, 1)$, where $\eta$ is again a diagonal metric with
$p$ (or $q$) eigenvalues equal to $-1$ and $q-1$ (or $p-1$)
eigenvalues equal to $+1$. Elements of ${\rm SO}(p,q)$ thus satisfy
\be
\label{p-orthogonal}
g^{-1} = \pmatrix{\eta&0\cr 0&1\cr} g^{\rm T} \pmatrix{\eta&0\cr
0&1\cr}\,,
\ee
so that the metric $\eta$ is obviously H-invariant.  
The coset representative \eqn{coset-representative-alpha} 
satisfies the condition \eqn{p-orthogonal}, provided that
\begin{equation}
  \label{eq:noncompact-S}
  \a_i = \eta_{ij}\,\a^j\,.
\end{equation}
When $\eta$ equals the unit matrix, we are dealing with a compact
space. When $\eta$ has negative eigenvalues the matrix
\eqn{coset-representative-alpha} is no longer orthogonal and the space
will be noncompact.\footnote{%%%%%%%%%%%%%%%%%%%%%%%%%%%%%%%%%%%%%%
   Noncompact groups are not fully covered by exponentiation, such as
   in the coset
   representatives \eqn{coset-representative-alpha},
   because there are disconnected components. We also note that, in
   the noncompact case, there are other decompositions than
   \eqn{H-decomposition}, which offer distinct advantages. We will not
   discuss these issues here. } %%%%%%%%%%%%%%%%%%%%%%%%%%%%%%%%%%%
%%%%%%%%%%%%%%%%%%%%%%%%%%%%%%%%%%%%%%%%%%%%%%%%%%%%%%%%%%%%%%%%%%%
Let us mention some examples. For $\eta=-{\bf 1}$ we have the
hyperbolic space\footnote{%%%%%%%%%%%%%%%%%%%%%%%%%%%%%%%%%%%%%%%%%
   We are a little cavalier here with our terminology. 
   Also the de Sitter and anti-de Sitter spaces are hyperbolic, as we
   shall see later (cf. \eqn{embedding-condition}), but 
   they are pseudo-Riemannian. We reserve the term hyperbolic for the
   Riemannian hyperbolic space. Unlike the de Sitter and anti-de
   Sitter spaces the former spaces are double-sheeded. In
   general, coset spaces where H 
   is the maximal compact subgroup of a noncompact group G, have a
   positive or negative definite metric and are thus Riemannian
   spaces. } %%%%%%%%%%%%%%%%%%%%%%%%%%%%%%%%%%%%%%%%%%%%%%%%%%%%%%
${\rm SO}(n,1)/{\rm SO}(n)$, for $\eta=
(-,+,\cdots,+)$ we have the
de Sitter space ${\rm SO}(n,1)/{\rm SO}(n-1,1)$, and for $\eta=
(-,\cdots,-,+)$, we have the anti-de Sitter space ${\rm
SO}(n-1,2)/{\rm SO}(n-1,1)$. In this way we can thus treat a variety
of spaces at the same time. However, note that $\eta$, which eventually
will play the role of the tangent-space metric, is `mostly plus' for
de Sitter, and `mostly minus' for the hyperbolic and the anti-de
Sitter space. This aspect will be important later on when comparing
the curvature for these spaces.

The elements $g(\alpha)\in {\rm G}$ define a 
{\it representative} of the ${\rm G}/{\rm H}$ coset space,
and therefore a parametrization of the corresponding space. Of course,
the coset representative is not 
unique. We have decomposed the generators of ${\rm G}$ into
generators ${\sf h}$ of H and generators ${\sf k}$  belonging to its 
complement; the latter have been used to generate the coset representative.
In our example, the latter are associated with the last row and 
column of \eqn{coset-representative-alpha}, so that they satisfy
(schematically), 
\bea
\label{h-k-algebra}
{[{\sf h},{\sf h}]} &=& {\sf h}\,,\nonumber \\
{[{\sf h},{\sf k}]} &=& {\sf k}\,,\nonumber \\
{[{\sf k},{\sf k}]} &=& {\sf h}\,.
\eea
The first commutation relation states that the ${\sf h}$ form a subalgebra. 
The second one implies that the generators ${\sf k}$ form a 
representation of H, which ensures that in an infinitesimal
neighbourhood of a point invariant under H, the coordinates $\alpha^i$
rotate under H according to that representation. 
In the more general case the third commutator may also yield the
generators ${\sf k}$. When they do not, the homogeneous 
space is {\it symmetric}.  Obviously
the above relations involve a choice of basis;  the generators ${\sf k}$ are
defined up to additive terms belonging to elements of the algebra
associated with H. 

Let us now proceed and evaluate  \eqn{coset-representative-alpha},
\be
g(\alpha) = \pmatrix{
\delta^i_j + \alpha^i \alpha_j 
 {\displaystyle \cos \alpha -1\over \displaystyle\alpha^2} &
 {\displaystyle \sin \alpha \over \displaystyle\alpha} \alpha^i \cr
\noalign{\vskip 8mm}
- {\displaystyle\sin \alpha\over \displaystyle\alpha}\alpha_j   & 
\cos \alpha\cr}\;,
\ee
where $\alpha^2 =\eta_{ij}\, \alpha^i\alpha^j$. Obviously the space is
compact when $\eta$ is positive, because in that case the parameter
space can be restricted to $0 \leq \alpha < \pi$. This corresponds to
the sphere $S^n$. In all other cases the parameter space is obviously
noncompact and the sine and cosine may change to the hyperbolic sine and
cosine in those parts of the space where $\alpha^2$ is
negative. Observe that the appearance of $\eta$ in the above
formulae is the result of the fact that the generators ${\sf k}$ are
normalized according to ${\rm tr}({\sf k}_i\,{\sf k}_j)= - 2
\,\eta_{ij}$. 

One may choose a different parametrization of the cosets by
making a different decomposition than in
\eqn{coset-representative-alpha}. Different  parametrizations
are generally related through (coordinate-dependent) ${\rm H}$
transformations acting from the right. We may also chose different
coordinates , such as, for
instance,\footnote{%%%%%%%%%%%%%%%%%%%%%%%%%%%%%%%%%%%%%%%%%%%% 
The coordinates $(y^i,\pm\sqrt{1-y^2})$ are sometimes called {\it
homogeneous} coordinates, because the G-transformations act linearly
on these coordinates. Inhomogeneous coordinates are the ratios
$y^i/\sqrt{1-y^2}$. } %%%%%%%%%%%%%%%%%%%%%%%%%%%%%%%%%%%%%%%%%% 
\be
y^i = \alpha^i \frac{\sin \alpha}{\alpha} \;,
\ee
so that the coset representative reads
\be
g(y) = 
\pmatrix{
\delta^i_j + y^iy_j \, 
{\displaystyle \pm \sqrt{1-y^2}-1\over\displaystyle y^2} &
y^i \cr 
\noalign{\vskip 8mm}
-y_j  & \pm \sqrt{1-y^2} \cr}\,. \label{coset-representative-y}
\ee
where, depending on the sign choice, we parametrize different parts of
the space (for the sphere $S^n$, the upper or 
the lower hemisphere). For the sphere the range of the coordinates is   
restricted by $y^2 = \Sigma_i (y^i)^2 \leq 1$. Note that the 
$n\times n$  submatrix in \eqn{coset-representative-y} equals the square
root of the matrix $\delta^i_j - y^i y_j$.

One may use the coset representative to sweep out the coset space from
one point ({\it i.e.} the `north pole') in the $(n+1)$-dimensional
embedding space. Acting with \eqn{coset-representative-y} on the point
$(0,\ldots,0,1)$ yields the following coordinates in the embedding
space, 
\be
Y^A = (y^i, \pm \sqrt{1-y^2})\;.
\ee
Using \eqn{p-orthogonal} one then shows that the coset space is
embedded in $n+1$ dimensions according to 
\be
\label{embedding-condition}
\eta_{ij} \,Y^iY^j +(Y^{n+1})^2 = 1\,.
\ee

Since the $g(y)$ are contained in G one may examine the
effect of G transformations acting on $g(y)$, which will induce
corresponding transformations in the coset space. To see this, we 
multiply $g(y)$ by a constant element $o_{\rm G}\in {\rm G}$ from the
left. After this multiplication the result is in general no longer
compatible with the coset representative $g(y)$, but by applying a
suitable $y$-dependent  ${\rm H}$ transformation, $o_{\rm H}(y)$, from
the right, we can again bring $g(y)$ in the desired form. In other words,
one has 
\be
g(y) \longrightarrow o_{\rm G}\, g(y) = g(y^\prime) \,o_{\rm H}(y)
\,.  \label{G-transf}
\ee
Hence the effect is a change of coordinates $y\to y^\prime$ in a way
that satisfies the group multiplication laws. The infinitesimal
transformation $y^i\to y^{\prime \,i} = y^i + \xi^i(y)$ defines the
so-called Killing vectors $\xi^i(y)$. Writing $o_{\rm G} \approx 
{\bf 1} + \hat{\sf g}$ and $o_{\rm h} \approx {\bf 1} + 
\hat{\sf h}(y)$, we find the relation,
\be
\label{killing-1} 
\xi^i(y) \,\pa_ig(y) = \hat {\sf g}\,g(y) - g(y)\,\hat{\sf h}(y)\,.
\ee
We return to this result in the next subsection. 

Applying this construction to the case at hand, one finds that there
are two types of isometries. One corresponding to the group $\rm H$,
which changes the coordinates $y^i$ by constant rotations. The other
corresponds to $n$ coordinate dependent shifts,
\be
\delta y^i =\epsilon^i \, \sqrt{1-y^2} \,,
\ee
where the $\epsilon^i$ are $n$ constant parameters. 
As the reader can easily verify, both types of transformations take
the form of a constant $\rm G$ transformation on the embedding
coordinates $Y^A$ which leaves the 
embedding condition \eqn{embedding-condition} invariant. 

Coset representatives can be defined in different representations of
the group G. The most interesting one is the spinor
representation. Assume that we have a representation of the Clifford 
algebra ${\cal C}(p,q)$. The representation transforms (not
necessarily irreducibly) under ${\rm
SO}(p,q)$ generated by the matrices $\ft12 \G_{ij}$, but in fact it
transforms also as a spinor under ${\rm SO}(p,q+1)$, as one can verify
by including extra generators equal to the matrices $\ft12
\G_i$. Consequently we can define a representative of ${\rm
SO}(p,q+1)/{\rm SO}(p,q)$ in the spinor representation, 
\begin{equation}
  \label{eq:spinor-coset-a}
 g(\a) = \exp[ \ft12  \G_i\, \a^i] = \cos (\a/2) \,{\bf 1} + i\,
 {\sin(\a/2) \over \a}\;\a^i\G_i \,,
\end{equation}
with $\a$ defined as before, $\a^2 =\sqrt{\a_i\a^i}$, and
$\{\G_i,\G_j\}= -2\eta_{ij}\,{\bf 1}$. This construction can applied
as well to cosets of other (pseudo-)orthogonal groups.  
In terms of the coordinates $y^i$ the representative reads 
\begin{equation}
  \label{eq:spinor-coset-y}
 g(y) = \ft12  \Big[\sqrt{1+y}+\sqrt{1-y}\,\Big]  \,{\bf 1} 
+ { y^i\G_i\over \sqrt{1+y} +\sqrt{1-y}}  \,.
\end{equation}
One can act with this representative on a constant spinor, specified
at the north pole, {\it i.e.} we define $\psi(y) = g^{-1}(y)
\,\psi(0)$. The resulting $y$-dependent spinor $\psi(y)$ is a so-called 
Killing spinor of the coset space. We shall exhibit this
below. Obviously, similar results can be obtained in 
other representations of $\rm G$.
%%%%%%%%%
%%%%%%%%%%%%%%%%%%%%%%%%%%%%%%%%%%%%%%%%%%%%%%%%%%%%%%%%%%%%%%%%%%%%%%
\subsection{Geometrical quantities}
\label{sec:geom-quant}
Geometrical quantities of the homogeneous space are defined from the
left-invariant one-forms $g^{-1}{\rm d}g$, where $g(y)\in {\rm G}$, so 
that the one-forms take their value in the Lie algebra associated with
$\rm G$. It 
is convenient to use the language of differential forms,but by no 
means essential.  The exterior derivative ${\rm  d}g(y)$
describes the change of $g$ induced by an infinitesimal variation of
the coset-space coordinates $y^i$. The one-forms $g^{-1}{\rm d} g$ are
called left-invariant, because they are invariant 
under left  multiplication of $g$ with constant elements of ${\rm
  G}$. The significance of this fact will be clear in a sequel. Because
the $g$'s themselves are elements of ${\rm G}$, the one-form
$g^{-1}{\rm d}g$ takes  its values in the Lie  algebra associated
with G. Therefore the one-forms can be decomposed into the generators
${\sf h}$ and ${\sf k}$, introduced earlier, {\it i.e.}, 
\be
g^{-1} {\rm d}g = \omega + e\, , \label{e-omega}
\ee
where $\omega$ is decomposable into the generators ${\sf h}$ and $e$
into the generators ${\sf k}$. Hence $e$ defines a square matrix, with
indices $i$ 
that label the coordinates and indices $a$ that label the generators
${\sf k}$. These one-forms $e$ are thus related to the vielbeine  
of the coset space, which define a tangent frame at each point of the
space. The one-forms $\omega$ define the spin
connection\footnote{%%%%%%%%%%%%%%%%%%%%%%%%%%%%%%%%%%%%%%%%%%%%%%
  Observe that in supergravity we have defined the spin connection field
  with opposite sign.}, %%%%%%%%%%%%%%%%%%%%%%%%%%%%%%%%%%%%%%%%%%%%
associated with tangent-space rotations that belong to the
group $\rm H$. Eq. \eqn{e-omega} is of central importance for the
geometry of the coset spaces.  As a first consequence we note that the
spinor $\psi(y)$, defined with the help of the representative
\eqn{eq:spinor-coset-y} at the end of the previous subsection,
satisfies the equation, 
\be
\Big({\rm d} + \omega + e\Big)\psi(y)=0\,.
\ee
Upon writing this out in terms of the gamma matrices, one recovers
precisely the so-called Killing spinor equation
({\it c.f.} \eqn{killing-spinor}). 

Let us now proceed and investigate the properties of the one-forms
$\omega$ and $e$. In general it is not necessary to specify the coset
representative, as different representatives are related by
$y$-dependent $\rm H$ transformation acting from the right on $g$,
{\it i.e.},  
\be
\label{local-H-transf} 
g(y)  \longrightarrow g(y) \,h(y)\,, \qquad h(y) \in {\rm H} \,. 
\ee
This leads to a different parametrization of the coset space. It 
is straightforward to see how $\omega$ and $e$ transform under
\eqn{local-H-transf}, 
\be
(\omega + e) \longrightarrow h^{-1} (\omega +e)h + h^{-1} {\rm d}h\, .
\ee
This equation can again be decomposed (using the first two relations
\eqn{h-k-algebra} in terms of the generators ${\sf h}$
and ${\sf k}$,  which yields  
\bea
\omega_i &{\longrightarrow}  & h^{-1}\omega_i h +  h^{-1}  
\partial_i h \,, \nonumber \\
e_i & {\longrightarrow} & h^{-1} e_i h\,.
\eea
Obviously, $\omega$ acts as a gauge connection for the local $\rm H$
transformations. 
Furthermore it follows from \eqn{e-omega-sphere} that $\omega_i$ and $e_i$
transform as covariant vectors under coordinate transformations, i.e.
\bea
y^i & {\longrightarrow} & y^i + \xi^i \,,\nonumber \\
\omega_i & {\longrightarrow} & \omega_i - \partial_i \xi^j \omega_j - \xi^j 
\partial_j \omega_i \,, \nonumber \\
e_i & {\longrightarrow} & e_i - \partial_i \xi^j e_j - \xi^j
\partial_j e_i \, .
\eea

We can also define Lie-algebra valued curvatures associated with
$\omega_i$ and $e_i$, 
\bea
\label{G/H-curvatures}
R_{ij}({\rm H})  & = & \partial_i \omega_j -\partial_j\omega_i 
+ [ \omega_i , \omega_j] \,, \nonumber \\
R_{ij}({\rm G/H}) & = & \partial_i e_j - \partial_j e_i + 
[\omega_i, e_j] -[\omega_j, e_i]\,.  \label{curv-omega-e}
\eea
Introducing ${\rm H}$-covariant derivatives, we note the relations 
\begin{eqnarray}
  \label{eq:cov-G/H}
  {[D_i,D_j]} &=& - R_{ij}({\rm H}) \,,\nonumber \\
  R_{ij}({\rm G/H}) &=&  D_i e_j - D_j e_i \,.  
\end{eqnarray}
The values of these curvatures follow from the
Cartan-Maurer equations. To derive these equations we take the
exterior derivative of the defining relation \eqn{e-omega}, 
\begin{equation}
{\rm d} (g^{-1} {\rm d}g) =- (g^{-1}{\rm d}g) \wedge (g^{-1} {\rm d}g) \,,  
\end{equation}
or, in terms of $\o$ and $e$,
\begin{equation}
{\rm d}(\o +e) = - (\o+e)\wedge (\o + e)\,. 
\end{equation}
Decomposing this equation in terms of the Lie algebra generators,
using the relations \eqn{h-k-algebra}, we find 
\be
\label{CM-omega-e}
R_{ij}({\rm H})   =   - [ e_i,e_j] \, ,\qquad  R_{ij}({\rm G/H}) = 0\,.  
\ee
Note that the vanishing of $R_{ij}(G/H)$ is a consequence of the fact
that we assumed that the coset space was {\it symmetric} (see the text
below \eqn{h-k-algebra}). 

As we already alluded to earlier, the fields $e_i$ can be decomposed
into the generators ${\sf k}$ and thus define a set of vielbeine
$e_i^{\,a}$ that specify a tangent frame at each point in the coset
space. In the context of differential geometry the indices $i$ are
called world indices, because they refer to the coordinates of a
manifold, whereas the indices $a, b, \ldots$ that label the generators
${\sf k}$ are called tangent-space indices (or local Lorentz indices  
in the context of general relativity). Because the generators ${\sf
k}$ form a representation of $\rm H$, this group rotates the tangent
frames. Usually the group $\rm H$ can be embedded into ${\rm SO}(n)$ (or a
noncompact version thereof) and leaves some target-space
metric invariant (we will see the importance of this fact shortly). 
The quantity $\omega_i$ thus acts as the connection associated
with rotations of the tangent frames, and therefore we call it the
spin connection. 

These aspects are easily recognized in the examples we are discussing,
because the group $\rm H$ was precisely the (pseudo)orthogonal
group. Hence, using the same matrix decomposition as before, we find
explicit expressions for the vielbein and the spin connection,
\be
g^{-1}{\rm d}g = \pmatrix{
\omega_{i}(y)\,{\rm d}y^i & e_{i}(y)\,{\rm d}y^i \cr 
\noalign{\vskip 6mm}
- e_i(y)\, {\rm d}y^i  & 0 \cr} \,. \label{e-omega-sphere}
\ee
From \eqn{coset-representative-y} one readily obtains
\bea
\label{e-omega-components}
\omega_i^{ab} & = & \Big( y^a\,\delta^b_i - y^b\,\delta^a_i \Big)
\frac{1 \mp \sqrt{1-y^2}}{y^2} \,, \nonumber \\
e_i^{\,a} & = & \delta^a_i + \frac{y_i\,y^a}{y^2} 
\Big( \pm \frac{1}{\sqrt{1-y^2}} - 1 \Big)\, , 
\eea
where, as before, indices are raised and lowered with $\eta$. Note
that $\omega^{ab}_i$ is antisymmetric in $a,b$, which follows from
the (pseudo)orthogonality of $\rm H$. The inverse vielbein reads
\be
e_a^{\,i} =  \delta_a^i + \frac{y_a\,y^i}{y^2} 
\Big(\pm \sqrt{1-y^2}-1\Big)\, , 
\ee
Furthermore, the curvatures
introduced before, are readily identified with the curvature of the
spin connection and with the torsion tensor. From the Cartan-Maurer
equations, explained above, we thus find in components,
\be
R_{ij}^{ab}(\omega) =   2\,e_{[i}^a \,e_{j]}^b \,, \qquad 
D^{~}_ie_j^{\,a}- D^{~}_je_i^{\,a} = 0 \,.
\ee
To define a metric $g_{ij}$ one contracts an ${\rm H}$-invariant
symmetric rank-2 tensor with the vielbeine. The obvious invariant
tensor is $\eta_{ab}$, so that 
\be
g_{ij} = \eta_{ab}\, e_i^{\,a} e_j^{\,b} \,. \label{coset-space-metric}
\ee
When there are several $\rm H$-invariant tensors, there is a more
extended class of metrics that one may consider, but in the case at
hand the metric is unique up to a proportionality factor. In the 
parametrization \eqn{e-omega-components} one obtains for the metric
and its inverse,
\be
g_{ij} = \eta_{ij} + \frac{y_i \,y_j}{1-y^2}\,,\qquad g^{ij} =
\eta^{ij} - y^i \,y^j \, .  
\ee
Given the fact that we have already made a choice for $\eta$
previously, a `mostly plus' metric requires to include a minus sign in
the definition \eqn{coset-space-metric} for the hyperbolic and anti-de
Sitter spaces. This sign is important when comparing to spheres
or de Sitter spaces.   

{From} the vielbein postulate, we know that the affine connection is
equal to $\Gamma_{ij}{}^{\!k} =  e_a{}^{\!k} \,D_i e_j{}^{\!a}$, from which we can
define the Riemann curvature. For the examples at hand, this leads to 
\be
\Gamma_{ij}{}^{\!k}   = y^k g_{ij} .
\ee
Because the torsion is zero, the connection coincides with Christoffel
symbol. 
Because $\omega_i$ differs in sign as compared
to the spin connection used in section~3.1, the Riemann tensor
is equal to minus the curvature $R_{ij}^{ab}(\omega)$, upon
contraction with  $\eta_{ac}\, e^c_k\, e_b^{\,l}$, and we find the 
following result for the Riemann curvature tensor,
\be
\label{R-curvature}
R_{ijk}{}^l =  -g^{~}_{ki}\, \d_{j}^l +g^{~}_{kj}\, \d_{i}^l \,,
\ee
where $g_{ij}$ is the metric tensor defined by
\eqn{coset-space-metric}. Thus the curvature is of definite sign,
but we stress that this is related to the signature choice that we made
for the metric, as we have discussed above. 
The Riemann curvature \eqn{R-curvature} is proportional to the
metric, which indicates that we are dealing with a maximally symmetric
space. This means that the maximal number of isometries (equal to
$\ft12 n(n+1)$) is realized for this space. 

All coset spaces have isometries corresponding to the
group $\rm G$. The diffeomorphisms associated with these 
isometries are generated by Killing vectors $\xi^i(y)$, which we
introduced earlier. Combining \eqn{killing-1} with \eqn{e-omega}, we
obtain,   
\be
\label{killing-2} 
\xi^i(y) \,(\o_i(y) + e_i(y)) = g^{-1}(y) \hat {\sf g}\,g(y) -
\hat{\sf h}(y)\,. 
\ee
Decomposing this equation according to the Lie algebra, we find
\bea
\label{killing-3}
\xi^i(y)\,e_i(y) &=& {\tilde {\sf g}}(y) \,,\nonumber \\ 
\hat{\sf h}(y) &=& -\xi^i(y) \,\o_i(y) + {\tilde {\sf h}}(y)  \,,
\eea
where
\bea 
{\tilde {\sf h}}(y) &=& \Big[g^{-1}(y) \,\hat {\sf
g}\,g(y)\Big]_{\rm H} \,, \nn\\
{\tilde {\sf g}}(y) &=& \Big[g^{-1}(y) \,\hat {\sf
g}\,g(y)\Big]_{\rm G/H} \,.
\eea
The contribution $\hat {\sf h}(y)$ is only relevant for those
quantities that live in the tangent space. 

Now we return to the observation that the left-invariant forms, from
which $e$ and $\o$ were constructed, are invariant
under the group G. Therefore, it follows that $e$ and $\o$ are both
invariant as well. Moreover, we established ({\it c.f.}
\eqn{G-transf}) that 
the G-transformation acting on the left can be decomposed into a
diffeomorphism combined with a coordinate-dependent H-transformation.
Therefore, the vielbein $e$ and the spin connection $\o$ are invariant
under these combined transformations. Since the metric is H-invariant
by construction, it thus follows that the metric is invariant under
the diffeomorphism associated with G. Hence,
\be
\label{killing-0}
\delta g_{ij} = D_i \xi_j (y) + D_j \xi_i (y)= 0\,,
\ee
where $\xi_i = g_{ij} \xi^j$ and $\xi^i$ is the so-called Killing
vector defined by \eqn{killing-3}. For the vielbein and spin
connection, which transform under $\rm H$, we find
\bea
\pa_i\xi^j \,e_j + \xi^j\,\pa_je_i + [e_i, \hat {\sf h}(y)] &=&
0\,,\nonumber \\ 
\pa_i\xi^j \,\o_j + \xi^j\,\pa_j\o_i  +\pa_i\hat {\sf h}(y) + [\o_i,
\hat {\sf h}(y)]&=& 0\,. 
\eea 
In terms of ${\tilde {\sf h}}(y)$ these results take a more
covariant form, 
\bea
\pa_i\xi^j \,e_j + \xi^j\,D_je_i + [e_i, {\tilde {\sf h}}(y)] &=&
0\,,\nonumber \\ 
D_i {\tilde {\sf h}}(y) &=& R_{ij}(H)\,\xi^j \,. 
\eea 
Combining the first equation with the first equation\eqn{killing-3}
yields
\be
R_{ij}(G/H) \,\xi^j + [e_i, {\tilde {\sf g}}(y)]_{G/H} = 0\,.
\ee
Observe that both terms vanish separately for a symmetric space.

The diffeomorphisms generated  by the Killing vector fields will
give rise to the group G. This follows from \eqn{killing-1}. Let us label
the generators of the group G by indices $\a,\b,\ldots,$ and introduce
structure constants by
\be 
[\hat {\sf g}_\a , \hat {\sf g}_\b]= f_{\a\b}{}^{\!\g}\,\hat {\sf
g}_\g\,.
\ee
The Killing vectors and the corresponding H-transformations then
satisfy corresponding group multiplication properties,
\bea
\xi_\b^j\, \pa_j\xi^i_\a - \xi_\a^j\, \pa_j\xi^i_\b &=&
f_{\a\b}{}^{\!\g} \,\xi^i_\g\,,  \nonumber \\
{}[{ \tilde{\sf h}}_\a ,{\tilde {\sf h}}_\b] &=& 
f_{\a\b}{}^{\!\g} \, {\tilde {\sf h}}_\g + \xi^i_\a\,\xi^j_\b\,
R_{ij}(H) \,.  
\eea
One can consider fields on the coset space, which are functions of the
coset space coordinates assigned to a representation of the group
H. On such fields the isometries are generated by the operators,
\be
\label{G-isometry}
-\xi^i_\a D_i + {\tilde {\sf h}}_\a\,.
\ee
On the basis of the results above one can show that these operators
satisfy the commutation relations of the Lie algebra associated with
the isometry group G. To show this we note the identity
\bea
\label{double-isometry}
(-\xi^i_\a D_i + {\tilde {\sf h}}_\a) (-\xi^i_\b D_i + {\tilde
{\sf h}}_\b) &\!\!=\!\!& \xi^i_\a \xi^j_\b (D_i D_j - R_{ij}(H) )
+ { \tilde{\sf h}}_\a\, { \tilde{\sf h}}_\b  \nn\\
&&
+ [\xi^i_\a (D_i\xi^j_\b) + { \tilde{\sf h}}^{~}_\a \,\xi^j_\b  +
{\tilde{\sf h}}^{~}_\b \,\xi^j_\a  ] D_j  \,.\;{~~~}
\eea
%%%%%%%%%%%%%%%%%%%%%%%%%%%%%%%%%%%%%%%%%%%%%%%%%%%%%%%%%%%%%%%%%%%%% 
%%%%%%%%%%%%%%%%%%%%%%%%%%%%%%%%%%%%%%%%%%%%%%%%%%%%%%%%%%%%%%%%%%%%%
\subsection{Nonlinear sigma models with homogeneous target space}
\label{sec:nonl-sigma-models}
%%%%%%%%%%%%%%%%%%%%%%%%%%%%%%%%%%%%%%%%%%%%%%%%%%%%%%%%%%%%%%%%%%%%%
It is now rather straightforward to describe a nonlinear sigma model
based on a homogeneous target space by making use of the above
framework. One starts from scalar fields which take 
their  values in the homogeneous space, so that the fields $\phi^i
(x)$ define a map from the spacetime to the coset space. Hence we may
follow the same procedure as before and define a coset representative
$\vv(\phi^i(x))\in{\rm G}$, which now depends on $n$
fields. Subsequently, one uses the analogue of \eqn{e-omega}, to
define Lie-algebra valued quantities ${\cal Q}_\m$ and ${\cal P}_\m$, 
\be
\vv^{-1}\partial_\mu\vv = {\cal Q}_\m + {\cal P}_\m \,,
\ee
where ${\cal Q}_\m$ is decomposable into the generators ${\sf h}$ and
${\cal P}_\m$ into the generators ${\sf k}$. Obviously one has the
relations 
\be
\label{Q-P-decomposition} 
{\cal Q}_\mu(\phi )  =  \omega_i(\phi )\, \partial_\mu \phi^i \,, 
\qquad 
{\cal P}_\mu (\phi ) =  e_i (\phi ) \, \partial_\mu \phi^i
\,. 
\ee
The above expressions
show that ${\cal Q}_\m$ and ${\cal P}_\m$ are just the pull backs of
the target space connection and vielbein to the spacetime.  

The local ${\rm H}$ transformations depend on the fields $\phi(x)$ and
thus indirectly on the spacetime coordinates. Therefore one may
elevate these transformations to transformations that
depend arbitrarily on $x^\m$. Under such transformations we have 
\be
\vv(\phi) \to \vv(\phi)\,h(x)\,. \label{eq:local-H}
\ee
By allowing ourselves to perform such local gauge
transformations, we introduced new degrees of freedom into $\vv$
associated with the group H. Eventually we will fix this gauge
freedom, but until that point $\vv$ will just be an unrestricted
spacetime dependent element of the group G. After imposing the gauge
condition on $\vv(x)$ one obtains the coset representative
$\vv(\phi(x))$. From \eqn{eq:local-H} we derive the following
local H-transformations, 
\bea
{\cal Q}_\mu(x) & \longrightarrow & h^{-1}(x) \,{\cal Q}_\mu(x)\, h(x)
+ h^{-1}(x) \,\partial_\mu h(x)\, , \nonumber \\ 
{\cal P}_\mu(x) & \longrightarrow& h^{-1}(x)\, {\cal P}_\mu(x)\, h(x) \, .
\eea
Hence ${\cal Q}_\mu$ acts as a gauge field associated with the local 
$\rm H$ transformations. Furthermore both ${\cal P}_\m$ and ${\cal
Q}_\m$ are invariant under rigid G-transformations. 
It is convenient to introduce a corresponding H-covariant derivative,
\be
D_\mu \vv = \partial_\mu \vv - \vv {\cal Q}_\m\,,
\ee
so that \eqn{Q-P-decomposition}  reads
\be
\vv^{-1}D_\m\vv ={\cal P}_\m  \,.  \label{Dvv}
\ee

Just as before, one derives the Cartan-Maurer equations \eqn{CM-omega-e}, 
\bea
F_{\mu\nu}({\cal Q})& = & \partial_\mu {\cal Q}_\nu - \partial_\nu 
{\cal Q}_\mu +[ {\cal Q}_{\mu}, {\cal Q}_{\nu }]  =  - [{\cal
P}_{\mu},  {\cal P}_{\nu }]  \,,  \nonumber \\
D_{[ \mu} {\cal P}_{\nu]} & = & \pa_{[\m}{\cal P}_{\n]} + [
{\cal Q}_{[\m}, {\cal P}_{\n]}] = 0 \,. \label{CM-Q-P}
\eea
Here we made use of the commutation relations \eqn{h-k-algebra}

There are several ways to write down the Lagrangian of the
corresponding nonlinear sigma model. Obviously the Lagrangian must be
invariant under both the rigid G transformations and the local H
transformations. Hence we write
\be
\lagr = \ft12 {\rm tr} \, \Big[D_\mu \vv^{-1}\, D^\mu \vv \Big]\, . 
\ee
One can interpret this result in a first- and in a second-order
form. In the first one regards the gauge field ${\cal Q}_\m$ as an 
independent field, whose field equations are algebraic and are 
solved by \eqn{Dvv}. After substituting the result one obtains the
second-order form, which presupposes \eqn{Dvv} from the beginning. The
result can be written as,
\be
\lagr =  - \ft{1}{2} {\rm tr}\,\Big[ {\cal P}_\mu \,{\cal P}^\mu\Big] \,.
\label{eq:AA-lagrangian} 
\ee
Clearly this Lagrangian is invariant under the group G. At this stage
one still has the full gauge invariance with respect to 
local ${\rm H}$ transformations and one can impose a gauge restricting
$\vv$ to a coset representative. When this is not done, the
theory is invariant under 
${\rm G}_{\rm rigid}\times {\rm H}_{\rm local}$  
with both groups acting {\it linearly}. However, as soon as one
imposes a gauge and restricts $\vv$ to a coset representative
parametrized by certain fields $\phi^i$, the
residual subgroup is such that the $\rm H$ transformations are linked
to the G transformations and depend on the fields
$\phi^i$. This combined subgroup still generates a representation
of the group G, but it is now realized in a nonlinear fashion. In order
to deal with complicated supergravity theories that involve
homogeneous spaces, the strategy is to postpone this gauge choice till
the end, so that one is always dealing with a manifest linearly
realized symmetry group 
${\rm G}_{\rm rigid}\times {\rm H}_{\rm local}$. As we intend to
demonstrate, this strategy allows for a systematic approach, whereas the
gauge-fixed approach leads to unsurmountable difficulties (at
least, for the spaces of interest). It is straightforward to demonstrate
that \eqn{eq:AA-lagrangian} leads to the standard form of the
nonlinear sigma model, 
\be
\label{target-space-form}
\lagr =   - \ft12  g_{ij}(\phi )\, \partial_\mu \phi^i
\,\partial^\mu\phi^j \,, 
\ee
where the target space metric is given by \eqn{coset-space-metric}. In
this form the local H invariance is absent, but the invariance under G
is still there and realized as target space isometries generated by
corresponding Killing vectors.

It is easy to see how to couple matter fields to the
sigma model in a way that the invariance under the isometries remains 
unaffected. Matter fields are assigned to a representation of the
local H group, so that they couple to the sigma model fields through
the connection 
${\cal Q}_\m$ that appears in the covariant derivatives. Usually the
fields will remain invariant under the group G as long as one does not
fix the gauge and choose a specific coset representative. Also here we
can proceed in first- 
or second-order formalism. In first-order form the equation 
\eqn{Q-P-decomposition} will acquire some extra terms that depend on
the matter fields. Upon choosing a gauge, the matter fields transform
nonlinearly under the group G with transformations that take the form
of $\phi$-dependent H-transformations, determined by \eqn{killing-3}. 
However, gauge fields cannot couple in this way as their gauge
invariance would be in conflict with the local invariance under the
group H. Therefore, gauge fields have to transform under the rigid
group G. 

We emphasize that the presentation that we followed so far was rather
general; the maximally symmetric spaces that we considered served
only as an example. In supergravity we are often dealing with sigma
models based on homogeneous, symmetric target spaces. These target spaces
are usually noncompact and Riemannian, so that H is the maximally
compact subgroup of G. Later in this chapter we will be dealing with 
the ${\rm E}_{7(7)}/{\rm SU}(8)$ and ${\rm E}_{6(6)}/{\rm USp}(8)$
coset spaces. The exceptional groups are noncompact and are divided by
their maximal compact subgroups. The corresponding spaces have
dimension 70 and 42, respectively. 

%%%%%%%%%%%%%%%%%%%%%%%%%%%%%%%%%%%%%%%%%%%%%%%%%%%%%%%%%%%%%%%%
\subsection{Gauged nonlinear sigma models}
\label{sec:gaug-nonl-sigma}
%%%%%%%%%%%%%%%%%%%%%%%%%%%%%%%%%%%%%%%%%%%%%%%%%%%%%%%%%%%%%%%%
Given a nonlinear sigma model with certain isometries, one can  gauge
some or all of these isometries in the usual way: one elevates the
parameters of the isometry group (of a subgroup thereof) to arbitrary
functions of the spacetime coordinates and introduces the necessary
gauge fields (with their standard gauge-invariant Lagrangian
containing a kinetic term) and corresponding covariant derivatives. As
explained above, for 
sigma models based on homogeneous target spaces one can proceed in a
way in which all transformations remain linearly realized. To adopt
this approach is extremely important for the construction of gauged
supergravity theories, as we will discuss in the next section. We will
always use the second-order formalism so that the H-connection ${\cal
Q}_\m$ will not be an independent field. 

Since these new gauge transformations involve the isometry group they
must act on the group element $\vv$ as a subgroup of G. Hence the
covariant derivative of $\vv$ is now changed by the addition of the
corresponding (dynamical) gauge fields $A_\m$ which take their
values in the corresponding Lie algebra (which is a subalgebra of the
Lie algebra associated with G). Hence,
\begin{equation}
  \label{eq:gauged-cov-der}
D_\mu \vv(x) = \partial_\mu \vv(x) - \vv(x)\, {\cal Q}_\m(x)  -
g\,A_\m(x)  \,\vv(x) \,,
\end{equation}
where we have introduced a coupling constant $g$ to keep track of the
new terms introduced by the gauging. 
With this change, the expressions for ${\cal Q}_\m$ and ${\cal P}_\m$
will change. They remain expressed by \eqn{Dvv}, but the
derivative is now covariantized and modified by the terms depending on
the new gauge fields $A_\m$. The consistency of this procedure is
obvious as \eqn{Dvv} is fully covariant. Of course, the original rigid
invariance under G transformations from the left is now broken by the
embedding of the new gauge group into G. 

The modifications caused by the new minimal couplings are minor and
the effects can be concisely summarized by the Cartan-Maurer
equations, 
\bea
 \label{CM-gauged} 
{\cal F}_{\mu\nu}({\cal Q})& = & \partial_\mu {\cal Q}_\nu -
\partial_\nu {\cal Q}_\mu + [{\cal Q}_{\mu}, {\cal Q}_{\nu} ]\,,
\nn\\ 
&=& 
[{\cal  P}_{\mu}, {\cal P}_{\nu}]  -g \Big[\vv^{-1}
F_{\m\n}(A)\vv\Big]_{\rm H} \,,   
\nonumber \\
D_{[ \mu} {\cal P}_{\nu]} & = & -\ft12 g \Big[\vv^{-1}
F_{\m\n}(A)\vv\Big]_{\rm G/H} \,.
\eea
Because ${\cal Q}_\m$ and ${\cal P}_\m$ now depend on the gauge
connections $A_\m$, according to
\begin{eqnarray}
  \label{eq:extra-delta-P/Q}
{\cal Q}_\m = {\cal Q}_\m^{(0)} - g[\vv^{-1} A_\m \vv]_{\rm H}\,,
\qquad  
{\cal P}_\m = {\cal P}_\m^{(0)} - g[\vv^{-1} A_\m \vv]_{\rm G/H}\,.
\end{eqnarray}
When imposing a gauge condition, the last result for ${\cal
P}_\m$ exhibits precisely the Killing 
vectors \eqn{killing-3} (in the gauge where $\vv$ equals the coset
representative). When gauging isometries in a generic nonlinear sigma
model ({\it c.f.} \eqn{target-space-form}), one replaces the derivatives
according to $\pa_\m\phi \to \pa_\m\phi^i  -A_\m \xi^i(\phi)$, where
for simplicity we assumed a single isometry. The modifications in the
matter sector arise through the order $g$ contributions to 
${\cal Q}_\m$.  
Note that ${\cal P}_\m$ and ${\cal Q}_\m$ are invariant under the
new gauge group (but transform under local H-transformations, as
before). In the next subsection we will discuss the application of
this formulation to gauged supergravity. 
%%%%%%%%%%%%%%%%%%%%%%%%%%%%%%%%%%%%%%%%%%%%%%%%%%%%%%%%%%%%%%%%%%%

%%%%%%%%%%%%%%%%%%%%%%%%%%%%%%%%%%%%%%%%%%%%%%%%%%%%%%%%%%%%%%%%
%%%%%%%%%%%%%%%%%%%%%%%%%%%%%%%%%%%%%%%%%%%%%%%%%%%%%%%%%%%%%%%%
%\newpage %%%%%%%%%%%%%%%%%%%%%%%%%%%%%%%%%%%%%%%%%%%%%%%%%%%%%%%
%%%%%%%%%%%%%%%%%%%%%%%%%%%%%%%%%%%%%%%%%%%%%%%%%%%%%%%%%%%%%%%%
%%%%%%%%%%%%%%%%%%%%%%%%%%%%%%%%%%%%%%%%%%%%%%%%%%%%%%%%%%%%%%%%
%%%%%%%%%%%%%%%%%%%%%%%%%%%%%%%%%%%%%%%%%%%%%%%%%%%%%%%%%%%%%%%%%
%%%%%%%%%%%% CHAPTER V %%%%%%%%%%%%%%%%%%%%%%%%%%%%%%%%%%%%%%%%%
%%%%%%%%%%%%%%%%%%%%%%%%%%%%%%%%%%%%%%%%%%%%%%%%%%%%%%%%%%%%%%%%%
\section{Gauged maximal supergravity in 4 and 5 dimensions}
\setcounter{equation}{0}
%%%%%%%%%%%%%%%%%%%%%%%%%%%%%%%%%%%%%%%%%%%%%%%%%%%%%%%%%%%%%%%%%
%%%%%%%%%%%%%%%%%%%%%%%%%%%%%%%%%%%%%%%%%%%%%%%%%%%%%%%%%%%%%%%%
\label{sec:maximal-gauged-45}
%%%%%%%%%%%%%%%%%%%%%%%%%%%%%%%%%%%%%%%%%%%%%%%%%%%%%%%%%%%%%%%%
The maximally extended supergravity theories introduced in chapter~3
were obtained by dimensional reduction from 11-dimensional
supergravity on a hypertorus. In these theories the scalar
fields parametrize a G/H coset space 
({\it c.f.} table~\ref{maximal-cosets}) and the group G is also
realized as a symmetry of the full theory. Generically the fields
transform as follows. The graviton is invariant, the (abelian) 
gauge fields transform linearly under G and the fermions transform
linearly under the group H. However, in some dimensions the
G-invariance is not realized at the level of the action, but at the
level of the combined field equations and Bianchi identities. For
example, in 4 dimensions the 28 abelian  
vector fields do not constitute a representation of the group ${\rm
E}_{7(7)}$. In this case the group G is realized by electric-magnetic
duality and acts on the field strengths, rather than on the vector
fields. We return to electric-magnetic duality in
section~\ref{elec-magn}. 

It is an obvious question whether these theories allow an extension in
which the abelian gauge fields are promoted to nonabelian ones. This
turns out to be 
possible and the corresponding theories are known as gauged
supergravities. They contain an extra parameter $g$, which is the
gauge coupling constant. Supersymmetry requires the presence of
extra terms of order $g$ and $g^2$ in the Lagrangian. Apart from the
gauge field interactions there are fermionic masslike terms of order
$g$ and a scalar potential of order $g^2$. The latter may give rise
to groundstates with nonzero cosmological constant. To explain the
construction of gauged supergravity theories, we concentrate on the
maximal gauged supergravities in $D=4$ and 5 spacetime dimensions. 
An obvious gauging in $D=4$ dimensions is based on the group 
${\rm SO}(8)$, as the Lagrangian has a manifest ${\rm SO}(8)$
invariance and there are precisely 28 vector fields
\cite{deWitNic}. This gauging has 
an obvious Kaluza-Klein origin, and arises when compactifying seven
coordinates of $D=11$ supergravity on the sphere $S^7$, which has an
${\rm SO}(8)$ isometry group. The group emerges as the gauge group
of the compactified theory formulated in 4 dimensions. In this theory
the ${\rm E}_{7(7)}$ invariance group is broken to a local 
${\rm SO}(8)$ group so that the resulting theory is 
invariant under ${\rm SU}(8)_{\rm local}\times {\rm
SO}(8)_{\rm local}$. In this compactification the four-index field
strength acquires a nonzero values when all its indices are in the
four-dimensional spacetime. However, it turns out that many  
other subgroups of ${\rm E}_{7(7)}$ can be gauged. 
 
%%%%%%%%%%%%%%%%%%%%%%%%%%%%%%%%%%%%%%%%%%%%%%%%%%%%%%%%%%%%%%
%%%%%%%%%%%%%%%%%%%%%%%%%%%%%%%%%%%%%%%%%%%%%%%%%%%%%%%%%%%%%%
\begin{table}
\begin{center}
\begin{tabular}{l c c c c c  }\hline
\\ ~&~&~&~&~&~  \\[-7mm]
 ~ &$e_\m^{\,a}$& $\psi_\m^i$ & ${\cal F}_{\m\n};{\cal
G}_{\m\n}; {\cal H}_{\m\n}$ & $\chi^{ijk}$ &
$u_{ij}{}^{\!IJ}; v^{ijIJ}$ \\[.5mm] \hline 
~&~&~&~&~&~  \\[-3mm]
${\rm SU}(8)$ &{\bf 1} &{\bf 8}  &{\bf 1} &{\bf 56} & ${\bf
28}+\overline{\bf 28}$   \\[.5mm]
${\rm E}_{7(7)}$ &{\bf 1} &{\bf 1}& ${\bf 56}$ &{\bf 1} &${\bf 56}$ 
\\ \hline
~&~&~&~&~&~  \\[-3mm]
${\rm USp}(8)$   & {\bf 1} & {\bf 8}&{\bf 1}&{\bf 48} &${\bf
27}+\overline{\bf 27}$   \\[.5mm]
${\rm E}_{6(6)}$ & {\bf 1} &{\bf 1} & ${\bf 27}$ & {\bf 1}&  ${\bf 27}+
\overline {\bf 27}$   
   \\ \hline
\end{tabular}
\end{center}
\caption{\small
Representation assignments for the various supergravity
fields with respect to the groups G and H. In $D=4$ dimensions these
groups are ${\rm E}_{7(7)}$ and ${\rm SU}(8)$, respectively. In $D=5$
dimensions they are ${\rm E}_{6(6)}$ and ${\rm USp}(8)$. Note that the
tensors ${\cal F}_{\m\n}$, ${\cal G}_{\m\n}$ and/or ${\cal H}_{\m\n}$
denote the field strengths of the vector fields and/or (for $D=5$)
possible tensor fields. }\label{G/H-max-gauged} 
\end{table}
%%%%%%%%%%%%%%%%%%%%%%%%%%%%%%%%%%%%%%%%%%%%%%%%%%%%%%%%%%%%%%
%
%%%%%%%%%%%%%%%%%%%%%%%%%%%%%%%%%%%%%%%%%%%%%%%%%%%%%%%%%%%%%%

In $D=5$ dimensions the possible gaugings are not immediately clear, as
there is no obvious 27-dimensional gauge group. Again the Kaluza-Klein
scenario can serve as a guide. While $D=11$ supergravity has
no obvious compactification to five dimensions, type-IIB supergravity 
has a compactification on the sphere $S^5$. In this
solution the five-index (self-dual) field strength acquires a nonzero
value whenever the five indices take all values in either $S^5$ or in
the five-dimensional spacetime. Type-IIB supergravity has a manifest 
${\rm SL}(2)$ invariance and the isometry group of $S^5$ is 
${\rm SO}(6)$, so that the symmetry group of the
Lagrangian equals the ${\rm SL}(2)\times {\rm SO}(6)$ subgroup of ${\rm
E}_{6(6)}$, 
where ${\rm SO}(6)$ is realized as a local gauge group. This implies
that 15 of the 27 gauge fields become associated with the nonabelian
group ${\rm SO}(6)$, which leaves 12 
abelian gauge fields which are charged with respect to the same
group. This poses an obvious problem, as the abelian 
gauge transformations of these 12 fields will be in conflict with
their transformations under the ${\rm SO}(6)$ gauge group. The
solutions is to convert these 12 gauge fields to antisymmetric tensor
fields. The Lagrangian can thus be written in a form that is invariant
under ${\rm USp}(8)_{\rm local}\times {\rm SO}(6)_{\rm local}\times
{\rm SL}(2)$ \cite{GunaRomansWarner}. Also in 5 dimensions other
gauge groups are possible. We will briefly comment on this issue at
the end of the chapter. 

Before continuing with supergravity we first discuss some basic
features of the two coset spaces ${\rm E}_{7(7)}/{\rm SU}(8)$ and
${\rm E}_{6(6)}/{\rm USp}(8)$. Both these exceptional
Lie groups can be introduced in terms of 56-dimensional
matrices.\footnote{%%%%%%%%%%%%%%%%%%%%%%%%%%%%%%%%%%%%%%%%%%%%%%%%
  Strictly speaking the isotropy groups are ${\rm SU}(8)/{\bf Z}_2$ 
  and ${\rm USp}(8)/{\bf Z}_2$. }  %%%%%%%%%%%%%%%%%%%%%%%%%%%%%%%%

%%%%%%%%%%%%%%%%%%%%%%%%%%%%%%%%%%%%%%%%%%%%%%%%%%%%%%%%%%%%%%%%%%%
%%%%%%%%%%%%%%%%%%%%%%%%%%%%%%%%%%%%%%%%%%%%%%%%%%%%%%%%%%%%%%%%%%%
%%%%%%%%%%%%%%%%%%%%%%%%%%%%%%%%%%%%%%%%%%%%%%%%%%%%%%%%%%%%%%%%%%%
\subsection{On ${\rm E}_{7(7)}/{\rm SU}(8)$ and ${\rm
    E}_{6(6)}/{\rm USp}(8)$ cosets}
\label{sec:E7,6-cosets}
%%%%%%%%%%%%%%%%%%%%%%%%%%%%%%%%%%%%%%%%%%%%%%%%%%%%%%%%%%%%%%%%%%%
We discuss the ${\rm E}_{7(7)}$ and ${\rm E}_{6(6)}$ on a par for
reasons that will become obvious. To define the groups we consider
the fundamental representation, acting on a pseudoreal vector $(z_{IJ},
z^{KL})$ with $z^{IJ} = (z_{IJ})^\ast$, where the indices are
antisymmetrized index pairs $[IJ]$ and $[KL]$ and $I,J,K,L =
1,\ldots,8$. Hence the $(z_{IJ},z^{KL})$ span a 56-dimensional vector
space. Consider now infinitesimal transformations of the form,
\begin{eqnarray}
  \label{eq:delta-z}
\d z_{IJ} &=& \Lambda_{IJ}{}^{\!KL} \,z_{KL} + \Sigma_{IJKL}
\,z^{KL}\,,\nn\\
\d z^{IJ} &=& \Lambda^{IJ}{}_{\!KL} \,z^{KL} + \Sigma^{IJKL}
\,z_{KL}\,.
\end{eqnarray}
where $\Lambda_{IJ}{}^{\!KL}$ and $\Sigma_{IJKL}$ are subject to the
conditions 
\be
  \label{eq:sp56-properties}
(\Lambda_{IJ}{}^{\!KL})^\ast =  \Lambda^{IJ}{}_{\!KL}=
-\Lambda_{KL}{}^{\!IJ} \,, \qquad
   (\Sigma_{IJKL})^\ast =  \Sigma^{KLIJ} \,.
\ee
The corresponding group elements constitute the group ${\rm
Sp}(56;{\bf R})$ in a pseudoreal basis. This group is the group of
electric-magnetic dualities of maximal supergravity in $D=4$
dimensions. The matrices $\Lambda_{IJ}{}^{KL}$ are associated with  
its maximal compact subgroup, which is equal to ${\rm U}(28)$. 
The defining properties of elements $E$ of ${\rm Sp}(56;{\bf R})$ are 
\be
\label{def-sp56}
E^\ast = \omega \,E\, \omega\,, \qquad E^{-1} = \Omega\, E^\dagger \,\Omega\,, 
\ee
where $\omega$ and $\Omega$ are given by 
\be
\omega= \pmatrix{0&{\bf 1} \cr {\bf 1} & 0\cr} \;,\qquad 
\Omega= \pmatrix{{\bf 1} &0 \cr 0& -{\bf 1}\cr} \;.
\ee
The above properties ensure that the sequilinear form, 
\be
(z_1,z_2) = z_1^{IJ}\,z_{2IJ} - z_{1IJ}\,z_2^{IJ}\,,
\ee
is invariant. In passing we note that the real subgroup (in this
pseudoreal representation) is equal to the group ${\rm GL}(28)$. 

Let us now consider the ${\rm E}_{7(7)}$ subgroup, for which the
$\Sigma^{IJKL}$ is fully antisymmetric and the generators are further
restricted according to  
\bea
  \label{eq:E7}
&& \Lambda_{IJ}{}^{\!KL} = \d_{[I}^{[K} \,\Lambda_{J]}{}^{\!L]}  \,, 
\qquad \Lambda_I{}^J =    -\Lambda^J{}_I\,, \nn\\
&&
\Lambda_I{}^I = 0\,,\qquad \Sigma_{IJKL} = \ft1{24}
\varepsilon_{IJKLMNPQ} \,\Sigma^{MNPQ}\,.
\end{eqnarray}
Obviously the matrices $\Lambda_{I}{}^{\!J}$ generate the
group ${\rm SU}(8)$, which has dimension 63; since $\Sigma_{IJKL}$
comprise 70 real parameters, the dimension of ${\rm E}_{7(7)}$ 
equals $63+70=133$. Because ${\rm SU}(8)$ is the maximal compact
subgroup, the number of the noncompact generators minus the number 
of compact ones is equal to $70-63=7$. It is
straightforward to show that these matrices close under commutation
and generate the group ${\rm E}_{7(7)}$. To show this one needs a
variety of identities for selfdual tensors \cite{dW-so8}; one of them
is that the contraction $\Sigma_{IKLM}\,\Sigma^{JKLM}$ is
traceless. 

However, ${\rm E}_{7(7)}$ has another maximal 63-dimensional subgroup,
which is not compact. This is the group ${\rm SL}(8)$. It is possible
to choose conventions in which the ${\rm E}_{7(7)}$ matrices have a
different block decomposition than \eqn{eq:delta-z} and where the
diagonal blocks correspond to the group ${\rm SL}(8)$, rather than
to ${\rm SU}(8)$. We note that the subgroup generated by \eqn{eq:E7}
with $\Lambda_I{}^{\!J}$ and $\Sigma^{IJKL}$ real, defines the group
${\rm SL}(8;{\bf R})$.

The group ${\rm E}_{7(7)}$ has a quartic invariant,
\bea
&&J_4(z) = z_{IJ}z^{JK}z_{KL}z^{LI} -\ft14 (z_{IJ}z^{IJ})^2
\\
&&
+ \ft1{96}\Big[ \varepsilon_{IJKLMNPQ}\,z^{IJ}z^{KL}z^{MN}z^{PQ} + 
\varepsilon^{IJKLMNPQ}\,z_{IJ}z_{KL}z_{MN}z_{PQ}\Big] \,,
\nonumber 
\eea
which, however, plays no role in the following. For further
information the reader is encouraged to read the appendices of
\cite{cremmer}.  

Another subgroup is the group ${\rm E}_{6(6)}$, for which the
restrictions are rather 
similar. Here one introduces a skew-symmetric tensor $\Omega_{IJ}$,
satisfying  
\begin{equation}
  \label{eq:Omega}
 \Omega_{IJ} = - \Omega_{JI}\,,\quad (\Omega_{IJ})^\ast = \Omega^{IJ}
 \,,\quad \Omega_{IK}\Omega^{KJ} = - \d^J_I\,.
\end{equation}
Now we restrict ourselves to the subgroup of ${\rm U}(8)$ that leaves
$\Omega_{IJ}$ invariant. This is the group ${\rm USp}(8)$. The other
restrictions on the generators concern $\Sigma_{IJKL}$. Altogether we
have the conditions,  
\begin{eqnarray}
  \label{eq:E6}
&&   \Lambda_{IJ}{}^{\!KL} = \d_{[I}^{[K} \,\Lambda_{J]}{}^{\!L]}  \,, 
\qquad \Lambda_I{}^J =    -\Lambda^J{}_I\,, \nn\\
&&\Lambda_{[I}{}^K \Omega_{J]K}=0 \,,\qquad \Omega_{IJ}
\Sigma^{IJKL}=0\,,\nn \\
&&\Sigma_{IJKL} = \Omega_{IM}\, \Omega_{JN} \,\Omega_{KP}
\,\Omega_{LQ} \,\Sigma^{MNPQ}\,.
\end{eqnarray}
The maximal compact subgroup ${\rm USp}(8)$ thus has dimension
$64-28=36$, while there are $70- 28=42$ generators associated with
$\Sigma_{IJKL}$. Altogether we thus have $36+42=78$
generators,  while the difference between the numbers of noncompact and
compact generators equals $42-36=6$. These numbers confirm that we are
indeed dealing with ${\rm E}_{6(6)}$ and its maximal compact subgroup
${\rm USp}(8)$. 
Because of the constraints \eqn{eq:E6} the 56-dimensional
representation defined by \eqn{eq:delta-z} is reducible and
decomposes into two singlets and a ${\bf 27}$ and a 
$\overline{\bf 27}$ representation. To see this we 
observe that the following restrictions are preserved by the group, 
\begin{equation}
  \label{eq:E6-27}
 \Omega_{IJ}z^{IJ}=0 \,,\qquad z_{IJ} = \pm \Omega_{IK} \Omega_{JL}
 z^{KL}\,.
\end{equation}
The first one suppresses the singlet representation and the second one
projects out the ${\bf 27}$ or the $\overline{\bf 27}$ representation.

The group ${\rm E}_{6(6)}$ has a cubic invariant, defined by
\be
\label{J3} 
J_3(z) = z^{IJ}z^{KL}z^{MN}\,\Omega_{JK}\Omega_{LM}\Omega_{NI}\,,
\ee
which plays a role in the ${\rm E}_{6(6)}$ invariant Chern-Simons
term in the supergravity Lagrangian. 

There is another  maximal subgroup of ${\rm E}_{6(6)}$, which is
noncompact,  that will be relevant in the following. This is the group
${\rm SL}(6)\times {\rm SL}(2)$, which has 
dimension $35+3=38$, and which plays a role in many of the
known gaugings, where the gauge group is embedded into the 
group ${\rm SL}(6)$, so that ${\rm SL}(2)$ remains as a rigid
invariance group of the Lagrangian.

%%%%%%%%%%%%%%%%%%%%%%%%%%%%%%%%%%%%%%%%%%%%%%%%%%%%%%%%%%%%%%%%
\subsection{On ungauged maximal supergravity Lagrangians}
\label{sec:feat-ungauged-sg}
%%%%%%%%%%%%%%%%%%%%%%%%%%%%%%%%%%%%%%%%%%%%%%%%%%%%%%%%%%%%%%%%
An important feature of pure extended supergravity theories is that
the spinless fields take their values in a homogeneous
target space ({\it c.f.} table~\ref{maximal-cosets}, where we have listed
these spaces). Because the spinless
fields always appear in nonpolynomial form, it is vital to exploit the
coset structure explained in the previous section in the construction
of the supersymmetric action and 
transformation rules, as well as in the gauging. We will not be
complete here but sketch a number of features of the maximal
supergravity theories in $D=4,5$ where the coset structure plays an
important role. We will be rather cavalier about numerical
factors, spinor conventions, etcetera. In this way we will, hopefully,
be able to bring out the main features of the G/H structure, without
getting entangled in issues that depend on the spacetime
dimension. For those and other details we refer to the original
literature \cite{deWitNic,GunaRomansWarner}. 

One starts by introducing a so-called 56-bein $\vv(x)$, which is a
$56\times 56$ matrix that belongs to the group ${\rm E}_{7(7)}$ or 
${\rm E}_{6(6)}$, depending on whether we are in $D=4$, or 5
dimensions. A coset representative is obtained by exponentiation of
the generators defined in \eqn{eq:delta-z}. Schematically,
\be
\label{exp-E}
\vv(x)= \exp \pmatrix{0& \overline \Sigma(x)\cr
   \noalign{\vskip 3mm}
\Sigma(x)& 0\cr}\,,
\ee
where the rank-4 antisymmetric tensor $\Sigma$ satisfies the algebraic
restrictions appropriate for the exceptional group. 
As explained in the previous section, the 56-bein is
reducible for ${\rm E}_{6(6)}$, but we will use the reducible
version in order to discuss the two theories on a par. Our notation will
be based on a description in terms of right cosets, just as in the
previous sections, which may differ from the notations used in the original
references where one sometimes uses left cosets. Hence, we assume that
the 56-bein transforms under  
the exceptional group from the left and under the local ${\rm SU}(8)$
(or ${\rm USp}(8)$) from the right. The 56-bein can be decomposed in
block form according to
\begin{equation}
  \label{eq:u-v}
\vv(x) = \pmatrix{ u^{ij}{}_{\!IJ}(x) & -v_{ kl  IJ}(x)\cr 
\noalign{\vskip 6mm}
-v^{ijKL}(x) &   u_{kl}{}^{\!KL}(x)\cr} \,,
\end{equation}
with the usual conventions $u^{ij}{}_{\!IJ}= (u_{ij}{}^{\!IJ})^\ast$
and  $v_{ijIJ}=(v^{ij IJ})^\ast$.  
Observe that the indices of the matrix are antisymmetrized index pairs
$[IJ]$ and $[ij]$. In the above the row indices are $([IJ],[KL])$, and
the column indices are $([ij],[kl])$. The latter are the indices
associated with the local ${\rm SU}(8)$ or ${\rm USp}(8)$. The
notation of the submatrices is chosen such as to make contact with
\cite{deWitNic}, where left cosets were chosen, upon interchanging $\vv$
and $\vv^{-1}$. Observe also that \eqn{exp-E} is a coset
representative, {\it i.e.} we have fixed the gauge with respect to
local ${\rm SU}(8)$ or ${\rm USp}(8)$, whereas in  \eqn{eq:u-v} gauge
fixing is not assumed. According to  \eqn{def-sp56} the inverse
$\vv^{-1}$ can be expressed in terms of the complex conjugates of the
submatrices of $\vv$, 
\begin{equation}
  \label{eq:u-v-1}
\vv^{-1}(x) = \pmatrix{ u_{ij}{}^{\!IJ}(x) &  v_{ijKL}(x)\cr 
\noalign{\vskip 6mm}
v^{klIJ}(x) &   u^{kl}{}_{\!KL}(x)\cr} \,.
\end{equation}
Consequently we derive the identities, for ${\rm E}_{7(7)}$,
\bea
u^{ij}{}_{\!IJ}\, u_{kl}{}^{\!IJ} -v^{ijIJ}\,v_{klIJ} &=& 
\d^{ij}_{kl} \,, \nonumber \\
u^{ij}{}_{\!IJ} \,v^{klIJ}- v^{ijIJ} \,u^{kl}{}_{\!IJ} &=& 0\,,
\eea
or, conversely,
\bea
u^{ij}{}_{\!IJ}\, u_{ij}{}^{\!KL} -v_{ijIJ}\, 
v^{ijKL} &=& \d^{IJ}_{KL} \,, \nonumber \\
u^{ij}{}_{\!IJ} v_{ijKL}- v_{ijIJ} u^{ij}{}_{\!KL} &=& 0\,.
\eea
The corresponding equations for ${\rm E}_{6(6)}$ are identical,
except that the antisymmetrized Kronecker symbols on the right-hand
sides are replaced according to  
\be
\d^{ij}_{kl}\to \d^{ij}_{kl} -\ft18\Omega_{kl}\, \Omega^{ij}\,,\qquad 
\d^{IJ}_{KL}\to \d^{IJ}_{KL} -\ft18\Omega_{KL}\, \Omega^{IJ}\,.
\ee
Furthermore the matrices $u$ and $v$ vanish when contracted with the
invariant tensor $\Omega$ and they are pseudoreal, {\it e.g.},
\be
u_{ij}{}^{\!IJ} \Omega_{IJ}=0\,,\qquad   u_{ij}{}^{\!KL}
\Omega_{IK}\Omega_{JL} = \Omega_{ik}\Omega_{jl} u^{kl}{}_{\!IJ}\,,
\ee
with similar identities for the $v^{ijIJ}$. 
In this case the (pseudoreal) matrices $u^{ij}{}_{IJ} \pm
\Omega_{IK}\,\Omega_{JL} \,v^{ijKL}$ and their complex conjugates
define (irreducible) 
elements of ${\rm E}_{6(6)}$ corresponding to the ${\bf 27}$ and
$\overline{\bf 27}$ representations. We note the identity
\bea
\Big(u^{ij}{}_{IJ} + \Omega_{IK}\,\Omega_{JL}\,v^{ijKL}\Big) 
\Big(u_{kl}{}^{IJ} - \Omega^{IM}\,\Omega^{JN} \,v_{klMN}\Big) = 
\d^{ij}_{kl} -\ft18 \Omega_{kl}\, \Omega^{ij}\,. 
\nn\\[-2mm] {~}
\eea
In this case we can thus decompose the 56-bein in terms of a 27-bein
and a $\overline{27}$-bein. 

Subsequently we evaluate the quantities ${\cal Q}_\m$ and ${\cal
P}_\m$, 
\be
\label{P-Q-max}
\vv^{-1} \pa_\m\vv = \pmatrix{{\cal Q}_{\m\, ij}{}^{\!mn} 
&{\cal P}_{\m\, ijpq} \cr \noalign{\vglue 6mm}
{\cal P}_\m^{klmn} & {\cal Q}_\m{}^{\!kl}{}_{\!pq} \cr}\,,
\ee
which leads to the expressions,
\bea
{\cal Q}_{\m \,ij}{}^{\!kl}&=& u_{ij}{}^{\!IJ}\, \pa_\m u^{kl}{}_{\!IJ}
-v_{ijIJ}\, \pa_\m v^{klIJ} \,, \nonumber\\
{\cal P}_\m^{ijkl} &=&  v^{ijIJ}\,\pa_\m u^{kl}{}_{\!IJ}-
u^{ij}{}_{\!IJ} \, \pa_\m v^{klIJ}  \,.
\eea
The important observation is that ${\cal Q}_{\m ij}{}^{\!kl}$ and
${\cal P}_\m^{ijkl}$ are subject to the same constraints as the
generators of the exceptional group listed in the previous
section. Hence, ${\cal P}_\m^{ijkl}$ is fully
antisymmetric and subject to a reality constraint. Therefore it
transforms according  to the 70-dimensional representation of ${\rm
SU}(8)$, with the reality condition,
\be
{\cal P}_\m^{ijkl} = \ft1{24}\,\varepsilon^{ijklmnpq}\, {\cal
 P}_{\m\,mnpq}\,, 
\ee
or, to the 42-dimensional representation of ${\rm USp}(8)$, with the
reality condition,
\be
{\cal P}_\m^{ijkl} = \Omega^{im} \Omega^{jn} \Omega^{kp} \Omega^{lq}
 \, {\cal P}_{\m\,mnpq}\,, 
\ee
Likewise ${\cal Q}_\m$ transforms as a connection associated with 
${\rm SU}(8)$ or ${\rm USp}(8)$, respectively. Hence ${\cal Q}_{\m
\,ij}{}^{\!kl}$ must satisfy the decomposition,  
\be 
{\cal Q}_{\m\, ij}{}^{\!kl}= \d^{[k}_{[i}\, {\cal Q}_{\m\, j]}{}^{\!l]}\,, 
\ee
so that ${\cal Q}_{\m \,i}{}^{\!j}$ equals
\be
{\cal Q}_{\m \,i}{}^{\!j} = \ft23 \Big[u_{ik}{}^{\!IJ}\, \pa_\m
u^{jk}{}_{\!IJ} - v_{ikIJ}\, \pa_\m v^{jkIJ}  \Big]\,.
\ee
Because of the underlying Lie algebra the connections ${\cal
Q}_{\m\,i}{}^{\!j}$ satisfy ${\cal Q}_{\m}{}^{\!i}{}_{\!j} = - {\cal
Q}_{\m j}{}^i$ and ${\cal Q}_{\m i}{}^i=0$, as well as an extra
symmetry condition in the case of ${\rm USp}(8)$ (cf. \eqn{eq:E6}). 

Furthermore we can evaluate the Maurer-Cartan equations \eqn{CM-Q-P}, 
\bea
F_{\mu\nu}({\cal Q})_i{}^j & = & \partial_\mu {\cal
Q}_{\nu\,i}{}^{\!j} - \partial_\nu {\cal Q}_{\mu\,i}{}^{\!j} + 
{\cal Q}_{[\mu\,i}{}^{\!k} {\cal Q}_{\nu]k}{}^{\!j} 
= -\ft43\,{\cal P}_{[\mu}{}^{\!jklm} \, {\cal P}_{\nu]iklm} \,,
\nonumber \\ 
D_{[ \mu}^{~} {\cal P}_{\nu]}^{ijkl} &=& \pa_{[\m}{\cal P}_{\n]}^{ijkl}
+ 2 {\cal Q}^{~}_{[\m \,m}{}^{\![i}\, {\cal P}_{\n]}^{jkl]m} = 0
\,. \label{ECM-Q-P} 
\eea
Observe that the use of the Lie algebra decomposition for G/H is
crucial in deriving these equations. Such decompositions
are an important tool for dealing with the spinless fields in this
nonlinear setting. Before fixing a  
gauge, we can avoid the nonlinearities completely and carry out the
calculations in a transparent way. Fixing the gauge prematurely and
converting to a specific coset representative for G/H would lead to
unsurmountable difficulties.  

Continuing along similar lines we turn to a number of other features
that are of interest for the Lagrangian and transformation rules. The
first one is the observation that {\it any} variation of the 56-bein
can be written, up to a local H-transformation, as 
\begin{equation}
  \label{eq:dvv}
  \d \vv= \vv\, \pmatrix{0& \overline\Sigma\cr \Sigma &0\cr}\,,
\end{equation}
or, in terms of submatrices,  
\begin{equation}
  \label{eq:duv}
  \d u_{ij}{}^{\!IJ} = -\Sigma_{ijkl} \,v^{klIJ}\,,\qquad
  \d v_{ijIJ} = -\Sigma_{ijkl} \,u^{kl}{}_{\!IJ}\,.
\end{equation}
where $\Sigma^{ijkl}$ is the rank-four antisymmetric tensor
corresponding to the generators associated with ${\rm G}/{\rm H}$
({\it i.e.}, the generators denoted by $\sf k$ in the previous
chapter).  Because $\Sigma$ takes the form of an H-covariant tensor,
the variation \eqn{eq:duv} is consistent with both groups G and H. 
Under this variation the quantities ${\cal Q}_\m$ and ${\cal
P}_\m$ transform systematically, 
\bea
 \label{dQP}
  \d Q_{\m\,i}{}^{\!j}  &=& \ft 23\Big(  \Sigma^{jklm} \,{\cal P}_{\m
  \,iklm} - \Sigma_{iklm} \,{\cal P}_\m^{jklm} \Big)  \,,\nn \\
\d {\cal P}^{ijkl}_\m &=& D_\m \Sigma^{ijkl} = \pa_\m \Sigma^{ijkl} + 2
  {\cal Q}_{\m \,m}{}^{\![i} \Sigma^{jkl]m} \,.  
\eea
Observe that this establishes that ${\cal Q}_\m$ and ${\cal P}_\m$
can be assigned to the adjoint representation of the group G, as is
already obvious from the decomposition  \eqn{P-Q-max}.

As was stressed above, any variation of $\vv$ can be decomposed into
\eqn{eq:dvv}, up to a local H-transformation. In particular this
applies to supersymmetry transformations. The supersymmetry variation
can be written in the form \eqn{eq:dvv}, where $\Sigma$ is an
H-covariant expression proportional to the supersymmetry 
parameter $\e^i$ and the fermion fields $\chi^{ijk}$. Hence it must be
of the form $\Sigma^{ijkl}\propto \bar \e^{[i}\chi^{jkl]}$, up to
complex conjugation and possible contractions with H-covariant
tensors, Furthermore 
$\Sigma$ must satisfy the restrictions associated with the exceptional
group, {\it i.e.} \eqn{eq:E7} or \eqn{eq:E6}. 

The supersymmetry
variation of the spinor $\chi^{ijk}$ contains the quantity ${\cal
P}_\m^{ijkl}$, which incorporates the spacetime derivatives of the
spinless fields, so that up to proportionality constants we must have
a variation, 
\be
\d \chi^{ijk} \propto {\cal P}_\m^{ijkl} \,\g^\m \e_l\,.
\ee

The verification of the supersymmetry algebra on $\vv$ is rather 
easy. Under two consecutive (field-dependent) variations \eqn{eq:duv}
applied in different orders on the 56-bein, we have  
\begin{equation}
  \label{eq:comm}
[\d_1,\d_2] \,\vv = \vv\, \pmatrix{0& 2 \,\d_{[1} \overline\Sigma_{2]}\cr 
\noalign{\vskip 2 mm} 
2\,\d_{[1} \Sigma_{2]} &0 \cr}  + \vv 
\left[\pmatrix{0&\overline\Sigma_1\cr \noalign{\vskip 2 mm}
\Sigma_1&0\cr} , 
\pmatrix{0&\overline \Sigma_2\cr \noalign{\vskip 2 mm}
\Sigma_2&0\cr}\right]    \,.  
\end{equation}
The last term is just an infinitesimal H-transformations. For the
first term we note that $\d_1 \Sigma_2$ leads to a term proportional
to ${\cal P}_\m^{ijkl}$, combined with two supersymmetry parameters,
$\e_1$ and $\e_2$, of the form $(\bar \e_1^{[i} \g^\m \e_{2\, m})\,{\cal
P}_\m^{jkl]m}$. Taking into account the various H-covariant
combinations in the actual expressions implied by \eqn{eq:E7} or
\eqn{eq:E6}, respectively, this contribution can be written in the form 
\be
[\d_1,\d_2] \,\vv \propto (\bar \e_1^i \g^\m \e_{2\, i}-\bar \e_2^i \g^\m
\e_{1\, i}  )\;\vv\,\pmatrix{0 &
\overline{\cal P}_\m \cr 
\noalign{\vskip 1mm} {\cal P}_\m &0\cr} \,.
\ee
This is precisely a spacetime diffeomorphism, up to a local
H-transformation proportional to ${\cal Q}_\m$, as follows from
\eqn{P-Q-max}. Hence up to a number of field-dependent
H-transformations, the supersymmetry commutator closes on $\vv$ into a
spacetime diffeomorphism (up to terms of higher-order in the spinors
that we suppressed).
 
Let us now turn to the action. Apart from higher-order spinor terms,
the terms in the Lagrangian pertaining to the graviton, gravitini,
spinors and scalars take the following form,
\bea
\label{lagr-1}
e^{-1} \lagr_1 & = & -  \ft{1}{2} R(e,\omega ) - \ft{1}{2} 
\bar{\psi}_\mu^i \gamma^{\mu\nu\rho} 
  \Big[(\pa_\n -\ft12 \omega^{ab}_\n \,\g_{ab})\d^j_i
+\ft12 {\cal Q}_{\n\,i}{}^{\!j} \Big]  \psi_{\rho j} 
\nonumber \\
& & - \ft{1}{12} \bar \chi^{ijk}\gamma^\m \Big[(\pa_\m -\ft12
\omega^{ab}_\m \,\g_{ab})\d^l_k 
+ \ft32{\cal Q}_{\m\,k}{}^{\!l}\Big]  \chi_{ijl} - 
\ft{1}{12} {\cal P}_\mu^{ijkl}\, {\cal P}^\mu_{ijkl} \nonumber \\
& & -\ft{1}{6}\sqrt{2} \bar{\chi}_{ijk} \gamma^\nu 
\gamma^\mu \psi_{\nu l} \,{\cal P}_\mu^{ijkl}  \,.
\eea
This Lagrangian is manifestly invariant with respect to ${\rm
E}_{7(7)}$ or ${\rm E}_{6(6)}$. 
Here we distinguish the Einstein-Hilbert term for gravity, the
Rarita-Schwinger Lagrangian for the gravitini, the Dirac Lagrangian
and the nonlinear sigma model associated with the G/H target
space. The last term represents the Noether coupling term for the
spin-0/spin-$\ft12$ system. For $D=4$ the fermion fields are chiral
spinors and we have to add 
the contributions from the spinors of opposite chirality; for $D=5$ we
are dealing with so-called symplectic Majorana spinors. Here we
disregard such details and concentrate on the symmetry issues.  

The vector fields bring in new features, which are different for
spacetime dimensions $D=4$ and 5. In $D=5$ dimensions the vector
fields $B_\m^{IJ}$ transform as the ${\bf 27}$ representation of ${\rm
E}_{6(6)}$, so that they satisfy the reality constraint $B_{\m\,IJ} =
\Omega_{IK} \Omega_{JL}\,B_\m^{KL}$, and the Lagrangian is manifestly
invariant under the corresponding transformations. It is impossible to
construct an 
invariant action just for the vector fields and one has to make use
of the scalars, which can be written in terms of the
$\overline{27}$-beine, 
$u^{ij}{}_{\!IJ} + v^{ij KL} \Omega_{IK} \Omega_{JL}$, and which
can be used to convert ${\rm E}_{6(6)}$ to ${\rm USp}(8)$
indices. Hence we define a ${\rm USp}(8)$ covariant field strength
for the vector fields, equal to 
\be
{\sf F}_{\m\n}^{ij} = (u^{ij}{}_{\!IJ}  - v^{ij KL} \Omega_{IK} \Omega_{JL})
(\pa_\m B^{IJ}_\n -  \pa_\n B^{IJ}_\m ) \, . 
\ee
The invariant Lagrangian of the vector fields then reads, 
\bea
\label{lagr-2}
 \lagr_2 &=& -\ft1{16}e\, {\sf F}_{\m\n}^{ij} \,{\sf F}^{\m\n kl}
 \,\Omega_{ik} \Omega_{jl} \nn \\
& & 
- \ft 1{12} \varepsilon^{\m\n\rho\sigma\lambda}\, B_\m^{IJ}\, \pa_\n^{~}
B_\rho^{KL} \,  \pa_\sigma^{~}  B_\lambda^{MN} \,\Omega_{JK}
\Omega_{LM} \Omega_{NI}   \nn\\
&& + \ft14e \, {\sf F}_{\m\n}^{ij}\,{\cal O}^{\m\n}_{ij}  \,,
\eea
where we distinguish the kinetic term, a Chern-Simons interaction
associated with the ${\rm E}_{6(6)}$ cubic invariant \eqn{J3} and a
moment coupling with the fermions. Here ${\cal O}^{\m\n} _{ij}$
denotes a covariant tensor antisymmetric in both spacetime and 
${\rm USp}(8)$ indices and quadratic in the
fermion fields, $\psi_\m^i$ and $\chi^{ijk}$. Observe that the
dependence on the spinless fields is completely implicit. Any
additional dependence would affect the
invariance under ${\rm E}_{6(6)}$. The result obtained by
combining the Lagrangians \eqn{lagr-1} and \eqn{lagr-2} gives the full
supergravity Lagrangian invariant under rigid ${\rm E}_{6(6)}$ and
local ${\rm USp}(8)$ transformations, up to terms quartic in the
fermion fields. We continue the discussion of the $D=4$ theory 
in the next section, as this requires to first introduce the concept
of electric-magnetic duality. 

%%%%%%%%%%%%%%%%%%%%%%%%%%%%%%%%%%%%%%%%%%%%%%%%%%%%%%%%%%%%%%%%
\subsection{Electric-magnetic duality and ${\rm E}_{7(7)}$ }
\label{elec-magn}
%%%%%%%%%%%%%%%%%%%%%%%%%%%%%%%%%%%%%%%%%%%%%%%%%%%%%%%%%%%%%%%%
For $D=4$ the Lagrangian is not invariant under ${\rm E}_{7(7)}$ but
under a smaller group, which acts on the vector fields (but not
necessarily on the 56-bein) according to a 28-dimensional subgroup of
${\rm GL}(28)$. However, the combined equations of motion and the
Bianchi identities are invariant under the group ${\rm E}_{7(7)}$. 
This situation is typical for $D=4$ theories with abelian 
vector fields, where the symmetry group of field equations and Bianchi
identities can be bigger than that of the Lagrangian, and where 
different Lagrangians not related by local field redefinitions, can
lead to an equivalent set of field equations and Bianchi
identities. However, the phenomenon is not restricted to 4 dimensions
and can occur for antisymmetric tensor gauge fields in any even number
of spacetime dimensions (see, {\it e.g.}, \cite{Tanii}). The
4-dimensional version has been known for a long time and is commonly
referred to as electric-magnetic duality  
(for a recent review of this duality in supergravity, see,
{\it e.g.}, \cite{susy30}). Its simplest form 
arises in Maxwell theory in four-dimensional 
(flat or curved) Minkowski space, where one can perform 
(Hodge)  duality rotations, which commute with the Lorentz group and
rotate the electric and magnetic fields and inductions according to
$ {\bf E} \,\leftrightarrow\, {\bf H}$ and ${\bf B}\leftrightarrow {\bf
D}$. 

This duality can be generalized to any $D=4$ dimensional field
theory with abelian vector fields and no charged fields, so that the
gauge fields enter the Lagrangian only through their 
(abelian) field strengths. These
field strengths (in the case at hand we have 28 of them, labelled by
antisymmetric index pairs $[IJ]$, but for the moment we will remain
more general and label the field strengths by $\a,\b,\ldots$) are
decomposed into selfdual and  anti-selfdual components $F_{\m\n}^{\pm
\a}$ (which, in Minkowski 
space, are related by complex conjugation) and so are the field
strengths $G^\pm_{\m\n\,\a}$ that appear in the field equations,
which are defined by  
\be
\label{def-G}
G_{\m\n\,\a}^\pm= \pm {4 i\over e}\,  \displaystyle{\pa \lagr\over \pa
F^{\pm\,\a\m\n} }\;.
\ee
Together $F_{\m\n}^{\pm \a}$ and $G^\pm_{\m\n\,\a}$ comprise the
electric and magnetic fields and inductions. The Bianchi identities
and equations of motion for the abelian gauge fields take the form 
\begin{equation}
\partial^\mu \big(F^{+} -F^{-}\big){}^{\a}_{\m\n} =
\partial^\mu \big(G^+ -G^-\big){}_{\m\n\, \a} =0\,,
\label{Maxwell}
\end{equation}
which are obviously invariant under {\it real}, {\it constant},
rotations of the field strengths $F^\pm$ and $G^\pm$, 
\be
\pmatrix{F^{\pm \a}_{\mu\nu}\cr \noalign{\vskip 2mm} G^\pm_{\mu\nu\,
\b}\cr}  \longrightarrow 
\pmatrix{U&Z\cr \noalign{\vskip 2mm} W&V\cr} \pmatrix{F^{\pm  
\a}_{\mu\nu}\cr  \noalign{\vskip 2mm} G^\pm_{\mu\nu \,\b}\cr}\,,
\label{FGdual}
\ee
where $U^\a_{\,\b}$, $V_\a^{\,\b}$, $W_{\a\b}$ and $Z^{\a\b}$ are 
constant, real,  $n\times n$ submatrices and $n$ denotes the number of
independent gauge potentials. In $N=8$ supergravity we have 56 such
field strengths of each duality, so that the rotation is associated
with a $56\times56$ matrix.  
The relevant question is whether the rotated equations
\eqn{Maxwell} can again follow from a Lagrangian. More precisely, does
there exist a new  
Lagrangian depending on the new, rotated, field strengths, such that
the new tensors $G_{\m\n}$ follow from this Lagrangian as in
\eqn{def-G}. This condition amounts to an integrability condition,
which can only have a solution (for generic Lagrangians) provided
that the matrix is an element of the group 
${\rm Sp}(2n;{\bf R})$.\footnote{%%%%%%%%%%%%%%%%%%%%%%%%%%%%%%%%%
%%%%%%%%%%%%%%%%%%%%%%%%%%%%%%%%%%%%%%%%%%%%%%%%%%%%%%%%%%%%%%%%%%
  Without any further assumptions, one can show that in Minkowski spaces
  of dimensions $D=4k$, the rotations of the field equations and Bianchi
  identities associated with $n$ rank-$(k-1)$ antisymmetric gauge fields that
  are described by a Lagrangian, constitute the group ${\rm Sp}(2n;{\bf
  R})$. For rank-$k$ antisymmetric gauge fields in $D=2k+2$
  dimensions, this group equals ${\rm SO}(n,n;{\bf R})$. Observe that
  these groups do not constitute an invariance of the theory, but merely
  characterize an equivalence class of Lagrangians. 
  The fact that the symplectic redefinitions of the field strengths
  constitute the group ${\rm Sp}(2n;{\bf R})$ was first derived in
  \cite{GZ}, but in the context of a duality {\it invariance} rather
  than of a {\it reparametrization}. In this respect our presentation
  is more in the spirit of a later treatment in \cite{CFG} for
  $N=2$ vector multiplets coupled to supergravity (duality
  invariances for these theories were introduced in
  \cite{dWVP}).} %%%%%%%%%%%%%%%%%%%%%%%%%%%%%%%%%%%%%%%%%%%%%%%%%
%%%%%%%%%%%%%%%%%%%%%%%%%%%%%%%%%%%%%%%%%%%%%%%%%%%%%%%%%%%%%%%%%%
This implies that the submatrices satisfy the constraint
\bea
\label{realdual}
&&U^{\rm T} V -W^{\rm T}Z = VU^{\rm T} - WZ^{\rm T} = {\bf 1}\,,
\nonumber\\ 
&&U^{\rm T}W = W^{\rm T}U \;,\qquad Z^{\rm T}V =V^{\rm T}Z \,.
\eea
We distinguish two subgroups of ${\rm Sp}(2n;{\bf R})$. One is the
invariance group of the combined field equations and Bianchi
identities, which usually requires the other fields in the Lagrangian
to transform as well. Of course, a generic theory does not have such
an invariance group, but maximal supergravity is known to have an ${\rm
E}_{7(7)}\subset {\rm Sp}(56;{\bf R})$ invariance. However, this
invariance group is not necessarily realized as a symmetry of the
Lagrangian. The subgroup that is a symmetry of the Lagrangian, is
usually smaller and 
restricted by $Z=0$ and $U^{-1}=V^{\rm T}$; the subgroup associated
with the matrices $U$ equals ${\rm GL}(n)$. Furthermore the Lagrangian is not 
uniquely defined (it 
can always be reparametrized via an electric-magnetic duality
transformation) and neither is its invariance group. More precisely,
there exist different Lagrangians with different symmetry groups,
whose Bianchi identities and equations of motion are the same (modulo
a linear transformation) and are invariant under the same group
(which contains the symmetry groups of the various
Lagrangians as subgroups). These issues are extremely important when
gauging a subgroup of the invariance group, as this requires the gauge
group to be contained in the invariance group of the Lagrangian.  

Given the fact that we can rotate the field strengths by
electric-magnetic duality transformations, we assign different indices
to the field strengths and the underlying gauge groups than to the
56-bein $\vv$. Namely, we label the fields strengths by independent
index pairs $[AB]$, which are related to the index pairs $[IJ]$ of the
56-bein ({\it c.f.} \eqn{eq:u-v}) in a way that we will discuss
below. Furthermore, to remain in the context of the pseudoreal basis
used previously, we form the linear combinations, 
\be 
{\sf F}_{1\m\n\,AB}^{+} = \ft12(i\, G^+_{\m\n\,AB}+ F_{\m\n}^{+ AB})\,,
\quad 
{\sf F}_{2\m\n}^{+AB} =\ft12(i\, G^+_{\m\n\,AB} - F_{\m\n}^{+ AB})\,.
\ee
Anti-selfdual field strengths 
$({\sf F}_{1\m\n}^{-AB},{\sf F}_{2\m\n\,AB}^{-})$ 
follow from complex conjugation. On this basis the field strengths rotate
under ${\rm Sp}(56;{\bf R})$ according to the matrices $E$ specified
in \eqn{def-sp56}; the real ${\rm GL}(28)$ subgroup is induced by
corresponding linear transformations of the vector fields. 

To exhibit how one can deal with a continuous variety of
Lagrangians, which are manifestly invariant under different subgroups
of ${\rm E}_{7(7)}$ , let us remember that the 
tensors $F^{AB}_{\m\n}$ and $G_{\m\n\,AB}$ are related by \eqn{def-G}
and this relationship must be consistent with ${\rm E}_{7(7)}$. In
order to establish this consistency, the 56-bein plays a crucial
role. The relation involves a constant ${\rm Sp}(56;{\bf R})$ matrix
${\sf E}$ (so that it satisfies the conditions \eqn{def-sp56}), 
\be
{\sf E}=\pmatrix{{\sf U}_{IJ}{}^{\!AB} & {\sf V}_{IJCD}\cr
\noalign{\vskip4mm}  {\sf V}^{KLAB}& {\sf U}^{KL}{}_{\!CD} \cr}\,.
\ee 
On the basis of ${\rm E}_{7(7)}$ and ${\rm SU}(8)$ covariance, the
relation among the field strengths must have the form, 
\be
\label{FGO}
\vv^{-1}\, \,{\sf E} \,\pmatrix{{\sf F}_{1\m\n\,AB}^{+}
\cr\noalign{\vskip2mm}  {\sf F}_{2\m\n}^{+AB} \cr}  =
\pmatrix{{\sf F}_{\m\n\,ij}^{+}\cr\noalign{\vskip3mm}
{\cal O}_{\m\n}^{+kl} \cr} \,,
\ee
where ${\cal O}_{\m\n}^+$ is an ${\rm SU}(8)$ covariant tensor
quadratic in the fermion fields and independent of the scalar
fields, which appears in the moment couplings in the
Lagrangian. Without going into the details we mention that the
chirality and duality of ${\cal O}_{\m\n}^+$ is severely restricted so
that the structure of \eqn{FGO} is unique ({\it c.f.} \cite{deWitNic}).  
The tensor ${\sf F}^+_{\m\n\,ij}$ is an ${\rm SU}(8)$ covariant field
strength that appears in the supersymmetry transformation rules of the spinors,
which is simply defined by the above condition.

Hence the matrix ${\sf E}$ allows the field strengths and the 56-bein
to transform under ${\rm E}_{7(7)}$ in an equivalent but nonidentical
way. One could consider absorbing this matrix into the 
definition of the field strengths $({\sf F}_1,{\sf F}_2)$, but such a
redefinition cannot be carried out at the level of the Lagrangian,
unless it belongs to a ${\rm GL}(28)$ subgroup which can act on
the gauge fields themselves. 
In the basis \eqn{def-sp56} the generators of ${\rm GL}(28)$
have a block decomposition with ${\rm SO}(28)$ generators in both 
diagonal blocks and identical real, symmetric, $28\times 28$ matrices
in the off-diagonal blocks. 
On the other hand, when ${\sf E}\in{\rm E}_{7(7)}$, it can be
absorbed into the 56-bein $\vv$. The various Lagrangians are thus encoded
in ${\rm Sp}(56;{\bf R})$ matrices ${\sf E}$, up to multiplication
by ${\rm GL}(28)$ from the right and multiplication by 
${\rm E}_{7(7)}$ from the left, {\it i.e.} in elements of ${\rm E}_{7(7)}
\backslash {\rm Sp}(56;{\bf R})/ {\rm GL}(28)$. 

From \eqn{FGO} one can straightforwardly determine the relevant terms
in the Lagrangian. For convenience, we redefine the 56-bein by
absorbing the matrix $\sf E$,
\be
\label{u-v-mod}
\hat\vv(x) = {\sf E}^{-1}\,\vv(x) = 
\pmatrix{ u^{ij}{}_{\!AB}(x) &  -v^{ kl\,AB}(x)\cr 
\noalign{\vskip 6mm}
- v^{ij\,CD}(x) &   u_{kl}{}_{\!CD}(x)\cr} \,,
\ee
where we have to remember that $\hat \vv$ is now no longer a group
element of ${\rm E}_{7(7)}$. Note, however, that the ${\rm
E}_{7(7)}$ tensors ${\cal Q}_\m$ and ${\cal P}_\m$ are
not affected by the matrix $\sf E$ and have identical expressions in
terms of $\vv$ and $\hat \vv$. This is not the case for the terms in
the Lagrangian that contain the abelian field
strengths,  
\be 
F^{AB}_{\m\n} = \pa_\m A_\n^{AB} - \pa_\n A_\m^{AB}\,,
\ee
and which take the form, 
\bea
\label{L3}
\lagr_3\!&=& \!-\ft18 e\, F^{+AB}_{\m\n} \, F^{+CD\,\m\n} \,
[(u+v)^{-1}]^{AB}{}_{\!\!ij} \, (u^{ij}{}_{\!CD} - v^{ijCD})  \nn \\ 
&& -\ft12 e\, F^{+AB}_{\m\n} \, [(u+v)^{-1}]^{AB}{}_{\!\!ij}\,  
{\cal O}^{+\m\n\,ij}    \nn\\[1mm]
&& + \mbox{ h.c.}\,,
\eea
where the $28\times 28$ matrices satisfy $[(u+v)^{-1}]^{AB}{}_{\!ij}\,
(u^{ij}{}_{\!CD} +v^{ijCD})= \d^{AB}_{CD}$.    
The ${\rm SU}(8)$ covariant field strength ${\sf F}_{\m\n ij}^+$ will
appear in the supersymmetry  transformation rules for the fermions,
and is equal to
\be
 F_{\m\n}^{+AB}= (u^{ij}{}_{\!AB}+ v^{ijAB}) \,{\sf F}_{\m\n ij}^{+}   
-  (u_{ij}{}^{\!AB}+ v_{ijAB}) \, {\cal O}_{\m\n}^{+ij}\,.
\ee
Clearly the Lagrangian depends on the matrix ${\sf E}$. Because the 
matrix ${\sf E}^{-1}\vv$ is an element of ${\rm Sp}(56;{\bf R})$, the
matrix multiplying the two field strengths in \eqn{L3} is symmetric
under the interchange of $[AB]\leftrightarrow[CD]$.\footnote{%%%%%%%%
%%%%%%%%%%%%%%%%%%%%%%%%%%%%%%%%%%%%%%%%%%%%%%%%%%%%%%%%%%%%%%%%%%%%%
  Such symmetry properties follow from the symmetry under 
  interchanging index pairs in the products
$(u^{ij}{}_{\!AB}- v^{ijAB})\, (u^{kl}{}_{\!AB}+ v^{klAB})$ and  
$(u^{ij}{}_{\!AB}+ v^{ijAB})\, (u_{ij}{}^{\!CD}+ v_{ijCD})$.} %%%%%%%
%%%%%%%%%%%%%%%%%%%%%%%%%%%%%%%%%%%%%%%%%%%%%%%%%%%%%%%%%%%%%%%%%%%%%   

In order that the Lagrangian be invariant under a certain subgroup of
${\rm E}_{7(7)}$, one has to make a certain choice for the matrix ${\sf
E}$. According to  the analysis leading to \eqn{FGdual} and
\eqn{realdual}, this subgroup 
is generated on $\hat\vv$ by matrices $\Lambda$ and $\Sigma$, just as in
\eqn{eq:delta-z}, but with indices $A,B,\ldots$, rather than with
$I,J,\ldots$, satisfying 
\be
\mbox{Im } \Big(\Sigma_{ABCD} + \Lambda_{AB}{}^{\!CD}\Big) =0\,.
\ee
In order to be a subgroup of ${\rm E}_{7(7)}$ as well, they must also
satisfy \eqn{eq:E7}, but only after a proper conversion of the 
$I,J,\ldots$ to $A,B,\ldots$ indices. The gauge fields transform under
the real  subgroup ({\it i.e.}, the imaginary parts of the generators  
act exclusively on the 56-bein). A large variety of symmetry groups
exists, as one can deduce 
from the symmetry groups that are realized in maximal supergravity in
higher dimensions. One such group whose existence can be directly inferred
in this way, is ${\rm E}_{6(6)}\times {\rm SO}(1,1)$.

%%%%%%%%%%%%%%%%%%%%%%%%%%%%%%%%%%%%%%%%%%%%%%%%%%%%%%%%%%%%%%%%
\subsection{Gauging maximal supergravity; the $T$-tensor}
\label{T-tensor}
%%%%%%%%%%%%%%%%%%%%%%%%%%%%%%%%%%%%%%%%%%%%%%%%%%%%%%%%%%%%%%%%
The gauging of supergravity is effected by switching on the gauge
coupling constant, after assigning the various fields to
representations of the gauge group embedded in ${\rm
E}_{7(7)}$ or ${\rm E}_{6(6)}$. Only the gauge fields
themselves and the spinless fields can transform under this gauge
group. Hence the abelian field strengths are changed to nonabelian
ones and derivatives of the scalars are covariantized according to 
\be
\label{gauge-bein}
\pa_\m \vv \to \pa_\m \vv - g A^{AB}_\m \,T_{AB} \vv \,,
\ee
where the gauge group generators $T_{AB}$ are $56\times56$ matrices
which span a subalgebra of dimension equal to at most the number of
vector fields, embedded in the Lie 
algebra of  ${\rm E}_{7(7)}$ or ${\rm E}_{6(6)}$. The
structure constants of the gauge group are given by 
\be
{[}T_{AB}, T_{CD}] = f_{AB,CD}{}^{\!\!EF}\;T_{EF}\,. 
\ee 
It turns out that the viability for a gauging depends sensitively on
the choice of the gauge group and its corresponding embedding. This
aspect is most nontrivial for the $D=4$ theory, in view of
electric-magnetic duality. Therefore, we will mainly concentrate on
this theory. In $D=4$ dimensions, one must start from a 
Lagrangian that is symmetric under the desired gauge group, which
requires one to make a suitable choice of the matrix $\sf E$. In $D=5$ 
dimensions, the Lagrangian is manifestly symmetric under ${\rm
E}_{6(6)}$, so this subtlety does not arise. When effecting the
gauging the vector fields may decompose into those associated with the
nonabelian gauge group and a number of remaining gauge fields. When
the latter are charged under the gauge group, then there is a
potential obstruction to the gauging as the gauge invariance of these
gauge fields cannot coexist with the nonabelian gauge
transformations. However, in $D=5$ this obstruction can be avoided,
because (charged) vector fields can alternatively be described as
antisymmetric rank-2 tensor fields. For instance, 
the gauging of ${\rm SO}(p,6-p)$ involves 15 nonabelian gauge fields
and 12 antisymmetric tensor fields. The latter can transform under the
gauge group, because they are not realized as tensor {\it gauge} 
fields. Typically this conversion of vector into tensor fields leads
to terms that are inversely proportional to the gauge coupling
\cite{TownPilcNieuw}. However, to write down a corresponding
Lagrangian requires an even number of tensor fields.

Introducing the gauging leads directly to a loss of supersymmetry,
because the new terms in the Lagrangian lead to new variations. For
convenience we now restrict ourselves to $D=4$ dimensions. The
leading variations are induced by the modification \eqn{gauge-bein} of
the Cartan-Maurer equations. This modification was already noted in
\eqn{CM-gauged} and takes the form
\bea
F_{\mu\nu}({\cal Q})_i{}^j & = &  -\ft43\,{\cal P}_{[\mu}{}^{\!jklm}
\, {\cal P}_{\nu]iklm} -g\,F_{\m\n}^{AB} \,{\cal Q}_{AB\,i}{}^j\,, 
\nonumber \\ 
D_{[ \mu}^{~} {\cal P}_{\nu]}^{ijkl} &=& - \ft12 g\,F_{\m\n}^{AB} \,{\cal
P}_{AB}^{ijkl}  
\,, \label{GECM-Q-P} 
\eea
where 
\be
\label{P-Q-gauged}
\vv^{-1} T_{AB} \vv = \pmatrix{{\cal Q}_{AB\, ij}{}^{\!mn} 
&{\cal P}_{AB\, ijpq} \cr \noalign{\vglue 6mm}
{\cal P}_{AB}^{klmn} & {\cal Q}_{AB}{}^{\!kl}{}_{\!pq} \cr}\,.
\ee
These modifications are the result of the implicit dependence of
${\cal Q}_\m$ and ${\cal P}_\m$ on the vector potentials
$A_\m^{AB}$. The fact that the matrices $T_{AB}$ 
generate a subalgebra of the algebra 
associated with ${\rm E}_{7(7)}$, in the basis appropriate for $\vv$,
implies that the quantities ${\cal Q}_{AB}$ and ${\cal P}_{AB}$
satisfy the constraints,
\bea
{\cal P}_{AB}^{ijkl} &=& \ft1{24}\,\varepsilon^{ijklmnpq}\, {\cal
 P}_{AB\,mnpq}\,, \nonumber \\
{\cal Q}_{AB\, ij}{}^{\!kl}&=& \d^{[k}_{[i}\, {\cal Q}_{AB\,
j]}{}^{\!l]}\,, 
\eea
while ${\cal Q}_{ABi}{}^{\!j}$ is antihermitean and traceless. 
It is straightforward to write down the explicit expressions for
${\cal Q}_{AB}$ and ${\cal P}_{AB}$, 
\bea
{\cal Q}_{AB \,i}{}^{\!j} &=& \ft23 \Big[u_{ik}{}^{\!IJ}\, (\Delta_{AB}
u^{jk}{}_{\!IJ}) - v_{ikIJ}\, (\Delta_{AB} v^{jkIJ})  \Big]
 \,, \nonumber\\ 
{\cal P}_{AB}^{ijkl} &=&  v^{ijIJ}\,(\Delta_{AB} u^{kl}{}_{\!IJ})-
u^{ij}{}_{\!IJ} (\Delta_{AB} v^{klIJ})  \,.
\eea
where $\Delta_{AB}u$ and $\Delta_{AB}v$ indicate the change of
submatrices in $\vv$ induced by multiplication with the generator
$T_{AB}$.  
Note that we could have expressed the above quantities in terms of the
modified 56-bein $\hat \vv$, on which the ${\rm E}_{7(7)}$
transformations act in the basis that is appropriate for the field
strengths, provided we change the generators $T_{AB}$ into 
\be 
\hat T_{AB} = {\sf E}^{-1} T_{AB}\,{\sf E}\,.
\ee
This is done below. 

When establishing supersymmetry of the action one needs the
Cartan-Maurer equations at an early stage to cancel variations from
the gravitino kinetic terms and the Noether term (the term in the
Lagrangian proportional to $\bar \chi\psi_\m {\cal P}_\n$). The
order-$g$ terms in the Maurer-Cartan equation yield the leading
variations of the Lagrangian. They are linearly proportional to the
fermion fields and read, 
\bea
\label{g-F-variation}
\d\lagr&=& \ft14 g(\bar \epsilon_j \g^\rho\g^{\m\n} \psi^i_{\rho}
-\bar \epsilon^i \g^\rho\g^{\m\n} \psi_{\rho j}) \,
 {\cal Q}_{AB\,i}{}^j\, (u^{kl}{}_{\!AB} + v^{klAB} ) \, 
{\sf F}^+_{\m\n kl} \nn\\
 &&+ \ft1{288} \varepsilon^{ijklmnpq}\, \bar \chi_{ijk}\g^{\m\n} \e_l
\,{\cal P}_{AB\,mnpq} \, (u^{rs}{}_{\!AB} + v^{rsAB} ) \, {\sf
F}^+_{\m\n rs}\nn\\[1.2mm]
&&+ {\mbox{ h.c.}}
\eea
The first variation is proportional to an ${\rm SU}(8)$ tensor
$T_i^{jkl}$, which is known as the $T$-tensor, 
\bea
\label{T1-tensor}
T_i^{jkl} &=& \ft34{\cal Q}_{AB\,i}{}^j\, (u^{kl}{}_{\!AB} + v^{klAB})
\\  
&=& \ft12 \Big[u_{im}{}^{\!CD}\, (\hat\Delta_{AB}
u^{jm}{}_{\!CD}) - v_{imCD}\, (\hat\Delta_{AB} v^{jmCD})  \Big] 
(u^{kl}{}_{\!AB} + v^{klAB})\,,\nn 
\eea
where $\hat\Delta_{AB} u$ and $\hat\Delta_{AB} v$ are the submatrices
of $\hat T_{AB}\hat\vv$. 
Another component of the $T$-tensor appears in the second variation
and is equal to  
\bea
\label{T2-tensor}
T_{ijkl}^{mn} &=&\ft12 {\cal P}_{AB\,ijkl} \, (u^{mn}{}_{\!AB} +
v^{mnAB} ) 
\\
&=& \ft12\Big[ v_{ijCD}\,(\hat\Delta_{AB} u_{kl}{}^{\!CD})- 
u_{ij}{}^{\!CD} (\hat\Delta_{AB} v_{klCD})\Big]
\, (u^{mn}{}_{\!AB} + v^{mnAB} )\,. \nn
\eea
The $T$-tensor is thus a cubic product of the 56-bein $\hat\vv$ which
depends in a nontrivial way on the embedding of the gauge group into
${\rm E}_{7(7)}$. It satisfies a number of important properties. Some of
them are rather obvious (such as $T_i^{ijk}=0$), and follow rather
straightforwardly from the definition. We will concentrate on
properties which are perhaps less obvious. Apart 
from the distinction between $\vv$ and $\hat\vv$, which is a special
feature of $D=4$ dimensions, these properties are generic. 

First we observe that ${\rm SU}(8)$ covariantized variations of the
$T$-tensor are again proportional to the $T$-tensor. These variations
are induced by \eqn{eq:dvv} and \eqn{eq:duv}. Along the same lines as 
before we can show that the ${\rm SU}(8)$ tensors ${\cal Q}_{AB}$ and
${\cal P}_{AB}$ transform  in the adjoint representation of ${\rm
E}_{7(7)}$, which allows one to derive,
\bea
\label{T-variations}
\d T_i^{jkl} &=&  \Sigma^{jmnp}_{~} \,T^{kl}_{imnp} -\ft1{24}
\varepsilon^{jmnpqrst} \,\Sigma_{imnp}^{~} \,T^{kl}_{qrst} +
\Sigma^{klmn}_{~}\, T^j_{imn} \,, \nn\\
\d T_{ijkl}^{mn} &=& \ft43 \Sigma_{p[ijk}^{~}\, T^{pmn} _{l]} -\ft1{24}
\varepsilon_{ijklpqrs}\, \Sigma^{mntu}_{~}\, T^{pqrs}_{tu} \,.
\eea
This shows that the ${\rm SU}(8)$ covariant $T$-tensors can be
assigned to a  
representation of ${\rm E}_{7(7)}$. This property will play an
important role below. 

Before completing the analysis leading to a consistent gauging we
stress that all variations are from now on expressed in terms of the
$T$-tensor, as its variations yield again the same tensor. This
includes the ${\rm SU}(8)$ covariant derivative of the $T$-tensor,
which follows directly  
from \eqn{T-variations} upon the substitutions $\d\to D_\m$ and
$\Sigma \to {\cal P}_\m$.  A viable gauging requires that the
$T$-tensor satisfies a number of rather nontrivial identities, as we
will discuss shortly, but the new terms in the Lagrangian and
transformation rules have a universal form, irrespective of the gauge
group. Let us first describe these new terms. First of all, to cancel
the variations \eqn{g-F-variation} we need masslike terms in the
Lagrangian, 
\bea
\label{g-masses}
\lagr_{\rm masslike} & = & g\, e\Big\{ \ft12 \sqrt{2}\,  A_{1ij}
\bar{\psi}\,^i_\mu  
\gamma^{\mu\nu} \psi^j_\nu + \ft{1}{6}  A_{2i}^{jkl} \,
\bar{\psi}^i_\mu \gamma^\mu \chi_{jkl} \nonumber \\
& & \hspace{7mm} + A_3^{ijk,lmn}\,
\bar{\chi}_{ijk}\chi_{lmn} + {\rm h.c.}\Big \} \, ,
\eea
whose presence necessitates corresponding modifications of the
supersymmetry transformations of the fermion fields, 
\bea 
\label{g-variations}
\d_g\bar\psi^i_\m &=& -\sqrt{2} g \, A_1^{ij}\,\bar\e_j \gamma_\m \,,\nn\\
\d_g\chi^{ijk}&=& - 2 g \,A_{2l}{}^{\!ijk}\,\bar\e^l\,. 
\eea
Finally at order $g^2$ one needs a potential for the spinless fields, 
\be
\label{g-potential}
P(\vv ) = g^2 \Big \{\ft{1}{24}  \vert A_{2i}{}^{\!jkl}\vert^2 -
\ft{1}{3} \vert A_1^{ij}\vert^2\Big\} \,.
\ee
These last three formulae will always apply, irrespective of the gauge
group. Note that the tensors $A_1^{ij}$, $A_{2i}{}^{\!jkl}$ and
$A_3^{ijk,lmn}$ have certain symmetry properties dictated by the way
they appear in the Lagrangian 
\eqn{g-masses}. To be specific, $A_1$ is symmetric in $(ij)$, $A_2$
is fully antisymmetric in $[jkl]$ and $A_3$ is antisymmetric in
$[ijk]$ as well as in $[lmn]$ and symmetric under the interchange 
$[ijk]\leftrightarrow[lmn]$. This implies that these tensors transform
under ${\rm SU}(8)$ according to the representations
\bea
\label{A:tosu8}
A_1&:& {\bf 36}\,,\nn \\
A_2&:& {\bf 28}+ {\bf 420}\,, \nn\\
A_3&:& \overline{\bf 28}+\overline{\bf 420}+ \overline{\bf 1176} +
\overline{\bf 1512} \,. 
\nn 
\eea

The three ${\rm SU}(8)$ covariant tensors, $A_1$, $A_2$ and $A_3$, 
which depend only on the spinless fields, must be linearly related to
the $T$-tensor, because they were introduced for the purpose of
cancelling the variations \eqn{g-F-variation}. To see how this can be
the case, let us analyze the ${\rm SU}(8)$ content of the
$T$-tensor. As we mentioned 
already, the $T$-tensor is cubic in the 56-bein, and as such is
constitutes a certain tensor that transforms under ${\rm
E}_{7(7)}$. The transformation properties were given in
\eqn{T-variations}, where we made use of the fact that the $T$-tensor
consists of a product of the
fundamental times the adjoint representation of ${\rm
E}_{7(7)}$. Hence the $T$-tensor comprises the representations,
\be
{\bf 56}\times{\bf 133} = {\bf 56} + {\bf 912} + {\bf 6480} \,.
\ee
The representations on the right-hand side can be decomposed under
the action of ${\rm SU}(8)$, with the result
\bea
\label{T:e7tosu8}
{\bf 56}&=& {\bf 28}+ \overline{\bf 28}\,, \nn\\
{\bf 912}&=& {\bf 36}+\overline{\bf 36} + {\bf 420}
+ \overline{\bf 420}\,, \\
{\bf 6480}&=& {\bf 28}+ \overline{\bf 28} +{\bf 420}+ 
\overline{\bf  420} +{\bf 1280}+ \overline{\bf 1280}
+{\bf 1512}+ \overline{\bf 1512} \,. \nn
\eea
Comparing these representations to the ${\rm SU}(8)$
representations to which the tensors $A_1-A_3$ (and their complex
conjugates) belong, we note that there is a mismatch between
\eqn{T:e7tosu8} and \eqn{A:tosu8}. In view of \eqn{T-variations} the
$T$-tensor must be restricted by suppressing complete representations
of ${\rm E}_{7(7)}$ in 
order that its variations and derivatives remain consistent. This
proves that the $T$-tensor cannot contain the entire ${\bf 6480}$
representation of ${\rm E}_{7(7)}$, so that it 
must consist of  the ${\bf 28}+{\bf 36}+{\bf 420}$ representation of
${\rm SU}(8)$ (and its  complex conjugate). This implies that the
$T$-tensor is decomposable into $A_1$ and $A_2$, whereas $A_3$ is not
an independent tensor and can be expressed in terms of $A_2$.  
Indeed this was found by explicit calculation, which gave rise to
\bea
T_i^{jkl} &=& - \ft{3}{4} A_{2i}{}^{\!jkl} + \ft{3}{2}  
A_1^{j[k}\,\delta_i^{l]} \,,\nn \\
T^{mn}_{ijkl} &=& - \ft 43 \d^{[m}_{[i} T^{n]}_{jkl]}\,,\nn\\
A_3^{ijk,lmn}&=&- \ft1{108}\sqrt{2} \, \varepsilon^{ijkpqr[lm}
T_{pqr}^{n]} \,.
\eea
Note that these conditions are necessary, but not sufficient as one
also needs nontrivial identities quadratic in the $T$-tensors in order
to deal with the variations of the Lagrangian of order $g^2$. One then
finds that there is yet another 
constraint, which suppresses the ${\bf 28}$ representation of the
$T$-tensor, 
\be
T_i^{[jk]i}=0\,.
\ee
Observe that a contraction with the first upper index is also equal to
zero, as follows from the definition \eqn{T1-tensor}. Hence the
$T$-tensor transforms under ${\rm E}_{7(7)}$ according to the 
${\bf 912}$ representation which decomposes into the ${\bf 36}$ and
${\bf 420}$ representations of ${\rm SU}(8)$ and their complex
conjugates residing in the tensors $A_1$ and $A_2$, respectively, 
\be
A_1^{ij} = \ft4{21} T_k^{ikj}\,,\qquad A_{2i}^{jkl} = -\ft 43
T_i^{[jkl]}\,.  
\ee

Although we concentrated on the $D=4$ theory, we should stress once more
that many of the above features are generic and apply in other 
dimensions. For instance, the unrestricted $T$-tensors in $D=5$ and 3
dimensions belong to the following representations of ${\rm E}_{6(6)}$  
and ${\rm E}_{8(8)}$, respectively\footnote{%%%%%%%%%%%%%%%%%%%%%
  The $D=3$ theory has initially no vector fields, but those can be
  included by adding Chern-Simons terms. These terms lose their
  topological nature when gauging some of the ${\rm E}_{8(8)}$
  isometries \cite{NicSam}. }%%%%%%%%%%%%%%%%%%%%%%%%%%%%%%%
%%%%%%%%%%%%%%%%%%%%%%%%%%%%%%%%%%%%%%%%%%%%%%%%%%%%%%%%%%%%%%%%%%
\bea
D=5&:&{\bf 27}\times{\bf 78} = {\bf 27} + {\bf 351} + {\bf 1728} \,,
\nn\\ 
D=3&:& {\bf 248}\times{\bf 248} = {\bf 1} + {\bf 248}+ {\bf 3875}
 +{\bf 27000} +  {\bf 30380} \,. 
\eea
In these cases a successful gauging requires the $T$-tensor to be
restricted to the ${\bf 351}$ and the ${\bf 1}+{\bf 3875}$ 
representations, respectively, which decompose as follows
under the action of ${\rm USp}(8)$ and ${\rm SO}(16)$, 
\bea
\label{T:e68to}
{\bf 351}&=& {\bf 36}+ {\bf 315} \,, 
\nn \\
{\bf 1}+{\bf 3875}&=& {\bf 1}+ {\bf 135}+{\bf 1820} + {\bf 1920} 
\,.
\eea
These representations correspond to the tensors $A_1$ and $A_2$; for
$D=5$ $A_3$ is again dependent while for $D=3$ there is an independent
tensor $A_3$ associated with the ${\bf 1820}$ representation of 
${\rm SO}(16)$.  

We close with a few comments regarding the various gauge groups that
have been considered. As we mentioned at the beginning of this
chapter, the first gaugings were to some extent motivated by
corresponding Kaluza-Klein compactifications. The $S^7$ and the $S^4$
\cite{PPvN} compactifications of 11-dimensional supergravity and the
$S^5$ compactification of IIB supergravity, gave rise to the gauge
groups ${\rm SO}(8)$, ${\rm SO}(5)$ and ${\rm SO}(6)$,
respectively. Noncompact gauge groups were initiated in \cite{Hull}
for the 4-dimensional theory; for the 5-dimensional theory they were
also realized in \cite{GunaRomansWarner}  
and in \cite{AndCordFreGual}. In $D=3$ dimensions
there is no guidance from Kaluza-Klein compactifications and one has
to rely on the group-theoretical analysis described above. In that
case there exists a large variety of gauge groups of rather high
dimension \cite{NicSam}. Gaugings can also be constructed via a  
so-called Scherk-Schwarz reduction from higher dimensions
\cite{AndDauFerrLle}. To give a really exhaustive classification
remains cumbersome. For explorations based on the group-theoretical
analysis explained above, see \cite{NicSam2,dWSamTrig}.
%%%%%%%%%%%%%%%%%%%%%%%%%%%%%%%%%%%%%%%%%%%%%%%%%%%%%%%%%%%%%%%%

%%%%%%%%%%%%%%%%%%%%%%%%%%%%%%%%%%%%%%%%%%%%%%%%%%%%%%%%%%%%%%%%
%%%%%%%%%%%%%%%%%%%%%%%%%%%%%%%%%%%%%%%%%%%%%%%%%%%%%%%%%%%%%%%%
%\newpage %%%%%%%%%%%%%%%%%%%%%%%%%%%%%%%%%%%%%%%%%%%%%%%%%%%%%%%
%%%%%%%%%%%%%%%%%%%%%%%%%%%%%%%%%%%%%%%%%%%%%%%%%%%%%%%%%%%%%%%%
%%%%%%%%%%%%%%%%%%%%%%%%%%%%%%%%%%%%%%%%%%%%%%%%%%%%%%%%%%%%%%%%
%%%%%%%%%%%%%%%%%%%%%%%%%%%%%%%%%%%%%%%%%%%%%%%%%%%%%%%%%%%%%%%%
%%%%%%%%%%%%%% CHAPTER VI %%%%%%%%%%%%%%%%%%%%%%%%%%%%%%%%%%%%%%
%%%%%%%%%%%%%%%%%%%%%%%%%%%%%%%%%%%%%%%%%%%%%%%%%%%%%%%%%%%%%%%%
\section{Supersymmetry in anti-de Sitter space}
\label{susy-ads}
\setcounter{equation}{0}
%%%%%%%%%%%%%%%%%%%%%%%%%%%%%%%%%%%%%%%%%%%%%%%%%%%%%%%%%%%%%%%%
In section~\ref{simple-supergravity} we presented the first steps
in the construction of a generic supergravity theory, starting with
the Einstein-Hilbert Lagrangian for gravity and the Rarita-Schwinger
Lagrangians for the gravitino fields. We established 
the existence of two supersymmetric gravitational backgrounds, namely
flat Minkowski space and anti-de Sitter space with a cosmological
constant proportional to $g^2$, where $g$ was some real coupling
constant proportional to the the inverse anti-de Sitter radius. The
two cases are clearly related and flat space is obtained in the limit
$g\to 0$, as can for instance be seen from the expression of the
Riemann curvature ({\it c.f.} \eqn{ads-curv}),
\be
R^{~}_{\m\n\rho}{}^{\!\sigma} = g^2(g^{~}_{\m\rho} \,\d_\n^\s -
g^{~}_{\n\rho} \,\d_\m^\s )\,.
\ee
Because both flat Minkowski space and  anti-de Sitter space
are maximally symmetric, they have $\ft12 D(D+1)$ independent
isometries which comprise the Poincar\'e group or the group 
${\rm SO}(D-1,2)$, respectively. 
The algebra of the combined bosonic and fermionic symmetries
is called the anti-de Sitter superalgebra. Note again that the
derivation in section~\ref{simple-supergravity} was incomplete and
in general one will need to introduce additional fields.  

In this chapter we will mainly be dealing with simple anti-de Sitter
supersymmetry and we will always assume that $3<D\leq7$. In that case
the bosonic subalgebra coincides with  
the anti-de Sitter algebra. In $D=3$ spacetime dimensions the anti-de
Sitter group ${\rm SO}(2,2)$ is not simple. There exist $N$-extended
versions where one introduces $N$ supercharges, each
transforming as a spinor under the anti-de Sitter group. These $N$
supercharges transform under a compact R-symmetry group, whose
generators will appear in the $\{Q,\bar Q\}$ anticommutator. As we
discussed in sect.~2.5, the R-symmetry 
group is in general not the same as in Minkowski space; according to 
Table~\ref{pq-spinors}, we have ${\rm H}_{\rm R} = {\rm SO}(N)$ for
$D=4$,  ${\rm H}_{\rm R} ={\rm U}(N)$ for $D=5$, and 
${\rm H}_{\rm R} ={\rm USp}(2N)$ for $D=6,7$. For $D>7$ the 
superalgebra is no longer simple \cite{nahm}; its bosonic subalgebra
can no longer be restricted to the sum of the anti-de Sitter 
algebra and the R-symmetry algebra, but one needs extra bosonic
generators that transform as high-rank antisymmetric tensors under the
Lorentz group (see also, \cite{vanHoltenVanProeyen}).  In contrast to
this, there exists an ($N$-extended)   
super-Poincar\'e algebra associated with flat Minkowski space of
any dimension, whose bosonic generators correspond to the
Poincar\'e group, possibly augmented with the R-symmetry 
generators associated with rotations of the supercharges.  

Anti-de Sitter space is isomorphic to ${\rm
SO}(D-1,2)/{\rm SO}(D-1,1)$ and thus belongs to the coset spaces that
were discussed extensively in chapter~4. According to
\eqn{embedding-condition} it is possible to describe anti-de Sitter
space as a hypersurface in a $(D+1)$-dimensional embedding
space. Denoting the extra 
coordinate of the embedding space by $Y^-$, so that we have
coordinates $Y^A$ with $A= -,0,1,2,\ldots,D-1$, this hypersurface is
defined by 
\be
- (Y^-)^2 - (Y^0)^2 + \vec Y{}^{\,2} = \eta_{AB} \,Y^AY^B = -g^{-2}\,.
\label{embedding}
\ee
Obviously, the hypersurface is invariant under linear transformations
that leave the metric $\eta_{AB}  ={\rm diag}\,(-,-,+,+,\ldots, +)$
invariant. These transformations constitute the group SO$(D-1,2)$. The 
$\ft12D(D+1)$ generators denoted by $M_{AB}$ act on the embedding
coordinates by 
\be 
M_{AB} = Y_A{\pa\over \pa Y^B} -Y_B{\pa\over \pa Y^A}\,,
\ee
where we lower and raise indices by contracting with $\eta_{AB}$ and
its inverse $\eta^{AB}$. It is now easy to evaluate the commutation
relations for the $M_{AB}$,
\be
{[}M_{AB}, M_{CD}] = \eta_{BC}\,M_{AD} - \eta_{AC}\,M_{BD} -\eta_{BD}\,M_{AC}
+\eta_{AD}\,M_{BC}\,. \label{ads-algebra}
\ee
Anti-de Sitter space has the topology of $S^1 \mbox{[time]}\times
{\bf R}^{D-1}\mbox{[space]}$ and has closed timelike curves. These
curves can be avoided by unwrapping  $S^1$, so that one finds the
universal covering space denoted by CadS, which has the topology of
${\bf R}^{D}$. There exist no Cauchy 
surfaces in this space. Any attempt to determine the outcome of some
evolution or wave equation from a spacelike surface requires fresh
information coming from a timelike infinity which takes a finite amount
of time to arrive \cite{HawkEll,AvisIshamStorey}. Spatial infinity is
a timelike surface which cannot be reached by timelike
geodesics. There are many ways
to coordinatize anti-de Sitter space, but we will avoid using explicit
coordinates.  

For later use we record the (simple) anti-de Sitter superalgebra, which in
addition to \eqn{ads-algebra} contains the (anti-)commutation relations, 
\bea
\{Q_\a, \bar Q_\b\} &=& - \ft12 (\G_{AB})_{\a\b} \, M^{AB} \,,\nn\\
{[}M_{AB},\bar Q_\a] &=& \ft 12 (\bar Q\, \G_{AB})_\a\,.
\label{ads-superalgebra} 
\eea
Here the matrices $\G_{AB}$, which will be defined later, are the
generators of ${\rm SO}(D-1,2)$ group in the spinor representation. 
As we alluded to earlier, this algebra changes its form when
considering $N$ supercharges which rotate under
R-symmetry, because the R-symmetry generators will appear on the 
right-hand side of the $\{Q,\bar Q\}$ anticommutator. 

The relation with the Minkowski case proceeds by means of a so-called
Wigner-In\"on\"u contraction. Here one rescales the generators
according to $M_{-A}\to g^{-1} P_A$, $Q\to g^{-1/2} Q$, keeping the
remaining generators $M_{AB}$ corresponding to the Lorentz subalgebra
unchanged. In the limit $g\to 0$, the generators $P_A$ will form a
commuting subalgebra and the full algebra contracts to the
super-Poincar\'e algebra.   

On spinors, the anti-de Sitter algebra can be realized by the
following combination of gamma matrices $\G_a$ in $D$-dimensional
Minkowski space, 
\be
M_{AB} = \ft 12 \Gamma_{AB} = \left\{ 
\begin{array}{lll}
\ft 12 \G_{ab}& \mbox{for} & A,B= a,b\,, \\[3mm]
\ft 12\G_a  &\mbox{for} & A=-\,, B= a \end{array}\right.
\ee
with $a,b= 0,1,\ldots,D-1$. Our gamma matrices satisfy the Clifford
property $\{\Gamma^a\,,\, \Gamma^b\} = 2\, \eta^{ab}\, {\bf 1}$, where
$\eta^{ab} = {\rm diag}\,(-,+,\ldots,+)$ is the $D$-dimensional
Lorentz-invariant metric.\footnote{%%%%%%%%%%%%%%%%%%%%%%%%%%%%%%%%
%%%%%%%%%%%%%%%%%%%%%%%%%%%%%%%%%%%%%%%%%%%%%%%%%%%%%%%%%%%%%%%%%%%
  Note that when the gravitino is a Majorana spinor, the quantities
  $\G_{AB}\e$   should satisfy the same Majorana constraint.} %%%%%
%%%%%%%%%%%%%%%%%%%%%%%%%%%%%%%%%%%%%%%%%%%%%%%%%%%%%%%%%%%%%%%%%%%
Concerning the R-symmetry group in anti-de Sitter space, the reader
is advised to consult section~2.5. 

Of central importance is the quadratic Casimir operator of the
isometry group ${\rm SO}(D-1,2)$, defined by 
\be 
{\cal C}_2 = -\ft12 \,M^{AB}\,M_{AB}\,. \label{Casimir}
\ee
The group SO$(D-1,2)$ has more Casimir operators when $D>3$, but these
are of higher order in the generators and will not play a role
in the following.  
To make contact between the masslike terms in the wave equations and
the properties of the irreducible representations of the anti-de
Sitter group, which we will discuss in section~\ref{masslike},  it
is important that we establish the relation between the wave operator
for fields that live in anti-de Sitter space, which involves the 
appropriately covariantized D'Alembertian $\Box_{\rm adS}$, and the
quadratic Casimir 
operator ${\cal C}_2$. We remind the reader that fields in anti-de
Sitter space are multi-component functions of the anti-de Sitter
coordinates that 
rotate irreducibly under the action of the Lorentz group ${\rm
SO}(D-1,1)$. The appropriate formulae were given at the end of
section~4.2 ({\it c.f.} \eqn{G-isometry} and \eqn{double-isometry})
and from them one can derive,
 \be
{\cal C}_2 = \Box_{\rm adS}\Big\vert_{g=1} +
{\cal C}_2^{\rm Lorentz}\,,
\label{casimir=box}
\ee
where ${\cal C}_2^{\rm Lorentz}$ is the quadratic Casimir operator
for the representation of the Lorentz group to which the fields have
been assigned. This result can be proven for any symmetric,
homogeneous, space (see, for example, \cite{Pilch-Schell}). 
For scalar fields, the second term in \eqn{casimir=box} vanishes and
the proof is elementary (see, {\it e.g.}, \cite{deWitHerger}).

Let us now briefly return to the supersymmetry algebra as it is
realized on the vielbein field. Using the transformation rules
\eqn{ads-susy-grav} 
The commutator of two supersymmetry transformations
yields an infinitesimal general-coordinate transformation and a
tangent-space Lorentz transformation. For example, we obtain for the
vielbein, 
\bea
{[}\d_1,\d_2]\,e_\m{}^a &=& \ft12 \bar\e_2\,\G^a\,\d_1\psi_\m -\ft12
\bar\e_1\,\G^a\,\d_2\psi_\m  \nn\\
&=& D_\m(\ft12\bar\e_2\,\G^a\e_1) + \ft12 g\,
(\bar\e_2\,\G^{ab}\e_1)\, e_{\m b}\,. \label{qq-comm}
\eea
The first term corresponds to a spacetime diffeomorphism and the
second one to a tangent space (local Lorentz) transformation. Here we
consider only the gravitational sector of the theory; for a 
complete theory there are additional contributions, but nevertheless
the above terms remain and \eqn{qq-comm} should be realized uniformly
on all the fields. In the anti-de Sitter background, were the
gravitino field vanishes, the parameters of the supersymmetry
transformations are Killing spinors satisfying \eqn{killing-spinor} so
that the gravitino field remains zero under supersymmetry. Therefore
both the gravitino and the vielbein are left invariant under
supersymmetry, so that the combination of symmetries on the   
right-hand side of \eqn{qq-comm} should vanish when $\e_1$ and $\e_2$
are Killing spinors. Indeed, the diffeomorphism
with parameter $\xi^\m = \ft12 \bar\e_2\,\G^\m\e_1$, is an
anti-de Sitter  Killing vector ({\it i.e.}, it satisfies 
\eqn{killing-0}),  because $D_\m
(g\,\bar\e_2 \G_{\n\rho}\e_1) = -g^2\, g_{\m[\rho}\xi_{\n]}$ is
antisymmetric in $\m$ and $\n$. As for all Killing
vectors, higher derivatives can be decomposed into the Killing vector
and its first derivative. Indeed, we find $D_\m  (g\,\bar\e_2
\G_{\n\rho}\e_1) = -g^2\, g_{\m[\rho}\xi_{\n]}$ in the case at
hand. The Killing vector 
can be decomposed into the $\ft12 D(D+1)$ Killing vectors of the
anti-de Sitter space. The last term in \eqn{qq-comm} is a compensating
target space transformation of the type we have been discussing
extensively in section~4.2 for generic coset spaces.
%%%%%%%%%%%%%%%%%%%%%%%%%%%%%%%%%%%%%%%%%%%%%%%%%%%%%%%%%%%%%%%%
%%%%%%%%%%%%%%%%%%%%%%%%%%%%%%%%%%%%%%%%%%%%%%%%%%%%%%%%%%%%%%%%
\subsection{Anti-de Sitter supersymmetry and masslike terms}
\label{masslike}
\setcounter{equation}{0}
%%%%%%%%%%%%%%%%%%%%%%%%%%%%%%%%%%%%%%%%%%%%%%%%%%%%%%%%%%%%%%%%
In flat Minkowski space all fields
belonging to a supermultiplet are subject to field equations with the
same mass, because the momentum operators commute 
with the supersymmetry charges, so that $P^2$ is a
Casimir operator. For supermultiplets in anti-de Sitter space this is
no longer the case, so that masslike terms will 
not necessarily be the same for different fields belonging to the same 
multiplet. We have already discussed the interpretation of masslike
terms for the gravitino, following \eqn{cosm-term-lagr}. This phenomenon
will be now illustrated below in a specific example, namely a scalar
chiral supermultiplet in $D=4$ spacetime dimensions. Further clarification
from an algebraic viewpoint will be given later in
section~\ref{superalgebra}. 

A scalar chiral supermultiplet in 4 spacetime dimensions consists of a
scalar field $A$,  
a pseudoscalar field $B$ and a Majorana spinor field $\psi$. 
In anti-de Sitter space the supersymmetry transformations of the fields are
proportional to a spinor parameter $\e(x)$, which is a
Killing spinor in the anti-de Sitter space, {\it i.e.}, $\e(x)$ must satisfy 
the Killing spinor equation \eqn{killing-spinor}. In the notation of
this section, this equation reads,
\be
\label{killing-spinor-4}
\Big(\pa_\m -\ft14\omega^{ab}\g_{ab} + \ft12 g\,e_\m^{\;a}\g_a\Big)\e
=0\,,
\ee
where we made the anti-de Sitter vierbein and spin connection
explicit. We allow for two constants $a$  
and $b$ in the supersymmetry transformations, which we parametrize as
follows,   
\bea
\d A &=& \ft14 \bar\e\psi\,, \qquad  \d B= \ft14 i\bar\e \g_5\psi\,,\nn\\
\d \psi &=& \rlap/\partial (A+ i\g_5B)\e - (a\,A + i b\,\g_5B)\e\,.
\eea
In this expression the anti-de Sitter vierbein field has been used to
contract the gamma matrix with the derivative.  
The coefficient of the first term in $\d\psi$ has been chosen
such as to ensure that $[\d_1,\d_2]$ yields the correct coordinate
transformation 
$\xi^\m D_\m$ on the fields $A$ and $B$.
To determine the constants $a$ and $b$ and the field equations of the
chiral multiplet, we consider the closure of the supersymmetry algebra
on the spinor field. After some Fierz reordering we obtain the result, 
\be
{[}\d_1,\d_2] \psi =  \xi^\m D_\m \psi + \ft1{16} (a-b)
\,\bar\e_2\gamma^{ab}\e_1 \, \g_{ab} \psi -\ft12  \xi^\rho \g_\rho
[\rlap/\!D \psi + \ft12 (a+b) \psi] \,.  \;{~}
\ee
We point out that derivatives acting on $\epsilon(x)$ occur in this
calculation at an intermediate stage and should not be suppressed in view of
\eqn{killing-spinor-4}. However, they produce terms proportional to $g$
which turn out to cancel in the above commutator. Now we note that the
right-hand side should constitute a coordinate transformation and 
a Lorentz transformation, possibly up to a field equation. Obviously,
the coordinate transformation coincides with \eqn{qq-comm} but
the correct Lorentz transformation is only reproduced  provided that
$a-b = 2g$. If we now define $m =\ft 12(a+b)$, so that the last term
is just the Dirac equation with mass $m$, we find
\be
a= m+g\,,\qquad b= m-g\,.
\ee
Consequently, the supersymmetry transformation of $\psi$ equals 
\be 
\d\psi = \rlap/\!D (A+i\g_5 B )\e - m (A+i\g_5 B ) \e - g(A-i\g_5 B )\, 
\e\,, \label{delpsi}
\ee
and the fermionic field equation equals 
$(\rlap/\!D + m)\psi =0$. The second term in \eqn{delpsi}, which is
proportional to $m$, can be accounted for by introducing an auxiliary
field to the supermultiplet. The third term, which is proportional to
$g$, can be understood as a compensating $S$-supersymmetry
transformation associated with auxiliary fields in 
the supergravity sector (see, {\it e.g.}, \cite{deWit82}). In order to
construct the corresponding field 
equations for $A$ and $B$, we consider the variation of the fermionic
field equation. Again we have to take into account that derivatives on
the supersymmetry parameter are not equal to zero.  This yields the
following second-order differential equations,  
\bea
{}[ \Box_{\rm adS} + 2g^2 - m(m-g) ]\, A&=&0\,, \nn\\
{}[ \Box_{\rm adS} + 2g^2 - m(m+g) ] \,B &=&0\,,\nn\\
{}[ \Box_{\rm adS} + 3 g^2 - m^2 ]\, \psi &=& 0 \,. \label{field-eqs}
\eea
The last equation follows from the Dirac equation. Namely, one
evaluates $(\rlap/\!D-m)(\rlap/\!D+m)\psi$, which gives rise to the
wave operator 
$ \Box_{\rm adS} +\ft12 [\rlap/\!D, \rlap/\!D] - m^2$. The commutator
yields the Riemann 
curvature of the  anti-de Sitter space. In an anti-de Sitter space of
arbitrary dimension $D$ this equation then reads,   
\be 
\label{Dirac}
{}[  \Box_{\rm adS} + \ft14 D(D-1) g^2 -m^2 ]\psi =0\,,
\ee
which, for $D=4$ agrees with the last equation of \eqn{field-eqs}. 
A striking feature of the above result is that the field equations
\eqn{field-eqs} all have different 
mass terms, in spite of the fact that they belong to the same
supermultiplet \cite{BreitFreed}. Consequently, the role of mass is
quite different in anti-de Sitter space as compared to flat Minkowski
space. This will be elucidated later in section~\ref{adSreps}. 

For future applications we also evaluate the Proca equation for a
massive vector field,
\be
\label{Proca-0}
D^\mu(\pa_\m A_\n - \pa_\n A_\m) - m^2 \,A_\n = 0 \,.
\ee
This leads to\footnote{%%%%%%%%%%
  When $m\not= 0$, otherwise one can impose this equation as a gauge
  condition. } %%%%%%%%%%%%%%%%%%
$D^\m A_\m=0$, so that \eqn{Proca-0} reads $D^2
A_\n -[D^\m,D_\n] A_\m - m^2 \,A_\n = 0$ or, in anti-de Sitter space,
\be
\label{Proca}
{}[ \Box_{\rm adS} + (D-1)g^2 - m^2]\, A_\m =0 \,. 
\ee
This can be generalized to an antisymmetric tensor of rank $n$, whose
field equation reads (antisymmetrizing over indices
$\n_1,\ldots,\n_n$), 
\be
\label{antisymm}
(n+1)\, D^\mu \pa_{[\m} C_{\n_1\cdots\n_n]} - m^2 \,C_{\n_1\cdots
\n_n}  = 0 \,. 
\ee
In the same way as before, this leads to 
\be
{}[\Box_{\rm adS} + n(D-n)g^2 - m^2] \,C_{\n_1\cdots\n_n} = 0\,.
\ee

The $g^2$ term in the field equations for the scalar fields can be
understood from the observation that the scalar D'Alembertian (in an
arbitrary gravitational background) can be 
extended to a conformally invariant operator (see, {\it e.g.}, 
\cite{deWit82}),
\be
 \Box  + {1\over 4}{D-2\over D-1}\, R =  \Box   + \ft14 D(D-2)\,
g^2 \, ,  \label{massless-scalar}
\ee
which seems the obvious candidate for a massless wave
operator for scalar fields. 
Indeed, for $D=4$, we do reproduce the $g^2$ dependence in the
first two equations \eqn{field-eqs}. Observe that 
the Dirac operator $\rlap /\!D$ is also conformally invariant and so is the
wave equation associated with the Maxwell field. 

Using \eqn{casimir=box} we can now determine the values for the
quadratic Casimir operator for the representations described by
scalar, spinor, vector and tensor fields. The quadratic Casimir operator of the
Lorentz group takes the values $0$, $\ft18 D(D-1)$, $D-1$ and $n(D-n)$
for scalar, spinor, vector and tensor fields respectively. Combining
this result with \eqn{massless-scalar}, \eqn{Dirac}, \eqn{Proca} and
\eqn{antisymm}, \eqn{casimir=box} yields the following values for the
quadratic Casimir operators, 
\bea
\label{conf-box-C}
{\cal C}_2^{\rm scalar} &=& - \ft 14 D(D-2) + m^2\,,\nn\\
{\cal C}_2^{\rm spinor} &=& - \ft 18 D(D-1) + m^2\,,\nn\\
{\cal C}_2^{\rm vector} &=&  m^2\,,\nn\\
{\cal C}_2^{\rm tensor} &=&  m^2\,.  
\eea
For spinless fields, $m^2$ is {\it not} the coefficient in the mass
term of the Klein-Gordon equation, as that is given by the value for
${\cal C}_2^{\rm scalar}$, while, for spinor and vector fields,
$m$ and $m^2$ do correspond to the mass terms in the Dirac and Proca 
equations, respectively. We thus derive specific values for
${\cal C}_2$ for massless scalar, spinor, vector and tensor fields upon
putting $m=0$. The fact that the value of ${\cal C}_2$ does not depend
on the rank of the tensor field is in accord with the fact that, for 
$m^2=0$, a rank-$n$ and a rank-$(D-n-2)$ tensor gauge field are
equivalent on shell (also in curved space). 

Hence, we see that the interpretation of the mass parameter is not
straightforward in the context of anti-de Sitter space. In the next
section we will derive a rather general lower bound on the value of
${\cal C}_2$ for the lowest-weight representations of the anti-de
Sitter algebra ({\it c.f.} \eqn{C2-bound}), which implies that the
masslike terms for scalar fields can have a negative coefficient
$\mu^2$ subject to the inequality,   
\be
\m^2 \geq - \ft14(D-1)^2\,.
\ee
This result is known as the the Breitenlohner-Freedman bound
\cite{BreitFreed}, which ensures the
stability of an anti-de Sitter background against small fluctuations
of the scalar fields. For spin-$\ft12$ the bound on ${\cal C}_2$
implies that $m^2\geq 0$, whereas for spin-$0$ we find that 
$m^2\geq-\ft14$, with $m^2$ as defined in \eqn{conf-box-C}. 

In the next section we study unitary representations of the anti-de
Sitter algebra and this study will confirm some of the results found
above. For massless representations of higher spin there is a
decoupling of degrees of freedom, 
which uniquely identifies the massless representations and their
values of ${\cal C}_2$. In a number of cases this decoupling is more
extreme and one obtains a so-called singleton representation which does
not have a smooth Poincar\'e limit. In those cases there is no
decoupling of a representation that 
could be identified as massless and therefore there remains a certain  
ambiguity in the definition of `massless' representations. This can
also be seen from the observation that (massless) 
antisymmetric tensor gauge fields of rank $n=D-2$ are on-shell
equivalent to massless scalar fields. While we concluded above that
these tensor fields lead to ${\cal C}_2 =0$, massless 
scalar fields have ${\cal C}_2 =-\ft14D(D-2)$ according to
\eqn{conf-box-C}. The difference may not 
be entirely surprising in view of the fact that the antisymmetric tensor
Lagrangian is not conformally invariant for arbitrary values of
$D$, contrary to the scalar field Lagrangian. At any rate, we have
established the existence of two different field representations
that describe massless, spinless states which correspond to different
values for the anti-de Sitter Casimir operator   
${\cal C}_2$. At the end of the next section, where we discuss
unitary representations of the anti-de Sitter algebra, we briefly return
to the issue of massless representations. The connection between the
local field theory description and the 
anti-de Sitter representations tends to be subtle. 
%%%%%%%%%%%%%%%%%%%%%%%%%%%%%%%%%%%%%%%%%%%%%%%%%%%%%%%%%%%%%%%%%%
%%%%%%%%%%%%%%%%%%%%%%%%%%%%%%%%%%%%%%%%%%%%%%%%%%%%%%%%%%%%%%%%%%
\subsection{Unitary representations of the anti-de Sitter algebra}
\label{adSreps}
%\setcounter{equation}{0}
%%%%%%%%%%%%%%%%%%%%%%%%%%%%%%%%%%%%%%%%%%%%%%%%%%%%%%%%%%%%%%%%%%
In this section we discuss unitary representations of the anti-de
Sitter algebra. We refer to \cite{Fronsdal} for some of the original
work, and to \cite{FreedNicolai,Nicolai,deWitHerger} where part of this
work was reviewed. In order to underline the general
features we will stay as much as possible in general spacetime
dimensions $D>3$.\footnote{%%%%%%%%%%%%%%%%%%%%%%%%%%%%%%%%%%%%%%%
   The case of $D=3$ is special because
   ${\rm SO}(2,2) \cong ({\rm SL}(2,{\rm R})
   \times  {\rm SL}(2,{\rm R}))/{{\bf Z}_2}$.%%%%%%%%%%%%%%%%%%%%%
} %%%%%%%%%%%%%%%%%%%%%%%%%%%%%%%%%%%%%%%%%%%%%%%%%%%%%%%%%%%%%%%%
The anti-de Sitter isometry group, ${\rm SO}(D-1,2)$, is 
noncompact, which implies that unitary representations will be infinitely
dimensional. For these representations the generators are anti-hermitean, 
\be
M_{AB}^\dagger = -M_{AB} \,.
\ee
Here we note that the cover group of SO$(D-1,2)$  has the
generators $\ft12 \Gamma_{\m\n}$ and $\ft12\Gamma_\m$, acting on spinors
which are finite-dimensional objects. These generators, however, have
different hermiticity properties.

The compact subgroup of the anti-de Sitter group is ${\rm SO}(2)\times {\rm
SO(}D-1)$ corresponding to rotations of the compact anti-de Sitter time
and spatial rotations. It is convenient to decompose the $\ft12
D(D+1)$ generators as follows. First, the generator $M_{-0}$ is related
to the energy 
operator when the radius of the anti-de Sitter space is taken to
infinity. The eigenvalues of this generator, associated with
motions along the circle, are quantized in integer units in order to
have single-valued functions, unless one passes to the covering space
CadS. The energy operator $H$ will thus be defined as
\be 
H = -i M_{-0} \,.
\ee
Obviously the generators of the spatial rotations are the operators
$M_{ab}$ with $a,b= 1,\ldots ,D-1$. Note that we have changed
notation: here and henceforth in this chapter the indices $a,b,\ldots$
refer to space indices.  The remaining generators $M_{-a}$ and
$M_{0a}$ are combined into $D-1$ pairs of mutually conjugate
operators,  
\be 
M_a^\pm= -i M_{0a} \pm  M_{-a}\,,
\ee
satisfying $(M_a^+)^\dagger = M_a^-$. The anti-de Sitter commutation
relations then read, 
\bea \label{ads-dec-algebra}
{[}H, M_a^\pm] &=& \pm M^\pm_a\,,\nonumber \\[1mm]
{[}M_a^\pm,M_b^\pm ]&=& 0\,,\nn\\[1mm]
{[} M_a^+ , M_b^-] &=& -2 (H\, \d_{ab} + M_{ab})\,.
\eea
Clearly, the $M_a^\pm$ play the role of raising and lowering operators:
when applied to an eigenstate of $H$ with eigenvalue $E$, application
of $M^\pm_a$ yields a state with eigenvalue $E\pm 1$. We also give the
Casimir operator in this basis,
\bea
{\cal C}_2&=& -\ft 12 M^{AB}M_{AB}\nn\\
&=& H^2 -\ft 12 \{M_a^+,M_a^-\} - \ft12 (M_{ab})^2\nn\\
&=& H(H-D+1)+ J^2 -M_a^+ M_a^-\,,
\label{newCasimir}
\eea
where $J^2$ is the total spin operator: the quadratic Casimir operator
of the rotation group ${\rm SO}(D-1)$, defined by 
\be
J^2 = - \ft12 (M_{ab})^2\,.
\ee
In simple cases, its value is well known. For $D=4$ it is expressed
in terms of the `spin' $s$ which is an integer for bosons and a
half-integer for fermions and the spin-$s$ representation has 
dimension $2s+1$ and $J^2=s(s+1)$. For $D=5$, the corresponding
rotation group ${\rm SO}(4)$ is
the product of two ${\rm SU}(2)$ groups, so that irreducible
representations are characterized by two spin values,
$(s_+,s_-)$. Their dimension is 
equal to $(2s_++1)(2s_-+1)$ and $J^2 = 2(J_+^{\,2} + J_-^{\,2})$ with
$J_\pm^{\,2}= s_\pm(s_\pm+1)$. Summarizing:
\be
J^2 = \left\{\begin{array}{l c l} 
   s(s+1) &\mbox{for}& D=4\,, \\
   2s_+(s_++1) + 2s_-(s_-+1) &\mbox{for}& D=5 \,.
\end{array} \right.
\ee
The ${\rm SO}(D-1)$ representations for $D>5$ are specified by giving the 
eigenvalues of additional (higher-order) ${\rm SO}(D-1)$ Casimir
operators. A restricted class of representations will be discussed in a
sequel; a more general discussion of all possible representations
requires a more technical set-up and is outside the scope of these
lectures.   

In this section we restrict ourselves to the bosonic case, but in
passing, let us already briefly indicate how some of the other
(anti-)commutators of the simple anti-de Sitter superalgebra decompose
(c.f. \eqn{ads-superalgebra}),  
\bea \label{ads-dec-superalgebra}
\{Q_\a\,, Q^\dagger_\b\} &=& H\,\d_{\a\b} - 
\ft12 iM_{ab}\,(\G^a\G^b\G^0)_{\a\b} 
 \nonumber \\[1mm]
&& + \ft12(M^+_a\,\G^a\,(1+i\G^0) + M^-_a\,\G^a \,(1-i\G^0))_{\a\b}
\,,\nonumber\\[1mm]
{[}H\,, Q_\a ] &=& -\ft12 i (\G^0\,Q)_\a\,, \nonumber \\[1mm]
{[}M^\pm_a\,, Q_\a ] &=& \mp \ft12  (\G_a(1\mp i\G^0)\,Q)_\a  \,. 
\eea
For the anti-de Sitter superalgebra, all the bosonic operators can be
expressed as bilinears of the supercharges, so that in principle one
could restrict oneself to fermionic operators only and employ the
projections $(1\pm i\G^0)Q$ as the basic lowering and raising
operators. This will be discussed later in
section~\ref{superalgebra}.   

We now turn to irreducible representations of the anti-de Sitter algebra
\eqn{ads-dec-algebra}. We start with the observation that the energy
operator can be diagonalized so that we can label the states according
to their eigenvalue $E$. 
Because application of $M_a^\pm$ leads to the states with
higher and lower eigenvalues $E$, we expect the representation to cover
an infinite range of eigenvalues, all separated by integers. 
For a unitary representation the
$M_a^+M^-_a$ term in \eqn{newCasimir} is positive, which implies that
the Casimir operator is bounded by 
\be
{\cal C}_2 \leq  -\ft14(D-1)^2 +\Big[J^2 +
\left(E-\ft12(D-1)\right)^2 \Big]_{\rm minimal}   \,,
\label{unitarity-bound}
\ee
where the subscript indicates that one must choose the minimal value
that $J^2+ (E-\ft12(D-1))^2$ takes in the
representation. Among other things, this number will depend on whether
the eigenvalues $E$ take integer or half-integer values. 

%%%%%%%%%%%%%%%%%%%%%%%%%%%%%%%%%%%%%%%%%%%%%%%%%%%%%%%%%
%%%%%%%%%%%%%%%%%%% spin 0 %%%%%%%%%%%%%%%%%%%%%%%%%%%%%%
%%%%%%%%%%%%%%%%%%%%%%%%%%%%%%%%%%%%%%%%%%%%%%%%%%%%%%%%%
\setlength{\unitlength}{0.5mm}
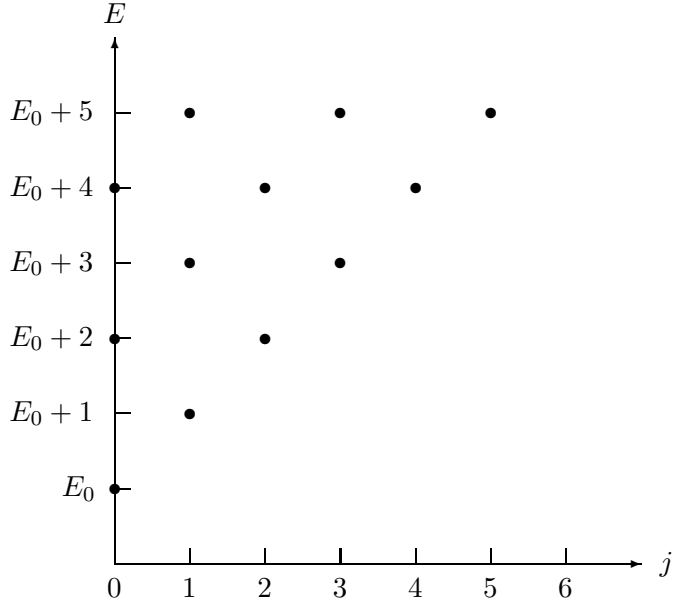
\begin{figure}[t]
\begin{picture}(190,160)(0,0)
\put(30,10){\vector(1,0){140}}
\put(175,8){$j$}
\put(30,10){\vector(0,1){140}}
\put(27,154){$E$}
\put(28,1){$0$}
\put(48,1){$1$}
\put(68,1){$2$}
\put(88,1){$3$}
\put(108,1){$4$}
\put(128,1){$5$}
\put(148,1){$6$}
\put(50,10){\line(0,1){4}}
\put(70,10){\line(0,1){4}}
\put(90,10){\line(0,1){4}}
\put(110,10){\line(0,1){4}}
\put(130,10){\line(0,1){4}}
\put(150,10){\line(0,1){4}}
\put(30,30){\line(1,0){4}}
\put(30,50){\line(1,0){4}}
\put(30,70){\line(1,0){4}}
\put(30,90){\line(1,0){4}}
\put(30,110){\line(1,0){4}}
\put(30,130){\line(1,0){4}}
\put(16,28){$E_0$}
\put(30,30){\circle*{3}}
\put(2,48){$E_0+1$}
\put(50,50){\circle*{3}}
\put(2,68){$E_0+2$}
\put(30,70){\circle*{3}}
\put(70,70){\circle*{3}}
\put(2,88){$E_0+3$}
\put(50,90){\circle*{3}}
\put(90,90){\circle*{3}}
\put(2,108){$E_0+4$}
\put(30,110){\circle*{3}}
\put(70,110){\circle*{3}}
\put(110,110){\circle*{3}}
\put(2,128){$E_0+5$}
\put(50,130){\circle*{3}}
\put(90,130){\circle*{3}}
\put(130,130){\circle*{3}}
%
%%%%%%%
%%%%%%%%%%%%%%%%%%%%%%%%%%%%%%%%%%%%%%%
%
\end{picture}
\caption{\small States of the spinless representation in terms of the
  energy   eigenvalues $E$ and the angular momentum $j$. Each point
  corresponds to the spherical harmonics of $S^{D-1}$: traceless,
  symmetric tensors $Y^{a_1\cdots a_l}$ of rank $l=j$. 
}
\label{spinless-ads-irrep} 
%\vspace{-4mm}
\end{figure}
%%%%%%%%%%%%%%%%%%%%%%%%%%%%%%%%%%%%%%%%%%%%%%%%%%%%%%%%%%%%%

Continuous representations cover the whole range of eigenvalues 
$E$ extending from $-\infty$ to $\infty$. However, when there is
a state with some eigenvalue $E_0$ that is annihilated 
by all the $M^-_a$, then only states with eigenvalues $E>E_0$ will
appear in the representation. This is therefore not a continuous
representation but a so-called lowest-weight representation. The
ground state of this representation (which itself transforms as an
irreducible representation of the rotation group and may thus be
degenerate) is denoted by $\vert E_0,J\rangle$ and satisfies
\be
\label{ground-state-condition}
M^-_a\,\vert E_0,J\rangle = 0\,. 
\ee
The unitarity upper bound \eqn{unitarity-bound} on ${\cal C}_2$ is
primarily useful for continuous representations. For unitary
lowest-weight representations one can derive various lower bounds, as
we shall see below. Substituting 
the condition \eqn{ground-state-condition} in the expression
\eqn{ads-dec-algebra} applied to the ground state  
$\vert E_0,J\rangle$, we derive at once the eigenvalue of the
quadratic Casimir 
operator associated with this representation in terms of $E_0$ and
$J^2$, 
\be
{\cal C}_2 = E_0(E_0-D+1) + J^2 \,.
\label{eq:casimir-repr}
\ee
Since ${\cal C}_2$ is a Casimir operator, 
this value holds for any state belonging to the corresponding
irreducible representation. For real values of $E_0$ the Casimir
operator is bounded by
\be
\label{C2-bound}
{\cal C}_2\geq J^2-\ft14(D-1)^2\,.
\ee
As we already discussed at the end of the previous section, for scalar
fields ($J^2=0$) this is just the Breitenlohner-Freedman bound
\cite{BreitFreed}. Additional restrictions based on unitarity will be
derived shortly. They generally lead to a lower bound for $E_0$ and
thus to a corresponding lower bound for ${\cal C}_2$. Unless this bound
supersedes \eqn{C2-bound} there can exist a
degeneracy in the sense that there are two possible, permissible
values for $E_0$ with the same value for ${\cal C}_2$. These two 
values correspond to two different solutions of the field equations
subject to different boundary conditions at spatial infinity.

In what follows we restrict ourselves to lowest-weight representations,
because those have a natural interpretation in the limit of large
anti-de Sitter radius in terms of Poincar\'e representations. 
Alternatively we can construct highest-weight representations, but those
will be similar and need not to be discussed separately. 

The full lowest-weight representation can now be constructed by acting
with the raising 
operators on the ground state $\vert E_0,J\rangle$. To be precise,
all states of energy $E=E_0+ n$ are constructed by an $n$-fold product
of creation operators $M^+_a$. In this way one
obtains states of higher eigenvalues $E$ with higher spin. The
simplest case is the one where the vacuum has no spin ($J=0$). For
given eigenvalue $E$, the states decompose into the state of the
highest spin generated by the traceless symmetric product of $E-E_0$
operators $M^+_a$ and a number of lower-spin descendants. These
states are all shown  in fig.~\ref{spinless-ads-irrep}.  

%%%%%%%%%%%%%%%%%%%%%%%%%%%%%%%%%%%%%%%%%%
\begin{table}[t]
\begin{center}
\begin{tabular}{l l l  }\hline
~&~&~ \\[-3.5mm]
${\rm representation}$ & $Y^{A_1\cdots A_l}$ & $Y^{B;A_1\cdots A_l}$
\\ \hline  
~&~&~ \\[-3.5mm]
$D$ & $l(l+D-3)$ & $(l+D-4) (l+1)$  \\
$D=4$ & $s=l$            & $s=l$     \\ 
$D=5$ & $s_\pm= \ft12 l$ & $s_\pm =s_\mp +1=\ft12(l+1)$  \\[1mm] \hline
\end{tabular}
\end{center}
\caption{\small Two generic ${\rm SO}(D-1)$ representations. One is 
  the symmetric traceless tensor representation (corresponding to the
  spherical harmonics on $S^{D-2}$)  denoted by $l$-rank tensors
  $Y^{A_1\cdots A_l}$, and the representation spanned by mixed tensors 
  $Y^{B;A_1\cdots A_l}$ of rank $l+1$ (which is not independent for
  $D=4$). We list the corresponding  
  eigenvalues of the quadratic Casimir operator $J^2$, for general
  $D$. For $D=4$ these representations are characterized by an integer
  spin $s$. For $D=5$ there are two such numbers, $s_\pm$, as we explained
  in the text.  
}\label{so-irreps-b}
\end{table}
%%%%%%%%%%%%%%%%%%%%%%%%%%%%%%%%%%%%%%%%%%%

In the following we consider a number of representations of ${\rm
SO}(D-1)$ that exist for any dimension. For the bosons we consider the
spherical harmonics, spanned by $l$-rank traceless, symmetric tensors
$Y^{a_1\cdots a_l}$.  Multiplying such tensors with the vector
representation gives rise to two of these representations with rank
$l\pm1$, and a `mixed' representation, spanned by 
mixed tensors $Y^{b;a_1\cdots a_l}$. Table~\ref{so-irreps-b} lists the
value of $J^2$ 
for these representations, for general $D$ and for the specific cases
of $D=4,5$. In a similar table~\ref{so-irreps-f} we list the value
of $J^2$ for the irreducible symmetric tensor-spinors, denoted by
$Y^{\a;a_1\cdots a_l}$. They are symmetric $l$-rank tensor spinors
that vanish upon contraction by a gamma matrix and appear when taking 
products of spherical harmonics with a simple spinor. 

Armed with this information it is straightforward to find the
decompositions of the spinor representation of the anti-de Sitter
algebra. One simply takes the direct product of the spinless
representation with a spin-$\ft12$ state. That implies that every
point with spin $j$ in fig.~\ref{spinless-ads-irrep} generates two points with spin
$j\pm\ft12$, with the exception of points associated with $j=0$, which
will simply move to $j=\ft12$. The result of this is shown in
fig.~\ref{spinor-ads-irrep}. 

%%%%%%%%%%%%%%%%%%%%%%%%%%%%%%%%%%%%%%%%%%%%%%%%%%%%%%%%%%%%%%%%%
\begin{table}[t]
\begin{center}
\begin{tabular}{l l }\hline
${\rm representation}$ & $Y^{\a; a_1\cdots a_l}$ \\ \hline 
~&~ \\[-3.3mm]
$J^2$ &  $l(l+D-2) + \ft18 (D-1)(D-2)$ \\
$D=4$ &  $s =l+\ft12$ \\ 
$D=5$ &  $s_\pm =s_\mp-\ft12 = \ft12 l$    \\[1mm] \hline
\end{tabular}
\end{center}
\caption{\small The eigenvalues of the quadratic ${\rm SO}(D-1)$ Casimir
  operator $J^2$ for the symmetric tensor-spinor representation
  spanned by tensors $Y^{\a;a_1\cdots a_l}$ for
  general dimension $D$ and for the specific cases $D=4,5$.
}\label{so-irreps-f}
\end{table}
%%%%%%%%%%%%%%%%%%%%%%%%%%%%%%%%%%%%%%%%%%%%%%%%%%%%%%%%%%%%%%%%%

However, the spinless and the spinor representations that we have
constructed so far are not necessarily irreducible. 
To see this consider the excited state that has the same spin content
as the ground state, but with an energy equal to $E_0+2$ or $E_0+1$,
for the scalar and spinor representations, respectively, and compare
their value for the Casimir operator with that of the corresponding
ground state. In this way we find for the scalar
\be
E_0 (E_0-D+1) = (E_0+2) (E_0-D+3) + \Big\vert M_a^- \vert
E_0+2,\mbox{spinless}\rangle 
\Big\vert^2\,. \label{equal-0-bound} 
\ee
This leads to 
\be
2E_0 +3-D = \ft12 \Big\vert M_a^- \vert E_0+2,\mbox{spinless}\rangle \Big\vert^2\,,
\ee
so that unitary of the representation requires the inequality, 
\be
E_0\geq \ft12(D-3)\,. 
\ee
For $E_0=\ft12(D-3)$ we have the so-called {\it singleton}
representation\footnote{%%%%%%%%%%%%%%%%%%%%%%%%%%%%%%%%%%%%%%%%%%%%
  The singleton representation was first found by Dirac
  \cite{Dirac2} in 4-dimensional anti-de Sitter space and was known as
  a `remarkable representation'. In the context of the oscillator
  method, which we will refer to later, singletons in anti-de
  Sitter spaces of dimension $D\not=4$ are called `doubletons' 
  \cite{GunaNieuWarn}. In these lectures we will only use the name 
  singleton to denote these remarkable representations.}, %%%%%%%%%%%
%%%%%%%%%%%%%%%%%%%%%%%%%%%%%%%%%%%%%%%%%%%%%%%%%%%%%%%%%%%%%%%%%%%%% 
where we have only one state for each given spherical harmonic. The Casimir
eigenvalue for this representation equals 
\begin{equation}
  \label{eq:spinless-singleton-C}
  {\cal C}_2(\mbox{spinless singleton}) = -\ft14 (D+1)(D-3) \,.
\end{equation}
The excited state then constitutes the ground state for a separate
irreducible spinless representation, but now with $E_0=\ft12(D+1)$,
which, not surprisingly, has the same value for ${\cal C}_2$. 
 
%%%%%%%%%%%%%%%%%%%%%%%%%%%%%%%%%%%%%%%%%%%%%%%%%%%%%%%%
%%%%%%%%%%%%%%%%%%%%%% spin 1/2 %%%%%%%%%%%%%%%%%%%%%%%%
%%%%%%%%%%%%%%%%%%%%%%%%%%%%%%%%%%%%%%%%%%%%%%%%%%%%%%%%
\setlength{\unitlength}{0.5mm}
\begin{figure}[t]
\begin{picture}(190,160)(0,0)
\put(30,10){\vector(1,0){140}}
\put(175,8){$j$}
\put(30,10){\line(0,1){19}}
\put(30,31){\line(0,1){38}}
\put(30,71){\line(0,1){38}}
\put(30,111){\vector(0,1){40}}
\put(27,154){$E$}
\put(28,1){$0$}
\put(48,1){$1$}
\put(68,1){$2$}
\put(88,1){$3$}
\put(108,1){$4$}
\put(128,1){$5$}
\put(148,1){$6$}
\put(50,10){\line(0,1){4}}
\put(70,10){\line(0,1){4}}
\put(90,10){\line(0,1){4}}
\put(110,10){\line(0,1){4}}
\put(130,10){\line(0,1){4}}
\put(150,10){\line(0,1){4}}
\put(31,30){\line(1,0){3}}
\put(30,50){\line(1,0){4}}
\put(31,70){\line(1,0){3}}
\put(30,90){\line(1,0){4}}
\put(31,110){\line(1,0){3}}
\put(30,130){\line(1,0){4}}
\put(16,28){$E_0$}
\put(30,30){\circle{2}}
\put(40,30){\circle*{3}}
\put(2,48){$E_0+1$}
\put(50,50){\circle{2}}
\put(40,50){\circle*{3}}
\put(60,50){\circle*{3}}
\put(2,68){$E_0+2$}
\put(30,70){\circle{2}}
\put(40,70){\circle*{3}}
\put(70,70){\circle{2}}
\put(60,70){\circle*{3}}
\put(80,70){\circle*{3}}
\put(2,88){$E_0+3$}
\put(50,90){\circle{2}}
\put(40,90){\circle*{3}}
\put(60,90){\circle*{3}}
\put(90,90){\circle{2}}
\put(80,90){\circle*{3}}
\put(100,90){\circle*{3}}
\put(2,108){$E_0+4$}
\put(30,110){\circle{2}}
\put(40,110){\circle*{3}}
\put(70,110){\circle{2}}
\put(60,110){\circle*{3}}
\put(80,110){\circle*{3}}
\put(110,110){\circle{2}}
\put(100,110){\circle*{3}}
\put(120,110){\circle*{3}}
\put(2,128){$E_0+5$}
\put(50,130){\circle{2}}
\put(40,130){\circle*{3}}
\put(60,130){\circle*{3}}
\put(90,130){\circle{2}}
\put(80,130){\circle*{3}}
\put(100,130){\circle*{3}}
\put(130,130){\circle{2}}
\put(120,130){\circle*{3}}
\put(140,130){\circle*{3}}
%
%%%%%%%%%%%%%%%%%%%%%%%%%%%%%%%%%%%%%%%%%%%%%%
%%%%%%%%%%%%%%%%%%%%%%%%%%%%%%%%%%%%%%%%%%%%%%
%
\end{picture}
\caption{\small 
  States of the spinor representation in terms of the energy
  eigenvalues $E$ and the angular momentum; the half-integer values
  for $j=l+\ft12$ denote that we are dealing with a symmetric
  tensor-spinor of rank $l$. The small circles
  denote the original spinless multiplet from which the spinor multiplet 
  has been constructed by a direct product with a spinor. } 
\label{spinor-ads-irrep}
%\vspace{-4mm}
\end{figure}
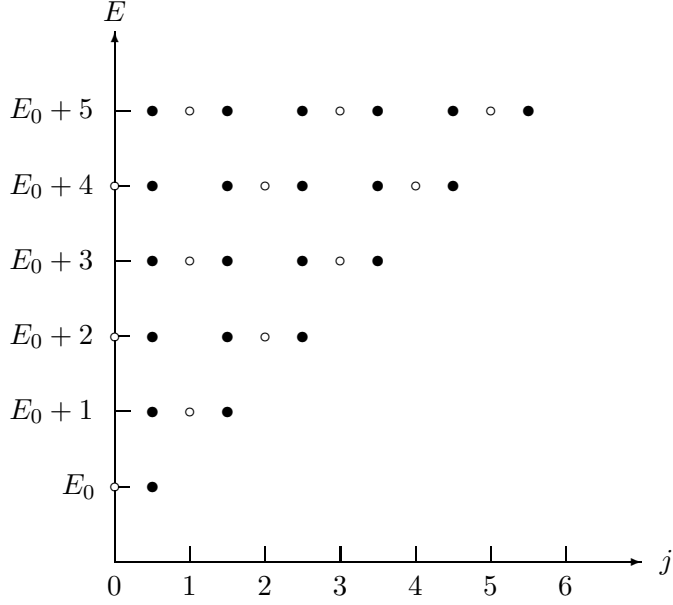
%%%%%%%%%%%%%%%%%%%%%%%%%%%%%%%%%%%%%%%%%%%%%%%%%%%%%%%%%%%%%%

For the spinor representation one finds a similar result,
\bea
 E_0 (E_0-D+1) +J^2  &=& (E_0+1) (E_0-D+2)+ J^2 \nonumber \\
&&+ \Big\vert M_a^- \vert E_0+1,\mbox{spinor}\rangle \Big\vert^2\,. 
\label{equal-1/2-bound}
\eea
As the value for $J^2$ are the same for the ground state and the
excited state one readily derives 
\be
2E_0 -D+3 = \Big\vert M_a^- \vert E_0+1,\mbox{spinor} \rangle
\Big\vert^2  \,, 
\ee
so that one obtains the unitarity bound
\be
E_0\geq \ft12(D-2) \,.
\ee
For $E_0=\ft12(D-2)$ we have the spinor singleton representation,
which again consists of just one state for every value of the total
spin. For the spinor representation the value of the Casimir operator
equals 
\be
{\cal C}_2(\mbox{spinor singleton}) = -\ft18(D+1)(D-2)\,.
\ee
Note that in $D=4$, both singleton representations have the same
eigenvalue of the Casimir operator.  

%%%%%%%%%%%%%%%%%%%%%%%%%%%%%%%%%%%%%%%%%%%%%%%%%%%%%%%%%%
%%%%%%%%%%%%%%%%%%%%%% singletons %%%%%%%%%%%%%%%%%%%%%%%%
%%%%%%%%%%%%%%%%%%%%%%%%%%%%%%%%%%%%%%%%%%%%%%%%%%%%%%%%%%
\setlength{\unitlength}{0.5mm}
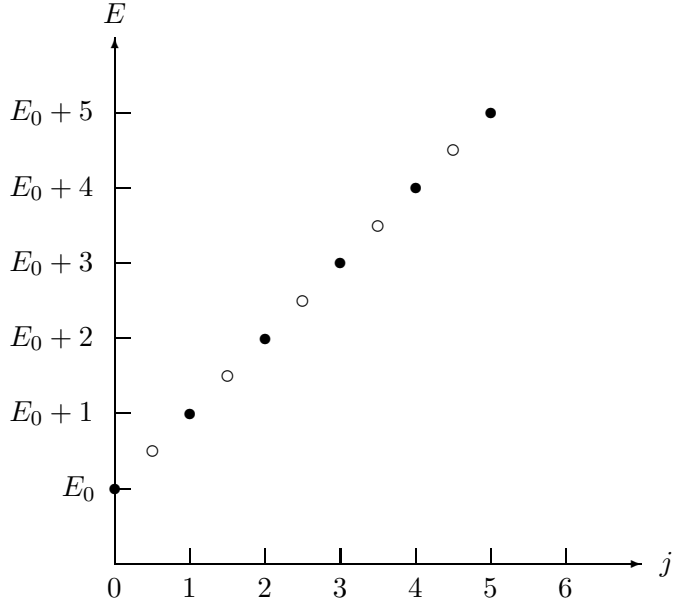
\begin{figure}[t]
\begin{picture}(190,160)(0,0)
\put(30,10){\vector(1,0){140}}
\put(175,8){$j$}
\put(30,10){\vector(0,1){140}}
\put(27,154){$E$}
\put(28,1){$0$}
\put(48,1){$1$}
\put(68,1){$2$}
\put(88,1){$3$}
\put(108,1){$4$}
\put(128,1){$5$}
\put(148,1){$6$}
\put(50,10){\line(0,1){4}}
\put(70,10){\line(0,1){4}}
\put(90,10){\line(0,1){4}}
\put(110,10){\line(0,1){4}}
\put(130,10){\line(0,1){4}}
\put(150,10){\line(0,1){4}}
\put(30,30){\line(1,0){4}}
\put(30,50){\line(1,0){4}}
\put(30,70){\line(1,0){4}}
\put(30,90){\line(1,0){4}}
\put(30,110){\line(1,0){4}}
\put(30,130){\line(1,0){4}}
\put(16,28){$E_0$}
\put(30,30){\circle*{3}}
\put(2,48){$E_0+1$}
\put(50,50){\circle*{3}}
\put(40,40){\circle{3}}
\put(2,68){$E_0+2$}
\put(70,70){\circle*{3}}
\put(60,60){\circle{3}}
\put(2,88){$E_0+3$}
\put(90,90){\circle*{3}}
\put(80,80){\circle{3}}
\put(2,108){$E_0+4$}
\put(110,110){\circle*{3}}
\put(100,100){\circle{3}}
\put(2,128){$E_0+5$}
\put(130,130){\circle*{3}}
\put(120,120){\circle{3}}
%
%%%%%%%%%%%%%%%%%%%%%%%%%%%%%%%%%%%%%%%%%%%%%%
%%%%%%%%%%%%%%%%%%%%%%%%%%%%%%%%%%%%%%%%%%%%%%
%
\end{picture}
\caption{\small 
The spin-0 and spin-$\ft12$ singleton representations. The
solid dots indicate the states of the spin-0 singleton, the circles the
states of the spin-$\ft12$ singleton. It is obvious that singletons
contain much less degrees of freedom than a generic local field. The
value of $E_0$, which denotes the spin-0  
ground state energy, is equal to $E_0=\ft12(D-3)$. The spin-$\ft12$
singleton ground state has an energy which is one half unit higher, as is
explained in the text. 
}
\label{singleton-irreps} 
%\vspace{-4mm}
\end{figure}
%%%%%%%%%%%%%%%%%%%%%%%%%%%%%%%%%%%%%%%%%%%%%%%%%%%%%%%%%%%%%%

The existence of the singletons was first noted by
Dirac \cite{Dirac2}. These representations are characterized by the
fact that they do not exist in the Poincar\'e limit. To see this, 
note that Poincar\'e representations correspond to plane waves which
are decomposable into an infinite number of spherical harmonics,
irrespective of the size of the spatial momentum (related to the
energy eigenvalue). That means that, for given spin, one is dealing
with an infinite, continuous tower of modes, which is just what one 
obtains in the limit of vanishing energy increments for the generic
spectrum shown in, {\it e.g.}, 
fig. \ref{spinless-ads-irrep}. In contradistinction, the singleton
spectrum is different as the states have a single energy eigenvalue
for any given value of the spin, as is obvious in
fig.~\ref{singleton-irreps}. Consequently, wave functions that
constitute singleton representations do not depend on the radius of the
anti-de Sitter spacetime and can be regarded as living on the
boundary.   

%%%%%%%%%%%%%%%%%%%%%%%%%%%%%%%%%%%%%%%%%%%%%%%%%%%%%%%%%
%%%%%%%%%%%%%%%%%%% spin 1 %%%%%%%%%%%%%%%%%%%%%%%%%%%%%%
%%%%%%%%%%%%%%%%%%%%%%%%%%%%%%%%%%%%%%%%%%%%%%%%%%%%%%%%%
\setlength{\unitlength}{0.5mm}
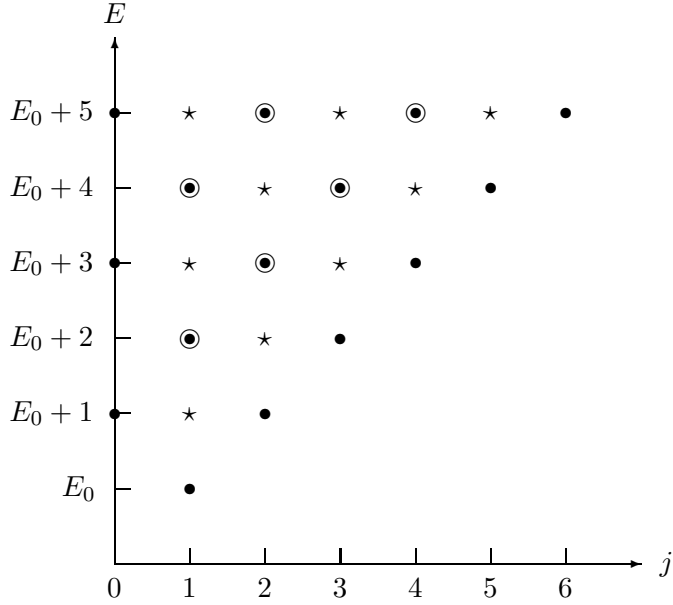
\begin{figure}[t]
\begin{picture}(190,160)(0,0)
\put(30,10){\vector(1,0){140}}
\put(175,8){$j$}
\put(30,10){\vector(0,1){140}}
\put(27,154){$E$}
\put(28,1){$0$}
\put(48,1){$1$}
\put(68,1){$2$}
\put(88,1){$3$}
\put(108,1){$4$}
\put(128,1){$5$}
\put(148,1){$6$}
\put(50,10){\line(0,1){4}}
\put(70,10){\line(0,1){4}}
\put(90,10){\line(0,1){4}}
\put(110,10){\line(0,1){4}}
\put(130,10){\line(0,1){4}}
\put(150,10){\line(0,1){4}}
\put(30,30){\line(1,0){4}}
\put(30,50){\line(1,0){4}}
\put(30,70){\line(1,0){4}}
\put(30,90){\line(1,0){4}}
\put(30,110){\line(1,0){4}}
\put(30,130){\line(1,0){4}}
\put(16,28){$E_0$}
\put(50,30){\circle*{3}}
\put(2,48){$E_0+1$}
\put(48,48){$\star$}
\put(30,50){\circle*{3}}
\put(70,50){\circle*{3}}
\put(2,68){$E_0+2$}
\put(50,70){\circle*{3}}
\put(50,70){\circle{5}}
\put(68,68){$\star$}
\put(90,70){\circle*{3}}
\put(2,88){$E_0+3$}
\put(30,90){\circle*{3}}
\put(48,88){$\star$}
\put(70,90){\circle*{3}}
\put(70,90){\circle{5}}
\put(88,88){$\star$}
\put(110,90){\circle*{3}}
\put(2,108){$E_0+4$}
\put(50,110){\circle*{3}}
\put(50,110){\circle{5}}
\put(68,108){$\star$}
\put(90,110){\circle*{3}}
\put(90,110){\circle{5}}
\put(108,108){$\star$}
\put(130,110){\circle*{3}}
\put(2,128){$E_0+5$}
\put(30,130){\circle*{3}}
\put(48,128){$\star$}
\put(70,130){\circle*{3}}
\put(70,130){\circle{5}}
\put(88,128){$\star$}
\put(110,130){\circle*{3}}
\put(110,130){\circle{5}}
\put(128,128){$\star$}
\put(150,130){\circle*{3}}
%
%%%%%%%
%%%%%%%%%%%%%%%%%%%%%%%%%%%%%%%%%%%%%%%
%
\end{picture}
\caption{\small 
States of the spin-$1$ representation in terms of the energy
  eigenvalues $E$ and the angular momentum $j$. Observe that there are
now points with double occupancy, indicated by the circle superimposed
on the dots and states transforming as mixed tensors (with $l=j$)
denoted by a $\star$. The double-occupancy points exhibit the
structure of a spin-0  multiplet with  
ground state energy $E_0+1$. This multiplet becomes
reducible and can be dropped when $E_0=D-2$, as is explained
in the text. The remaining points then constitute a massless spin-$1$
multiplet, shown in fig.~\ref{massless-spin1-irrep}. 
}
\label{spin1-ads-irrep} 
%\vspace{-4mm}
\end{figure}
%%%%%%%%%%%%%%%%%%%%%%%%%%%%%%%%%%%%%%%%%%%%%%%%%%%%%%%%%%%%%

To obtain the spin-1 representation one can take the direct product of
the spinless multiplet with a
spin-1 state. Now the situation is more complicated, however,  as the
resulting multiplet contains states of spin lower than that of the
ground state. In principle, each point in fig.~\ref{spinless-ads-irrep} 
now generates three points, associated with two
spherical harmonics, associated with rank-$j^\pm1$ tensors as well as
mixed tensors of rank $j+1$ (so that $l=j$). An exception are the
spinless points, which simply 
move to $j=1$. The result of taking the product is depicted in
fig.~\ref{spin1-ads-irrep}.  This procedure can be extended directly
to ground states that transform as a spherical harmonic $Y^{a_1\cdots
a_l}$.

%%%%%%%%%%%%%%%%%%%%%%%%%%%%%%%%%%%%%%%%%%%%%%%%%%%%%%%%%%%%%%
%%%%%%%%%%%%%%%%%%% massless spin 1 %%%%%%%%%%%%%%%%%%%%%%%%%%
%%%%%%%%%%%%%%%%%%%%%%%%%%%%%%%%%%%%%%%%%%%%%%%%%%%%%%%%%%%%%%
\setlength{\unitlength}{0.5mm}
\begin{figure}[t]
\begin{picture}(190,160)(0,0)
\put(30,10){\vector(1,0){140}}
\put(175,8){$j$}
\put(30,10){\vector(0,1){140}}
\put(27,154){$E$}
\put(28,1){$0$}
\put(48,1){$1$}
\put(68,1){$2$}
\put(88,1){$3$}
\put(108,1){$4$}
\put(128,1){$5$}
\put(148,1){$6$}
\put(50,10){\line(0,1){4}}
\put(70,10){\line(0,1){4}}
\put(90,10){\line(0,1){4}}
\put(110,10){\line(0,1){4}}
\put(130,10){\line(0,1){4}}
\put(150,10){\line(0,1){4}}
\put(30,30){\line(1,0){4}}
\put(30,50){\line(1,0){4}}
\put(30,70){\line(1,0){4}}
\put(30,90){\line(1,0){4}}
\put(30,110){\line(1,0){4}}
\put(30,130){\line(1,0){4}}
\put(16,28){$E_0$}
\put(50,30){\circle*{3}}
\put(2,48){$E_0+1$}
\put(48,48){$\star$}
\put(70,50){\circle*{3}}
\put(2,68){$E_0+2$}
\put(50,70){\circle*{3}}
\put(68,68){$\star$}
\put(90,70){\circle*{3}}
\put(2,88){$E_0+3$}
\put(48,88){$\star$}
\put(70,90){\circle*{3}}
\put(88,88){$\star$}
\put(110,90){\circle*{3}}
\put(2,108){$E_0+4$}
\put(50,110){\circle*{3}}
\put(68,108){$\star$}
\put(90,110){\circle*{3}}
\put(108,108){$\star$}
\put(130,110){\circle*{3}}
\put(2,128){$E_0+5$}
\put(48,128){$\star$}
\put(70,130){\circle*{3}}
\put(88,128){$\star$}
\put(110,130){\circle*{3}}
\put(128,128){$\star$}
\put(150,130){\circle*{3}}
%
%%%%%%%
%%%%%%%%%%%%%%%%%%%%%%%%%%%%%%%%%%%%%%%
%
\end{picture}
\caption{\small 
  States of the massless $s=1$ representation in terms of the energy
  eigenvalues $E$ and the angular momentum $j$. Now $E_0$ is no longer
arbitrary but it is fixed to $E_0=D-2$. 
}
\label{massless-spin1-irrep} 
%\vspace{-4mm}
\end{figure}
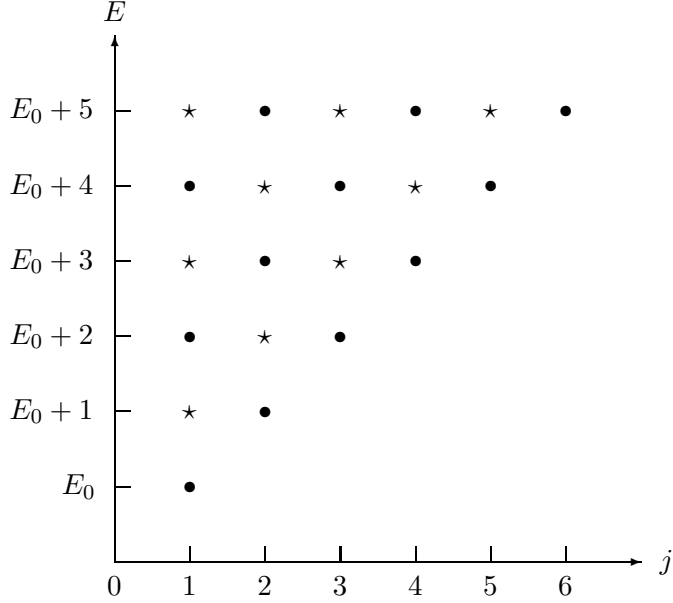
%%%%%%%%%%%%%%%%%%%%%%%%%%%%%%%%%%%%%%%%%%%%%%%%%%%%%%%%%%%%%

Along the same lines as before, we investigate whether this
representation can become reducible for special values of the ground
state energy.  We compare the value of the Casimir operator for the
first excited states with minimal spin to the value for the ground
state specified in \eqn{eq:casimir-repr}. Hence we consider the states
with  $E=E_0+1$ and $j=l-1$, assuming that the ground state  
has $l\geq1$. In that case we find 
\bea
{\cal C}_2 &=& (E_0+1)(E_0-D+2) + (l-1) (l+D-4) - \Big\vert M_a^-  \vert
E_0+1,l-1\rangle \Big\vert^2 \nn\\
&=& E_0(E_0-D+1) + l(l+D-3)\,,
\eea
so that 
\be
E_0-l -D+3 = \ft12 \Big\vert M_a^-  \vert
E_0+1,l-1\rangle \Big\vert^2 \,.
\ee
Therefore we establish the unitarity bound 
\be 
\label{D-unitarity-l}
E_0 \geq l+D-3\,,\qquad(l\geq 1)
\ee 
When $E_0= l+D-3$, however,  the state $\vert
E_0+1,l-1\rangle$ is itself the ground state of an irreducible
multiplet, which decouples from the original multiplet 
together with its corresponding excited states. This can be
interpreted as the result of a gauge symmetry. Because these
representations have a smooth Poincar\'e limit they are not singletons
and can therefore be regarded as {\it massless}
representations. Hence massless representations with spin $l\geq1$ are
characterized by  
\be 
\label{massless-l}
E_0= l +D-3 \,. \qquad (l\geq 1)
\ee 
For these particular values the quadratic Casimir operator acquires a
minimal value equal to 
\be
{\cal C}_2({\rm massless})  = 2(l-1)(l+D-3) \,. \qquad (l\geq 1)
\label{massless-casimir}
\ee
We recall that this result is only derived for $l\geq1$. For certain
other cases, the identification of massless representation is somewhat
ambiguous, as we already discussed.  We return to this issue at the
end of this section.

The above arguments can be easily extended to other grounds states,
but this requires further knowledge of the various
representations of the rotation group, at least for general
dimension. This is outside the scope of these lectures.  However, in
$D=4,5$ dimensions this information is readily available. For 
a spin-$s$ ground state in $4$ spacetime dimensions we immediately
derive the unitarity bound (for $s>\ft12$),
\be
\label{4D-unitarity-s}
E_0\geq s+1\,,
\ee
by following the same procedure as leading to \eqn{D-unitarity-l}.
When the bound is saturated we obtain a massless representation. The
equation corresponding to \eqn{massless-casimir} becomes 
\begin{equation}
  \label{eq:D4-massless-casimir}
  {\cal C}_2= 2 (s^2-1)\,,
\end{equation}
It turns out that this result applies to all spin-$s$ representations,
even to $s=0,\ft12$ conformal fields, for which we cannot use this
derivation. This is a special property for $D=4$ dimensions. 
 
The case of $D=5$ requires extra attention, because here the
rotation group factorizes into two ${\rm SU}(2)$ groups. We briefly
summarize some results. First let us assume the the groundstate has
spin $(s_+,s_-)$ with $s_\pm\geq\ft12$. In that case we find
that the ground state energy satisfies the unitarity bound,
\begin{equation}
  \label{eq:s-pm-bound}
 E_0 \geq s_+ + s_- +2\,. \qquad(s_\pm\geq \ft12)
\end{equation}
This bound is saturated for massless states, for which the $E=E_0+1$
states with spin $(s_+-\ft12,s_--\ft12)$ decouples. The corresponding
value for the  Casimir operator is equal to
\begin{equation}
  \label{eq:s-pm-casimir}
  {\cal C}_2 = (s_++s_-)^2 + 2 s_+(s_++1) +2 s_-(s_-+1) -4\,.
\end{equation}
For $s_\pm = \ft12 l$ these values are in agreement with earlier
result.

What remains to be considered are the ground states with spin
$(0,s)$. Here we find 
\begin{equation}
  \label{eq:0-s-bound}
 E_0 \geq 1+s\,.
\end{equation}
When the bound is saturated we have again a singleton
representation. The corresponding values for the  Casimir operator are  
\begin{equation}
  \label{eq:0-s-casimir}
  {\cal C}_2({\rm singleton}) = 3(s^2-1)\,.
\end{equation}
The singleton representations for $s=0,\ft12$ were already found
earlier. Note that for $D=5$ there are thus infinitely many singleton
representations, unlike in 4 dimensions, with a large variety of spin
values. This is generically the case for arbitrary dimensions $D\not
=4$ and is thus related to the fact that the rotation group is of
higher rank. 

{From} the above it is clear that we are dealing with the phenomenon 
of multiplet shortening for specific values of the energy and spin of
the representation, just as in the earlier discussions on BPS multiplets in
previous chapters. This phenomenon can be understood 
from the fact that the $[M_a^+,M_b^-]$ commutator acquires zero or
negative eigenvalues for certain values of $E_0$ and $J^2$. When
viewed in this way, the shortening of the representation is
qualitatively similar to the shortening of BPS multiplets based on the
anticommutator of the supercharges. Our discussion of the shortening
of anti-de Sitter supermultiplets in section~\ref{superalgebra} will
support this point of view. The same phenomenon of multiplet shortening
is well known and relevant in conformal field theory in $1+1$ dimensions.
 
The purpose of this section was to elucidate the various principles
that underlie the anti-de Sitter representations and their relation
with the field theory description. Here we are not striving for
completeness. There are in fact alternative and often more
systematic techniques for constructing the lowest-weight
representations. A powerful method to construct the unitary 
irreducible representations of the anti-de Sitter algebra, is
known as the oscillator method \cite{GunaSacl}, which is
applicable in any number of spacetime dimensions and which can also be
used for supersymmetric extensions of the anti-de Sitter algebra.  
There is an extensive literature on this; for a recent elementary
introduction to this method we refer to  \cite{deWitHerger}. 

We close this section with a number of comments regarding `massless'
representations and their field-theoretic description. As we
demonstrated above, certain representations can, for a specific value 
of $E_0$, decouple into different irreducible representations. 
This phenomenon takes place when some unitarity bound is saturated. In
that case one has representations that contain fewer degrees of
freedom. When these `shortened' representations have a smooth
Poincar\'e limit, they are called massless; when they do not, they are
called singletons. For the case of spin-0 or spin-$\ft12$
representations, for example, the spectrum of states is qualitatively
independent of the value for $E_0$, as long as $E_0$ does not 
saturate the unitarity bound and a singleton representation
decouples. Therefore the concept of mass remains ambiguous. We have 
already discussed this in section~\ref{masslike}, where we emphasized
that the absence of mass terms in the field equations is also not a
relevant criterion for masslessness. In  
table~\ref{massless-low-spin} we have collected a number of examples
of spin-0 and spin-$\ft12$ representations with the criteria
according to which they can be regarded as massless. One of them is
tied to the fact that the corresponding field equation is conformally
invariant, as we discussed at the end of
section~\ref{masslike}. Another one follows from the fact that we are
dealing with a gauge field. Here the example is an antisymmetric
rank-$(D-2)$ gauge field, which is on-shell equivalent to a
scalar. 

We also invoke a criterion introduced by G\"unaydin (see \cite{GunTrieste}
and the discussion in \cite{GunaMini}), according to which
every representation should be regarded as massless that appears in
the tensor product of two singleton representations. For instance, it
is easy to verify that the product of two spinless singletons leads to
an infinite series of higher spin representations that are all massless
according to \eqn{massless-l}. However, it also contains the $l=0$
representation with $E_0=D-3$, to which \eqn{massless-l} does not
apply so that the interpretation as a massless representation is less
obvious. It is interesting to consider this criterion for masslessness 
in $D=5$ dimensions. The tensor product of the singleton
representations with spin $(0,s_-)$ and $(s_+,0)$ leads to a ground
state with spin $(s_+,s_-)$ and $E_0=2+s_++s_-$, which are obviously
massless in view of \eqn{eq:0-s-bound} and \eqn{eq:s-pm-bound}. Taking
the product of two singleton representations, one with spin $(s_1,0)$
and another one with spin $(s_2,0)$ leads to ground states with spin
$(s,0)$ and energy $E_0= 2 + s+n$, where $n$ is an arbitrary positive
integer. Hence these representations should be regarded as massless.

This interpretation can be tested as follows. In maximal gauged
supergravity in 5 dimensions with gauge group ${\rm SO}(6)$, one of
these representations appears as part of the `massless' supergravity
multiplet. This anti-de Sitter representation is described by a
(complex) tensor field, whose field equation takes the form,
\be
\label{D-5-tensor}
e^{-1} \varepsilon^{\m\n\rho\sigma\lambda} \,D_\rho B_{\sigma\lambda}
+ 2i m \,B^{\m\n} = 0\,,
\ee
where $m=\pm g$. From this equation one can show that $B_{\m\n}$
satisfies \eqn{antisymm} so that ${\cal C}_2 = 1$ ({\it c.f.}
\eqn{conf-box-C}). On shell the equation \eqn{D-5-tensor} projects out
the degrees of freedom corresponding to spin $(1,0)$ or $(0,1)$, depending
on the sign of $m$. From this one derives that $E_0=3$ (a
second solution with $E_0=1$ violates the unitarity bound
\eqn{eq:0-s-bound}). 

%%%%%%%%%%%%%%%%%%%%%%%%%%%%%%%%%%%%%%%%%%%%%%%%%%%%%%%%%%%%%%%%
%%%%%%%%%%%%%%%%%%%%%% massless low spin %%%%%%%%%%%%%%%%%%%%%%%
\begin{table}[t]
\begin{center}
\begin{tabular}{l c c l }\hline
${\rm spin}$ & $E_0$ & ${\cal C}_2$ & {\rm type} \\ \hline 
~&~&~&~\\[-3mm]
$0$ &  $\ft12 D-1$ & $ -\ft14D(D-2)$& conformal
scalar \\[.5mm]
$0$ &  $\ft12 D$ & $ -\ft14D(D-2)$ & {\rm conformal
scalar} \\[.5mm]
$0$ &  $D-3$ & $ -2(D-3)$ & $\in {\rm singleton}\times
{\rm singleton}$  \\[.5mm]
$0$ &  $D-1$ & $ 0$  & ($D-2$)-rank gauge
field 
\\[.5mm]
$\ft12$ &  $\ft12 D-\ft12$ & $-\ft18D(D-1)$ &  conformal
spinor 
\\[.5mm] 
$\ft12$ &  $D-\ft52$ & $\ft18(D^2-15D +32) $ & 
$\in {\rm singleton}\times{\rm singleton}$
\\[.5mm]
\hline
\end{tabular}
\end{center}
\caption{\small Some unitary anti-de Sitter representations of spin 0
and $\ft12$ which are massless according to various criteria, and the
corresponding values for $E_0$ and ${\cal C}_2$. } 
\label{massless-low-spin}
\end{table}
%%%%%%%%%%%%%%%%%%%%%%%%%%%%%%%%%%%%%%%%%%%%%%%%%%%%%%%%%%%%%%
%%%%%%%%%%%%%%%%%%%%%%%%%%%%%%%%%%%%%%%%%%%%%%%%%%%%%%%%%%%%%%

%%%%%%%%%%%%%%%%%%%%%%%%%%%%%%%%%%%%%%%%%%%%%%%%%%%%%%%%
%%%%%%%%%%%%%%%%%%%%%%%%%%%%%%%%%%%%%%%%%%%%%%%%%%%%%%%%
\subsection{The superalgebras ${\rm OSp}(N\vert 4)$} 
\label{superalgebra} 
%%%%%%%%%%%%%%%%%%%%%%%%%%%%%%%%%%%%%%%%%%%%%%%%%%%%%%%%
In this section we return to the anti-de Sitter superalgebras. 
We start from the (anti-)commutation relations
already presented in \eqn{ads-dec-algebra} and
\eqn{ads-dec-superalgebra}. For definiteness we discuss the case of
4 spacetime dimensions with a Majorana supercharge $Q$. This
allows us to make contact with the material discussed in
section~\ref{masslike}. These anti-de Sitter supermultiplets were
first discussed in \cite{BreitFreed,Heidenreich,FreedNicolai,Nicolai}.  
In most of the section we discuss simple supersymmetry ({\it i.e.}, 
$N=1$), but at the end we turn to more general $N$. 
We choose conventions where the $4\times 4$ gamma matrices are given
by 
\be
\G^0= \pmatrix{-i {\bf 1}& 0 \cr \noalign{\vskip 2mm}
                0& i {\bf 1} \cr} \,,\qquad 
\G^a= \pmatrix{0 & -i\sigma^a \cr \noalign{\vskip 2mm}
                i\sigma^a & 0 \cr} \,, \quad a=1,2,3 \,,
\ee
and write the Majorana spinor $Q$ in the form
\be
Q= \pmatrix{q_\a \cr \noalign{\vskip 2mm} \varepsilon_{\a\b} \,q^\b
\cr}\,, 
\ee
where $q^\a \equiv q_\a^\dagger$, the indices $\a,\b,\ldots$ are
two-component spinor indices and the $\s^a$ are the Pauli spin matrices.
We substitute these definitions into \eqn{ads-dec-superalgebra} and obtain
\bea
{[}H\,,q_\a]  &=& -\ft12 q_\a    \,,\nn\\[1mm]
{[}H\,,q^\a]  &=& \ft12 q^\a    \,,\nn\\[1mm]
\{q_\a\,,q^\b \}  &=& (H\,{\bf 1}+ \vec J\cdot \vec\sigma)_\a{}^\b
\,,\nn\\[1mm] 
\{q_\a\,,q_\b \}  &=& M^-_a \,(\s^a\s^2)_{\a\b}  \,, \nn\\[1mm]
\{q^\a\,,q^\b \}  &=& M^+_a \,(\s^2\s^a)^{\a\b}   \,,
\label{q-sup-alg}
\eea
where we have defined the (hermitean) angular momentum operators  
$J_a = -\ft12\, i\, \varepsilon_{abc} M^{bc}$. We see that the operators
$q_\a$ and $q^\a$ are lowering and raising operators, respectively. They
change the energy of a state by half a unit. Observe that the relative sign
between $H$ and $\vec J\cdot \s$ in the third (anti)commutator is not
arbitrary but fixed by the closure of the algebra. 

In analogy to the bosonic case, we study unitary irreducible representations 
of the ${\rm OSp}(1\vert 4)$ superalgebra. We assume that there exists
a lowest-weight state $|E_0,s\rangle$, characterized by the fact that it is
annihilated by the lowering operators $q_\a$,
\be
q_\a |E_0,s\rangle = 0 \,.
\ee
In principle we can now choose a ground state and build the whole
representation upon it by applying products of raising operators $q^\a$.
However, we only have to study the {\it antisymmetrized} products of
the $q^\a$, because the symmetric ones just yield products of the
operators $M^+_a$ by virtue of (\ref{q-sup-alg}). Products of the
$M^+_a$ simply lead to the higher-energy states in the anti-de
Sitter representations 
of given spin that we considered in section~\ref{adSreps}. By
restricting ourselves to the antisymmetrized products of the $q^\a$ we 
thus restrict ourselves to   
the ground states upon which the full anti-de Sitter
representations are build. These ground states are   
$|E_0,s\rangle$, $q^\a |E_0,s\rangle$ and
$q^{[\a}q^{\b]}|E_0,s\rangle$. 
Let us briefly discuss these representations for different $s$.

The $s=0$ case is special since it contains less anti-de Sitter
representations than the
generic case. It includes the spinless states $|E_0,0\rangle$ and
$q^{[\a}q^{\b]}|E_0,0\rangle$ with ground-state energies $E_0$ and $E_0+1$,
respectively. There is one spin-$\ft12$ pair of ground states
$q^\a|E_0,0\rangle$, with energy 
$E_0+\ft12$. As we will see below, these states can be described by
the scalar field $A$, the pseudo-scalar field $B$ and the spinor field
$\psi$ of the scalar chiral 
supermultiplet, that we studied in section \ref{masslike}.
Obviously, the bounds for $E_0$ that we derived in the previous
sections should be respected, so that $E_0>\ft12$. For $E_0=\ft12$ the
multiplet degenerates and decomposes into a super-singleton,
consisting of a spin-0 and a spin-$\ft12$ singleton, and another spinless
supermultiplet with $E_0=\ft32$. 

For $s\geq\ft12$ we are in the generic situation. We obtain the ground
states 
$|E_0,s\rangle$ and $q^{[\a}q^{\b]}|E_0,s\rangle$
which have both spin $s$ and which have energies $E_0$ and $E_0+1$,
respectively. There are two more (degenerate) ground states,
$q^\a|E_0,s\rangle$, 
both with energy $E_0+\ft12$, which decompose into the ground states
with spin $j=s-\ft12$ and $j=s+\ft12$.

As in the purely bosonic case of section \ref{adSreps}, 
there can be situations in which states decouple so that we are
dealing with multiplet shortening associated with gauge
invariance in the corresponding field theory. The corresponding 
multiplets are then again called massless. We now discuss this in a
general way analogous to the way in which one discusses BPS multiplets
in flat space. Namely, we consider the 
matrix elements of the operator $q_\a\,q^\b$ between the
$(2s+1)$-degenerate ground states $|E_0,s\rangle$,  
\bea
\langle E_0,s| \,q_\a q^\b\,|E_0,s\rangle & = & 
\langle E_0,s| \{q_\a\,, q^\b\}|E_0,s\rangle \nn \\
& = & \langle E_0,s| (E_0\, {\bf 1}+\vec J \cdot \vec \sigma)_\a{}^\b
|E_0,s\rangle \,. \label{matrix} 
\eea
This expression constitutes an hermitean  matrix in both the
quantum numbers of the 
degenerate groundstate and in the indices $\a$ and $\b$, so that it
is $(4s+2)$-by-$(4s+2)$. Because we 
assume that the representation is unitary, this matrix must be
positive definite, as one can verify by inserting a complete set of
intermediate states between the operators $q_\a$ and $q^\b$ in the
matrix element on the left-hand side. 
Obviously, the right-hand side is manifestly hermitean as well, but in
order to be positive definite the eigenvalue $E_0$ of $H$ must be
big enough to compensate for possible negative eigenvalues of $\vec
J\cdot\vec\sigma$, where the latter is again  
regarded as a $(4s+2)$-by-$(4s+2)$ matrix. To determine its 
eigenvalues, we note that 
$\vec J\cdot\vec\sigma$ satisfies the following identity, 
\be
\label{J.sigma}
(\vec J\cdot \vec\sigma)^2 + (\vec J\cdot \vec\sigma) = s(s+1) {\bf 1}
\,, 
\ee
as follows by straightforward calculation.  
This shows that $\vec J\cdot\vec\sigma$ has only two (degenerate)
eigenvalues (assuming $s\not =0$, so that the above equation is not
trivially satisfied), namely $s$ and $-(s+1)$. Hence in
order for \eqn{matrix} to be positive definite, $E_0$ must satisfy the
inequality 
\be
E_0 \geq s+1 \,, \quad \mbox{for } s \geq \ft12 \,. 
\ee
If the bound is saturated, i.e.\ if $E_0 = s+1$, the 
expression on the right-hand side of \eqn{matrix} has zero eigenvalues
so that there are zero-norm states in the multiplet which
decouple. In that case we must be dealing with a massless
multiplet. This is the bound \eqn{4D-unitarity-s}, whose applicability
is extended to spin $\ft12$. The ground state with $s=\ft12$ and
$E_0=\ft32$ leads to the massless vector supermultiplet
in 4 spacetime dimensions. 

As we already mentioned one can also use the oscillator method to 
construct the irreducible representations. There is an extended
literature on this. The reader may consult, for instance,
\cite{GunaNieuWarn,GunaMarWarn}. 

Armed with these results we return to the masslike terms of section
\ref{masslike} for the chiral supermultiplet. The ground-state energy
for anti-de Sitter multiplets corresponding to the scalar field $A$, the
pseudo-scalar field $B$ and the Majorana spinor field  $\psi$, are
equal to $E_0$, $E_0+1$ and
$E_0+\ft12$, respectively. The Casimir operator therefore takes the values
\bea
{\cal C}_2{(A)} & = & E_0(E_0-3) \,,\nonumber \\
{\cal C}_2{(B)} & = & (E_0+1)(E_0-2)\,,\nonumber \\
{\cal C}_2{(\psi)} & = & (E_0+\ft12)(E_0-\ft52) +\ft34 
\label{chiralcasimir}\,.
\eea
For massless anti-de Sitter multiplets, we know that the quadratic
Casimir operator is given by \eqn{eq:D4-massless-casimir}, so we present the
value for ${\cal C}_2 - 2(s^2-1)$ for the three multiplets, i.e
\bea
{\cal C}_2{(A)} +2 & = & (E_0-1)(E_0-2)\,,\nonumber\\
{\cal C}_2{(B)} +2& = & E_0(E_0-1)\,,\nonumber\\
{\cal C}_2{(\psi)} +\ft32 & = & (E_0-1)^2 \,.
\eea
The terms on the right-hand side are not present for massless fields
and we should therefore identify them somehow with the common mass
parameter $m$ of the supermultiplet. Comparison with the field
equations \eqn{field-eqs} shows (for $g=1$) 
that we obtain the correct contributions provided we make the
identification $E_0=m+1$. Observe that we could have made a slightly
different identification here; the above result remains the same under
the interchange of $A$ and $B$ combined with a change of sign in $m$
(the latter is accompanied by a chiral redefinition of $\psi$). 

When $E_0=2$ there exists, in principle, an alternative field
representation for describing this supermultiplet. The spinless 
representation with $E_0=2$ can be described by a scalar field, the
spin-$\ft12$ representation with $E_0=\ft52$ by a spinor field, 
and the second spinless representation with $E_0=3$ by
a rank-2 {\it tensor} field. The Lagrangian for the tensor
supermultiplet is not conformally invariant in 4 dimensions, and this
could account for the unusual ground state energy for the spinor
representation. We have not constructed this supermultiplet in anti-de
Sitter space; in view of the fact that it contains a tensor gauge
field, it should be regarded as massless. 

From Kaluza-Klein compactifications of supergravity one can deduce
that there should also exist shortened {\it massive} supermultiplets. The
reason is that the underlying supergravity multiplet in higher
dimensions is shortened because it is
massless. When compactifying to an anti-de Sitter ground 
state with supersymmetry the massless supermultiplets remain
shortened by the same mechanism, but also the infinite tower of
massive Kaluza-Klein states should comprise shortened supermultiplets. For
toroidal compactifications the massive Kaluza-Klein states belong to
BPS multiplets whose central charges are the momenta associated with
the compactified dimensions. For nontrivial compactifications that 
correspond to supersymmetric anti-de Sitter ground states, the massive
Kaluza-Klein states must be shortened according to the mechanism
exhibited in this section. The singleton multiplets decouple from the
Kaluza-Klein spectrum. Therefore it follows that there must exist
shortened massive representations of the extended supersymmetric
anti-de Sitter algebra. 

To exhibit this we generalize the previous analysis to the
$N$-extended superalgebra, denoted by ${\rm OSp}(N,4)$. As it turns
out, the analysis is rather similar. The supercharges now carry an
extra ${\rm SO}(N)$ index and are denoted by $q_{\a i}$ and $q^{\a
i}$, with $q^{\a i} = (q_{\a i})^\dagger$ with $i=1,\dots,N$. The most 
relevant change to the (anti)commutators \eqn{q-sup-alg} is in the
third one, which reads
\be
\{q_{\a i}\,,q^{\b j} \}  = \d_i^j \,\d_\a^\b\,H+  \d_i^j\,\vec J\cdot
\vec\sigma_\a{}^\b +\d_\a^\b\;  \vec T\cdot \vec\Sigma_i{}^j \,,
\label{ex-sup-alg1}
\ee
where $\vec T$ are the hermitean $\ft12 N(N-1)$ generators of 
${\rm SO}(N)$ which act on the supercharges in the fundamental
representation, generated by the hermitean matrices $\vec \Sigma$. The
last two anticommutators are given by
\bea
\{q_{\a i}\,,q_{\b j}\}  &=& M^-_a \,(\s^a\s^2)_{\a\b}\;\d_{ij} \,,
\nn\\[1mm] 
\{q^{\a i}\,,q^{\b j}\}  &=& M^+_a \,(\s^2\s^a)^{\a\b}\;\d^{ij} \,.
\label{ex-sup-alg2}
\eea
The construction of lowest-weight representations proceeds in the same
way as before. One starts with a ground state of energy $E_0$ which
has a certain spin and transforms according to a representation of
${\rm SO}(N)$ which is annihilated by the $q_{\a i}$. Denoting the
${\rm SO}(N)$ representation by $t$ (which can be expressed in terms
of the eigenvalues of the Casimir operators or Dynkin labels), we have
\be 
q_{\a i}\vert E_0,s,t\rangle =0\,.
\ee
Excited states are generated by application of the $q^{\a i}$, which
are mutually anticommuting, with exception of the combination that
leads to the operators $M^+_a$ which will generate the full anti-de Sitter 
representations. Hence the generic $N$-extended representations
decompose into ordinary anti-de Sitter representation whose ground
states have energy $E_0+\ft12 n$ and which can be written as 
\be 
q^{[\a_1 i_1}\,\cdots\, q^{\a_n i_n]}\, \vert E_0,s,t\rangle \,.
\ee
Here the antisymmetrization applies to the combined $(\a i)$
labels. As before the unitarity limits follow from the separate limits
on the anti-de Sitter representations and from the right-hand side of
the anticommutator \eqn{ex-sup-alg1}, which decomposes into three
terms, namely the Hamiltonian, the rotation generators and the
R-symmetry generators, taken in the space of ground state
configurations ({\it c.f.} \eqn{matrix}). We have already determined
the possible eigenvalues of $\vec J\cdot\vec\sigma$ which are equal to
$s$ or $-(s+1)$. In a similar way one can determine the eigenvalues for 
$\vec T\cdot\vec\Sigma$ by noting that it satisfies a 
polynomial matrix equation such as \eqn{J.sigma} with coefficients
determined by the Casimir operators. For instance, for $N=3$ we
derive,  
\be
-(\vec T\cdot\vec\Sigma)^3+ 2 (\vec T\cdot\vec\Sigma)^2 +
(t^2+t-1) (\vec T\cdot\vec\Sigma) = t(t+1)\, {\bf 1}\,,
\ee
where $\vec T^2 = t(t+1)\,{\bf 1}$. This equation shows that the
eigenvalues of $\vec T\cdot\vec\Sigma$ take the values $-t$,
1 or $t+1$, unless $t=0$. Combining these results we find that the 
right-hand side of \eqn{ex-sup-alg1} in the space of degenerate ground
state configurations has the following
six eigenvalues: $E_0+s-t$, $E_0+s+1$, $E_0+s+t+1$, $E_0-s-t-1$,
$E_0-s$ or $E_0+t+1$. All these eigenvalues must be positive, so that
in the generic case where $s$ and $t$ are nonvanishing, we derive the 
unitarity bound, $E_0\geq 1+s+t$. Incorporating also the
possibility that $s$ or $t$ vanishes, the combined result takes the
following form,
\be
\begin{array}{l c l}
E_0\geq 1+s+t   &\mbox{for}& s\geq \ft12\,,\; t\geq \ft12\,,\\[.5mm]
E_0\geq 1+s     &\mbox{for}& s\geq\ft12\,,\; t = 0\,, \\[.5mm]
E_0\geq t       &\mbox{for}& s=0\,,\; t\geq\ft12  \,.
\end{array}
\ee
Whenever one of these bounds is saturated, certain anti-de Sitter
representations must decouple. The ground states with $s=0$ and $E_0=t$
define massive shortened representations of the type that appear in
Kaluza-Klein compactifications \cite{FreedNicolai}. In the Poincar\'e 
limit these representations become all massless. 

Obviously these techniques can be extended to other cases, either by 
changing the number of supersymmetries or by changing the spacetime
dimension. There is an extended literature to which we refer the
reader for applications and further details.  

Before closing the chapter we want to return to the remarkable
singleton representations. Long before the formulation of the AdS/CFT
correspondence it was realized that supersingleton representations
could be described by conformal supersymmetric field theories on a
boundary. Two prominent examples were noted (see, {\it e.g.},
\cite{GunaMarWarn,GunaNieuWarn}), namely the singleton
representations  in $D=5$ and 7 anti-de Sitter space, which correspond
to $N=4, D=4$ supersymmetric gauge theories and the chiral $(2,0)$ tensor
multiplet in $D=6$ dimensions. The singletons decouple from the
Kaluza-Klein spectrum, precisely because they are related to 
boundary degrees of freedom. Group-theoretically they are of
interest because their products lead to the massless and massive
representations that one encountered in the Kaluza-Klein context. 
Another theme addresses the connection between singletons and higher-spin
theories. Here the issue is whether the singletons play only a
group-theoretic role or whether they have also a more dynamical 
significance. We refrain from speculating about these questions and
just refer to some recent papers \cite{FerrFrons,SezginSun,Segal}.  
In \cite{SezginSun} the reader may also find a summary of some useful
results about singletons as well as an extensive list of
references.  

In the next chapter we will move to a discussion of superconformal
symmetries, which are based on the same anti-de Sitter algebra. We
draw the attention of the reader to the fact that in
chapter~\ref{superconformal}, $D$ will always denote the 
spacetime dimension of the superconformal theory. The corresponding
superalgebra is then the anti-de Sitter superalgebra, but in spacetime
dimension $D+1$. 
%%%%%%%%%%%%%%%%%%%%%%%%%%%%%%%%%%%%%%%%%%%%%%%%%%%%%%%%%%%%%%%%

%%%%%%%%%%%%%%%%%%%%%%%%%%%%%%%%%%%%%%%%%%%%%%%%%%%%%%%%%%%%%%%%
%%%%%%%%%%%%%%%%%%%%%%%%%%%%%%%%%%%%%%%%%%%%%%%%%%%%%%%%%%%%%%%%
%\newpage %%%%%%%%%%%%%%%%%%%%%%%%%%%%%%%%%%%%%%%%%%%%%%%%%%%%%%%
%%%%%%%%%%%%%%%%%%%%%%%%%%%%%%%%%%%%%%%%%%%%%%%%%%%%%%%%%%%%%%%%
%%%%%%%%%%%%%%%%%%%%%%%%%%%%%%%%%%%%%%%%%%%%%%%%%%%%%%%%%%%%%%%%
%%%%%%%%%%%%%%%%%% CHAPTER 7 %%%%%%%%%%%%%%%%%%%%%%%%%%%%%%%%%%%
%%%%%%%%%%%%%%%%%%%%%%%%%%%%%%%%%%%%%%%%%%%%%%%%%%%%%%%%%%%%%%%%
\section{Superconformal symmetry}
\setcounter{equation}{0}
\label{superconformal}
%%%%%%%%%%%%%%%%%%%%%%%%%%%%%%%%%%%%%%%%%%%%%%%%%%%%%%%%%%%%%%%%
Invariances of the metric are known as isometries. Continuous
isometries are generated by so-called Killing vectors, satisfying 
\be
D_\m\xi_\n +D_\n\xi_\m = 0\,. \label{killing-vector}
\ee
The maximal number of linearly independent Killing vectors is equal to
$\ft12 D(D+1)$. A space that has the maximal number of isometries is
called maximally symmetric. 
A weaker condition than \eqn{killing-vector} is, 
\be
D_\m\xi_\n +D_\n\xi_\m = {2\over D}\, g_{\m\n}
D_\rho\xi^\rho\,. \label{ckilling-vector} 
\ee
Solutions to this equation are called {\it conformal} Killing
vectors. Note that the above equation is the traceless part of
\eqn{killing-vector}. The conformal Killing vectors that are not
isometries are thus characterized by a nonvanishing
$\xi=D_\m\xi^\m$. For general dimension $D>2$ there  
are at most $\ft12(D+1)(D+2)$ conformal Killing vectors. For $D=2$
there can be infinitely many conformal Killing vectors. These result
can be derived as follows. First one shows that 
\begin{equation}
  \label{eq:1}
D_\m D_\n\xi_\rho = R_{\n\rho\m}{}^\s \,\xi_\s - {1\over D}
\Big[g_{\m\n} \,D_\rho\xi - g_{\rho\m} \,D_\n\xi - g_{\rho\n}
\,D_\m\xi \Big]\,. 
\end{equation}
For Killing vectors (which satisfy $\xi=0$) this result implies that
the second derivatives of Killing vectors  are determined by the
vector and its first derivatives. When expanding about a certain point on the
manifold, the Killing vector is thus fully determined by its value at
that point and the values of its first derivatives (which are antisymmetric
in view of \eqn{killing-vector}). Altogether there are
thus $\ft12D(D+1)$ initial conditions to be fixed and they parametrize
the number of independent Killing vectors. For conformal Killing
vectors, where $\xi\not= 0$ one then proves that $(D-2)D_\m D_\n \xi$
and $D^\m D_\m\xi$ are determined in terms of lower derivatives. This
suffices to derive the maximal number of conformal Killing vectors
quoted above for $D>2$. Both ordinary and conformal Killing vectors
generate a group.  

In what follows we choose a Minkowski signature for the
$D$-dimensional space, a restriction that is mainly relevant when
considering supersymmetry. Flat Minkowski spacetime has the maximal
number of conformal Killing vectors, which decompose as follows, 
\be
\xi^\m \propto \left \{ \begin{array}{l l }
\xi_{\rm P}^\m        &\mbox{spacetime translations ($P$)}\\[.5mm]
\e^{\m}{}_{\!\n}\,x^\n    &\mbox{Lorentz transformations ($M$)}\\[.5mm]
\Lambda_{\rm D} \,x^\m &\mbox{scale transformations ($D$)}\\[.5mm]
(2\, x^\m x^\n -  x^2 \,\eta^{\m\n} )\Lambda_{{\rm K} \n} 
&\mbox{conformal boosts ($K$)} 
\end{array} \right. \label{flat-conf-killing}
\ee
Here $\xi_{\rm P}^\m$, $\e^{\m\n}= - \e^{\n\m}$, $\Lambda_{\rm D}$ and
$\Lambda_{\rm K}^\m$ are constant parameters. Obviously $\xi =
D(\Lambda_{\rm D} + x_\m \Lambda^\m_{\rm K})$. The above  conformal
Killing vectors generate the group ${\rm SO}(D,2)$. This is the same
group as the anti-de Sitter group in $D+1$ dimensions. The case
of $D=2$ is special because in that case the above transformations
generate a semisimple group, ${\rm SO}(2,2) \cong ({\rm SL}(2,{\rm R})
\times  {\rm SL}(2,{\rm R}))/{Z_2}$. This follows directly by writing out
the infinitesimal transformations \eqn{flat-conf-killing} for the
linear combinations $x\pm t$, 
\be 
\d(x \pm t) = (\xi_{\rm P}^x\pm\xi_{\rm P}^t) 
+ (\Lambda_{\rm D} \mp \e^{xt} ) 
(x\pm t) + \ft12(\Lambda_{\rm K}^x \mp \Lambda_{\rm K}^t) (x\pm
t)^2\,.  
\ee
However, for $D=2$ there are infinitely many conformal Killing
vectors, corresponding to two copies of the Virasoro algebra. The 
corresponding diffeomorphisms can be characterized in terms of two
independent functions $f_\pm$ and take the form,
\be
\d x= f_+(x+t) +f_-(x-t)\,,\qquad \d t = f_+(x+t) - f_-(x-t)\,.
\ee

The fact that, for $D\geq3$ the anti-de Sitter and the conformal group
coincide for dimensions $D+1$ and $D$, respectively, can be clarified
by extending the $D$-dimensional spacetime parametrized by coordinates
$x^\m$ with an 
extra (noncompact) coordinate $y$, assuming the line element,
\be
\label{bulk-metric}
{\rm d}s^2 = {g_{\m\n}\,{\rm d}x^\m{\rm d}x^\n + {\rm d}y^2 \over y^2}
\;, 
\ee
so that the right-hand side of \eqn{ckilling-vector}, which is
responsible for the lack of invariance of the line element of the
original $D$-dimensional space, can be cancelled by a scale
transformation of extra 
coordinate $y$. It is straightforward to derive the nonvanishing
Christoffel symbols for this extended space,
\be 
\{{}_\m{}^{\!y}{}_{\!\n}\} = y^{-1}\,g_{\m\n}\,,\quad
\{{}_\m{}^{\!\n}{}_{\!y}\} = -y^{-1}\, \d^\n_\m\,,\quad
\{{}_y{}^{\!y}{}_{\!y}\} = -y^{-1} \,,   
\ee
where $\{{}_\m{}^{\!\rho}{}_{\!\n}\}$ remains the same for both spaces and all
other components vanish. The corresponding expressions for the
curvature components are 
\bea
R_{\m\n\rho}{}^\s  &=&R^D_{\m\n\rho}{}^\s + 2 y^{-2}\, g_{\rho[\m}
\d_{\n]}^\s \,,\nn \\  
R_{\m y\rho}{}^y &=& y^{-2}\,g_{\m\rho}\,,\nn \\
R_{y\n y}{}^\s  &=& y^{-2} \d_\n^\s \,.
\eea
With these results one easily verifies that the curvature tensor of the
${(D\!+\!1)}$-dimensional extension of a flat $D$-dimensional Minkowski
space is that
of an anti-de Sitter spacetime with unit anti-de Sitter radius
(i.e. $g=1$ in \eqn{ads-curv}). This was the reason why we adopted a
positive signature in the line element \eqn{bulk-metric} for the
coordinate $y$. 

Subsequently one can show that 
the $D$-dimensional conformal Killing vectors satisfying $D_\m
D_\n\xi=0$ can be extended to Killing vectors of the
$(D+1)$-dimensional space,
\begin{equation}
  \label{eq:D+1isometry}
  \xi^\m(x,y) =  \xi^\m(x) - {y^2\over 2D} \,\pa^\m\xi(x) \,,\qquad
  \xi^y(x,y)={y\over D} \,\xi(x) \,.
\end{equation}
The condition $D_\m D_\n\xi=0$ holds for the conformal Killing vectors
\eqn{flat-conf-killing}. For $D=2$ these vectors generate a finite
subgroup of the infinite-dimensional conformal group, and only this
group can be extended to isometries of the ($D+1$)-dimensional
space. Nevertheless, near the boundary \cite{HawkEll} of 
the space ($y\approx 0$), the conformal Killing vectors generate
asymptotic symmetries. Such a phenomenon was first analyzed in
\cite{BrownHenneaux}. 

This setting is relevant for the adS/CFT correspondence
and there exists an extensive literature on this (see, {\it e.g.}, 
\cite{Mald,GubKlePol,Witten-holo,BalaKraLawr,dHSkenSolo,Aharony}, and
also the lectures presented at this school).   
Also the relation between the D'Alembertians of the
extended and of the original $D$-dimensional spacetime is
relevant in this context. Straightforward calculation yields,
\begin{equation}
  \label{eq:conf-box}
  \Box_{D+1} = y^{2} \,\Box^D + (y \,\pa_y)^2 - D\,
  y\, \pa _y  \,.
\end{equation}
Near the boundary where $y$ is small, the fields can be approximated by
$y^\Delta\,\phi(x)$.  
We may compare this to solutions of the Klein-Gordon equation in 
the anti-de Sitter space, for which we know that the D'Alembertian
equals the quadratic Casimir operator ${\cal C}_2$. In terms of the
ground state energy $E_0$ of the anti-de Sitter representation, we
have ${\cal C}_2= E_0(E_0-D)$ (observe that we must replace
$D$ by $D+1$ in \eqn{eq:casimir-repr}), which shows that we have the
identification $\Delta=E_0$ or $\Delta=D-E_0$. This identification is
somewhat remarkable in view of the fact that $E_0$
is the energy eigenvalue associated with the ${\rm SO}(2)$ generator
of the anti-de Sitter algebra and not with the noncompact scale
transformation of $y$, which associated with the ${\rm SO}(1,1)$
eigenvalue. The identification of the generators is discussed in more
detail in the next section. 

%%%%%%%%%%%%%%%%%%%%%%%%%%%%%%%%%%%%%%%%%%%%%%%%%%%%%%%%%%%%%%%%%%%%%%%%
\subsection{The superconformal algebra}
\label{sec:superc-algebra}
From the relation between the conformal and the anti-de Sitter algebra
one can determine the superextension of the
conformal algebra generated by the above conformal Killing vectors. In 
comparison to the anti-de Sitter algebra and superalgebra
({\it c.f.} \eqn{ads-algebra} and \eqn{ads-superalgebra}) 
we make a different decomposition than the one that led to
\eqn{ads-dec-algebra} and \eqn{ads-dec-superalgebra}. We start from a
$D$-dimensional 
spacetime of coordinates carrying indices $a= 0,1,\ldots,D-1$, which
we extend with {\it two} 
extra index values, so that $A= -,0,1,\ldots, D-1, D$. For the
bosonic generators which generate the group ${\rm SO}(D,2)$ we have 
\bea
M_{D-}  &\longrightarrow& D \,,  \nn\\
M_{ab}  &\longrightarrow& M_{ab} \,, \nn\\
M_{Da}  &\longrightarrow& \ft12( P_{a}-K_a) \,,  \nn\\
M_{-a}  &\longrightarrow& \ft12(P_a + K_{a}) \,,
\eea
Here we distinguish the generator $D$ of the dilatations, $\ft12D(D-1)$
generators $M_{ab}$ of the Lorentz transformations, $D$ generators
$P_a$ of the translations, and $D$ generators $K_a$ of the conformal 
boosts. 

The algebra associated with ${\rm SO}(D,2)$ was given in
\eqn{ads-algebra} and corresponds to the following commutation relations,
\be
\label{eq:conf-boson}
\begin{array}{rcl}
{[}D,P_a] &\!=\!& -P_a\,,\\[.5mm]
{[}M_{ab},P_c] &\!=\!& -2 \,\eta_{c[a} \,P_{b]}\,,\\[.5mm]
{[}M_{ab},M_{cd}] &\!=\!& 4\,\eta_{[a[c}\,M_{d]b]}\,,\\[.5mm]
{[}D, M_{ab}] &\!=\!& 0 \,,
\end{array}
\quad 
\begin{array}{rcl}
{[}D,K_a] &\!=\!&  K_a \,, \\[.5mm]
{[}M_{ab},K_c] &\!=\!& -  2\,\eta_{c[a}\,K_{b]} \,,\\[.5mm]
{[}P_a,P_b] &\!=\!&{[}K_a,K_b]=0  \,, \\[.5mm] 
{[}K_a,P_b] &\!=\!& 2(M_{ab} + \eta_{ab} \,D)\,.
\end{array}
\ee
To obtain the superextension (for $D\leq 6$) one must first extend the 
spinor representation associated with the $D$-dimensional spacetime to
incorporate two extra gamma matrices $\G_D$ and $\G_-$. According to the
discussion in section~\ref{R-symmetry} (see, 
in particular, table~\ref{pq-spinors}) this requires a doubling of the 
spinor charges,   
\begin{equation}
  \label{eq:double-Q}
Q\to {\cal Q}  = \pmatrix{S_\a\cr Q_\a\cr}  \,,\qquad \bar Q\to
\bar{\cal Q}= (\bar Q_\a, \bar S_\a)\,,
\end{equation}
and we define an extended set of gamma matrices $\G_A$ by, 
\begin{equation}
  \label{eq:extra-gamma}
\Gamma_a = \pmatrix{ \G^a &0 \cr \noalign{\vskip 1mm}0 &-\G^a \cr} 
\qquad 
\Gamma_D= \pmatrix{ 0 &{\bf 1} \cr\noalign{\vskip 1mm} {\bf 1}&0 \cr} 
\qquad 
\Gamma_- = \pmatrix{ 0 &{\bf 1} \cr\noalign{\vskip 1mm}  -{\bf 1}&0
\cr}  \,. 
\end{equation}
The new charges $S_\a$ generate so-called {\it special} supersymmetry
transformations \cite{FKTvN}. The decomposition of the conjugate spinor is
somewhat subtle, to make contact with the Majorana condition employed
for the anti-de Sitter algebra. 

The anticommutation relation for the spinor charges follows from
\eqn{ads-superalgebra} and can be written as  
\begin{eqnarray}
  \label{eq:susy}
  \{{\cal Q},\bar {\cal Q} \} &=& \pmatrix{ \{S,\bar Q\} & \{S,\bar
    S\}\cr\noalign{\vskip 3mm}  \{Q,\bar Q\} & \{Q,\bar S\}\cr}    \nn\\[2mm]
&=& \pmatrix{-\ft12 \G^{ab}M_{ab} - D &  -\G^a K_a \cr \noalign{\vskip 5mm}  
-\G^a P_a & -\ft12 \G_{ab}M_{ab} + D \cr}\,,
\end{eqnarray}
or,
\bea
\label{conf-anticomm}
\{ Q_\a,\bar Q_\b\} &\!=\!& -\G^a_{\a\b}\, P_a \,, \nn \\[.5mm]
\{ S_\a,\bar S_\b\} &\!=\!& -\G^a_{\a\b}\, K_a \,,\nn\\[.5mm]
\{ Q_\a, \bar S_\b\} &\!=\!&-\ft12 \G^{ab}_{\a\b}\,M_{ab}+
\eta_{\a\b} \,D \,. 
\eea
The nonvanishing commutators of the spinor charges with the bosonic
generators read, 
\be
\label{eq:conf-fermion}
\begin{array}{rcl}
{[}M_{ab} ,\bar Q_\a] &\!=\!& \ft12 (\bar Q \G_{ab})_\a\,,\\[.5mm]
{[}D,\bar Q_\a] &\!=\!& -\ft12 \bar Q_a \,,\\[.5mm]
 {[}K_a, \bar Q_\a] &\!=\!&  -(\bar S \G_a)_\a \,,
\end{array}
\quad
\begin{array}{rcl}
{[}M_{ab} ,\bar Q_\a] &\!=\!& \ft12 (\bar Q \G_{ab})_\a\,,\\[.5mm]
{[}D,\bar S_\a] &\!=\!& \ft12 \bar S_\a \,, \\[.5mm]
{[}P_a, \bar  S_\a] &\!=\!& - (\bar Q \G_a)_\a \,.
\end{array}
\ee
Here we are assuming the same gamma matrix conventions as in the
beginning of chapter~\ref{supergravity}. From the
results quoted in the previous chapter, we know that, up to $D=6$, the
bosonic subalgebra will be the sum of the conformal algebra and the
R-symmetry algebra. The R-symmetry can be identified from 
table~\ref{pq-spinors} and the corresponding generators will appear on
the righ-hand side of the $\{Q,S\}$ anticommutator; the other
(anti)commutation relations listed above remain unchanged. In
addition, commutators with the R-symmetry generators must be
specified, but those follow from the R-symmetry assignments of the
supercharges. The above (anti)commutators satisfy the Jacobi
identities that are at most quadratic in the fermionic generators. The
validity of the remaining Jacobi identities, which are cubic in the
fermionic generators, requires in general the presence of the R-symmetry
charges. The results given so far suffice to discuss the most salient
features of the superconformal algebra and henceforth we will be
ignoring the contributions of the R-symmetry generators. Note also that
the numbers of bosonic and fermion generators do not match; this
mismatch will in general remain when including the R-symmetry
generators. 

As before, the matrix on the right-hand side of \eqn{eq:susy} may have
zero eigenvalues, leading to shortened supermultiplets. Those
multiplets are in one-to-one correspondence with the anti-de Sitter
supermultiplets. Its eigenvalues are subject to certain positivity
requirements in order that the algebra is realized in a
positive-definite Hilbert space.
 
The abstract algebra can be connected to the spacetime transformations
\eqn{flat-conf-killing} in flat spacetime introduced at the beginning
of this chapter. To see this we derive how the conformal
transformations act on generic fields. In principle, this is an 
application of the theory of homogeneous spaces discussed in
chapter~\ref{homogeneous} and we will demonstrate this
for the bosonic transformations \cite{MackSalam}; a supersymmetric
extension can be given in superspace. Let us assume that the action of
these spacetime transformations denoted by $g$ takes the following
form on a generic multicomponent field $\phi$,  
\begin{equation}
  \label{eq:field-transf}
\phi(x) {\longrightarrow} \phi_{g}(x) = {\rm S}(g, x) \, 
\phi(g^{-1} x)\,, 
\end{equation}
where $\rm S$ is some matrix acting on the components of $\phi$. Observe
that there exists a subgroup of the conformal group that leaves a
point in spacetime invariant and choose, by a suitable translation,
this point equal to $x^a=0$. From \eqn{flat-conf-killing} it then
follows that the corresponding stability group of this point is 
generated by the generators $M$ of the Lorentz 
group, the generator $D$ of the scale transformations and the
generators $K$ of the conformal boosts.  Hence we conclude that the
matrices ${\rm S}(g,0)$ 
must form a representation of this subgroup, whose generators
are denoted by the matrices $\hat M_{ab}$, $\hat D$ and
$\hat K_a$. Generic fields are thus assigned to representations of
this subgroup.  

On the other hand, we want the translation operators to act
exclusively on the coordinates $x^a$, so that according to
\eqn{eq:field-transf} (the generators are antihermitean),
\begin{equation}
  \label{eq:P-phi}
P_a \,\phi(x)= - {\pa\over\pa x^a} \, \phi(x)\,.
\end{equation}
Hence we may write $\phi(x)=\exp(- x^aP_a)\, \phi(0)$. Subsequently we
define the infinitesimal variation $\phi_g(x)\approx \phi(x) +\d
\phi(x)$, where $\d\phi(x)$ is generated by 
\be
\d\phi(x) = \Big[\xi_{\rm P}^aP_a + \ft12 \e^{ab}M_{ab} +\Lambda_{\rm
D} D + \Lambda_{\rm K}^aK_a\Big]\phi(x)\,. 
\ee
This variation can be converted to the basis defined by the fields at
the origin, by sandwiching between 
$\exp(x^aP_a)$ and $\exp(-x^aP_a)$. The result is then related to the
infinitesimal variation of ${\rm S}(g,0) \approx {\bf 1} +  \ft12 
\hat\e^{ab}\hat M_{ab} + \hat\Lambda_{\rm D} \hat D + \hat\Lambda_{\rm
K}^a \hat K$ and terms proportional to $P_a$. Using the commutation
relations \eqn{eq:conf-boson} and using \eqn{eq:field-transf} and 
\eqn{eq:P-phi}, it follows that 
%
%\be
%{\rm S}(g,x) = \exp(-y^a P_a)\,{\rm S}(g,0)\,\exp(x^b P_b)\,, \;\mbox{
%with  } y= g\,x\,. 
%\ee
%Writing this out for infinitesimal transformations with $y^a = x^a +
%\xi^a$, and ${\rm O}(g,0)\approx {\bf 1} + \ft12 \e^{ab}\hat M_{ab} +
%\Lambda_{\rm D} \hat D + \Lambda_{\rm K}^a \hat K$, we derive that
%conformal transformations act infinitesimally on $\phi$ according to  
\bea
  \label{eq:d-phi-conf}
\d\phi(x) &\!\!=\!\!& - \Big[\xi^a_{\rm P} - \e^{ab}x_b +\Lambda_{\rm D}\,
x^a -2 x^a\,x_b\Lambda_{\rm K}^b  +x^2\,\Lambda_{\rm K}^a\Big] \pa_a
\phi(x) \\ 
&& + \Big[\Big(\ft12 \e^{ab} - 2 \Lambda_{\rm K}^{[a}
x^{b]} \Big) \hat M_{ab}   + \Big(\Lambda_{\rm D} - 2\Lambda_{\rm K}^a
x_a \Big) \hat D 
+ \Lambda_{\rm K}^a \hat K_a\Big] \phi(x)\,, \nn
\eea
where the first term represents the conformal Killing vectors parametrized
in \eqn{flat-conf-killing}. Note that the combination of sign
factors is dictated by the algebra \eqn{eq:conf-boson}. 

The procedure applied above is just a simple example of the
construction of induced representations on a ${\rm G}/{\rm H}$
coset manifold. Indeed, we are describing flat space as a coset
manifold, where the conformal group plays the role of the isometry
group G and the stability group plays the role of the isotropy group
H. The coset representative equals $\exp(- x^aP_a)$, from which it
follows ({\it c.f.} \eqn{e-omega}) that the vielbein is constant and
diagonal and the connections associated with 
the stability group are zero. Hence the metric is invariant under the
conformal transformations, as established earlier, while the vielbein
is invariant after including the compensating transformations
represented by the second line of \eqn{eq:d-phi-conf}. Explicit
evaluation then shows that the invariance of the flat vielbein requires
the compensating tangent-space transformations,
\be
\d e_\m{}^{\!a} \propto \e^{ab} \, e_{\m, b} - \Lambda_{\rm D}
\,e_\m{}^{\!a}\,,
\ee
with parameters specified by \eqn{eq:d-phi-conf}. 
Note that the special conformal boosts do not act on the tangent space
index of the vielbein. 

In the next two sections we will discuss how one can deviate from flat
space in the context of the conformal group. There are two approaches
here which lead to related results. One is to start from a gauge
theory of the conformal group. This conformal group has a priori
nothing to do with spacetime transformations and the resulting theory
is described in some unspecified spacetime. Then one imposes 
constraints on certain curvatures. This is similar to what we described
in section~\ref{supergravity}, where we imposed a constraint on the
torsion tensor ({\it c.f.} \eqn{eq:torsion-constraint}), so that the
spin connection becomes a dependent field and the Riemann tensor
becomes proportional to the curvature of the spin connection
field. This approach amounts to imposing the maximal number of
conventional constraints. The second approach starts from the coupling
to superconformal matter and the corresponding superconformal
currents. 

%%%%%%%%%%%%%%%%%%%%%%%%%%%%%%%%%%%%%%%%%%%%%%%%%%%%%%%%%%%%%%%%%%
\subsection{Superconformal gauge theory and supergravity}
\label{superconf-gauge}
%%%%%%%%%%%%%%%%%%%%%%%%%%%%%%%%%%%%%%%%%%%%%%%%%%%%%%%%%%%%%%%%%%
In principle it is straightforward to set up a gauge theory associated
with the superconformal algebra. We start by associating a gauge field
to every generator,
\be
\begin{array}{lcccccc}
\mbox{generators:} & P          & M          & D    & K          
    & Q & S \\[.5mm] 
\mbox{gauge fields:} & e_\m^{\;a} & \o^{ab}_\m & b_\m & f_\m^{\;a} 
    & \psi_\m & \phi_\m \\[.5mm]
\mbox{parameters:} & \xi_{\rm P}^a & \e^{ab} & \Lambda_{\rm D} &
    \Lambda_{\rm K}^a & \e & \eta 
\end{array}
\ee
%%%%%%%%%%%%%%%%%%%%%%%%%%%%%%%%%%%%%%%%%%%%%%%%%%%%%%%%%%%%%%%%
Up to normalization factors, the transformation rules for the
gauge fields, which we specify below, follow directly from the
structure constants of the superconformal algebra,  
\begin{eqnarray}
  \label{eq:conf-gauge-fields}
\d e_\m{}^{\!a} &=& {\cal D}_\m \xi_{\rm P}^a  -\Lambda_{\rm
  D}e_\m^{\;a} +\ft12 \bar\e\G^a\psi_\m \,, 
\nn\\
\d\o_\m^{ab}  &=& {\cal D}_\m \e^{ab} + \Lambda_{\rm K}^{[a}
e_\m^{\;b]} - \xi_{\rm P}^{[a} f_\m{}^{\!b]}-  \ft14 \bar \e \G^{ab}
\phi_\m +\ft14 \bar \psi_\m  \G^{ab} \eta  \,, 
\nn \\[1mm] 
\d b_\m   &=& {\cal D}_\m \Lambda_{\rm D} +\ft12 \L_{{\rm
K}a}e_\m{}^{\!a} -\ft12 \xi_{{\rm P}a} f_\m{}^{\!a} 
+ \ft14 \bar \e \phi_\m  -\ft14 \bar \psi_\m\eta  \,,  
\nn\\[1mm]
\d f_\m^{\;a} &=& {\cal D}_\m \Lambda_{\rm K}^a  +\Lambda_{\rm
  D}e_\m^{\;a} + \ft 12 \bar\eta \G^a\phi_\m \,,
\nn\\[1mm]
\d\psi_\m     &=& {\cal D}_\m \e -\ft12 \Lambda_{\rm
  D}\psi_\m  - \ft12 e_\m{}^{\!a} \G_a \eta +\ft12   \xi_{\rm P}^a \G_a
  \phi_\m \,,  
\nn\\[1mm]
\d\phi_\m     &=& {\cal D}_\m \eta +\ft12  \Lambda_{\rm
  D} \phi_\m - \ft12 f_\m^{\;a}\G_a \e + \ft12 \Lambda_{\rm K}^a\G_a
  \psi_\m    \,.
\end{eqnarray}
Here we use derivatives that are covariantized with respect to
dilatations and Lorentz transformations, {\it i.e.}, 
\begin{eqnarray}
  \label{eq:MD-cov-derivative}
{\cal D}_\m \xi_{\rm P}^a &=& \pa_\m\xi_{\rm P}^a + b_\m
\xi_{\rm P}^a - \omega_\m{}^{ab}\xi_{{\rm P}b}\,,   
\nn\\[.5mm]
{\cal D}_\m \Lambda_{\rm K}^a &=& \pa_\m\Lambda_{\rm K}^a - b_\m
\Lambda_{\rm K}^a-\omega_\m{}^{ab}\Lambda_{{\rm K}b}\,,   
\nn\\[1mm]
{\cal D}_\m \Lambda_{\rm D} &=& \pa_\m\Lambda_{\rm D}\,,   
\nn\\[1mm]
{\cal D}_\m \e  &=&  (\pa_\m +\ft12 b_\m -\ft14
\o_\m^{ab} \G_{ab} )\e \,,
\nn\\
{\cal D}_\m \eta  &=&  (\pa_\m -\ft12 b_\m -\ft14
\o_\m^{ab} \G_{ab} )\eta \,. 
\end{eqnarray}
Again we suppressed the gauge fields for the R-symmetry generators.
 
The above transformation rules close under commutation, up to the
commutators of two supersymmetry transformations acting on the
fermionic gauge fields. In that case, one needs
Fierz reorderings to establish the closure of the algebra, which
depend sensitively on the dimension and on the presence of additional
generators (for $D=4$, see, for example, \cite{FKTvN}). As an example
we list some of the commutation relations that can be obtained from
\eqn{eq:conf-gauge-fields}, 
\bea
{[} \d_{\rm P}(\xi_{\rm P}),\d_{\rm K}(\Lambda_{\rm K})]&=&  \d_{\rm
D}(\ft12 \Lambda_{\rm K}^a\xi_{\rm P}^b\,\eta_{ab})+ \d_{\rm
M}(\Lambda_{\rm K}^{[a}\xi_{\rm P}^{b]}) \,,
\nn\\[.5mm]
\{ \d_{\rm Q}(\e_1),\d_{\rm Q}(\e_2)\}&=&  \d_{\rm P} 
(\ft12 \bar\e_2\G^a\e_1)\,, 
\nn\\[.5mm]
\{ \d_{\rm S}(\eta_1),\d_{\rm S}(\eta_2)\}&=&  \d_{\rm K}
(\ft12 \bar\eta_2\G^a\eta_1) \,,
\nn\\
\{ \d_{\rm Q}(\e),\d_{\rm S}(\eta)\}&=&  \d_{\rm M}(\ft14
\bar\e \G^{ab}\eta) + \d_{\rm D}(-\ft14 \bar\e \eta) \,, 
\nn\\[.5mm]
{[} \d_{\rm Q}(\e),\d_{\rm K}(\Lambda_{\rm K})]&=&  \d_{\rm S}(\ft12
\Lambda^a_{\rm K} \G_{a}\e)\,,
\nn\\[.5mm]
{[} \d_{\rm S}(\eta),\d_{\rm P}(\Lambda_{\rm P})]&=&  \d_{\rm Q}(\ft12
\xi_{\rm P}^a \G_{a}\eta)\,.
\eea

For completeness we also present the corresponding curvature tensors
of the superconformal gauge theory,
\begin{eqnarray}
  \label{eq:sc-curvatures}
R_{\m\n}^a (P)   &=& 2 \,{\cal D}_{[\m} e^a_{\n]} -\ft12 \bar
\psi_{[\m}\G^a\psi_{\n]}    \,, 
\nn \\
R_{\m\n}^{ab}(M) &=& 2 \,\pa_{[\m}\, \o^{ab}_{\n]} -2\, \o_{[\m}^{ac}\,
\o^{~}_{\n]\,c}{}^b -  2\,f^{[a}_{[\m}\, e_{\n]}^{b]} + \ft12\bar
\psi_{[\m}\G^{ab} \phi_{\n]}\,,
\nn \\ 
R_{\m\n} (D)     &=& 2 \,{\cal D}_{[\m} b_{\n]}-  f^{a}_{[\m}\,
e^{~}_{\n]\,a}  -\ft12  \bar \psi_{[\m} \phi_{\n]}  \,,
\nn \\
R_{\m\n}^a (K)   &=& 2\, {\cal D}_{[\m} f^a_{\n]} -\ft12 \bar
\phi_{[\m}\G^a\phi^{~}_{\n]}\,,  
\nn \\[1mm]
R_{\m\n} (Q)     &=& 2 \,{\cal D}_{[\m} \psi_{\n]} - e_{[\m}^{\;a}
\,\G_a \phi_{\n]}  \,,
\nn \\[1mm]
R_{\m\n} (S)     &=& 2\, {\cal D}_{[\m} \phi_{\n]}  - f_{[\m}^{\:a} 
\,\G_a \psi^{~}_{\n]}\,. 
\end{eqnarray}
These curvature tensors transform covariantly and their
transformation rules follow from the structure constants of the
superconformal algebra. They also satisfy a number of Bianchi
identities which are straightforward to write down. As an example and
for future reference we list the first three identities,
\bea
\label{sc-bianchi}
{\cal D}_{[\m} R_{\rho]}^a (P) + R^{ab}_{[\m\n}(M)\, e_{\rho]\,b} -
  R_{[\m\n}(D)\, e_{\rho]}^{\;a}  %\quad &&
%\nn\\
-\ft12 \bar\psi_{[\rho} \G^a   R_{\m\n]} (Q) 
  \!\!&=&\!\! 0  \,, 
\nn \\
{\cal D}_{[\m} R_{\n\rho]}^{ab}(M) + R^{[a}_{[\m\n}(K)\, e_{\rho]}^{\,b]}
+ R^{[a}_{[\m\n}(P)\, f_{\rho]}^{\,b]} \quad &&
\nn\\
+\ft14 \bar\phi_{[\rho} \G^{ab} R_{\m\n]} (Q) 
+\ft14 \bar\psi_{[\rho} \G^{ab} R_{\m\n]} (S) 
\!\!&=&\!\! 0 \,,
\nn \\ 
{\cal D}_{[\m} R_{\n\rho]}(D) + \ft12 R^{a}_{[\m\n}(K)\, e_{\rho]\,a}
- R^{a}_{[\m\n}(P)\, f_{\rho]\,a} \quad &&
\nn\\
+ \ft14 \bar\phi_{[\rho} R_{\m\n]} (Q) 
- \ft14 \bar\psi_{[\rho} R_{\m\n]} (S) 
\!\!&=& \!\!0  \,.\quad{~}
\end{eqnarray}

At this stage, the superconformal algebra is not related to symmetries
of spacetime. Of course, the gauge fields independently transform as
vectors under general coordinate transformations but these
transformations have no intrinsic relation with the gauge
transformations. This is the reason why, at this stage, there is no
need for the bosonic and fermionic degrees of freedom to match, as one
would expect for a conventional supersymmetric theory. 

There is a procedure to introducing a nontrivial entangling between the
spacetime diffeomorphisms and the (internal) symmetries associated
with the superconformal gauge algebra, based on curvature 
constraints. Here one regards the $P$ gauge field $e_\m{}^{\!a}$ as a
nonsingular vielbein field, whose inverse will be denoted by
$e_a{}^{\!\m}$. This interpretation is in line with the interpretation
presented in the previous section, where flat space was viewed as a
coset space. In that case, the curvature $R(P)$ has the interpretation
of a torsion tensor, and one can impose a constraint $R(P)=0$, so
that the $M$ gauge field $\o_\m^{ab}$ becomes a dependent field, just
as in \eqn{eq:torsion-constraint}. The effect of this constraint is
also that the $P$ gauge transformations are effectively replaced by
general-coordinate transformations. To see this, let us rewrite a 
$P$-transformation on $e_\m^{\;a}$, making use of the fact that there
exists an inverse vielbein $e_a^{\;\m}$,
\begin{equation}
  \label{eq:P-GCT}
\d e_\m^{\;a} = {\cal D}_\m\xi_{\rm P}^a = \pa_\m\xi^\n \,e_\n^{\;a} -
\xi^\n \,D_\n e_\m^{\;a} + \xi^\n  R_{\m\n}^a(P) \,,
\end{equation}
where $\xi^\m = \xi_{\rm P}^a \, e_a^{\;\m}$. Hence, when imposing the
torsion constraint $R(P)=0$, a $P$-transformation takes the form of a
(covariant) general coordinate transformation. This is completely in
line with the field transformations \eqn{eq:d-phi-conf}, where the
P-transformations were also exclusively represented by coordinate
changes, except that we are now dealing with arbitrary
diffeomorphisms. 

A constraint such as $R(P)=0$ is called a {\it conventional}
constraint, because it algebraically expresses some of the 
gauge fields in terms of the others. Of course, by doing so, the
transformation rules of the dependent fields are determined and they
may acquire extra terms beyond the original ones presented 
in \eqn{eq:conf-gauge-fields}. Because $R(P)=0$ is consistent with
spacetime diffeomorphisms, and the bosonic conformal transformations,
the field the field $\o_\m{}^{\!ab}$ will still transform under these
symmetries according to \eqn{eq:conf-gauge-fields}. This is also the
case for $S$-supersymmetry, but not for $Q$-supersymmetry, because the
constraint $R(P)=0$ is inconsistent with $Q$-supersymmetry. Indeed,
under $Q$-supersymmetry, the field $\o_\m{}^{\!ab}$ acquires an extra
term beyond what was presented in \eqn{eq:conf-gauge-fields},
which is proportional to $R(Q)$. We will not elaborate on the 
systematics of this procedure but concentrate on a number of
noteworthy features.  One of them is that there are potentially more
conventional constraints. Inspection of \eqn{eq:sc-curvatures} shows
that constraints on $R(M)$, $R(D)$ and $R(Q)$ can be conventional and
may lead to additional dependent gauge fields $f_\m{}^{\!a}$ and
$\phi_\m$ associated with special conformal boosts and special
supersymmetry transformations. A maximal set of conventional
constraints that achieves just that, takes the form
\begin{eqnarray}
  \label{eq:sc-constraints}
R_{\m\n}^a(P) &=& 0\,,\nn\\
e_b^{\;b} R_{\m\n}^{\;ab}(M) &=& 0\,,\nn\\
\G^\m R_{\m\n}(Q) &=& 0\,,   
\end{eqnarray}
where, for reasons of covariance, one should include possible
modifications of the curvatures due to the changes in the
transformation laws of the dependent fields. Other than that,
the precise form of the 
constraints is not so important, because constraints that differ by
the addition of other covariant terms result in the addition of
covariant terms to the dependent gauge fields, which can easily be
eliminated by a field redefinition. Note that $R_{\m\n}(D)$ is not
independent as a result of the first Bianchi identity on
$R_{\m\n}^a(P)$ given in \eqn{sc-bianchi} and should not be
constrained.  

At this point we are left with the vielbein field $e_\m{}^{\!a}$,
the gauge field $b_\m$ associated with the scale transformations, and
the  gravitino field $\psi_\m$ associated with $Q$-supersymmetry. All
other gauge fields have become dependent. The gauge transformations
remain with the exception of the $P$ transformations;  we have
diffeomorphisms, local Lorentz transformations ($M$), local scale
transformations ($D$), local conformal boosts ($K$), $Q$-supersymmetry
and $S$-supersymmetry. Note that $b_\m$ is the only field that
transforms nontrivially on special conformal boosts and therefore acts
as a compensator which induces all the $K$-transformations for the
dependent fields. Because the constraints are consistent with all
the bosonic transformations, those will not change and will still
describe a closed algebra. The superalgebra will, however, not close,
as one can verify by comparing the numbers of bosonic and fermionic
degrees of freedom. In order to have a consistent superconformal
theory one must add additional fields (for a review, see \cite{DW}). A
practical way to do this makes use of the superconformal multiplet of
currents \cite{BdRdW}, which we will discuss in the next section. This
construction is limited to theories with $Q=16$ supercharges and leads
to consistent conformal supergravity theories \cite{KTvN,dWvHVP,BdRdW}. 

We close this section with a comment regarding the number of degrees
of freedom described by the above gauge fields. The independent
bosonic fields, $e_\m{}^{\!a}$ and $b_\m$, comprise $D^2+D$ degrees of
freedom, which are subject to the $\ft12 D^2 -\ft32 D -1$ independent,
bosonic, gauge invariances of the conformal group. This 
leaves us with $\ft12D(D-1)-1$ degrees of freedom,  corresponding 
to the independent components of a 
symmetric, traceless, rank-2 tensor in $D-1$ dimensions, which
constitutes an irreducible representation of the Poincar\'e
algebra. This representation is the minimal representation 
that is required for an off-shell description of gravitons in $D$
spacetime dimensions. A similar off-shell counting argument applies to
the fermions, which comprise $(D-2)_{\rm s}$ degrees of freedom after
subtracting the gauge degrees of freedom associated with $Q$- and
$S$-supersymmetry. Here $n_{\rm s}$ denotes the spinor
dimension. Hence, the conformal framework is set up to reduce the
field representation to the smallest possible one 
that describes the leading spin without putting the fields on shell. The
fact that the fields can exist off the mass shell, implies that they
must constitute massive representations of the Lorentz
group. Similarly, the supermultiplet of fields 
on which conformal supergravity is based, comprise the smallest
{\it massive} supermultiplet whose highest spin coincides with the
graviton spin.
%%%%%%%%%%%%%%%%%%%%%%%%%%%%%%%%%%%%%%%%%%%%%%%%%%%%%%%%%%%%%%%%
\subsection{Matter fields and currents}
\label{superconf-matter}
%%%%%%%%%%%%%%%%%%%%%%%%%%%%%%%%%%%%%%%%%%%%%%%%%%%%%%%%%%%%%%%%
In the previous section we described how to set up a consistent gauge
theory for conformal supergravity. This theory has an obvious rigid
limit, where all the gauge fields are equal to zero, with the
exception of the vielbein which is equal to the flat vielbein,
$e_\m{}^{\!a} = \d_\m^a$. This is the background we considered in
section~\ref{sec:superc-algebra}. In this background we may have
(matter) theories that are superconformally invariant under rigid
transformations, described by \eqn{eq:d-phi-conf}. Suppose that we
couple such a rigidly superconformal matter theory in first order to
the gauge fields of conformal supergravity. Hence we write, 
\bea
\label{first-action}
\lagr\!\! &=&\!\! \lagr_{\rm matter} \nn\\
&& 
+ h_\m^{\,a} \, \theta_{a}{}^{\m} +
\ft12 \omega_\m^{ab} \,S_{ab}^\m + b_\m \, T^\m + f_\m{}^{\!a} \,
U_{a}{}^{\m} + \bar \psi_\m J^\m + \bar\phi_\m J^\m_{\rm S} \,,
\quad{~} 
\eea
where $h_\m^{\;a}$ denotes the deviation of the vielbein from its flat
space value, {\it i.e.},  $e_\m{}^{\!a} \approx \d_\m^{\,a} +
h_\m{}^{\!a}$.   
The first term denotes the matter Lagrangian in flat space. The
current $\theta_a{}^\m$ is the energy-momentum tensor. In linearized
approximation the above Lagrangian is invariant under {\it local}
superconformal transformations. To examine the consequences of this we
need the leading (inhomogeneous) terms in the transformations of the
gauge fields ({\it c.f.} \eqn{eq:conf-gauge-fields}), 
\bea
\begin{array}{ rrc l rcl }
\mbox{translations:}\!\!& \d h_\m^{\;a}&\!\!\!=\!\!\!& \pa_\m \xi_{\rm P}^a
\,, & & & \\[1mm] 
\mbox{Lorentz:}\!\!  &\!\! \d\o^{ab}_\m &\!\!\!=\!\!\!&\pa_\m \e^{ab}\,, 
   &\! \d h_\m^{\;a} &\!\!\!=\!\!\!& \e^{ab}\,\d_{\m\,b} \,, \\[1mm]
\mbox{dilatations:}\!\!  & \d b_\m &\!\!\!=\!\!\!& \pa_\m\Lambda_{\rm
D} \,, & \!\d h_\m^{\;a} &\!\!\!=\!\!\!&- \d_\m^a \,\Lambda_{\rm D}\,,
\\[.5mm] 
\mbox{conformal boosts:}\!\!  & \d f_\m^{\;a} &\!\!\!=\!\!\!&\!
\pa_\m\Lambda_{\rm K}^a\,,   
   &\!\!\! \d\o^{ab}_\m &\!\!\!=\!\!\!& \Lambda_{\rm K}^{[a}\, \d^{b]}_\m  \,,
   \;\, \d b_\m =\ft12\Lambda_{{\rm K}\m} \,, \\[1mm] 
\mbox{$Q$-supersymmetry:}\!\! & \d\psi_\m &\!\!\!=\!\!\!& \pa_\m \e\,,
   & &    \\[1mm] 
\mbox{$S$-supersymmetry:}\!\! & \d\phi_\m &\!\!\!=\!\!\!&\!
\pa_\m\eta \,, & \d\psi_\m &\!\!\!=\!\!\!& -\ft12 \G_\m\eta \,,
\end{array}
\eea

The variations of the action corresponding to \eqn{first-action} under
the superconformal transformations, ignoring variations that are
proportional to the superconformal gauge fields and assuming that the
matter fields satisfy their equations of motion, must vanish. One can
verify that this leads to a number of conservation equations for the
currents, 
\be
\label{current-cons}
\begin{array}{rcl}
\pa_\m\theta_{a}{}^{\m} &=&0 \,,          \\
\pa_\m S^\m_{ab} -2\, \theta_{[ab]} &=&0 \,, \\
\pa_\m T^\m + \theta_\m{}^\m &=&0  \,,   
\end{array}
\qquad
\begin{array}{rcl}
\pa_\m U_{a}{}^{\m}- \ft12 S^\m_{a\m} - \ft12 T_a &=&0\,, \\
\pa_\m J^\m  &=& 0\,,                       \\
\pa_\m J^{\m}_{\rm S} +\ft12 \G_\m J^\m &=&0\,, 
\end{array} 
\ee
where we used the flat vielbein to convert world into tangent space
indices and vice versa; for instance, we employed the notation  
$\theta_{ab} = \theta_a{}^{\m}\,e_{\m\,b}$ 
and $\theta_\m{}^{\m} = \theta_a{}^{\m}\,e_{\m}{}^{\!a}$.
Obviously, not all currents are conserved, but we can define a set of
conserved currents by allowing an explicit dependence on the
coordinates,
\begin{eqnarray}
  \label{eq:modified-currents}
\pa_\m \theta_{a}{}^\m &=& 0 \,, 
\nn\\
\pa_\m \left(S^\m_{ab} - 2\, \theta_{[a}{}^{\!\m}  x^{~}_{b]}  \right) &=&0 \,,
\nn\\
\pa_\m \left(T^\m + \theta_a{}^\m x^a \right) &=&0 \,,
\nn\\
\pa_\m \left(U_a{}^\m -\ft12 S^\m_{ab} x^b -\ft12 T^\m x^a - \ft12
\theta_b{}^\m \,(x_a x^b -\ft12 x^2 \d_a^b) \right) &=&0 \,,
\nn\\
\pa_\m J^\m &=&0 \,,
\nn\\
\pa_\m \left(J_{{\rm S}}^\m +\ft12 \G_\n J^\m x^\n  \right) &=&0 \,.
\end{eqnarray}
In this result one recognizes the various components
in \eqn{eq:d-phi-conf} and in \eqn{flat-conf-killing}. For
$S$-supersymmetry one can understand the expression for the current by
noting that the 
following combination of a constant $S$ transformation with a
spacetime dependent $Q$-transformation with $\e= \ft12 x^\m \G_\m \eta$
leaves the gravitino field $\psi_\m$ invariant. Observe that the terms
involving the energy-momentum tensor take the form
$\theta_a{}^\m\,\xi^a$, where  $\xi^a$ are the conformal Killing
vectors defined in \eqn{flat-conf-killing}. 

So far we have assumed that the gauge fields in \eqn{first-action} are 
independent. However, we have argued in the previous section that it
is possible to choose the gauge fields associated with the generators
$M$, $K$ and $S$, to depend on the other fields. At the linearized
level, the fields $\o_\m{}^{\!ab}$, $f_\m{}^{\!a}$ and $\phi_\m$ can
then be written as linear combinations of curls of the independent
gauge fields. After a partial integration, the currents
$\theta_a{}^\m$, $T^\m$ and $J^\m$ are modified by improvement terms:
terms of the form $\pa_\n\,A^{[\n\m]}$, which can be included into the
currents without affecting their divergence. Hence, the currents
$S^\m_{ab}$, $U_a{}^\m$ and $J_{\rm S}^\m$ no longer appear explicitly
but are absorbed in the remaining currents as improvement terms. We
don't have to work out their explicit form, because we can simply
repeat the analysis leading to \eqn{current-cons}, suppressing
$S^\m_{ab}$, $U_a{}^\m$ and $J_{\rm S}^\m$. We then obtain the
following conditions for the {\it improved} currents,
\bea
\label{imp-current-cons}
\pa^\m\theta^{\rm \scriptscriptstyle imp}_{\m\n} =
 \theta^{\rm \scriptscriptstyle imp}_{[\m\n]}= 
\theta^{\rm \scriptscriptstyle imp}_{\;\m}{}^\m =0  \,,&&   
\nn\\
\pa^\m J^{\rm \scriptscriptstyle imp}_\m =
\G^\m J^{\rm \scriptscriptstyle imp}_\m =0\,.&& 
\eea
Observe that these equations reduce the currents to  irreducible
representations of the Poincar\'e group, in accord with the earlier
counting arguments given for the gauge fields.

To illustrate the construction of the currents, let us consider a
nonlinear sigma model in flat spacetime with Lagrangian,
\be
\lagr = 
\ft12 g_{AB} \,\pa_\m\phi^A \,\pa^\m\phi^B\,.
\ee
Its energy-momentum operator can be derived by standard methods and is
equal to 
\be
\theta_{\m\n} = \ft 12 g_{AB} \left(\pa_\m\phi^A\,\pa_\n\phi^B - \ft12
\eta_{\m\n}\, \pa_\rho\phi^A\pa^\rho\phi^B \right)\,.
\ee
It is conserved by virtue of the field equations; moreover it is
symmetric, but not traceless. It is, however, possible to introduce an
improvement term,
\bea
\label{imp-e-m}
\theta^{\rm \scriptscriptstyle imp}_{\m\n} &=& 
\ft 12 g_{AB} (\pa_\m\phi^A\,\pa_\n\phi^B - \ft12
\eta_{\m\n}\, \pa_\rho\phi^A\pa^\rho\phi^B ) \nn \\
&&
+ {D-2\over 4(D-1)} \,\left(\eta_{\m\n}\,\pa^2 -\pa_\m\pa_\n\right)
\chi(\phi)\,. 
\eea
When $\chi(\phi)$ satisfies 
\be
\label{homothety}
D_A \pa_B \chi(\phi) = g_{AB}\,,
\ee
the improved energy-momentum tensor is conserved, symmetric and
traceless (again, upon using the field equations). This implies
that, $\chi_A= \pa_A\chi$ is a homothetic
vector.\footnote{%%%%%%%%%%%%%%%%%%%%%%%%%%%%%%%%%%%%%%%%%%%%
 A homothetic vector satisfies $D_A \chi_B+ D_A \chi_B = 2 g_{AB}$
 Here we are dealing with an {\it exact} homothety, for which $D_A
 \chi_B= D_B \chi_A$, and which can be solved by a potential $\chi$. 
} %%%%%%%%%%%%%%%%%%%%%%%%%%%%%%%%%%%%%%%%%%%%%%%%%%%%%%%%%%%%
From this result it follows that locally in the target space, $\chi$ can
be written as 
\be 
\chi =\ft12 g^{AB} \,\chi_A\,\chi_B\,, 
\ee
up to an integration constant. Spaces that have such a homothety are
cones. To see this, we decompose the target-space coordinates $\phi^A$
into $\phi$ and remaining coordinates $\varphi^a$, where 
$\phi$ is defined by  
\be
\chi^A{ \pa\over \pa\phi^A} = {\pa\over \pa\phi}\;.
\ee
It then follows that $\chi(\phi,\varphi) = \exp[2\phi] \,\hat
\chi(\varphi)$, where 
$\hat\chi$ is an undetermined function of the coordinates $\varphi^a$. 
In terms of these new coordinates we have $\chi^A= (1,0,\cdots,0)$ and
$g_{A\phi} = \chi_A = (2\chi, g_{a\phi})$.  From this result one
proves directly that the metric takes the form,
\be
\label{cone-metric}
({\rm d}s)^2 = {({\rm d}\chi)^2\over 2\chi}  + \chi  h_{ab}(\varphi)
\,{\rm d}\varphi^a {\rm d}\varphi^b \,,
\ee
where the $\phi$-independence of $h_{ab}$ can be deduced directly from 
\eqn{homothety}. This result shows that the target space is a cone
over a base manifold ${\cal M}_{\rm B}$ parametrized in terms of the
coordinates  $\varphi^a$ with metric $h_{ab}$ \cite{GibbonsRych}. In
the supersymmetic context it is important to note that, when the cone
is a K\"ahler or hyperk\"ahler space, it must be invariant under
${\rm U}(1)$  or ${\rm SU}(2)$. These features play an important role
when extending to the supersymmetric case. In that case ${\rm U}(1)$  or
${\rm SU}(2)$ must be associated with the R-symmetry of the
superconformal algebra. 

Coupling the improved energy-momentum tensor \eqn{imp-e-m} to
gravity must lead to a conformally  invariant theory of the nonlinear
sigma model and gravity. The relevant Lagrangian reads, 
\be 
\label{lagr-conf}
e^{-1}\,\lagr = \ft12 g_{AB} \,\pa_\m\phi^A \,\pa^\m\phi^B - {D-2\over
4(D-1)}\, \chi(\phi) \,R\,.
\ee
Indeed, this Lagrangian is invariant under local scale transformations
characterized by the functions $\Lambda_{\rm D}(x)$, 
\be
\d_{\rm D}\phi^A = w \Lambda_{\rm D} \chi^A\,, \qquad 
\d_{\rm D} g_{\m\n} = -2 \Lambda_{\rm D}\, g_{\m\n} \,.
\ee
where $w$ is the Weyl weight of the scalar fields which is equal to
$w=\ft12(D-2)$. The transformation of $g_{\m\n}$ is in accord with the
vielbein scale transformation written down in
section~\ref{superconf-gauge}. 
We should also point out that the coupling with the Ricci scalar can
be understood in the context of the results of the previous section. 
Using the gauge fields of the
conformal group, the Lagrangian reads, 
\be
e^{-1} \,\lagr = \ft12 g_{AB}\,g^{\m\n} (\pa_\m \phi^A
-w\,b_\m\,\chi^A )\,(\pa_\n \phi^B -w\,b_\n\,\chi^B ) - \ft12 w\,
f_\m{}^\m\,\chi \,.
\ee
As one can easily verify from the transformation rules
\eqn{eq:conf-gauge-fields}, this Lagrangian is invariant under local
dilatations, conformal boosts and spacetime diffeomorphisms. Upon
using the second constraint \eqn{eq:sc-constraints} for the gauge
field $f_\m{}^a$ associated with the conformal boosts and setting
$b_\m=0$ as a gauge condition for the conformal boosts, the
Lagrangian becomes equal to 
\eqn{lagr-conf}, which is still invariant under local dilatations. 
This example thus demonstrates the relation
between improvement terms in the currents and constraints on the gauge
fields.

It is possible to also employ a gauge condition for the dilatations. 
An obvious one amounts to putting $\chi$ equal to
a constant $\chi_0$, with the dimension of $[{\rm mass]}^{D-2}$, 
\be
\chi = \chi_{0}\,.
\ee
Substituting the metric \eqn{cone-metric} the Lagrangian then acquires the
form,
\be 
e^{-1}\,\lagr \propto  \ft12 h_{ab} \,\pa_\m\varphi^a
\,\pa^\m\varphi^b - {D-2\over 4(D-1)}\,\,R\,.
\ee
This Lagrangian describes a nonlinear sigma model with the base
manifold ${\cal M}_{\rm B}$ of the cone as a target space, coupled to
(nonconformal) gravity. The constant $\chi_0$ appears as an overall
constant and is inversely proportional to Newton's constant in $D$
spacetime dimensions. Observe that in order to obtain positive kinetic
terms, the metric $h_{ab}$ should be negative definite and
$\chi_0$ must be positive. 

The above example forms an important ingredient in the so-called
superconformal multiplet calculus that has been used extensively in
the construction of 
nonmaximal supergravity couplings. There is an extensive literature on
this. For an introduction to the 4-dimensional $N=1$ multiplet calculus,
see, {\it e.g.}, \cite{deWit82}, for 4-dimensional $N=2$ vector
multiplets and hypermultiplets, we refer to \cite{dWVP,dWRVD}. 
%%%%%%%%%%%%%%%%%%%%%%%%%%%%%%%%%%%%%%%%%%%%%%%%%%%%%%%%%%%%%%%%%%%

%%%%%%%%%%%%%%%%%%%%%%%%%%%%%%%%%%%%%%%%%%%%%%%%%%%%%%%%%%%%%%%%
%%%%%%%%%%%%%%%%%%%%%%%%%%%%%%%%%%%%%%%%%%%%%%%%%%%%%%%%%%%%%%%%
\vspace{5mm}
\noindent
{\bf Acknowledgements}\\  
{I am grateful to Sergio Ferrara, Murat G\"unaydin, N.D. Hari Dass,
Jan Louis, Francisco Morales, Hermann Nicolai, Peter van
Nieuwenhuizen, Soo-Jong Rey, Henning Samtleben, Ergin Sezgin,
Kostas Skenderis, Mario Trigiante, Stefan Vandoren and Toine Van
Proeyen for many helpful and stimulating discussions.} 
%%%%%%%%%%%%%%%%%%%%%%%%%%%%%%%%%%%%%%%%%%%%%%%%%%%%%%%%%%%%%%%%
%\newpage %%%%%%%%%%%%%%%%%%%%%%%%%%%%%%%%%%%%%%%%%%%%%%%%%%%%%%%
%%%%%%%%%%%%%%%%%%%%%%%%%%%%%%%%%%%%%%%%%%%%%%%%%%%%%%%%%%%%%%%%

%%%%%%%%%%%%%%%%%%%%%%%%%%%%%%%%%%%%%%%%%%%%%%%%%%%%%%%%%%%%%%%%
%%%%%%%%%%%%%%%%%%%%%%%%%%%%%%%%%%%%%%%%%%%%%%%%%%%%%%%%%%%%%%%%
%%%%%%%%%%%%%%%%%%%%%%%%%%%%%%%%%%%%%%%%%%%%%%%%%%%%%%%%%%%%%%%%
%%%%%%%%%%%%%%%%%%%%%%%%%%%%%%%%%%%%%%%%%%%%%%%%%%%%%%%%%%%%%%%%

%%%%%%%%%%%%%%%%%%%%%%%%%%%%%%%%%%%%%%%%%%%%%%%%%%%%%%%%%%%%%%%%%
%%%%%%%%%%%%%%%%%%%%%%%%%%%%%%%%%%%%%%%%%%%%%%%%%%%%%%%%%%%%%%%%%
%\newpage
%%%%%%%%%%%%%%%%%%%%%%%%%%%%%%%%%%%%%%%%%%%%%%%%%%%%%%%%%%%%%%%%%
%\input{lhapp}
%%%%%%%%%%%%%%%%%%%%%%%%%%%%%%%%%%%%%%%%%%%%%%%%%%%%%%%%%%%%%%%%%
%%%%%%%%%%%%%%%%%%%%%%%%%%%%%%%%%%%%%%%%%%%%%%%%%%%%%%%%%%%%%%%%
\end{document}